\documentclass[twoside,        
               BCOR=12mm,       
      	       DIV=10,
             ]{scrbook}

\usepackage{graphicx} 
\usepackage{tikz} 
\usepackage{pgfplots} 
\usepackage{color} 
\graphicspath{{./figures/}}
\definecolor{butter1}{rgb}{0.98,0.91,0.31}
\definecolor{butter2}{rgb}{0.93,0.83,0}
\definecolor{butter3}{rgb}{0.77,0.63,0}
\definecolor{skyblue1}{rgb}{0.45,0.62,0.81}
\definecolor{skyblue2}{rgb}{0.2,0.39,0.64}
\definecolor{skyblue3}{rgb}{0.13,0.29,0.53}
\definecolor{scarlet1}{rgb}{0.93,0.16,0.16}
\definecolor{scarlet2}{rgb}{0.8,0,0}
\definecolor{scarlet3}{rgb}{0.64,0,0}
\definecolor{chameleon1}{rgb}{0.54,0.88,0.2}
\definecolor{chameleon2}{rgb}{0.45,0.82,0.09}
\definecolor{chameleon3}{rgb}{0.3,0.6,0.02}
\definecolor{orange1}{rgb}{0.98,0.68,0.24}
\definecolor{orange2}{rgb}{0.96,0.47,0}
\definecolor{orange3}{rgb}{0.8,0.36,0}
\definecolor{plum1}{rgb}{0.68,0.5,0.66}
\definecolor{plum2}{rgb}{0.46,0.31,0.48}
\definecolor{plum3}{rgb}{0.36,0.21,0.4}
\definecolor{chocolate1}{rgb}{0.91,0.72,0.43}
\definecolor{chocolate2}{rgb}{0.75,0.49,0.07}
\definecolor{chocolate3}{rgb}{0.56,0.35,0.01}

\definecolor{aluminium1}{rgb}{0.93,0.93,0.92}
\definecolor{aluminium2}{rgb}{0.82,0.84,0.81}
\definecolor{aluminium3}{rgb}{0.73,0.74,0.71}
\definecolor{aluminium4}{rgb}{0.53,0.54,0.52}
\definecolor{aluminium5}{rgb}{0.33,0.34,0.32}
\definecolor{aluminium6}{rgb}{0.18,0.2,0.21}  

  \usepackage{csquotes}
  \usepackage{calc}      
  \usepackage{pdfpages} 

  \usepackage[utf8]{inputenc} 
  \usepackage[T1]{fontenc} 

  \usepackage{fourier} 
  \usepackage{fix-cm} 
  \usepackage{microtype} 

  \usepackage{amsfonts,amssymb} 
  \usepackage{amsmath,mathtools} 

  \usepackage[amsmath,hyperref,thmmarks]{ntheorem} 

  \theoremstyle{margin}
  \theoremheaderfont{\normalfont\bfseries}
  \theorembodyfont{\slshape}
  \theoremseparator{}
  \theoremsymbol{}
  \newtheorem{theorem}{Theorem}[chapter] 
  \newtheorem{proposition}[theorem]{Proposition}
  \newtheorem{lemma}[theorem]{Lemma}
  \newtheorem{corollary}[theorem]{Corollary}

  \theoremstyle{nonumberplain}

  \theoremstyle{margin}
  \theoremheaderfont{\normalfont\bfseries}
  \theorembodyfont{\rmfamily}
  \theoremseparator{}

  \theoremheaderfont{\scshape}
  \theorembodyfont{\upshape}
  \theoremstyle{nonumberplain}
  \theoremseparator{}
  \theoremsymbol{\ensuremath{\square}}
  \newtheorem{proof}{Proof}

  \theoremstyle{margin}
  \theoremheaderfont{\normalfont\bfseries}\theorembodyfont{\upshape}
  \theoremseparator{}
  \theoremsymbol{}
  \newtheorem{definition}[theorem]{Definition}
  
  \newtheorem{conjecture}[theorem]{Conjecture}
  \newtheorem{remark}[theorem]{Remark}

  \usepackage[backend=bibtex,natbib,isbn=false,doi=false,eprint=true,url=false,firstinits=true,style=alphabetic,backref=true]{biblatex}
  \AtEveryBibitem{\clearfield{month}}
  \AtEveryBibitem{\clearfield{day}}
  \AtEveryBibitem{\clearfield{note}}    

  \usepackage{setspace}  
  \usepackage[ngerman,english]{babel} 
  \usepackage{booktabs}                      
  \usepackage{longtable}                     
  \usepackage{array}                         
  \usepackage{float}

  \usepackage{xspace} 
  \usepackage[expproduct=cdot]{siunitx} 

  \usepackage{caption}
  \usepackage{subcaption}
\makeatletter
\DeclareFontFamily{OMX}{MnSymbolE}{}
\DeclareSymbolFont{MnLargeSymbols}{OMX}{MnSymbolE}{m}{n}
\SetSymbolFont{MnLargeSymbols}{bold}{OMX}{MnSymbolE}{b}{n}
\DeclareFontShape{OMX}{MnSymbolE}{m}{n}{
    <-6>  MnSymbolE5
   <6-7>  MnSymbolE6
   <7-8>  MnSymbolE7
   <8-9>  MnSymbolE8
   <9-10> MnSymbolE9
  <10-12> MnSymbolE10
  <12->   MnSymbolE12
}{}
\DeclareFontShape{OMX}{MnSymbolE}{b}{n}{
    <-6>  MnSymbolE-Bold5
   <6-7>  MnSymbolE-Bold6
   <7-8>  MnSymbolE-Bold7
   <8-9>  MnSymbolE-Bold8
   <9-10> MnSymbolE-Bold9
  <10-12> MnSymbolE-Bold10
  <12->   MnSymbolE-Bold12
}{}

\let\llangle\@undefined
\let\rrangle\@undefined
\DeclareMathDelimiter{\llangle}{\mathopen}%
                     {MnLargeSymbols}{'164}{MnLargeSymbols}{'164}
\DeclareMathDelimiter{\rrangle}{\mathclose}%
                     {MnLargeSymbols}{'171}{MnLargeSymbols}{'171}
\makeatother

  \usepackage{anysize}
  \marginsize{3cm}{4cm}{1cm}{2cm}

  \makeatletter
  \defbibenvironment{bibliography}
  {\list{\printtext[labelalphawidth]{
      \printfield{prefixnumber}%
      \printfield{labelalpha}}} 
      {\setlength{\bibhang}{0pt}%
      \setlength{\leftmargin}{\bibhang}%
      \setlength{\itemindent}{-\leftmargin}%
      \setlength{\itemsep}{\bibitemsep}%
      \setlength{\parsep}{\bibparsep}}}
  {\endlist}
  {\item}
  \makeatother

  \cleardoublepage
\addtokomafont{disposition}{\rmfamily}

\renewcommand*{\othersectionlevelsformat}[1]{%
  \makebox[0pt][r]{\csname the#1\endcsname\autodot\enskip}}

\usepackage[pdfstartview=FitH,      
            breaklinks=true,        
            bookmarksopen=true,     
            bookmarksnumbered=true,  
      	    colorlinks=false,
	          pdfborder={0 0 0},
            pdftitle={PhD Thesis Bernhard Altaner 2014},
            pdfauthor={Bernhard Altaner},
            pdfsubject={Foundations of Stochastic Thermodynamics},
            pdfkeywords={Entropy, Information, Statistical Physics, Stochastic Thermodynamics, Dynamical Systems, Ergodic Theory},
	          pdftex,
            colorlinks=true,
            linktocpage=true,
            linkbordercolor=scarlet1,
            citebordercolor=chameleon1,
            urlbordercolor=skyblue1,
            linkcolor=scarlet3,
            citecolor=chameleon3,
            urlcolor=skyblue3,
            ]{hyperref}

\usepackage[
        xindy,
        nomain,
        nonumberlist, 
        acronym,      
        toc,          
        ]
        {glossaries}

\usepackage{xcolor}
\definecolor{quotemark}{gray}{0.7}
\makeatletter
\def\fquote{%
    \@ifnextchar[{\fquote@i}{\fquote@i[]}
           }%
\def\fquote@i[#1]{%
    \def\tempa{#1}%
    \@ifnextchar[{\fquote@ii}{\fquote@ii[]}
                 }%
\def\fquote@ii[#1]{%
    \def\tempb{#1}%
    \@ifnextchar[{\fquote@iii}{\fquote@iii[]}
                      }%
\def\fquote@iii[#1]{%
    \def\tempc{#1}%
    \vspace{1em}%
    \noindent%
    \begin{list}{}{%
         \setlength{\leftmargin}{0.1\textwidth}%
         \setlength{\rightmargin}{0.1\textwidth}%
                  }%
         \item[]%
         \begin{picture}(0,0)%
         \put(-15,-5){\makebox(0,0){\scalebox{3}{\textcolor{quotemark}{``}}}}%
         \end{picture}%
         \begingroup\itshape}%
 \def\endfquote{%
 \endgroup\par%
 \makebox[0pt][l]{%
 \hspace{0.8\textwidth}%
 \begin{picture}(0,0)(0,0)%
 \put(15,15){\makebox(0,0){%
 \scalebox{3}{\color{quotemark}''}}}%
 \end{picture}}%
 \ifx\tempa\empty%
 \else%
    \ifx\tempc\empty%
       \hfill\rule{100pt}{0.5pt}\\\mbox{}\hfill\tempa,\ \emph{\tempb}%
   \else%
       \hfill\rule{100pt}{0.5pt}\\\mbox{}\hfill\tempa,\ \emph{\tempb},\ \tempc%
   \fi\fi\par%
   \vspace{0.5em}%
 \end{list}%
 }%
 \makeatother

\newcommand{\etc}{\textit{etc}}
\newcommand{\tsub}[1]{\ensuremath{_{\mathrm{#1}}}} 
\newcommand{\tsup}[1]{\ensuremath{^{\mathrm{#1}}}} 

\newcommand{\eq}[1]{ Eq.~(\ref{eq:#1})\,}

\newcommand{\eg}{\textit{e.g.~}}
\newcommand{\ie}{\textit{i.e.~}}
\newcommand{\cf}{\textit{cf.~}}

\newcommand{\rlind}[1]{^{\left(#1\right)}} 
\newcommand{\tind}[1]{_{#1}} 

\newcommand{\ifam}[1]{\ensuremath{\left(#1\right)}} 

\newcommand{\spc}{\ensuremath{X}} 
\newcommand{\proj}{\ensuremath{\pi}} 

\newcommand{\parti}{\ensuremath{\mathcal{Q}}} 
\newcommand{\powset}{\ensuremath{\mathcal{P}}} 
\newcommand{\fsets}{\ensuremath{\mathcal{F}}} 
\newcommand{\alg}{\ensuremath{\mathcal{A}}} 

\newcommand{\palg}{\ensuremath{\mathcal{B}}} 
\newcommand{\salg}{\ensuremath{\mathcal{A}}} 

\newcommand{\pset}{\ensuremath{B}} 
\newcommand{\sset}{\ensuremath{A}} 

\newcommand{\borel}{\ensuremath{\mathcal{B}}} 
\newcommand{\topo}{\ensuremath{\mathcal{T}}} 

\newcommand{\ptopo}{\ensuremath{\mathcal{T}_\pspace}} 

\newcommand{\cyl}{\ensuremath{Z}} 
\newcommand{\pcyl}{\ensuremath{\mathcal{C}}} 

\newcommand{\cyla}[2]{\ensuremath{\cyl\tind{#1}\left[#2\right]}} 
\newcommand{\pcyla}[2]{\ensuremath{\pcyl\tind{#1}\left[#2\right]}} 

\newcommand{\cyls}{\ensuremath{\mathcal Z}} 
\newcommand{\pcyls}{\ensuremath{\zeta}} 

\newcommand{\pms}{\ensuremath{\mu}} 
\newcommand{\ms}{\ensuremath{\nu}} 
\newcommand{\lms}{\ensuremath{\lambda}} 

\newcommand{\pden}{\ensuremath{\varrho}} 
\newcommand{\den}{\ensuremath{f}} 

\newcommand{\pspace}{\ensuremath{\Gamma} } 
\newcommand{\ospace}{\ensuremath{\Omega} } 
\newcommand{\cell}{\ensuremath{\mathcal{C}} } 
\newcommand{\vcell}{\ensuremath{\Pi} } 

\newcommand{\alphabet}{\Omega} 
\newcommand{\partmap}{\ensuremath{ M} } 
\newcommand{\shiftmap}{\ensuremath{\hat{s}}} 
\newcommand{\subshift}{\ensuremath{\mathbb S}} 

\newcommand{\factormap}{\ensuremath{K}} 
\newcommand{\mtraj}{\ensuremath{\traj \partmap}} 

\newcommand{\tsum}[2]{\ensuremath{\sum}_{\traj #1 (#2)}}
\newcommand{\tcup}[2]{\ensuremath{\bigcup}_{\traj #1 (#2)}}

\newcommand{\raover}[1]{\ensuremath{\overset{ {\, }_{\rightarrow}}{#1}}}

\newcommand{\lraover}[1]{\ensuremath{\overset{ {\, }_{\leftrightarrow}}{#1}}}

\newcommand{\tmx}{\ensuremath{t_{\mathrm{max}}}}
\newcommand{\tmn}{\ensuremath{t_{\mathrm{min}}}}

\newcommand{\mms}{\ensuremath{\mu}} 

\newcommand{\imms}{\ensuremath{\mu_{\infty}}} 
\newcommand{\fmms}{\ensuremath{\raover{\mu}}} 
\newcommand{\fbmms}{\ensuremath{\lraover{\mu}}} 

\newcommand{\revop}{\ensuremath{\mathrm{R}}} 
\newcommand{\back}[1]{\ensuremath{\widetilde{#1}}}

\newcommand{\reals}{\ensuremath{\mathbb{R}}} 
\newcommand{\naturals}{\ensuremath{\mathbb{N}}} 
\newcommand{\integers}{\ensuremath{\mathbb{Z}}} 

\newcommand{\graph}{\ensuremath{{G}}} 
\newcommand{\verts}{\ensuremath{{V}}} 
\newcommand{\edges}{\ensuremath{{E}}} 

\newcommand{\edge}{\ensuremath{{e}}} 
\newcommand{\adjm}{\ensuremath{\mathbb A}} 

\newcommand{\tprob}[2]{\ensuremath{w^{#1}_{#2}}} 
\newcommand{\btprob}[2]{\ensuremath{\back{w}^{#1}_{#2}}} 

\newcommand{\tmat}{\ensuremath{\mathbb W}} 
\newcommand{\btmat}{\ensuremath{\back{\tmat}}}

\newcommand{\ptprob}[2]{\ensuremath{q^{#1}_{#2}}} 
\newcommand{\ptmat}{\ensuremath{\mathbb Q}} 

\newcommand{\traj}[1]{\ensuremath{\underline{#1}} } 

\newcommand{\p}[2]{\ensuremath{p\tind{#1}\left( #2 \right)}}
\newcommand{\prob}{\ensuremath{\mathbb P}}
\newcommand{\proba}[3]{\ensuremath{\prob\tind{#1}\left[ \traj{#2}\rlind{#3} \right]}}

\newcommand{\trav}[1]{\ensuremath{\left\llangle #1 \right\rrangle}} 
\newcommand{\eav}[1]{\ensuremath{\left\langle #1 \right\rangle} } 

\newcommand{\trava}[3]{\ensuremath{\left\llangle #1 \right\rrangle}\tind{#2}\rlind{#3}} 
\newcommand{\eava}[2]{\ensuremath{\left\langle #1 \right\rangle}\tind{#2} } 

\newcommand{\topent}{\ensuremath{h_{\mathrm{top}}}} 
\newcommand{\Topent}{\ensuremath{H_{\mathrm{top}}}} 
\newcommand{\ksent}{\ensuremath{h}} 
\newcommand{\KSent}{\ensuremath{H}} 

\newcommand{\ent}{\ensuremath{S}} 
\newcommand{\dkl}{\ensuremath{D_{\mathrm{KL}}}} 

\newcommand{\icms}{\ensuremath{\pms^{\mathrm{ic}}}}
\newcommand{\priorms}{\ensuremath{\pms^{\mathrm{pr}}}}

\newcommand{\fgms}{\ensuremath{\mu^{\mathrm{fg}}}}
\newcommand{\cgms}{\ensuremath{\mu^{\mathrm{cg}}}}

\newcommand{\icden}{\ensuremath{\pden^{\mathrm{ic}}} } 
\newcommand{\priorden}{\ensuremath{\pden^{\mathrm{pr}}}} 

\newcommand{\fgden}{\ensuremath{\pden^{\mathrm{fg}}} } 
\newcommand{\cgden}{\ensuremath{\pden^{\mathrm{cg}}} } 

\newcommand{\visent}{\ensuremath{\ent^{\mathrm{vis}}}} 
\newcommand{\priorent}{\ensuremath{\ent^{\mathrm{pr}}}} 

\newcommand{\fgent}{\ensuremath{\ent^{\mathrm{fg}}}} 
\newcommand{\cgent}{\ensuremath{\ent^{\mathrm{cg}}}} 
\newcommand{\relent}{\ensuremath{\ent^{\mathrm{rel}}}} 

\newcommand{\crossent}{\ensuremath{\ent^\times}} 

\newcommand{\entrv}{\ensuremath{s}} 

\newcommand{\visentrv}{\ensuremath{\entrv^{\mathrm{vis}}}} 
\newcommand{\priorentrv}{\ensuremath{\entrv^{\mathrm{pr}}}} 

\newcommand{\fgentrv}{\ensuremath{\entrv^{\mathrm{fg}}}} 
\newcommand{\cgentrv}{\ensuremath{\entrv^{\mathrm{cg}}}} 
\newcommand{\relentrv}{\ensuremath{\entrv^{\mathrm{rel}}}} 

\newcommand{\crossentrv}{\ensuremath{\entrv^\times}} 

\newcommand{\contentrv}{\ensuremath{\lambda}} 

\newcommand{\dissentrv}{\entrv\tsup{flow}} 
\newcommand{\hiddentrv}{\entrv\tsup{hidd}} 

\newcommand{\ham}{\ensuremath{\mathcal{H}}}
\newcommand{\invo}{\ensuremath{{\mathcal{I}}}}

\newcommand{\obs}{\ensuremath{\varphi}} 
\newcommand{\invms}{\ensuremath{\pms_\infty}} 

\newcommand{\ja}{\ensuremath{\hat{J}}}
\newcommand{\jac}[2]{\ensuremath{\ja^{\left( #1 \right)}(#2)}}

\newcommand{\disfun}{\ensuremath{\Omega}} 

\newcommand{\sysent}{\ensuremath{\ent^{\mathrm{sys}}}} 
\newcommand{\medent}{\ensuremath{\ent^{\mathrm{med}}}} 
\newcommand{\totent}{\ensuremath{\ent^{\mathrm{tot}}}} 

\newcommand{\sysentrv}{\ensuremath{\entrv^{\mathrm{sys}}}} 
\newcommand{\medentrv}{\ensuremath{\entrv^{\mathrm{med}}}} 
\newcommand{\totentrv}{\ensuremath{\entrv^{\mathrm{tot}}}} 

\DeclareMathOperator{\Df}{D} 

\newcommand{\iverson}[1]{\ensuremath{\left\llbracket#1\right\rrbracket}}
\newcommand{\timdom}{\ensuremath{\mathbb{T}}} 

\newcommand{\tmic}{\ensuremath{\tau_{\mathrm{mic}}} } 
\newcommand{\tmes}{\ensuremath{\tau_{\mathrm{mes}}} } 
\newcommand{\tmom}{\ensuremath{\tau_{\vec p}} } 
\newcommand{\tconf}{\ensuremath{\tau_{\mathrm{conf}}} } 

\newcommand{\indset}{\ensuremath{I}} 

\newcommand{\flux}{\ensuremath{\phi}} 
\newcommand{\curr}{\ensuremath{J}} 
\newcommand{\fluxa}[2]{\ensuremath{\flux^{#1}_{#2}}} 
\newcommand{\curra}[2]{\ensuremath{\curr^{#1}_{#2}}} 
\newcommand{\aff}{\ensuremath{A}} 
\newcommand{\affa}[2]{\ensuremath{\aff^{#1}_{#2}}} 
\newcommand{\mot}{\ensuremath{\mathcal{E}}} 
\newcommand{\mota}[2]{\ensuremath{\mot^{#1}_{#2}}} 

\newcommand{\res}{\ensuremath{R}} 
\newcommand{\resa}[2]{\ensuremath{\res^{#1}_{#2}}} 
\newcommand{\pot}{\ensuremath{U}} 
\newcommand{\pota}[2]{\ensuremath{\pot^{#1}_{#2}}} 
\newcommand{\pow}{\ensuremath{P}} 
\newcommand{\powa}[2]{\ensuremath{\pow^{#1}_{#2}}} 
\newcommand{\vola}[2]{\ensuremath{\volt^{#1}_{#2}}} 

\newcommand{\tredges}{\ensuremath{ {\mathcal{T}} }}  

\newcommand{\diredges}{\ensuremath{ {\edges^\mathrm{d}} }}
\newcommand{\udiredges}{\ensuremath{ {\edges^\mathrm{u}} }}
\newcommand{\oredges}{\ensuremath{ {\edges^\mathrm{o}} }}

\newcommand{\currents}{\ensuremath{\mathcal{O}}}
\newcommand{\potentials}{\ensuremath{\mathcal{U}}}
\newcommand{\chords}{\ensuremath{\mathcal{H}}}

\newcommand{\cycles}{\ensuremath{\mathcal{Z}}}
\newcommand{\cocycles}{\ensuremath{\Upsilon}}

\newcommand{\transpose}{\ensuremath{^\top}}

\newcommand{\sprod}[2]{\ensuremath{\left\langle #1,#2\right\rangle}}

\DeclareMathOperator{\img}{im}
\DeclareMathOperator*{\var}{var} 
\DeclareMathOperator*{\mean}{\mathbb E} 
\newcommand{\assure}{\text{a.\,s.\@{}}}

\newcommand{\onsmat}{\ensuremath{\mathbb{L}}} 
\newcommand{\mobmat}{\ensuremath{\mathbb{M}}} 

\newcommand{\fcyls}[1][]{\ifthenelse{\equal{#1}{}}{\ensuremath{\cyls^+}}{\ensuremath{\mathcal Z^{+\left( #1\right)}}}} 
\newcommand{\bcyls}[1][]{\ifthenelse{\equal{#1}{}}{\ensuremath{\cyls^-}}{\ensuremath{\mathcal Z^{-\left( #1\right)}}}} 
\newcommand{\fbcyls}[1][]{\ifthenelse{\equal{#1}{}}{\ensuremath{\cyls^\pm}}{\ensuremath{\mathcal Z^{\pm\left( #1\right)}}}} 

\newcommand{\sent}[1]{\ensuremath{H}\left[#1\right] } 
\newcommand{\sentx}[1]{\ensuremath{H^{\times}\!\left[#1\right] }} 

\newcommand{\Molar}{\textup{M}}		

\newcommand{\df}[1]{\ensuremath{\,\mathrm{d}#1\,}}

\renewcommand{\bar}{\ensuremath{\overline}} 

\newcommand{\set}[1]{{\ensuremath{\left\{#1\right\}}}}

\newcommand{\audretsch}[1]{\ldots} 

\newcommand{\bvec}[1]{\ensuremath{\bold{\boldsymbol{#1}}}} 

\newcommand{\shouldbe}{\stackrel{!}{=}}

\renewcommand{\epsilon}{\varepsilon} 

\newcommand{\diff}[2]{\ensuremath{\frac{\df{#1}}{\df{#2}}}}

\newcommand{\tdiff}[2]{\ensuremath{\tfrac{\df{#1}}{\df{#2}}}}

\newcommand{\del}{\ensuremath{\partial}}

\newcommand{\kb}{\ensuremath{k_\mathrm{B}}}

\DeclareMathOperator*{\vgrad}{grad}
\DeclareMathOperator*{\vdiv}{div}
 
\DeclareMathOperator*{\sgn}{sgn} 

\newcommand{\abs}[1]{\ensuremath{\vert #1 \vert}}

\renewcommand{\abs}[1]{\ensuremath{\left\vert #1 \right\vert} }

\newglossary[slg]{symbolslist}{syi}{syg}{List of frequently used symbols}
 
\makeglossaries
 
\newglossaryentry{symb:ent-avg}{
name=\protect{\ent},
description={Entropy obtained as an ensemble average},
sort=S, type=symbolslist
}

\newglossaryentry{symb:ent-avg-var}{
name=\protect{\ensuremath{\Delta\!\ent}},
description={(Temporal) variation of an ensemble entropy},
sort=S-var, type=symbolslist
}

\newglossaryentry{symb:ostate}{
name=\protect{\ensuremath{\omega}},
description={Observable state, letter, symbol},
sort=omega, type=symbolslist
}

\newglossaryentry{symb:time-series}{%
name=\underline{$\omega$},
description={Observable time-series, symbolic sequence},
sort=omega-traj, type=symbolslist
}

\newglossaryentry{symb:ospace}{
  name=\protect{\ospace},
description={Space of observations, sample space, alphabet},
sort=Omega, type=symbolslist
}

\newglossaryentry{symb:prob-vec}{
  name=\protect{\ensuremath{\bvec{p}}},
description={Stochastic vector, discrete probability distribution},
sort=prob-discrete, type=symbolslist
}

\newglossaryentry{symb:sent}{
  name=\protect{\ensuremath{\sent{\,\cdot\,}}},
description={Shannon entropy},
sort=H-Shannon, type=symbolslist
}

\newglossaryentry{symb:DKL}{
  name=\protect{\ensuremath{\dkl}},
description={Kullback--Leibler divergence, relative entropy},
sort=DKL, type=symbolslist
}

\newglossaryentry{symb:pden}{
name=\protect{\ensuremath{\pden}},
description={Probability density},
sort=rho, type=symbolslist
}

\newglossaryentry{symb:affinity}{
  name=\protect{\ensuremath{\aff}},
description={Affinity (generalized thermodynamic force)},
sort=affinity, type=symbolslist
}

\newglossaryentry{symb:measurement}{
  name=\protect{\ensuremath{\partmap}},
description={Observable for the measurement process},
sort=Measurement, type=symbolslist
}

\newglossaryentry{symb:microstate}{
  name=\protect{\ensuremath{x}},
description={Microscopic state},
sort=x, type=symbolslist
}

\newglossaryentry{symb:pspace}{
  name=\protect{\ensuremath{\pspace}},
description={Phase space of a deterministic dynamical system},
sort=gamma, type=symbolslist
}

\newglossaryentry{symb:motance}{
  name=\protect{\ensuremath{\mot}},
description={Motance, \ie logarithmic ratio of forward and backward transition rates},
sort=Electromotance, type=symbolslist
}

\newglossaryentry{symb:tmat}{
  name=\protect{\ensuremath{\tmat}},
description={Transition matrix, rate matrix},
sort=Wmatrix, type=symbolslist
}

\newglossaryentry{symb:flux}{
  name=\protect{\ensuremath{\flux}},
description={Probability flux},
sort=Phi, type=symbolslist
}

\newglossaryentry{symb:current}{
  name=\protect{\ensuremath{\curr}},
description={(Probability) current},
sort=J, type=symbolslist
}

\newglossaryentry{symb:adjacency-matrix}{
  name=\protect{\ensuremath{\adjm}},
description={Adjacency matrix},
sort=Adjacency, type=symbolslist
}

\newglossaryentry{symb:sign}{
  name=\protect{\ensuremath{\sgn}},
description={Sign function},
sort=sgn, type=symbolslist
}

\newglossaryentry{symb:graph}{
  name=\protect{\ensuremath{\graph}},
description={Graph},
sort=Graph, type=symbolslist
}

\newglossaryentry{symb:transition-probability}{
  name=\protect{\ensuremath{\tprob{\omega}{\omega'}}},
description={Probability or rate for the transition $\omega \to \omega'$},
sort=w, type=symbolslist
}

\newglossaryentry{symb:involution}{
  name=\protect{\ensuremath{\invo}},
description={(Measure-preserving) time-reversal involution},
sort=Invo, type=symbolslist
}

\newglossaryentry{symb:prob}{
  name=\protect{\ensuremath{\prob}},
description={Probability measure, probability of an event},
sort=P, type=symbolslist
}

\newglossaryentry{symb:trav}{
  name=\protect{\ensuremath{\trav{\,\cdot\,}}},
description={Time-series average},
sort=<<>>, type=symbolslist
}

\newglossaryentry{symb:eav}{
  name=\protect{\ensuremath{\eav{\,\cdot\,}}},
description={Ensemble average},
sort=<>, type=symbolslist
}

\newglossaryentry{symb:observable}{
  name=\protect{\ensuremath{\obs}},
description={(Current-like) observable},
sort=phi-var, type=symbolslist
}

\newglossaryentry{symb:expansion-rate}{
  name=\protect{\ensuremath{\Lambda}},
description={Phase space expansion (rate)},
sort=Lambda, type=symbolslist
}

\newglossaryentry{symb:jacobian}{
  name=\protect{\ensuremath{\ja}},
description={Absolute value of the Jacobian determinant},
sort=omega, type=symbolslist
}

\newglossaryentry{symb:strobo-map}{
  name=\protect{\ensuremath{\Phi}},
description={Deterministic microscopic dynamics},
sort=Phi-strobe, type=symbolslist
}

\newglossaryentry{symb:flow}{
  name=\protect{\ensuremath{\Psi}},
description={Flow of a continuous dynamical system},
sort=Psi, type=symbolslist
}

\newglossaryentry{symb:reals}{
  name=\protect{\ensuremath{\reals}},
description={Real numbers},
sort=R-eals, type=symbolslist
}

\newglossaryentry{symb:integers}{
  name=\protect{\ensuremath{\integers}},
description={Integers},
sort=Z-integers, type=symbolslist
}

\newglossaryentry{symb:naturals}{
  name=\protect{\ensuremath{\naturals}},
description={Natural numbers (including zero)},
sort=N-aturals, type=symbolslist
}

\newglossaryentry{symb:power-set}{
  name=\protect{\ensuremath{\powset}},
description={Power set},
sort=P-owerset, type=symbolslist
}

\newglossaryentry{symb:topology}{
  name=\protect{\ensuremath{\topo}},
description={Topology},
sort=T-opo, type=symbolslist
}

\newglossaryentry{symb:empty-set}{
  name=\protect{\ensuremath{\emptyset}},
description={Empty set},
sort=\\emptyset, type=symbolslist
}

\newglossaryentry{symb:probability-measure}{
  name=\protect{\ensuremath{\pms}},
description={Probability measure},
sort=mu, type=symbolslist
}

\newglossaryentry{symb:time-domain}{
  name=\protect{\ensuremath{\timdom}},
description={Time domain},
sort=T-ime-domain, type=symbolslist
}

\newglossaryentry{symb:subshift}{
  name=\protect{\ensuremath{\subshift}},
description={Subshift},
sort=S-ubschift, type=symbolslist
}

\newglossaryentry{symb:shiftmap}{
  name=\protect{\ensuremath{\shiftmap}},
description={Shift map},
sort=s-hiftmap, type=symbolslist
}

\newglossaryentry{symb:cell}{
  name=\protect{\ensuremath{\cell_\omega}},
description={Phase space cell, partition element},
sort=C-ell, type=symbolslist
}

\newglossaryentry{symb:partition}{
  name=\protect{\ensuremath{\parti}},
description={Partition of phase space},
sort=, type=symbolslist
}

\newglossaryentry{symb:entropy-rv}{
  name=\protect{\ensuremath{\entrv}},
description={Entropic $\tau$-chain, \ie a time-series dependent random variable with an entropic interpretation},
sort=s-ent-rv, type=symbolslist
}

\newglossaryentry{symb:entropy-rv-var}{
  name=\protect{\ensuremath{\delta\!\entrv}},
description={(Temporal) variation of an entropic $\tau$-chain},
sort=s-ent-rv-var, type=symbolslist
}

\newglossaryentry{symb:physical-observables}{
  name=\protect{\ensuremath{\currents}},
description={Space of current-like observables},
sort=O, type=symbolslist
}

\newglossaryentry{symb:rate-function}{
  name=\protect{\ensuremath{I_\obs(x)}},
description={Rate function for the current-like observable $\obs$},
sort=I-rate-function, type=symbolslist
}

\newglossaryentry{symb:scgf}{
  name=\protect{\ensuremath{\lambda_\obs(q)}},
description={Scaled cumulant generating function for the current-like observable $\obs$},
sort=lambda-scfg, type=symbolslist
}

\begin{document}

  

\newlength{\indnt}
\setlength{\indnt}{3em}


\begin{titlepage}
\setlength{\parskip}{10pt}
\setlength{\topmargin}{30pt}

\begin{center}
{\Huge\bf
Foundations of Stochastic Thermodynamics\\
}
\vspace{\stretch{.2}}
{\Large%
Entropy, Dissipation and Information in Models of Small Systems\\
}
\vspace{\stretch{.3}}
{\huge \bf Bernhard Altaner\\}
\vspace{\stretch{.4}}
{\Large \bf Abstract\\}
\end{center}%
  Small systems in a thermodynamic medium --- like colloids in a suspension or the molecular machinery in living cells --- are strongly affected by the thermal fluctuations of their environment.
  Physicists model such systems by means of stochastic processes.
  Stochastic Thermodynamics (ST) defines entropy changes and other thermodynamic notions for individual realizations of such processes.
  It applies to situations far from equilibrium and provides a unified approach to stochastic fluctuation relations.
  Its predictions have been studied and verified experimentally.

  This thesis addresses the theoretical foundations of ST.
  Its focus is on the following two aspects:
  (i)~The stochastic nature of mesoscopic observations has its origin in the molecular chaos on the microscopic level.
  Can one derive ST from an underlying reversible deterministic dynamics?
  Can we interpret ST's notions of entropy and entropy changes in a well-defined information-theoretical framework?
  (ii)~Markovian jump processes on finite state spaces are common models for bio-chemical pathways.
  How does one quantify and calculate fluctuations of physical observables in such models?
  What role does the topology of the network of states play?
  How can we apply our abstract results to the design of models for molecular motors?

  The thesis concludes with an outlook on dissipation as information written to unobserved degrees of freedom --- a perspective that yields a consistency criterion between dynamical models formulated on various levels of description.

\vspace{\stretch{.4}}
\centering
{\large\bf
G\"ottingen, 2014}
\vfill

\end{titlepage}

\newcommand{\footlabel}[2]{%
    \addtocounter{footnote}{1}%
    \footnotetext[\thefootnote]{%
        \addtocounter{footnote}{-1}%
        \refstepcounter{footnote}\label{#1}%
        #2%
    }%
    $^{\ref{#1}}$%
}

\renewcommand{\footref}[1]{%
    $^{\ref{#1}}$%
}
\newpage
\thispagestyle{empty}
\section*{About this work}
This work was submitted as a dissertation for the award of the degree ``Doctor rerum naturalium'' of the Georg--August--Universit\"at G\"ottingen according to the regulations of the International Max Planck Research School ``Physics of Biological and Complex Systems'' (IMPRS PBCS) of the ``G\"ottingen Graduate School for Neurosciences, Biophysics, and Molecular Biosciences'' (GGNB).

\subsection*{Thesis Committee}
\begin{list}{}%
{\leftmargin=\indnt}
\item {Prof.~Dr.~J\"urgen Vollmer}, \textit{Supervisor} \footlabel{DCF}{Department ``Dynamics of Complex Fluids'', Max Planck Institute for Dynamics and Self-Organization, G\"ottingen}$^{,}$\footlabel{INLD}{Institute for Nonlinear Dynamics, Derpartment of Physics, Georg\--August\--Universit\"at G\"ottingen}
\item {Prof.~Dr.~Marc Timme}, \textit{Co-supervisor} \footlabel{ND}{Independent Research Group ``Network Dynamics'',  Max Planck Institute for Dynamics and Self-Organization, G\"ottingen}$^{,}$\footref{INLD}
\item {Prof.~Dr.~Eberhard Bodenschatz} \footlabel{NBC}{Department ``Fluid Dynamics, Pattern Formation and Nanobiocomplexity'', Max Planck Institute for Dynamics and Self-Organization}$^{,}$\footref{INLD}
\end{list}

\subsection*{Thesis referees}
\begin{list}{}%
  {\leftmargin=\indnt }
\item {Prof.~Dr.~J\"urgen Vollmer} \footref{DCF}$^{,}$\footref{INLD}
\item {Prof.~Dr.~Stefan Kehrein} \footlabel{CMT}{Condensed Matter Theory, Institute for Theoretical Physics, Department of Physics, Georg\--August\--Universit\"at G\"ottingen}
\end{list}

\subsection*{Examination committee}
\begin{list}{}%
{\leftmargin=\indnt}
\item {Prof.~Dr.~J\"urgen Vollmer} \footref{DCF}$^{,}$\footref{INLD}
\item {Prof.~Dr.~Marc Timme} \footref{ND}$^{,}$\footref{INLD}
\item {Prof.~Dr.~Stefan Kehrein} \footref{CMT}
\item {Prof.~Dr.~Stephan Herminghaus} \footref{DCF}$^{,}$\footref{INLD}
\item {Dr.~Marco G.\,Mazza} \footref{DCF}
\end{list}

\subsection*{Examination date}
\medskip
\begin{list}{}%
{\leftmargin=\indnt}
\item July, 31st 2014
\end{list}
\newpage

  \tableofcontents


  \chapter{Introduction}
  \label{chap:intro}
  \begin{fquote}[Aristotle][Art of Rhetoric][4th century BC]
  Tell them what you are going to say; say it; then tell them what you said.
\end{fquote}

\section{Motivation}
\label{sec:motivation}
Finding an appropriate title for a doctoral thesis is a difficult task.
Usually, one starts with a working title.
As research progresses and the doctoral candidate's knowledge deepens, a working title feels increasingly shallow.
Often, a good title only emerges when the thesis is almost ready --- at a time when it might be impossible to change it any more.

The title of the present thesis is ``Foundations of Stochastic Thermodynamics''.
Admittedly, such a title sounds rather like the title of a review article than a work of original research.
Also the subtitle ``Entropy, Dissipation and Information in Effective Models of Small Systems'' only slightly specifies the topic of this thesis.

Therefore, as a motivation and introduction to what follows, let us quickly go through the title before we formulate our research question.

\subsection{Stochastic thermodynamics}
\emph{Stochastic thermodynamics} (ST) is a modern paradigm for the treatment of small systems in thermodynamic environments \cite{Seifert2008,Seifert2012}.
In particular, ST studies \emph{non-equilibrium} situations, \ie conditions where a system is actively driven out of equilibrium by some force.
Examples include colloids in solution which are driven by external fields \cite{Speck_etal2007,Toyabe_etal2010,Horowitz+Parrondo2011}, complex fluids under flow \cite{Grmela+Oettinger1997}, actively moving micro-swimmers \cite{Astumian1997,Romanczuk_etal2012,Ganguly+Chaudhuri2013} as well as small electric devices \cite{Esposito_etal2012,Ciliberto_etal2013}.
Arguably, the most active field in ST is the study of biologically relevant macro-molecules, ranging from relatively simple molecules like RNA/DNA \cite{Liphardt_etal2001} to the complex molecular machinery of life \cite{Qian2005,Lipowsky+Liepelt2008,Seifert2011,Boon+Hoyle2012}.

The above examples show that the mechanisms of driving a system away from equilibrium are as diverse as the systems themselves \cite{Ciliberto_etal2010,Seifert2012}. 
Experiments on colloids often use optical tweezers, \ie \emph{external} electrical fields to drive the system.
In rheological experiments on soft matter, pressure gradients induce flows.
Actively moving particles often carry their own fuel, whereas enzymes and molecular motors reside in a solution of various chemical compounds, which are not in equilibrium with each other.
In the latter case, an enzyme's active site acts as a catalyst for otherwise kinetically hindered reactions.

At first sight it seems challenging to capture this variety of systems in one generalized framework.
However, for more than one hundred years, \emph{thermodynamics} has been very successful in describing a plethora of different phenomena \cite{deGroot+Mazur1984}.
The key for this success is the abstraction of a thermodynamic \emph{system} and the thermodynamic forces exerted on it by its surrounding \emph{medium}.
In this thesis we define a system as the degrees of freedom which are observed in experiments.
Hence, the state of a system is defined by the information accessible from a measurement.
For the colloid example the state of the system specifies the position of the particle's centre of mass and possibly its velocity and/or rotational degrees of freedom.
Similarly, for a complex biological macromolecule one is usually more interested in its tertiary or quaternary structure, \ie its overall geometric shape rather than the position of each atom.
Hence, the state of the system may be defined by a set of \emph{coarse-grained} degrees of freedom. 
All other unresolved degrees of freedom constitute the ``medium''.

The effect of driving and drag forces, which are mediated by the medium, are observable thermodynamic \emph{currents}.
In addition to these \emph{macroscopic} effects, \emph{small} (sometimes called \emph{mesoscopic}) systems also feel erratic forces.
The latter originate in the essentially random motion of the medium's constituents.
Usually these effects are collectively summarized as ``thermal noise''.
For small systems thermal noise manifests in fluctuations of physical observables.
For large systems the typical energy scales are well above the thermal energy of about  $\kb T \approx 4 \cdot 10^{-11} J$.
Consequently, fluctuations are not relevant and usually negligible on the macroscopic scale.
In order to observe these fluctuations experiments require a very high degree of precision. 
Hence, it is not surprising that the development of the \emph{theoretical framework} of ST in the last twenty year went hand in hand with the refinement of experimental techniques \cite{Ciliberto_etal2010}.

To account for the apparently random behaviour observed for small systems, the \emph{models} used in ST include fluctuating forces.
Thus, the system's \emph{trajectory} is obtained as a random process, rather than given by a deterministic evolution rule.
A realization of the fluctuating forces is called the \emph{noise history} of the system.
The mathematical framework of \emph{stochastic processes} allows the assignment of probabilities to noise histories.
Consequently, one assigns probabilities to fluctuation trajectories and other dynamical observables \cite{vKampen1992,Sekimoto1998}.

Stochastic thermodynamics obtains its name from its goal to generalize \emph{thermodynamic notions} like heat, work, dissipation and efficiency to this stochastic setting.
A big emphasis is put on the \emph{molecular machinery of life}, \ie the \emph{molecular motors} performing work within living cells.
The key innovation of modern ST is the definition of entropy changes in the system and its medium for single stochastic trajectories \cite{Sekimoto1998,Kurchan1998,Lebowitz+Spohn1999,Maes2004,Seifert2005}.
In this new approach, one considers both the properties of a single trajectory and of the entire \emph{ensemble}, which specifies the probability of finding the system in a specific state.
It was recently realized that this approach leads to a unification of stochastic \emph{fluctuations relations} \cite{Seifert2005}.
The latter are detailed versions of the \emph{second law of thermodynamics}.
They are statements about the probability of finding individual trajectories that yield a decrease rather than an increase of entropy.
In fact they are examples of the few exact generally applicable results for thermodynamic systems far from equilibrium \cite{Maes2004,Seifert2005,Seifert2012}.

Besides statements about the entropy, ST also aims to quantify noise-driven fluctuations in other physical observables.
Often one is interested in the probability of \emph{rare events} in small systems.
For instance, as a result of a fluctuation molecular machines may run in reverse or particles may move against an external field.
Note that such events are \emph{not} in contradiction with either the first or the second law of thermodynamics.
If a particle moves against an external field, the energy necessary is provided by its medium.
However, such a behaviour is atypical, \ie it occurs with a low \emph{probability}.
Upon \emph{averaging} over the entire ensemble, we still find that work is dissipated into heat and not the other way round, as guaranteed by the second law.

A well-established mathematical tool for the treatment of rare events is the theory of \emph{large deviations} (\cf for instance Ref.~\cite{Ellis2005}).
Large-deviations theory has been unified formally in 1966 by Varadhan \cite{Varadhan1966}.
It formalizes the heuristic ideas of the convergence of probability measures.
With its applications in statistical physics in general \cite{Touchette2009} and ST in particular \cite{Andrieux+Gaspard2007,Faggionato+dPietro2011}, large-deviations theory has become a prime example for the application of an abstract mathematical theory in a very interdisciplinary context.

\subsection{Foundations}
\begin{figure}[t]
  \centering
  \includegraphics{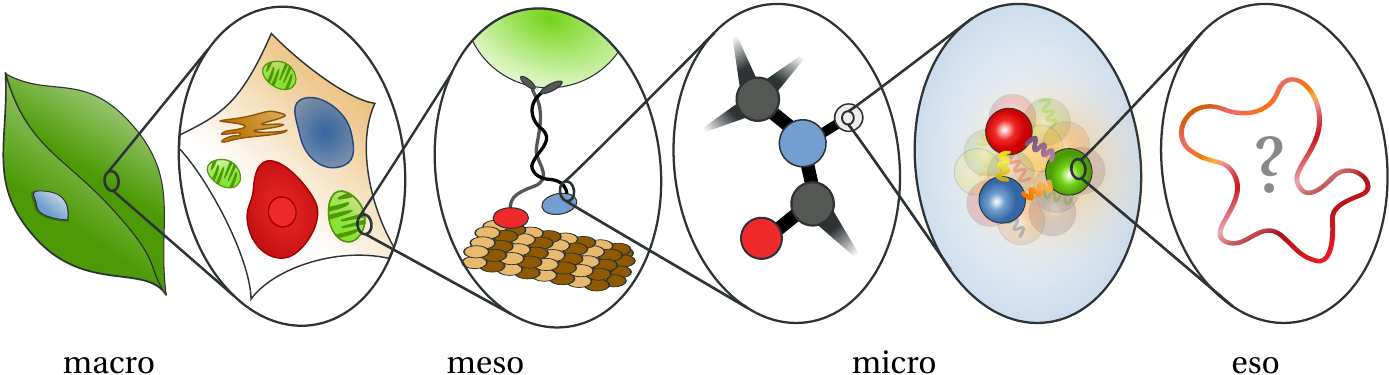}
  \caption{Different levels of description.
  The distinction between the macroscopic, mesoscopic and microscopic levels is not unambiguous.
  In this work, we make the following distinction:
  The macroscopic level is described using deterministic, irreversible laws like hydrodynamics.
  For the mesoscopic level, thermal noise plays a major role and stochastic models are used.
  The microscopic level refers to \emph{any} underlying deterministic and reversible description.
  The esoteric level comprises more fundamental theories which cannot be falsified (yet).
  }
  \label{fig:zoomlevels}
\end{figure}
Stochastic processes and large deviations theory provide the \emph{mathematical foundations} of ST.
Consequently, they will play a major role in the present work.
However, the ``Foundations'' appearing in the title of the present thesis also refer to another, more physical, aspect.
Stochastic thermodynamics is a framework for the treatment of stochastic behaviour observed in \emph{mesoscopic} systems.
In that sense it is an \emph{effective theory} with a validity for the description on a certain scale of observation.
Besides the mathematical foundations, this thesis is mainly concerned with the \emph{microscopic foundations} of ST, \ie the relation of ST to an underlying microscopic dynamics.

Admittedly, the distinction between the macroscopic, mesoscopic and microscopic scale of description is ambiguous.
Often typical length scales are used as a distinction.
However, there are no definite boundaries between, say, the microscopic and the mesoscopic level.
Hence, in the present thesis, we distinguish the scales of description by their model paradigms.
More precisely, we call a model or a theory \emph{macroscopic}, if its dynamical equations are deterministic and irreversible, \ie not symmetric upon reversing the direction of time.
\emph{Mesoscopic} theories, like ST, are based on stochastic models.
In analogy to Hamiltonian mechanics, we say that a system is described by a \emph{microscopic} theory, if it evolves according to time-reversible, deterministic laws, \cf Figure~\ref{fig:zoomlevels}.
With this terminology, the microscopic foundations of ST are concerned with a deterministic level of description underlying the stochastic mesoscopic description.

One of the fundamental assumptions of statistical mechanics is the \emph{Markovian postulate} regarding the dynamics of observable states \cite{Penrose1970}.
It states that the system's trajectory is generated by a \emph{memoryless} (so-called Markovian) process.

For ST, the Markovian postulate is understood as a consequence of the assumption of \emph{local equilibrium} (LE) \cite{Seifert2011}.
Local equilibrium is a consistency assumption that relates the statistics of the degrees of freedom of the medium to the statistics of the stochastic terms used in mesoscopic models.
More precisely, one assumes that \emph{on the time scale of mesoscopic (or macroscopic) observations}, the distribution of the unobserved degrees of freedom are well-described by \emph{equilibrium} probability densities.
Equilibrium distributions are asymptotic distributions, which are encountered in a non-driven system in the long-time limit.
They act as \emph{attractors}:
Under \emph{equilibrium conditions}, any initial distribution will converge to an equilibrium distribution.
In that process, the distribution loses the memory of its past, \ie the memory of its previous interactions with the system.

From this point of view, the Markovian postulate is a prerequisite for LE:
The random forces exerted by the medium on the system are assumed to be sampled from an \emph{equilibrium} distribution.
As a result, they are uncorrelated with the past of the system or medium.

The \emph{separation of time scales} between the microscopic and mesoscopic levels is also known as an \emph{adiabatic approximation} \cite{vKampen1992}:
From the perspective of the medium, the system evolves slowly enough for viewing the medium as being at a (constrained) thermodynamic equilibrium at any time.
Assuming an underlying microscopic dynamics in continuous time, the Markovian postulate can only hold in the limit of an \emph{infinite} separation of time scales.
Such an infinite separation is itself either an unphysical or uninteresting limit:
If the microscopic time scale is finite, the limit implies that \emph{nothing} ever changes on the observable level.
On the other hand, if we let the microscopic time scale approach zero we might run into relativistic problems.

Local equilibrium should thus be understood as a useful approximation for \emph{practical purposes} instead than a strict assumption.
Additionally, it is desirable to have a proper \emph{dynamical picture} of LE.
A major part of this is concerned with the relation between a microscopic deterministic dynamics and a stochastic description of observable, \ie experimentally accessible, states.

Classically, the microscopic-deterministic equations of motion are Hamiltonian.
However, modern computer simulations also use non-Hamiltonian, \emph{effective} deterministic-reversible equations of motion.
The microscopic character of such an approach is also implicit in the term ``Molecular dynamics'' (MD), which is often used synonymously with deterministic computer simulations \cite{Hoover1983,Evans+Morriss1990}.
In spite of their name, such models do not treat all molecules of a system individually.
For instance, MD is used to model the behaviour of single molecules in solution, without explicitly treating the dynamics of the solvent molecules.
Rather, the action of the solvent molecules is reduced to their role as a \emph{heat bath}, \ie the absorption and release of energy from and into the system.
Consequently, one speaks of \emph{thermostated} MD.

If microscopic is understood as ``from first principles'' or ``fundamental'', one could (rightfully) argue that effective models like thermostated MD are not microscopic theories.
However, in the present work we treat thermostated MD on the same level as Hamilton's equations of motion.
Our argument can be understood with regard to Figure~\ref{fig:zoomlevels}:
If there is no \emph{objective}, physical distinction in the terminology, the distinction must be made elsewhere.
The present work is theoretical in its nature.
Hence, it is only natural that we use the paradigms for the mathematical modelling to distinguish between different levels of description.

\subsection{Entropy, dissipation and information}
Let us now discuss the subtitle ``Entropy, Dissipation and Information in Models of Small Systems'' of the present thesis.
First, note that besides implying a separation of time scales, LE is also a statement about thermodynamic \emph{consistency}.
More precisely, the assumption of an equilibrium distribution for the medium allows for a definition of the thermodynamic \emph{entropy} of an observable state.
In fact, the term ``local'' in LE is a remnant of the formulation in its original context, \ie thermodynamic transport theory.
The latter is a continuum theory formulated in physical space.
In transport theory, LE is the assumption that at any point in space, the fundamental thermodynamic relations are obeyed by density fields for internal energy, entropy, temperature \etc~\cite{deGroot+Mazur1984}.

The notion of entropy first appeared in the work of Clausius \cite{Clausius1865}.
His intuition of entropy was that of energy exchanged with the medium as heat.
Building on Carnot's notion of a reversible process, he arrived at the system's entropy as a state variable.
Reversible processes are infinitely slow.
In practice, any real process is \emph{irreversible}.

Upon the completion of an irreversible \emph{cyclic} process, which brings the system back to its original state, the state of the medium has changed.
Though some energy might have been converted into the potential energy of a work reservoir (\eg a weight lifted against gravity), the heat in the medium has increased.\footnote{This is also the case for a \emph{heat pump} which uses the energy stored in a work reservoir to cool one heat bath while heating up another.
The net heat balance in the medium comprising all reservoirs and heat baths is still positive.}
Alternatively, we can say the entropy of the medium has increased.
This phenomenon is usually referred to as \emph{dissipation}.

With the introduction of statistical mechanics by Gibbs, entropy obtained a \emph{statistical interpretation}.
The Gibbs entropy formula
\begin{align*}
  \gls{symb:ent-avg} = -\kb \sum_\omega p_\omega \log p_\omega
\end{align*}
defines entropy with respect to the \emph{probability distribution} $p_\omega$.
In Gibbs' considerations, this probability distribution is interpreted as an ensemble with a \emph{frequentist} interpretation:
It specifies the sampling probability of observing a certain state when picking a system from a large number of identical copies.

At the same time, Boltzmann introduced entropy as
\begin{align*}
  S = \kb \log \vcell 
\end{align*}
where $\vcell$ is the number of microscopic states compatible with a given macroscopic state.
Using the framework of Hamiltonian mechanics together with the assumption of \emph{ergodicity}, a \emph{microscopical} relation between the two concepts of entropy can be established.

In the first half of the twentieth century, statistical mechanics was mostly discussed following Gibbs' and Boltzmann's lines of thought.
Ergodic theory \cite{Hopf1948,Cornfeld_etal1982}, which is concerned with probability and the evolution of dynamical systems, was originally perceived within this context.
At the same time, scientists started to formalize the notion of deterministic chaos, \ie situations where small changes in the initial state of the system grow exponentially fast with time.
Consequently, the ergodic theory for chaotic systems became the major field of study regarding the mathematical foundations of statistical mechanics \cite{Sinai1972,Bowen+Chazottes1975,Beck+Schloegl1995,Ruelle2004,Khinchin2013}.

In the 1940s, Shannon discovered the importance of Gibbs' formula in his theory of \emph{communication} \cite{Shannon1948}.
More precisely, he found that the entropy formula for probability distributions has all the desired properties of a quantity which characterizes the \emph{uncertainty} of the content of (statistically generated) messages.
Nowadays, one refers to the subject founded by Shannon as \emph{information theory}.
It constitutes the basis of \emph{all} digital communication, coding and information storage.

Realizing the importance of entropy for applied statistics in general, Jaynes argued that there is no \emph{conceptional difference} which distinguishes entropy in information theory from entropy in statistical mechanics \cite{Jaynes1957}.
Based on this premiss, he advocated a view of statistical physics (and science in general) as a theory of logical \emph{statistical inference} \cite{Jaynes2003}.
He claims that, if viewed in that way, statistical mechanics can be \emph{logically derived} from the structure of the \emph{underlying fundamental laws} \cite{Jaynes1957}. 
In that approach, the principle of \emph{maximum entropy} replaces the more technical ergodic requirements demanded by the usual treatment from the perspective of mathematical physics, \cf \eg Ref.~\cite{Hopf1948}.
As such it might help us to understand why classical thermodynamic concepts are --- perhaps unexpectedly --- useful in describing systems whose microscopic dynamics are vastly different from what is usually assumed.
An example is provided by the physics of wet granular media as described in Ref.~\cite{Herminghaus2014}.

Jaynes' approach has been both celebrated and rejected by parts of the physics community, partly due to his (physical) interpretation being applied outside of its original context.
After all, probability distributions (and thus the corresponding entropies) arise naturally at various levels of and within several different paradigms for the descriptions of physical and mathematical systems, \cf also Ref.~\cite{Frigg+Werndl2011}.
However, the \emph{thermodynamic} interpretation of the information/entropy associated with an arbitrary probability distribution has to be attempted \emph{cum grano salis}: %
In order to avoid logical fallacies, it is crucial to carefully review the framework in which these probabilistic notions arise.%
\footnote{Examples of common misconceptions of entropy that lead to apparent paradoxes are the ``constant entropy paradox'' for Hamiltonian dynamics (\cf \eg \cite{Ruelle1999}) and the interpretation of entropy as ``disorder''.}

In spite of the criticism of Jaynes' ideas by parts of the physics community, his premiss of a deep conceptional connection between statistical thermodynamics and information theory has been developed further.
With the advent of digital computers, Landauer and later Bennett discussed the ``thermodynamics of computation'' \cite{Landauer1961,Bennett1982,Bennett2003}.
Landauer's principle states that the erasure of an elementary unit of binary information, a \emph{bit}, from a storage medium in a computer comes at the price of at least $Q = \kb T \log2$ of dissipated heat~\cite{Landauer1961}.
Bennett put this result in the context of the old problem of Maxwell's or Szilard's demons \cite{Szilard1929,Bennett2003}.
He stresses that the \emph{information} that such an imaginary demon \emph{processes} equals the maximal amount of work that can be extracted by the demon.
Further thoughts in that direction have recently lead to a general framework of ``information thermodynamics'' \cite{Sagawa+Ueda2010,Sagawa2012}.
Conceptionally, a demon can be thought of as a feedback protocol --- a point of view that has proven useful for the optimal design of small thermodynamic engines~\cite{Horowitz+Parrondo2011}.
In light of the work discussed above, it should not be surprising that the predictions of information thermodynamics have been confirmed by recent experiments on small systems \cite{Toyabe_etal2010,Berut_etal2012}.
This research as well as other work in the same direction 
\cite{Hartich_etal2014}
strongly support the information-theoretical perspective on statistical mechanics.

In light of the examples given above, we consider it only natural to look at stochastic thermodynamics from Jaynes' point of view, \ie as a (dynamical) theory of statistical inference.
In fact, one can go a step further and generally understand the statistical mechanics of non-equilibrium situations as the study of models of \emph{information processing systems}.
The emphasis on \emph{models} is important; it stresses that information (and thus entropy) needs to be formulated in an operational or descriptive context.
At the very end of the present work, we return to these ideas and discuss them in more detail.

\clearpage
\subsection{Research questions}
After having motivated the context of this thesis, we formulate its research questions.
The work splits into two parts.

\paragraph{Microscopic foundations}
Within the framework of having a microscopic-deterministic and a coarse-grained, mesoscopic-stochastic level of description, we formulate two questions:

\begin{itemize}
  \item What are the implications of the \emph{Markovian postulate} on the mesoscopic level of description for the microscopic dynamics?
  \item Can, and if yes how, stochastic thermodynamics be obtained in an \emph{information-theoretical framework}?
\end{itemize}

Both questions point towards a dynamical or information-theoretical picture of local equilibrium.
Hence, in our investigations we will point out when certain physical assumptions appear as logical-probabilistic consistency relations between different models.

\paragraph{Mathematical foundations}
In the second part of the present thesis, we deal with the mathematical foundations of ST formulated on discrete state spaces.
The network of states, which we use to describe a mesoscopic system, is represented as a graph.
Using concepts from graph theory and the theory of large deviations we address the following questions:
\begin{itemize}
  \item What is the general structure of discrete ST and how can we use it in order to \emph{characterize fluctuations} of physical observables?
  \item How can we use such concepts in order to compare different \emph{mesoscopic models} for real physical systems with each other?
\end{itemize}

In the context of the first question, we see how the results of Kirchhoff on electrical circuits reappear in the present setting.
More precisely, we discuss the importance of \emph{cycles} for small systems driven away from equilibrium.
As a solution to the second question we propose to consider the statistics of \emph{dissipation}, which we interpret as information written to unobservable degrees of freedom.
We illustrate our results using models for a system which plays a huge role for the function of living cells: The molecular motor \emph{kinesin}.

\clearpage
\section{The structure of this thesis} 

\subsection{How to read this thesis}
The initial quote, in some form or another, is usually attributed to Aristotle and his teachings on rhetorics.
Admittedly, I have never studied the alleged ``Master of Rhetorics'' himself nor heard him speak.
Thus, I cannot say whether the quote is original.
However, it seems equally good advice for both writing a thesis and for giving an oral presentation.
I mention the advice at this point, because it may serve as a guide on how to \emph{read} the present work.

In the spirit of Aristotle's suggestion,  the multiple hierarchical levels of this thesis also show some amount of intended redundancy.
On the highest level, the outline presented in the next subsection will tell the reader what and what not to expect from the story told by this thesis.
Similarly, the discussion in the final chapter comes back to the general picture presented here.

The central Chapters~\ref{chap:entropy}--\ref{chap:fluctuations} are written in the same spirit.
Each chapter starts with an initial quote followed by a short introduction in order to give an idea of ``What is this about?''.
After the introduction, a presentation of the methods and results precedes a detailed discussion of the latter.
Finally, we give a short summary and  motivate the connection to the contents of the subsequent chapter.

\subsection{Outline}
%
Chapter~\ref{chap:entropy} reviews different notions of entropy and entropy changes as they occur in different physical and mathematical settings.
Consequently, that chapter should be considered as an extended introduction, providing the necessary mathematical and physical terminology needed in what follows.
In particular, we focus on entropy and dissipation in both stochastic and deterministic models of complex systems in thermodynamic environments.

The main part of the thesis is divided into two parts.
The first part starts with Chapter~\ref{chap:marksymdyn}, which revisits the above-mentioned Markovian postulate.
More precisely, we make explicit the requirements on dynamics, observables and ensembles such that the Markovian postulate holds.
For this formal treatment, we introduce an abstract framework for the process of recording mesoscopic time series on a system evolving according to deterministic microscopic laws.
Eventually, the mathematical results are put into the context of ergodic theory and we equip them with \emph{operational} interpretations.

In Chapter~\ref{chap:information-st} we attempt an information-theoretical interpretation of the framework introduced in Chapter~\ref{chap:marksymdyn}.
However, we will \emph{not} make use of the Markovian postulate or the concept of local equilibrium.
Instead we try to base our argument purely on information-theoretical aspects.
In order to make our considerations more transparent in examples, we introduce a versatile, yet analytically tractable, microscopic model dynamics.
We will see that the Markovian postulate holds rigorously for that model, and ST emerges as an information-theoretical interpretation.
Based on this central result, we conjecture a general mechanism for the emergence of ST from an underlying microscopic dynamics.

The second part of the thesis starts with Chapter~\ref{chap:cycles}, where we deal with the mathematical theory of Markovian dynamics on a finite state space.
Finiteness ensures that the topology induced by the stochastic dynamics on state space can be represented as a graph.
Viewing the graph as an electrical circuit, we present an electro-dynamical analogy of ST.
The rationale behind this analogy are algebraic-topological considerations, pioneered already in the nineteenth century by Kirchhoff.
In analogy to Kirchhoff's ``mesh'' or ``circuit law'', we see how \emph{cycles} play a fundamental role in non-equilibrium situations.
This in turn gives an intuition of the intimate connection between cycles and the thermodynamic (macroscopic) forces that drive the system.

Building on the electro-dynamical analogy, we investigate the structure of Markovian jump processes from the theory of algebraic topology.
We establish an analytical way to quantify fluctuations in these processes, \ie any behaviour that deviates from ensemble expectations.
Our results stress that the topology of the network is extremely important:
Fluctuations of \emph{any} physical observable are shown to depend only on the fluctuation statistics of currents associated with a set of fundamental cycles.

Chapter~\ref{chap:fluctuations} is concerned with fluctuations in models of ST.
This is particularly relevant for models of the molecular machinery of living cells.
In the light of evolution it is not surprising that their are many cases where fluctuations are important for the function of an organism.

We explicitly discuss the design and structure of chemo-mechanical models using the molecular motor \emph{kinesin} as an example.
As a main result, we present a fluctuation-sensitive model reduction procedure and investigate its heuristic motivation from the topological perspective established in Chapter~\ref{chap:cycles}.

In addition, we demonstrate how minimal models can be designed in a systematic way.
With our methods we give a detailed account of kinesin's phase diagram, which is spanned by chemical and mechanical driving forces.
In contrast to previous characterizations using approximations or numerics, our results are completely analytic.
Moreover, we find that the fluctuation statistics found in our simplified models agree very well with the prediction of a more complex model known in the literature.
The relative mismatches amount to only few percent in the majority of the phase diagram --- for values ranging over twenty logarithmic decades.
Finally, we show how our method unifies previous approaches to the exact calculation of dynamic properties of molecular machines, like drift and diffusion.

Chapter~\ref{chap:discussion} provides a summary and an outlook on interesting future research.
We finish with a personal perspective on non-equilibrium thermodynamics as the study of information processing devices.

\subsection{Notation, abbreviations and conventions}
A Ph.D.~thesis is always composed of work which has been obtained over an extended period of time.
During that time, preliminary results are being generalized and new definitions or formulations are constantly being created at the expense of older ones.
Consequently, it is fair to say that the general notation has evolved quite a bit during both research for and the formulation of a thesis.

In the optimal case, this evolution leads to a consistent presentation of the results.
As in so many cases, this optimum is hardly ever reached.
The current thesis is no exception to that rule.
Still, the reader might benefit from the following remarks.

\paragraph{Language}
We tried to use British English as a convention throughout the entire thesis.
Abbreviations are usually introduced in the context where they first appear.
The most commonly used ones are: stochastic thermodynamics (ST), local equilibrium (LE), [non-equilibrium] molecular dynamics ([NE]MD), subshift of finite type (SFT),  network multibaker map (NMBM), [scaled] cumulant-generating function ([S]CGF) and adenosine triphosphate (ATP).

\paragraph{Mathematical notation}
In the present work, ``$\log$'' denotes the \emph{natural} logarithm.
The natural numbers \gls{symb:naturals}$\,=\ifam{0,1,\cdots}$ always include zero as the neutral element of addition.
Ensemble averages $\gls{symb:eav}_t$ are denoted by chevrons and a subscript indicates that the probability density reflects an ensemble at time $t$.
\emph{Time series} \gls{symb:time-series} and orbits $\traj x$ are discrete or continuous successions of values and exhibit an under-bar to distinguish them from values $\omega_t$ or $x_t$ at a specific point in time.
Time series $\traj \omega\rlind{\tau}$ of finite run length $\tau$ are equipped with a superscript.
Similarly, averages $\gls{symb:trav}\tind{t}\rlind{\tau}$ which are taken over an ensemble of trajectories that start at time $t$ and extend until time $t+\tau$ carry both decorators. 
The \emph{time average} $\bar\obs\tind{t}\rlind{\tau}$ of an observable $\obs$ along a single trajectory $\traj \omega\rlind{\tau}$ is denoted with an over-bar.
Generally, a single point in time is indicated by a subscript $t$ while a run length is indicated by a superscript $(\tau)$. 

\paragraph{Figures}
All of the sketches were designed using the free software \textit{Inkscape}.
Contour plots were rendered using \textit{Mathematica}\texttrademark.

\paragraph{Copyright}
This work is licensed under the Creative Commons Attribution-ShareAlike 4.0 International License. To view a copy of this license, visit~ \url{http://creativecommons.org/licenses/by-sa/4.0/}.

\vfill
\par{\hfill\includegraphics[scale=.7]{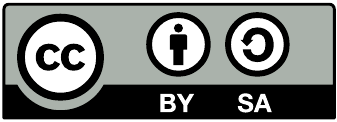}}


  \chapter{Notions of entropy and entropy production}
  \label{chap:entropy}
  
\begin{fquote}[J.~v.~Neumann to C.~E.~Shannon][1940--1941]
  You should call it entropy, for two reasons.
  In the first place, your uncertainty function has been used in statistical mechanics under that name, so it already has a name.
  In the second place, and more important, nobody knows what entropy really is, so in a debate you will always have the advantage.
\end{fquote}

\section*{What is this about?}
The introductory quote (or slightly different formulations thereof) has its origin in a conversation between John von Neumann and Claude E.~Shannon attributed to a period of time between autumn 1940 and spring 1941 \cite{Tribus+McIrvine1971}.
At that time, Shannon was working on his post-doctoral studies at the Institute for Advanced Study in Princeton, New Jersey, where von Neumann was one of the faculty members.
Previous to the conversation Shannon had realized the importance of the expression
\begin{align*}
  -\sum_{i} p_i \log p_i
\end{align*}
for his statistical formulation of signal transmission (\cf Section~\ref{sec:info-theory}).
He thought about calling it  ``uncertainty'' rather than ``information'', because he  was concerned that the latter term is already overly used and might be misleading.
The quote above is Neumann's alleged answer to Shannon when he was asked about the naming issue.

The present chapter picks up on the second part of the quote which regards the nature and meaning of entropy.
More precisely, we present different notions of entropy and entropy production that arise in different branches of physics and mathematics.
A main goal of this thesis is to outline and discuss connections between these notions.
The review character of this chapter sets the stage for the original results presented in Chapters~\ref{chap:marksymdyn}--\ref{chap:fluctuations}.

In the present chapter we introduce the notation and terminology for the rest of this work.
In contrast to von Neumann's suggestion, we aim to disentangle the different meanings of entropy.
If we are successful in that task, the reader of this thesis should know exactly what entropy \emph{is} --- at least from the perspective of the following investigations.

This chapter is structured as follows:
In Section~\ref{sec:entro-classical} we review entropy in the classical thermodynamics of the mid-nineteenth century.
After that, Section~\ref{sec:info-theory} reviews Shannon's and related notions of entropy as \emph{uncertainty} or information of data.
In Section~\ref{sec:stat-phys} we use the latter notion to define the entropy of a \emph{system} as the uncertainty in its observed configurations.
Consequently, we assign the entropy of a system's environment (which we will refer to as its \emph{medium})  to the (dynamics of) unobservable degrees of freedom.
Section~\ref{sec:stochastic-models} makes the distinction explicit for stochastic models and introduces the basic idea of \emph{stochastic thermodynamics}.
In Section~\ref{sec:md} we investigate this distinction in the context of deterministic models of complex systems in thermodynamic environments.
Finally, Section~\ref{sec:measurable-ds} returns to mathematical notions of entropy (production), which characterize the \emph{complexity} of abstract dynamical systems.

\section{Entropy in classical thermodynamics}
\label{sec:entro-classical}
In classical thermodynamics, the \emph{variation of the entropy of a thermodynamic system} is defined by the relation
\begin{align*}
  \Delta \sysent := -\int\frac{\delta Q^{\mathrm{med}}_{\mathrm{rev}}}{T}.
\end{align*}
In this definition, $T$ is the thermodynamic temperature and $Q^{\mathrm{med}}_{\mathrm{rev}}$ is the (integrated) heat flow into%
\footnote{Note that we define the heat flow from the perspective of the medium rather of the system. Hence, our sign convention differs from Clausius' classical work~\cite{Clausius1854}.}
the \emph{medium} for a so-called \emph{reversible} process.
A reversible process is \emph{defined} to be a sequence of changes to the system's state, such that the integral on the right-hand side depends only on the initial and final state of the system.
For a \emph{cyclic} process, the system state is the same both at the beginning and at the end of the process.
Hence, irrespective of its specific nature, a reversible cyclic process (in particular, a \emph{Carnot process}) obeys:
\begin{align*}
  -\oint\frac{\delta Q^{\mathrm{med}}_{\mathrm{rev}}}{T} = \gls{symb:ent-avg-var}\tsup{sys} = 0.
\end{align*}
This path-independence ensures that the \emph{entropy of the system} $\sysent$ is well-defined and obeys the differential relationship $T\df \sysent = \delta Q^{\mathrm{med}}_{\mathrm{rev}}$ for such reversible processes.
The Clausius inequality states that \emph{any} cyclic process obeys \cite{Clausius1854} 
\begin{align*}
  -\oint\frac{\delta Q^{\mathrm{med}}}{T} \leq 0.
\end{align*}
This is one of the many formulations of the \emph{second law of thermodynamics}.
Note that this equation does not imply that there has been no heat exchange with the medium.
Rather, it states that the integrated \emph{ratio} of a heat flux and a (generally varying) temperature vanishes.
Combining a reversible with an irreversible process yields
\begin{align*}
  \Delta \sysent \geq   -\int\frac{\delta Q^{\mathrm{med}}}{T} =: -\Delta\medent,
\end{align*}
where the right-hand side \emph{defines} the \emph{entropy variation in the medium}.
With that, we arrive at a formulation of the second law, where heat $Q^{\mathrm{med}}$ and temperature $T$ do not appear explicitly any more:
\begin{align}
  \Delta \totent := \Delta \sysent + \Delta \medent \geq 0.
  \label{eq:td-second-law}
\end{align}
This is the famous formulation of the second law that states that the \emph{total entropy} of a system together with its environment never decreases.\footnote{
  We refrain from using a statement referring to the \emph{universe}, as we do not divert into a discussion of entropy and information in cosmology.
  The interested reader is referred to Ref.~\cite{Bekenstein2003} for the general idea and to Ref.~\cite{Bousso2002} and the references therein for a detailed treatment.}

\section{Entropy as information or uncertainty}
\label{sec:info-theory}
Information theory is the branch of mathematics that deals with the quantification of information.
It was developed in 1948 by C.E.\,Shannon  as a theoretical framework for the processing of electrical signals.  
At that time Shannon was working at Bell labs, and his seminal work ``A Mathematical Theory of Communication'' appeared in the \emph{Bell Labs Technical Journal} \cite{Shannon1948}.
The main goal of the paper was to lay out the central elements of communication and to formalize them mathematically (\cf figure \ref{fig:shannon-communication}).
\begin{figure}[ht]
  \centering
  \includegraphics{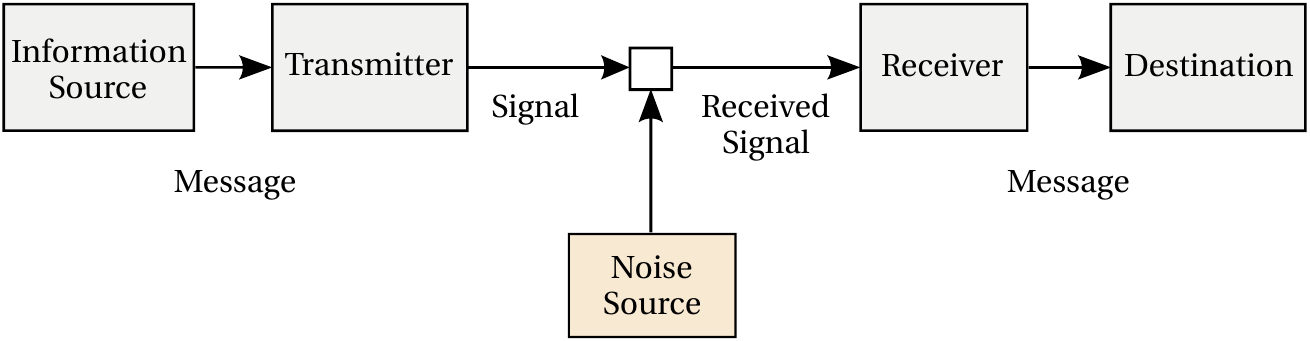}
  \caption{The elements of communication according to Shannon's original paper \cite{Shannon1948}.}
  \label{fig:shannon-communication}
\end{figure}

Information theory is a framework developed to make \emph{quantitative} statements about the information content of \emph{messages}.
In information theory, a message $\traj \omega$ is a string of \emph{letters} $\gls{symb:ostate}\in\alphabet$ composed from a finite \emph{alphabet} $\gls{symb:ospace}$.
More precisely, information theory is concerned with the \emph{probability} of a certain letter appearing in a message.
One can rephrase that statement as follows:
Information theory deals with strings of letters which are generated by a random source.
In that regard it can make statements about uncertainty, redundancy and encoding of messages.
However, it does not refer to \emph{qualitative} properties such as their meaning or their relevance.

In the following we will motivate information theory in the original setting of a discrete random variable $X$ taking values on a finite set $\alphabet = \set{1,2,\cdots,N}$.
We denote the probability to find a letter $\omega$ by $p^X_\omega$.
The probabilities of all possible letters are summarized in the stochastic vector $\gls{symb:prob-vec}^X = \ifam{p^X_\omega}_{\omega \in \alphabet}$.
Entropy is a scalar quantity that characterizes the average \emph{uncertainty} of a letter (or more abstractly, an \emph{event}) to occur.
Hence, entropy quantifies the amount of additional information obtained by \emph{observing} a letter in a message generated by a source solely characterized by $\bvec p^X$.

The requirements on such an entropy have been formalized mathematically in the so-called Khinchin axioms \cite{Khinchin1957}:
\begin{definition}[Shannon entropy]
  Let $X$ be a random variable taking values $\omega \in \alphabet$ on a finite set $\alphabet = \set{1,2,\cdots,N}$ with a probability distribution $\bvec p^X := \ifam{p^X_\omega}_{\omega\in\alphabet}$.
  Then, we call a scalar function $\sent{X}$ the \emph{entropy} (or \emph{uncertainty} or \emph{Shannon information}) of $X$ if it obeys the following axioms:
  \begin{enumerate}
    \item $\sent{X}$ depends only on $\bvec p^X$, \ie the enumeration of its entries must not matter.
    \item $\sent{X}$ takes its maximum value for the uniform distribution.
    \item Let $Y$ be a random variable taking values $y$ on a larger set $\alphabet^Y = \set{1,2,\cdots,M}\supset \alphabet^X$ such that its distribution $\bvec p^Y$ obeys $p^Y_\omega = p^X_\omega$  for all $\omega$ in $\alphabet^X$.
      Then, $\sent{X} = \sent{Y}$.
    \item For any two random variables $X$ and $Y$ with values in $\alphabet^X$ and $\alphabet^Y$, respectively, we have
      \begin{align*}
        \sent{X,Y}  = \sent{X} + \sum_{\omega \in \alphabet^X}p^X_\omega \sent{Y\vert X=\omega},
      \end{align*}
      where $\sent{X,Y}$ is the entropy of the joint distribution for the tuple $(X,Y)$ and $\sent{Y\vert X=\omega}$ is the entropy of the distribution of $Y$ conditioned on  $X=\omega$.
  \end{enumerate}
\end{definition}
It can be shown \cite{Khinchin1957} that the only functional $\gls{symb:sent}$ form satisfying these axioms is
\begin{align}
  \sent{X} = \sent{\bvec p^X} = -\sum_{\omega\in \alphabet}\left[ p_\omega \log_b p_\omega \right],
  \label{eq:shannon-entropy}
\end{align}
where $\log_b$ denotes the logarithm with respect to base $b$.
The dependence on the base can also be understood as choosing the \emph{unit} of entropy.
For instance, if $b=2$ the unit of entropy is called a \emph{bit}.
In statistical mechanics, often the natural logarithm is used and entropy is measured in units of the Boltzmann constant $\kb$.
In the remainder of this thesis we will use the natural logarithm and set $\kb\equiv1$.

To see that this definition of entropy appropriately captures the notion of the uncertainty of $X$, let us take a closer look at the first three axioms:
The first one says that $\sent{X}$ must be independent of the specific nature or enumeration of the events $\omega \in \alphabet$, \ie $\sent{(\frac{1}{3},\frac{2}{3})} = \sent{(\frac{2}{3},\frac{1}{3})}$ .
Hence, entropy is well-defined for \emph{any} random variable and we can compare arbitrary random variables with each other.
This certainly is a useful thing to demand of a \emph{generally applicable} concept of uncertainty.
The second axiom specifies that entropy should be maximal if no event is more probable than any other, in agreement with the informal meaning of uncertainty.
The third axiom states that adding zero-probability events to the possible values of a random variable does not change its uncertainty.

Finally, the fourth axiom specifies the additivity of uncertainty.
More precisely, it says that the uncertainty of conditional events averages to the uncertainty of the joint distribution.
Indeed, this axiom is necessary in order to obtain equation (\ref{eq:shannon-entropy}).
However, relaxing or dropping this axiom gives rise to a whole class of generalized entropies, with applications in contexts where a weaker form of additivity is sufficient or desired \cite{Renyi1961,Tsallis1988,Beck+Schloegl1995}.

Because a discrete probability vector has entries in the interval $[0,1]$, the entropy (\ref{eq:shannon-entropy}) is always positive.
This is not true for the \emph{differential entropy} of a probability density $\gls{symb:pden}\colon \pspace \to [0,\infty)$ on a continuous space $\pspace$:
\begin{align}
  \sent{\pden} := -\int_\pspace \pden(x) \log \pden(x) \df {x}
  \label{eq:differential-entropy}
\end{align}
As the integral is a generalized sum, we will usually use the differential notion of entropy, even if $\pden$ is actually a probability distribution $\bvec p$ on a discrete space.
Despite the fact that the expression \eqref{eq:differential-entropy} can take negative values (and hence without the direct interpretation as ``uncertainty''), the differential entropy is readily used in physics, especially in statistical mechanics.

Another important quantity is the \emph{relative entropy} or \emph{Kullback--Leibler divergence}.
For two probability distributions $\pden$ and $\pden'$ on a state space $\pspace$ such that $\pden' =0$ implies $\pden=0$, it is defined as
\begin{align}
  \dkl[\pden\Vert\pden'] := \int_\pspace \pden(x) \log\frac{\pden(x)}{\pden'(x)} \df x. 
  \label{eq:kullback-leibler-divergence}
\end{align}
By using the concavity of the logarithm, it is straightforward to show that $\gls{symb:DKL} \geq 0$ in general and that $\dkl = 0$ implies measure-theoretic equality of the distributions.

Another quantity we encounter in this work is the cross-entropy of two distributions.
It is a measure for the error one makes if a distribution $\pden'$ is assumed for a random variable with real distribution $\pden$:
\begin{align}
  \sentx{\pden;\pden'} :=-\int_\pspace\pden(x)\log{\pden'(x)}\df x = \sent{\pden} + \dkl[\pden\Vert\pden'],
  \label{eq:cross-entropy}
\end{align}
where the second equality requires that $\dkl$ is defined.

\section{Statistical physics and the distinction between system and medium}
\label{sec:stat-phys}
In this section we review the fundamental aspects of statistical physics we will need in the remainder of this work.
Classical statistical physics has been developed in order to provide a microscopic background for thermodynamics.
It is based on Hamiltonian dynamics, which is a deterministic evolution rule for microscopic states.
A microscopic state contains information about the degrees of freedom of all the particles that make up a macroscopic system.
The number of such degrees of freedom is very large.
Thus, computing the dynamics of individual configurations is cumbersome.
Moreover, for several reasons which we will analyse in more detail later, such calculations are also not effective in order to obtain physical statements.
Hence, rather than focussing on individual microscopic configurations, statistical physics makes probabilistic statements.
For instance, it features a statistical derivation of the second law of thermodynamics (\ref{eq:td-second-law}).

\subsection{The second law in statistical physics}
In classical thermodynamics, the second law is a \emph{macroscopic} statement about macroscopic states.
Similarly, the fundamental equations of  \emph{thermodynamic transport theory} are continuity equations for macroscopically defined quantities \cite{deGroot+Mazur1984}.
In both cases, matter is treated as a continuum and one neglects the existence of individual atoms or molecules.
At macroscopic scales, the granularity of matter is not visible and the continuum approximation is sufficient.
For smaller systems, however, \emph{fluctuations} due to finite particle numbers play a role.
For electrical systems, this effect is referred to as \emph{shot noise}~\cite{Blanter+Buettiker2000}.

In classical statistical physics, one relies on the notion of a \emph{thermodynamic limit}, where the number of particles goes to infinity.
In this limit, fluctuations are negligible.
In contrast, modern statistical physics does not necessarily assume this limit.
Consequently, fluctuations in non-macroscopic systems become relevant and should be included in the theory.
Modern generalizations of the second law are thus detailed probabilistic statements, rather than statements about (macroscopic) averages.
However, consistency requires that the second law of thermodynamics as formulated in (\ref{eq:td-second-law}) must emerge in the macroscopic limit.

The recent years have seen a multitude of such generalizations of the second law for different (non-thermodynamic) models of complex systems.
Amongst the most famous of such statements are the inequalities of C.\,Jarzynski \cite{Jarzynski1997} and G.\,Crooks \cite{Crooks1999}.
Even more recently, these relations have been understood as being consequences of the so-called fluctuation relations for finite systems in thermodynamic environments \cite{Maes2004,Seifert2012}.
Moreover, they have been tested and verified numerically and experimentally \cite{Ciliberto_etal2010}.

For the formulation of fluctuation relations, one defines entropy changes associated with the \emph{system} and its surrounding \emph{medium}, similar to equation (\ref{eq:td-second-law}).
While this distinction is quite clear for macroscopic thermodynamic systems like engines or refrigerators, for small systems it becomes more subtle.
In this work, we identify the system with the set of \emph{observed} degrees of freedom.
Consequently, the medium contains the \emph{unobserved} degrees of freedom.

This distinction based on observability has the advantage that there is no need for a \emph{spatial} separation of the system and the medium.
This is already an implicit feature of any hydrodynamic theory.
For instance, in the Navier--Stokes equation, viscosity acts as a \emph{transport coefficient} for an energy flow from observable hydrodynamic to unobservable internal degrees of freedom.

Other examples are systems in chemical environments.
In particular, we are interested in biological macromolecules which are often surrounded by different chemical compounds.
In biology, a macromolecular system often acts as a \emph{catalyst} which enables (or at least strongly accelerates) reactions between the chemical species.
If such a catalytic reaction additionally triggers an (observable) conformal change on the level of the system itself, one also speaks of \emph{molecular motors}.
In these examples, the medium is composed of the molecules of the solvent and the solutes as well as unobservable microscopic degrees of freedom of the macromolecule.
Even in a well-mixed environment, the solute concentrations need not be in equilibrium with each other.
Hence, the medium provides a \emph{heat bath} as well as different \emph{chemical reservoirs}, which are not spatially separated.

Although a distinction between system and environment based on \emph{observability} seems useful, it comes at the price of subjectivity:
Observability is always an operational, and thus a subjective quality, which is determined by the choice or capability of an observer performing \emph{measurements} on the system.
One goal of this thesis is to shed light on physical implications of that type of subjectivity.

\subsection{Entropy changes in statistical physics}
\label{sec:entropy-statphys}
Keeping the issue of subjectivity discussed in the last subsection in mind, we look for definitions of the entropy changes $\Delta\sysent$ and $\Delta\medent$ in modern statistical physics.
We begin with some general considerations here and then explicitly define these quantities for modern model paradigms.
In particular, we will look at stochastic (jump) processes and molecular dynamics simulations in sections \ref{sec:stochastic-models} and \ref{sec:md}, respectively.

A concept common to all models in statistical physics is the notion of an \emph{ensemble}.
An ensemble specifies the probability of picking a system at a certain microscopic state from a large number of copies of a system.
Mathematically, ensembles are probability densities%
\footnote{In this section, we only consider situations where such a density exists.
  We do not yet discuss the measure-theoretic formulation.
  For an account of the latter \cf Chapter \ref{chap:marksymdyn} or Ref~\cite{Altaner2014}.}
  $\pden\tsup{sys}\colon X\to[0,\infty)$ defined on the \emph{state space} $X$ of a \emph{model}.
The \emph{system's entropy} $\sysent$ is defined to be the (differential) entropy of the distribution $\pden\tsup{sys}$ of the observed degrees of freedom
\begin{align}
  \sysent := \sent{\pden\tsup{sys}} \equiv -\int_X\pden\tsup{sys}\log\pden\tsup{sys} \df{x}.
  \label{eq:system-entropy}
\end{align}

Subjectivity also enters into purely theoretical considerations of mathematical \emph{models} for physical systems, even without the reference to a measurement:
It manifests in the degrees of freedom we choose to make up the state space of a model.
A \emph{dynamical} model specifies an evolution rule on the state space.
Consequently, the dynamics prescribes an evolution operator $U\tind{t}\rlind{\tau}\colon \pden_t \mapsto \pden_{t+\tau}$ for the ensemble $\pden\tind{t}$.\footnote{
The evolution operator for deterministic dynamics is often called the Frobenius--Perron operator, whereas for stochastic systems it is often called the Smoluchowski or Fokker--Planck operator.}
Hence, the system's entropy becomes a time-dependent quantity $\sysent\tind{t} := \sent{\pden\tsup{sys}\tind{t}}$.
The \emph{temporal variation} of the system entropy in the interval $[t,t+\tau]$  is defined as
\begin{align*}
  \Delta\tind{t}\rlind{\tau} \sysent := \sysent\tind{t+\tau} -\sysent\tind{t}.
\end{align*}

As for classical thermodynamics, the entropy change in the medium is related to the \emph{irreversibility} of a process.
Let us denote the evolution operator of a suitable \emph{reversed process} by $\revop U\rlind{\tau}\tind{t}$.
Often, the term or operator responsible for the \emph{temporal variation of the entropy in the medium} has the form
\begin{align}
  \Delta\tind{t}\rlind{\tau}\medent \sim \int \log\left( \frac{U\rlind{\tau}\tind{t}}{\revop U\rlind{\tau}\tind{t+\tau}} \right)\pden\tind{t} \df x.
  \label{eq:medep-general}
\end{align}
Various examples of this relation can be found in \cite{Maes2004}.

In Sections \ref{sec:stochastic-models} and \ref{sec:md} we will be more concrete and give the expressions for $\Delta\tind{t}\rlind{\tau}\sysent $ and $ \Delta\tind{t}\rlind{\tau}\medent $ for some common models of complex systems in thermodynamic environments.
Beforehand, we revisit the microscopic theory of \emph{isolated systems}, namely Hamiltonian dynamics.

\subsection{Hamiltonian dynamics}
\label{sec:hamiltonian-dynamics}
Classical statistical mechanics is formulated based on \emph{Hamiltonian dynamics} \cite{Gibbs1948,Callen+Scott1998,Khinchin2013}.
In Hamiltonian dynamics, a point $\gls{symb:microstate}= \left( \vec q, \vec p \right) \in \gls{symb:pspace}$ fully represents the state of a system.
The dynamics is deterministic, \ie the state $x\tind{t}$ after some time $t$ is fully determined by the \emph{initial condition} $x_0$.
The \emph{phase space} $\pspace$ of Hamiltonian dynamics is the state space of an \emph{isolated system}.\footnote{
Closed and open systems can be obtained by considering only subsets of the phase space as the system, whereas the rest is identified with the medium.}
The degrees of freedom $x$ split into the (generalized) coordinates $\vec q$ and (generalized) momenta $\vec p$ of all $N$ particles that constitute the system.
For brevity, here and in the following we use the notation $\vec q = \set{\vec q_k }_{k=1}^N$, $\vec p = \set{ \vec p_k }_{k=1}^N$ where no ambiguity can arise.

The \emph{Hamiltonian}\footnote{ In this notation, the term $\vec p^2 /2m$ (\ref{eq:Hamiltonian}) is short for $\sum_{k=1}^N \frac{\vec{p}_k^2}{2 m_k}$ including the (possibly different) masses $m_k$.}
\begin{subequations}
\begin{equation}
  \ham(x) =  V(\vec q)+\frac{\vec {p} ^2}{2m}.
  \label{eq:Hamiltonian}
\end{equation}
  \label{eq:HamiltonianEOM}
is the dynamic variable that represents the total energy $E$ of the system.
It determines the equations of motion
  \begin{alignat}{2}
  \dot{ \vec{ q}} &=  &&{}\nabla_{\vec p} \ham= \frac{\vec p}{m},\\
  \dot{ \vec{ p}} &= -&&{}\nabla_{\vec q} \ham = -\nabla_{\vec q}V(\vec q).
\end{alignat}  
\end{subequations}
In the above equations \eqref{eq:HamiltonianEOM}, $\dot x := \frac{\df x}{\df t}$ denotes the total derivative with respect to time and $\nabla_{\set{x}}(\cdot)$ denotes the vector gradient (also denoted $\vgrad_{\set{x}}(\cdot)$) of a scalar function with respect to the set of coordinates $\set{x}$.

The first term in the Hamiltonian, $V(\vec q)$, is a potential that gives rise to (conservative) forces $F\tsup{cons}(\vec q) := -\nabla_{\vec q} V(\vec q)$.
The second term denotes the total kinetic energy.
Moreover, the Hamiltonian is a constant of motion, \ie $\ham(x\tind{t}) = \ham(x_0) = E$ does not change over time.
Hence, energy is conserved as we would expect it from an isolated system.

Hamiltonian dynamics are a standard example of deterministic-chaotic systems.
Its equations are usually non-linear and high dimensional, and thus generically show a sensitive dependence on initial conditions:
The distance of infinitesimally separated points in phase space $\delta\! x$ shows an (initial) exponential growth with time.
In contrast, detailed information on microscopic initial conditions is never accurately available for real systems.
Hence, it must be specified in a probabilistic way --- which lead to the notion of Gibbs' statistical ensembles.%
\footnote{
  A collection of Gibbs' pioneering work can be found in Ref.\,\cite{Gibbs1948}}
Moreover, Gibbs was the first to write down the functional form of the (differential) entropy (\ref{eq:shannon-entropy}) associated with a phase-space ensemble $\pden\tind{t}$.

A probability density for a dynamics $\partial_t x = f(x)$ satisfies a continuity equation, because probability is conserved.
Henceforth, we denote the partial derivative with respect to time $t$ by $\del_t$ and the divergence of a vector field $f$ by $\vdiv(f)\equiv \nabla \cdot f$.
The continuity equation then reads:
\begin{align}
  \partial_t \pden\tind{t} &= -\vdiv(f(x) \,\pden\tind{t})\nonumber\\
  &=- \pden\tind{t}\vdiv(f(x)) -f(x)\cdot\vgrad(\pden\tind{t}) .
  \label{eq:continuity-equation}
\end{align}
Rearranging this equation, we find for the total derivative of the probability density:
\begin{align*}
  \tdiff{\pden\tind{t}}{t} =  \partial_t \pden\tind{t} + f(x)\cdot\vgrad(\pden\tind{t}) = - \pden\tind{t}\vdiv(f(x)).
\end{align*}
Note that this equation can be rewritten as
\begin{align}
  \tdiff{\left(- \log \pden\tind{t}\right)}{t} = \vdiv(f) =: \Lambda,
  \label{eq:phase-space-expansion}
\end{align}
where $\gls{symb:expansion-rate}(x)$ is called the \emph{phase space expansion rate}.
For Hamiltonian dynamics, phase space volume is conserved, \ie the expansion rate identically vanishes:
\begin{align}
  \Lambda \equiv \vdiv (f) := \diff{(\partial_t q)}{q} +  \diff{(\partial_t p)}{p} = \frac{\mathrm{d}^2 \ham}{\df q \df p} - \frac{\mathrm{d}^2\ham}{\df p \df q}  = 0
  \label{eq:uniformly-conservative-system}
\end{align}
and thus
\begin{align}
  \frac{\df \pden\tind{t}}{\df t} = 0.
  \label{eq:liouville-equation}
\end{align}
This statement, usually known as the ``Liouville theorem'', was first written down by Gibbs in the context of his statistical ensembles.
He soon realized that conservation of phase space volumes implies the conservation of the entropy:
\begin{align}
  \frac{\df{\sent{\pden\tind{t}}}}{\df t} = 0.
  \label{eq:const-entropy}
\end{align}
This fact is often referred to as the ``paradox of constant entropy'' in Hamiltonian systems,  as it seems to be in contradiction with observations.
However, this problem is remedied if one accepts that one never has access to the microscopic density.
All that we can hope for in real observations is to find a distribution for some coarser, \emph{effective} degrees of freedom.
Indeed, the apparent paradox is resolved if one adapts our initial point of view, in which the system consists of observable and thus operationally accessible degrees of freedom, \cf Ref.~\cite{Penrose1970,Ruelle1999}.

In the following, we reserve the term ``Gibbs entropy'' $\ent\tsup{G}$ for the entropy obtained by a \emph{maximum entropy principle}.
More precisely, we say that $\pden \equiv \pden(\set{(a_i,\obs_i)})$ is \emph{compatible} with the macroscopic constraints $\set{(a_i,\obs_i)}$, if for the observables $\set{\obs_i\colon\pspace\to \gls{symb:reals}}$ one has
\begin{align}
  \eav{\obs_i} := \int_\pspace \obs_i \pden \df x = a_i,\quad \forall i.
  \label{eq:compatible}
\end{align}
In that case, the Gibbs entropy specified by $\set{(a_i,\obs_i)}$ is defined as
\begin{align}
  \ent\tsup{G} := \sup_{\pden'}\sent{\pden'(\set{a_i,\obs_i})},
  \label{eq:gibbs-entropy}
\end{align}
where the supremum is taken with respect to all compatible ensembles $\pden'(\set{(a_i,\obs_i)})$.
Often the supremum is given by a unique ensemble $\pden\tsup{G}(\set{(a_i,\obs_i)})$, which we will call the \emph{Gibbs ensemble} or \emph{Gibbs distribution}.

Hamilton's equations of motion are appealing because they constitute a microscopic theory derived from first principles.
However, besides the paradox of constant entropy they suffer another huge practical problem:
For macroscopic physical systems the number of particles, $N\sim 10^{23}$, is very large and makes computations hard.
The problem also does not vanish if we consider much smaller,  \emph{mesoscopic}%
\footnote{
  Usually, for the \emph{mesoscopic range} one considers typical molecular scales (less than \SI{10}{\nano\meter}) and typical macroscopic scales (larger than \SI{10}{\micro \meter}) as lower and upper boundaries, respectively.
  Because of this wide range, we prefer to define the term with respect to the modelling paradigm, \cf Sec.~\ref{sec:motivation}.
}
systems.
Such systems are usually immersed in some solvent (\eg water) and Hamiltonian dynamics requires us to treat this environment explicitly.

Thus, treating meso- or macroscopic systems in thermodynamic environments with the microscopic equations of motion (\ref{eq:HamiltonianEOM}) is a challenging task.
Even with state-of-the-art supercomputers, simulations of no more than a few (\numrange{e4}{e6}) particles on small time scales (\SIrange{e2}{e4}{\nano\second}) are possible.
Hence, developing and applying effective dynamical models with fewer degrees of freedom is a major subject of modern physics.
In the next two sections, we review  modelling paradigms for systems in thermodynamic environments.
We start with models based on stochastic processes, which have their origins already in the beginning of the twentieth century.
After that, we focus on deterministic models used in modern molecular dynamics simulations.

\section{The models of stochastic thermodynamics}
\label{sec:stochastic-models}

The first stochastic models were introduced as a theoretical framework to study the phenomenon of \emph{Brownian motion} in systems at or close to equilibrium.
Brownian motion provides an archetypal example of the dynamics of systems in thermodynamic environments.
As we will see shortly, already the study of the thermodynamic aspects of such a simple system yields important physical results.
The most famous one is the so-called Einstein relation which connects microscopic fluctuations to macroscopic dissipation.

\emph{Stochastic thermodynamics} is the area of statistical physics that seeks such relations for increasingly complex systems in non-equilibrium environments \cite{Seifert2008}.
Already in the middle of 20th century, stochastic models were formulated for a variety of (non-equilibrium) phenomena in many disciplines of science \cite{vKampen1992}.
They all have in common that the statistically random forces on the system exerted by the environment are modelled using stochastic terms.
For small systems like biomolecules in solution, these forces lead to notable \emph{fluctuations} in the system's dynamics.

Hill and Schnakenberg pioneered a thermodynamic interpretation of non-equilibrium steady states of master equations \cite{Hill1977,Schnakenberg1976}.
In particular, they proposed a general relation between abstract notions of entropy production for stochastic processes and thermodynamic dissipation.
These early considerations were based on the temporal evolution of an ensemble as specified by the master equation.
More recently, authors started to discuss notions of entropy and entropy production for individual \emph{realizations} of stochastic processes \cite{Kurchan1998,Lebowitz+Spohn1999}.
This idea led to the unification of a variety of fundamental non-equilibrium \emph{fluctuation relations} (FR) concerning the probability distributions of heat, work and entropy production \cite{Maes2004,Seifert2005}.
Here, we only briefly discuss stochastic FR in Section~\ref{sec:stochastic-fr}.
For a review on the general theory, we refer the reader to Ref.~\cite{Seifert2012}.

\subsection{Langevin and Fokker--Planck equations}
\label{sec:langevin}
The first stochastic models were introduced in the beginning of the twentieth century by Einstein \cite{Einstein1905}, Langevin \cite{Langevin1908} and Smoluchowski \cite{Smoluchowski1915}.
Their goal was to model the diffusion of a relatively heavy \emph{tracer particle} surrounded by a large number of much lighter particles.
One usually refers to the tracer particle as performing \emph{Brownian motion} in its fluid environment.
Today we know that every fluid, though it might appear as continuous, is made out of particles.
Further, we understand Brownian motion as the result of the irregular forces that the lighter particles exert on the tracer.
Hence, Brownian motion can be understood as a kind of shot noise, \ie an erratic behaviour that has its origin in the granularity of matter.
However, at the end of the 18th century the atomistic view had not been generally accepted.
Einstein emphasized that the success of the theory of Brownian motion gives an estimation of Avogadro's number and thus confirms the existence of molecules \cite{Einstein1905}.

\subsubsection{Brownian motion}
We start by illustrating the ideas of stochastic models in the framework of Brownian motion.
The mathematics are essentially the same for more general situations. 
For a comprehensive review of stochastic thermodynamics, we direct the reader to Ref.~\cite{Seifert2012}.

Consider a particle with position $q$ and velocity $\dot{q}$ in a fluid environment.
The particle is subject to conservative forces $F\tsup{cons} = -\partial_q V$ and a (Stokes) drag force $F\tsup{drag} = - \zeta \dot{q}$, where $\zeta$ denotes a phenomenological drag coefficient.
Further, we consider a microscopic noise term $\xi$ to model the collisions of the tracer with the fluid molecules.

In the \emph{overdamped limit} one assumes that accelerations are immediately damped away by the environment.
Hence, the macroscopic forces balance, \ie $F\tsup{cons} + F\tsup{drag}=0$ and thus $\dot{q} = F\tsup{cons}/\zeta$.
To this \emph{macroscopic} equation of motion we add the noise $\xi$ to obtain the \emph{overdamped Langevin equation}:

\begin{align}
  \dot{q} = \frac{-\partial_q V}{\zeta} + \xi
  \label{eq:langevin-overdamped}
\end{align}

A common assumption (which we will adopt here) is that $\xi$ obeys the statistics of \emph{white noise}.
White noise is uncorrelated with zero mean and variance $2D$.
More precisely, the averages realization of the stochastic force at time $t$ obey
\begin{equation}
  \eav{ \xi(t) }\tind{t} = 0, \qquad \eav{ \xi(t)\xi(0) }\tind{t} = 2 D\delta(t),
  \label{eq:WhiteNoise}
\end{equation}
where $\delta(t)$ denotes the Dirac $\delta$-distribution.

For Langevin dynamics, the average of an observable $\gls{symb:observable}\colon\pspace\to\reals$ can be written as an integral over a probability density $\pden\tind{t}$:
\begin{align*}
  \eav{\obs}\tind{t} := \int_\pspace \obs\, \pden\tind{t} \df x.
\end{align*}
The density $\pden\tind{t}$ specifies a time-dependent ensemble.
For white noise, its evolution is governed by the \emph{Smoluchowski equation} \cite{Smoluchowski1915}:
\begin{subequations}
\begin{align}
  \partial_t \pden\tind{t} = -\partial_q j\tind{t}.
\end{align}
Probability conservation is guaranteed by this equation, as the right-hand side amounts to the divergence of the \emph{instantaneous probability current}
\begin{align}
  j\tind{t} :=  \frac{-\partial_qV}{\zeta} \pden\tind{t} - D\partial_q\pden\tind{t} .
  \label{eq:smoluchowski-current}
\end{align}  
  \label{eq:smoluchowski}
\end{subequations}
The first contribution to the probability current is associated with the macroscopic force balance.
It thus expresses the macroscopic \emph{drift}.
The second term is an undirected diffusive current which is determined by the strength of the noise $D$.
For dilute systems, the probability density $\pden$ can also be understood as a particle density.
If the current $j\tind{t}$ in Equations~\eqref{eq:smoluchowski} is interpreted in that way, then $D$ is called a \emph{diffusion constant}.

Equilibrium is defined as a steady state ($\partial_t \pden\tind{t} = 0$) where probability currents vanish:
\begin{align}
 j\tind{t} \equiv 0.
  \label{eq:equilibrium}
\end{align}
In that case one also says that the system obeys \emph{detailed balance}.
For the current in Equation~\eqref{eq:smoluchowski-current}, the equilibrium condition \eqref{eq:equilibrium} yields
\begin{align*}
  0 = -\left( \frac{\partial_q V}{\zeta} + D\partial_q \right)\pden\tind{t}.
\end{align*}
Consistency with statistical mechanics requires that the equilibrium probability density amounts to a Boltzmann-distribution\footnote{Note that $\kb \equiv 1$.}, \ie $\pden(q) \propto \exp\frac{-V(q)}{T}$.
Hence, we get
\begin{equation}
  D = \frac{T}{\zeta},
  \label{eq:FDR}
\end{equation}
where $T$ is the temperature of the (isothermal) system.
In the context of Brownian motion one usually uses a Stokes drag constant $\zeta = 6 \pi \eta R$, where $R$ denotes the radius of the particle and $\eta$ is the dynamic viscosity of the fluid.
In that case, (\ref{eq:FDR}) is the so-called Smoluchowski--Einstein fluctuation-dissipation relation (FDR)
\begin{align}
  D = \frac{T}{6 \pi \eta R}.
  \label{eq:einstein-relation}
\end{align}
It thus relates the erratic motion of the tracer particle in equilibrium (diffusion) to the \emph{linear response} of the system to an externally applied force (drag).

A general connection between equilibrium fluctuations and the response to externally applied (small) forces is the \emph{fluctuation-dissipation theorem} \cite{Callen+Welton1951}.
For systems close to equilibrium, this theorem implies a linear response, which results in the purely exponential decay of fluctuations.

Another example of a linear response result close to equilibrium are the Onsager relations \cite{Onsager1931}.
They are statements about the \emph{thermodynamic current} $\curr_\alpha$ induced by a (small) \emph{thermodynamic force} or \emph{affinity} $\gls{symb:affinity}_\alpha$.
The index $\alpha$ distinguishes between the different driving mechanisms, because there may be multiple forces acting on the same system.
The driving forces are either external forces (like electric fields) or spatial gradients of intrinsic thermodynamic variables (like temperature or chemical potential).
With the  matrix of transport coefficients $\onsmat_{\alpha\beta}$, the \emph{Onsager relations}
\begin{align}
  \aff_\alpha = \sum_\beta \onsmat_{\alpha\beta} \curr_\alpha
  \label{eq:onsager}
\end{align}
provide a prime example of a linear-response relation.
In most cases we also have \emph{reciprocity}, which means that the Onsager coefficients are symmetric, \ie $ \onsmat_{\alpha\beta} = \onsmat_{\beta\alpha} $.

Above, we have derived the Smoluchowski-Einstein FDR, from the \emph{thermodynamic consistency argument}, namely the assumption of a Boltzmann distribution.
In general, linear response theory close to equilibrium follows from a more general thermodynamic consistency assumption called \emph{local equilibrium}.
We will discuss local equilibrium in more detail below.

\subsubsection{Entropies for the system and the medium}
In order to identify entropies and entropy changes in the system and the medium we follow Seifert's work \cite{Seifert2005,Seifert2012}.
In agreement with the  general prescription (\ref{eq:system-entropy}), the system's entropy is the differential entropy of the ensemble:
\begin{align*}
  \sysent\tind{t} := -\int_\pspace \pden\tind{t} \log \pden\tind{t} \df q.
\end{align*}
The instantaneous entropy change of the system is its time-derivative
\begin{align*}
  \delta\tind{t}\sysent := \partial_t \sysent\tind{t}.   
\end{align*}
Denoting the change in the medium by $\delta\tind{t}\medent$ and the total change by $\delta\tind{t}\totent$, it splits into two contributions:
\begin{subequations}
\begin{align}
  \delta\tind{t}\sysent  = \delta\tind{t}\totent - \delta\tind{t}\medent.
\end{align}
With the introduction of the velocity distribution, $v\tind{t} := \frac{j\tind{t}}{\pden\tind{t}}$ one finds that \cite{Sekimoto1998,Seifert2012}
  \begin{align}
    \delta\tind{t}\medent &= - \int_\pspace \frac{ v\tind{t}F\tsup{cons}}{T} \pden\tind{t} \df q,    \label{eq:fp-medent}\\
    \delta\tind{t}\totent &= \int_\pspace \frac{j\tind{t}^2}{D \pden\tind{t}} \df q  
    = \frac{\eav{v\tind{t}^2}\tind{t}}{D} \geq 0     \label{eq:fp-totent}.
  \end{align}
  \label{eq:fp-evs}
\end{subequations}
The thermodynamic interpretation is straightforward:
In the overdamped limit, any work performed in a potential $V$ is immediately dissipated.
The ensemble average of the instantaneous dissipated heat $\delta_t Q\tsup{med}$ is thus the associated power $\delta_t Q\tsup{med} = v\tind{t}F\tsup{cons} $.
Under isothermal conditions, the entropy change in the medium is the heat $Q$ divided by temperature $T$.
The total entropy is always positive and can be written in a form which is well-known from transport theory \cite{deGroot+Mazur1984}.

In this interpretation, the relations \eqref{eq:fp-evs} yield a differential form of the second law~\eqref{eq:td-second-law}:
\begin{align*}
\delta\tind{t}\totent = \delta\tind{t}\sysent+\delta\tind{t}\medent\geq0.
\end{align*}

\subsubsection{Underdamped motion and generalizations}
The Langevin equation is easily formulated for more general situations.
In fact, the original Langevin equation was formulated as an underdamped equation \cite{Langevin1908}.
In that case, the macroscopic equation is Newton's second law $\dot{p} = F\tsup{tot} = F\tsup{drag} + F\tsup{cons}$, where $p = m\dot{q}$ is the momentum of the particle with mass $m$.
Again, by adding a noise term to model the irregular microscopic forces we obtain:
\begin{subequations}
  \begin{align}
    \dot{q} &= \frac{p}{m},\\
    \dot{p} &= -\partial_q V -\frac{\zeta}{m} p + \xi.
  \end{align}
  \label{eq:langevin-underdamped}
\end{subequations}
Here, the strength of the noise fulfils a different fluctuation-dissipation relation, which can be found from demanding a Maxwell--Boltzmann distribution for the momenta.
Further generalizations consider multiple interacting particles in more spatial dimensions.
Because the evolution equation for the probability density retains the form of a linear advection-diffusion equation similar to \eq{smoluchowski}, one can at least formally solve it.

In practice, one is often interested in observable \emph{collective degrees of freedom}, like the hydrodynamic modes of a continuous density.
In order to obtain equations for these \emph{effective} degrees of freedom, one applies approximations at some point which turn the high-dimensional linear equation for the probability distribution into a lower dimensional form.
Unfortunately, the linear form of the evolution equation is usually lost \cite{Zwanzig1961,Mori1965}.

However, there are situations where a linear description of the evolution equation is still appropriate.
In that case, one can formulate a generalized version of the Langevin equation also for the collective degrees of freedom.
Common examples of such collective degrees of freedom are \emph{reaction coordinates} in biochemical systems or \emph{order parameters} in the physics of condensed matter \cite{Chaikin+Lubensky2000}.

The most general formulation of a classical Langevin equation for an arbitrary set of degrees of freedom $\omega$ reads \cite{Seifert2012}:
\begin{align}
  \partial_t \omega = \mobmat\left( -\nabla_\omega V(\omega) + F\tsup{diss}(\omega) \right) + \xi.
  \label{eq:langevin-general}
\end{align}
The mobility tensor $\mobmat$ summarizes the appropriate phenomenological transport coefficients.
If $\mobmat = \zeta^{-1} \mathbb{1}\,$ has scalar form, then it is the inverse of the drag coefficient $\zeta$.

Dissipative forces $F\tsup{diss}$ may include drag and other non-conservative forces.
Again, the microscopic noise term has white noise statistics and needs to be connected with macroscopic transport properties.
More precisely, the noise correlations obey
\begin{align}
  \eav{ \xi_i(t)\xi_j(0) }\tind{t} = 2 T \mobmat_{ij}\delta(t),
  \label{eq:white-noise-general}
\end{align}
 where $\xi_i(t)$ denotes the component of the noise associated with the $i$th component of $\omega$ and the numbers $\mobmat_{ij}$ are the entries of the (positive semi-definite) mobility tensor.

White noise ensures a linear evolution equation for the probability densities $\pden\tind{t}(\omega)$.
The resulting partial differential equation is called the \emph{Fokker--Planck equation}:
\begin{align}
  \partial_t \pden\tind{t}(\omega) &= -\nabla_\omega \cdot j\tind{t}(\omega) \nonumber\\
  &:= -\nabla_\omega \cdot\left( \mobmat\left( -\nabla_\omega V(\omega) + F\tsup{diss}(\omega) \right)\pden\tind{t}(\omega) - T \mobmat \nabla_\omega\pden\tind{t}(\omega) \right).
  \label{eq:fp-general}
\end{align}
In (experimental) applications, often time-dependent (\eg oscillatory) forces are used to probe the response of the system and hence determine the transport coefficients $\mobmat_{ij}$ \cite{Chaikin+Lubensky2000}.
Stochastic thermodynamics with explicitly time-dependent forces is thoroughly reviewed in Ref.~\cite{Seifert2012}.

\subsubsection{Local equilibrium}
One of the crucial assumptions in stochastic thermodynamics is \emph{local equilibrium} (LE).
In order to appreciate its meaning in the present context, consider the following situation:
An experimenter takes measurements on a many-particle system.
The possible measurement outcomes are the values $\omega\in\ospace$ of an \emph{observable} $\gls{symb:measurement}\colon\pspace \to \ospace$ characterizing the measurement process.
Let us further assume that the dynamics on the level of the collective variables $\omega = \partmap(x)$ are modelled by a generalized Langevin equation \eqref{eq:langevin-general}.
On that level of description, one ignores the hidden structure and dynamics of the microstates $x \in \pspace$.

In order to connect the stochastic dynamics with statistical physics, one (implicitly) assumes a distribution $\pden_\omega(x)$ for those microstates $x$ which yield a certain measurement result $\partmap(x) = \omega$.
The energy associated with a state $\omega$ is thus an \emph{internal energy} obtained as the conditioned expectation value of a (Hamiltonian) energy function $\ham(x)$.
Besides the average internal energy, the distribution $\pden_\omega$ also specifies an internal entropy.
The forces $F\tsup{cons} = -\nabla_\omega V(\omega)$ must therefore include \emph{entropic forces} as well.
The latter arise from the fact that different values of $\omega$ might be compatible with a different number of microscopic states.

In order to make the notion of compatibility precise, we formulate an ansatz for the constrained microscopic distribution $\pden_\omega(x)$.
The \emph{internal} entropy is then given as $\ent\tsup{int}_\omega = \sent{\pden_\omega}$.
Local equilibrium specifies the value of this entropy, by making assumptions about the distribution $\pden_\omega(x)$. 
Similar to the constraints on the strength of the noise in the formulation of Langevin equations, local equilibrium is a consistency assumption:
It demands consistency with the treatment of microstates in the statistical physics of equilibrium systems.
Further, it ensures that quantities like the potential $V(\omega)$ of a collective variable $\omega$ are well defined.
Note that here ``local'' refers to the association with a certain value $\omega$ rather than a spatial localization.

One of the necessary requirements for local equilibrium is the existence of a separation of time scales between effective mesoscopic degrees of freedom $\omega$ and the microscopic degrees of freedom $x$:
The dynamics of the collective variables are assumed to evolve on a typical time scale $\tmes$ which is much larger than the microscopic time scale $\tmic$.
Consequently, one expects that the distribution $\pden_\omega$ of the microstates $x$ has effectively relaxed to an equilibrium distribution.
More precisely, $\pden_\omega = \pden\tsup{G}(\omega,M)$ is assumed to be the Gibbs distribution compatible with the mesoscopic value $\omega$ and any further thermodynamic constraints.
In particular, the internal entropy of a state $\omega$ is assumed to have the value of the corresponding Gibbs entropy.
Then, the potential $V(q)$ should be understood as \emph{thermodynamic potential} like a \emph{free energy}.

Note that the separation of time scales implicitly enters the Langevin equation \eqref{eq:langevin-general} through the assumption of white noise \eqref{eq:white-noise-general}.
For the example of Brownian motion, the $\delta$-correlations are approximations to the real collision statistics, which have a small (but finite) relaxation time $\tmic$.
We assume that the temporal resolution for the observation of the motion of the heavy colloid is much larger than a microscopic time scale $\tmic$.
The time scale $\tmic$ also determines the decay time of microscopic--mesoscopic correlations.

In the context of bio-chemical systems, collective variables are usually chosen to reflect experimentally accessible observations.
Such variables may describe the configuration and chemical composition of macromolecules.
In that context, they are also known as reaction coordinates.
As an example, consider the configurational changes associated with the folding of proteins.
They occur on time scales $\tconf \approx \SIrange{e-4}{e0}{\second}$, which is much larger than the time scales of the microscopic constituents and the solvent.

This separation of time scales and the resulting quick loss of correlations, is called the \emph{Markov property}.
It means that the future of a state only depends on its current state.
Neither the system nor its environment keeps memory of the system's evolution in the past. 
This is explicitly visible in the Fokker--Planck equation (\ref{eq:fp-general}):
Both the conservative and the dissipative forces only depend on the instantaneous value of $\omega$.
The Fokker--Planck (or Smoluchowski) equation is a certain time- and space-continuous form of a \emph{master equation}.
It is appropriate if $\omega\in\ospace$ takes continuous values.
However, often it is enough to consider a discrete space of observations $\ospace$.

\subsection{Master equations}
\label{sec:discrete-st}
The master equation is the discrete-time version of the Fokker--Planck equation.
We have already seen that the Fokker--Planck equation specifies a Markovian (\ie memoryless) evolution on a continuous phase space.
In this section, we consider time-homogeneous Markov processes on a finite state space $\ospace = \set{1,2,\dots,N}$.
The ensemble at time $t$ is specified by a probability vector $\bvec p \tind{t}$.

The \emph{master equation} for its evolution in continuous time reads
\begin{align}
  \partial_t \bvec p \tind{t} = \bvec p \tind{t} \tmat,
  \label{eq:master-continuous}
\end{align}
where $\tmat$ is a \emph{rate matrix} containing the transition rates \gls{symb:transition-probability}.
For $\omega \neq \omega'$ they obey $\tprob{\omega}{\omega'}\geq0$.
The diagonal entries of $\tmat$ amount to $\tprob{\omega}{\omega} := -\sum_{\omega'\neq\omega}\tprob{\omega}{\omega'}$, because probability needs to be conserved.

In Chapters \ref{chap:marksymdyn} and \ref{chap:information-st}, we are more interested in the time-discrete case of \emph{Markov chains}.
The master equation for Markov chains reads:
\begin{align}
  \bvec p \tind{t+1} = \bvec p \tind{t} \tmat,
  \label{eq:master}
\end{align}
where the entries \gls{symb:tmat} obey $0\leq\tprob{\omega}{\omega'}\leq 1 $ and $\sum_{\omega'}\tprob{\omega}{\omega'}=1$.
A matrix that satisfies these conditions is called a \emph{stochastic} or \emph{transition matrix}.

For both continuous and discrete-time Markov chains we define the \emph{probability flux} \gls{symb:flux} from state $\omega$ to state $\omega'$ as
\begin{equation}
  \fluxa{\omega}{\omega'}(t) := \p{t}{\omega} \tprob{\omega}{\omega'},\,~\omega\neq \omega'.
  \label{eq:flux-defn-nonstat}
\end{equation}
The difference of the forward and backward flux is the \emph{probability current} \gls{symb:current} with entries
\begin{align}
  \curra{\omega}{\omega'}(t):= \fluxa{\omega}{\omega'}(t)- \fluxa{\omega'}{\omega}(t).
  \label{eq:currents-nonstat}
\end{align}

\subsubsection{The graph of the network of states}
\begin{figure}[th]
  \centering
  \includegraphics{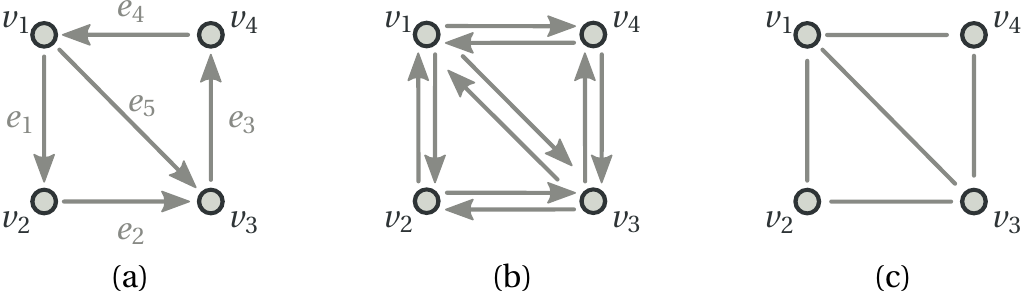}
  \caption{The network of states as a graph. 
  a) A directed graph $\graph\tsup{d}$ consisting of four vertices $v_i$ and five edges $e_i$.
  b) The directed graph $\graph\tsup{d}$ of a dynamically reversible Markovian jump process.
  c) By identifying the forward and backward edges $(\omega,\omega')$ and $(\omega',\omega)$ in b) with the unordered pair $\set{\omega,\omega'}$ we obtain an undirected graph $\graph\tsup{u}$.
  Note that the directed graph in a) is an \emph{oriented} version of $\graph\tsup{u}$, where we pick an arbitrary orientation for each undirected edge.}
  \label{fig:graph-intro}
\end{figure}
For both continuous and discrete time dynamics on finite state spaces, we have a visual representation of the \emph{network of states} as a \emph{graph} \gls{symb:graph}.
A directed graph $\graph\tsup{d} = (\verts,\edges)$ is a finite collection of \emph{vertices} $\verts$ and a set of tuples $\edges \subset \verts \times \verts$, which are called \emph{edges}.
In our case, we choose $\verts = \ospace$ and demand that $(\omega,\omega') \in \edges$ if and only if $\tprob{\omega}{\omega'} >0$.
Its adjacency matrix \gls{symb:adjacency-matrix} is a Boolean matrix (\ie a matrix consisting of zeros and ones) with an entry $a^{\omega}_{\omega'}=1$ whenever $(\omega,\omega')\in \edges$.

A graph can be easily drawn, \cf figure \ref{fig:graph-intro}a.
Vertices $\omega \in \verts$ are represented by points and an arrow pointing from $\omega$ to $\omega'$ is drawn if $(\omega,\omega') \in \edges$.
The visual representation of the state space as a graph provides an intuition for \emph{paths} and \emph{cycles}.
A directed path $\gamma\rlind{\tau}$ of length $\tau$ is a sequence of successive states $\gamma\rlind{\tau} = (\omega_0,\omega_1,\cdots,\omega_{\tau-1},\omega_{\tau})$ that obey
\begin{align*}
  (\omega_{i-1},\omega_i)\in\edges,~0<i\leq\tau.
\end{align*}
Further, we require that no \emph{internal state} $\omega_i$ with $0<i<\tau$ appears multiple times in a path $\gamma\rlind{\tau}$.
A cycle $\alpha\rlind{\tau}$ of length $\tau$ is a closed path where $\omega_0 =\omega_\tau$.

A graph is said to be \emph{connected}, if it does not consist of multiple disconnected parts.
We say that it is \emph{strongly connected}, if there exists a directed path between any two vertices.
If a graph is strongly connected, its adjacency matrix is called \emph{irreducible}.
We use the same terminology to refer to the transition matrix $\tmat$, which yields an adjacency matrix $\adjm = \abs{\gls{symb:sign}{\,\tmat}}$.
A Markov chain or a Markovian jump process on a finite state space with an irreducible adjacency matrix is also called \emph{ergodic}.

For continuous-time Markov processes, ergodicity implies the existence of a \emph{unique} invariant distribution $\bvec p\tind{\infty}$.
For time-discrete Markov chains uniqueness of $\bvec p\tind{\infty}$ additionally requires that the transition matrix $\tmat$ is \emph{aperiodic}.
Aperiodicity means that returns to any state are possible at arbitrary, but sufficiently large time-differences $t$.
Formally, we demand that there is a $\tau \in \naturals$ such that for all $t > \tau$ we have
\begin{align}
  (\tmat^t)\vert_{\omega,\omega} > 0, \quad \forall \omega \in \ospace,
  \label{eq:aperiodic}
\end{align}
where $(\tmat^t)\vert_{\omega,\omega}$ denotes the diagonal entries of the $t$th power of $\tmat$.
One also says that an irreducible and aperiodic Markov chain is \emph{mixing}.

For the rest of this section we assume \emph{dynamical reversibility}, \ie $\tprob{\omega}{\omega'}> 0\Leftrightarrow\tprob{\omega'}{\omega}> 0$.
This means, that the network of states is a simply connected, bi-directional graph.
In that case, we can also draw an undirected graph $\graph\tsup{u} = (\verts,\edges\tsup{u})$, where $\edges\tsup{u}$ contains sets of pairs $\set{\omega,\omega'}$ rather than tuples (\cf figure \ref{fig:graph-intro}c).
The physical motivation of dynamic reversibility has its origin in the reversibility of the microscopic (Hamiltonian) dynamics:
Assume that the value $\omega$ of a collective dynamical variable does not depend on the momenta of an underlying microscopic system.
Now consider a microscopic trajectory that takes the system from state $\omega$ to $\omega'$.
By flipping the momenta, we reverse the direction of time and hence obtain a microscopic trajectory that takes the system from  $\omega'$ to $\omega$.

\subsubsection{Stochastic thermodynamics for master equations}

For dynamically reversible Markov processes, we define the logarithmic ratio
\begin{align}
  \mota{\omega}{\omega'} := \log\frac{\tprob{\omega}{\omega'}}{\tprob{\omega'}{\omega}}.
  \label{eq:motance}
\end{align}
for both continuous- and discrete-time dynamics.

Because of an analogy with electrical networks (which we will discuss in Chapter \ref{chap:cycles}), we refer to $\mota{\omega}{\omega'}$ as the \emph{motance} of a transition ${\omega}\to{\omega'}$.
The integrated motance $\mot_\alpha$ of a cycle~$\alpha = \alpha\rlind{\tau}$ is defined as
\begin{align}
  \gls{symb:motance}_\alpha := \sum_{i=1}^\tau \mota{\omega_{i-1}}{\omega_{i}}.
  \label{eq:cycle-motance}
\end{align}
By analogy to vector calculus we say that the motance is \emph{conservative}, if it vanishes along any cycle:
\begin{align*}
  \mot_\alpha = 0,~\forall \alpha.
\end{align*}
Then, the integrated motance depends only on the initial and final state of any path $\gamma\rlind{\tau}$.
As a consequence one can write the motance between two states as the difference $\Delta\pota{\omega}{\omega'}$ of a potential $\pot_\omega$ defined on the states.
A formal treatment of this analogy will be the topic of Chapter~\ref{chap:cycles}.
An equivalent condition for a conservative motance is that the so-called \emph{Kolmogorov cycle criterion} holds for every closed path $\traj \omega\rlind{\tau}$ with $\omega_0 = \omega_\tau$:
\begin{align}
  \prod_{i = 1}^{\tau} \tprob{\omega_{i-1}}{\omega_{i}} = \prod_{i = 1}^{\tau} \tprob{\omega_{i}}{\omega_{i-1}}.
  \label{eq:kolmogorov-crit}
\end{align}

For the moment, let us focus on the steady state where we have an invariant probability distribution $\bvec p \tind{t} = \bvec p\tind{\infty}$ with $\del_t \bvec{p}_\infty = 0$.
In that case, fluxes $\fluxa{\omega}{\omega'}$ and currents $\curra{\omega}{\omega'}$ are time-independent.
For the steady state, we define the \emph{edge affinity}
\begin{align}
  \affa{\omega}{\omega'} := \log\frac{\fluxa{\omega}{\omega'}}{\fluxa{\omega'}{\omega}}\equiv \log\frac{\p{\infty}{\omega}\tprob{\omega}{\omega'}}{\p{\infty}{\omega'}\tprob{\omega'}{\omega}} 
  \label{eq:edge-affinity}
\end{align}
as the logarithmic ratio of forward and backward fluxes in the steady state.
Because the steady state probabilities cancel in the logarithmic ratio, the affinity $\aff_\alpha$ of a cycle of length $\tau$ corresponds to the motance of that cycle:
\begin{align}
  \aff_\alpha  &:= \sum_{t=1}^{\tau} \log\frac{\p{\infty}{\omega}\tprob{\omega}{\omega'}}{\p{\infty}{\omega'}\tprob{\omega'}{\omega}}  \label{eq:cycle-affinity}\\
  &= \sum_{t=1}^{\tau} \log\frac{\tprob{\omega}{\omega'}}{\tprob{\omega'}{\omega}} = \mot_\alpha\nonumber
\end{align}

The affinities and motances of cycles and edges play an important role in the early formulations of stochastic thermodynamics by T.\,Hill \cite{Hill1977} and J.\,Schnakenberg \cite{Schnakenberg1976}.
In analogy to the discussion of the Fokker--Planck equation, we say that the system is in \emph{equilibrium} if all steady state currents vanish identically:
\begin{align*}
  \curra{\omega}{\omega'} = 0,~ \forall \omega,\omega'.
\end{align*}
The detailed balance condition can be written using the fluxes:
\begin{align*}
  \fluxa{\omega}{\omega'} = \fluxa{\omega'}{\omega}.
\end{align*}
In a steady state, the master equation can be rewritten as
\begin{align*}
  \sum_{\omega'} \curra{\omega}{\omega'} = 0.
\end{align*}
We discuss an interpretation of this equation in more detail in Chapter \ref{chap:cycles}.
Whether a system will exhibit an equilibrium or a non-equilibrium steady state (NESS), depends on the elements of $\tmat$.
The Kolmogorov criterion \eqref{eq:kolmogorov-crit} provides a necessary and sufficient condition for a system to relax to an equilibrium steady state \cite{Schnakenberg1976}.

Initially, let us discuss continuous-time master equations.
Hill was the first one to attempt a thermodynamic interpretation of the master-equation framework.
He noticed the importance of cycles for non-equilibrium steady states \cite{Hill1977} which motivated Schnakenberg's network theory \cite{Schnakenberg1976}.
These authors also realized that the cycle affinities $A_\alpha$ obtain the values of (linear combinations) of the thermodynamic forces acting on the system.
Moreover, cyclic currents $\curr_\alpha$ are the analogues of the thermodynamic currents for these systems.
Schnakenberg also found that cycle currents and affinities close to equilibrium obey
\begin{align}
  \aff_\alpha = \sum_\beta \onsmat_{\alpha\beta} \curr_\alpha,
  \label{eq:schnakenberg-onsager}
\end{align}
with positive symmetric coefficients $\onsmat_{\alpha\beta} = \onsmat_{\beta\alpha}$.
Hence, \eq{schnakenberg-onsager} is another manifestation of Onsager's linear response relations \eqref{eq:onsager}.

This in turn motivates the identification of the bilinear expression
\begin{align}
  \delta\totent\tind{\infty} = \sum_{\alpha} \aff_\alpha \curr_\alpha
  \label{eq:total-ep-schnakenberg-cycle}
\end{align}
as the total entropy production in the steady state.
The sum in \eq{total-ep-schnakenberg-cycle} runs over a set of \emph{fundamental cycles} defined in Schnakenberg's network theory.
We will discuss the abstract algebraical features of master equations in more detail in Chapter \ref{chap:cycles}.
For now, we just mention that we can formulate \eq{total-ep-schnakenberg-cycle} also using edge currents and edge affinities:
\begin{align}
  \delta\totent\tind{\infty} = \frac{1}{2}\sum_{\omega,\omega'} \affa{\omega}{\omega'}\curra{\omega}{\omega'}\geq 0
  \label{eq:total-ep-schnakenberg}
\end{align}
The factor $\frac{1}{2}$ is needed to avoid double-counting of edges.
Positivity immediately follows from the fact that $\sgn(\affa{\omega}{\omega'})=\sgn(\curra{\omega}{\omega'})$, \cf Equations~\eqref{eq:edge-affinity} and \eqref{eq:currents-nonstat}.

Schnakenberg's considerations for the steady state can be generalized to the transient case.
We canonically identify the system's entropy as 
\begin{align}
  \sysent\tind{t} := -\sum_{\omega \in \ospace} \p{t}{\omega}\log\p{t}{\omega}.
  \label{eq:defn-sysent}
\end{align}
With the master equation \eqref{eq:master-continuous}, the time-derivative of the system's entropy reads
\begin{subequations}%
\begin{align}
  \partial_t\sysent\tind{t} &= \sum_{\omega,\omega'}\p{t}{\omega}\tprob{\omega}{\omega'}\log{\frac{\p{t}{\omega}}{\p{t}{\omega'}}}
  \label{eq:defn-sysent-var-continuous}
\end{align}
It can be split into two contributions $\partial_t\sysent\tind{t} = \delta\totent\tind{t} - \delta\medent\tind{t}$:
  \begin{align}
    \delta\medent\tind{t} &:=\sum_{\omega,\omega'}\p{t}{\omega}\tprob{\omega}{\omega'}\log{\frac{\tprob{\omega}{\omega'}}{\tprob{\omega'}{\omega}}},\label{eq:transient-medent}\\
   \delta\totent\tind{t} &:= \sum_{\omega,\omega'}\p{t}{\omega}\tprob{\omega}{\omega'}\log{\frac{\p{t}{\omega}\tprob{\omega}{\omega'}}{\p{t}{\omega'}\tprob{\omega'}{\omega}}}\label{eq:transient-totent}\\
   &= \frac{1}{2}\sum_{\omega,\omega'}\left[ \left( \fluxa{\omega}{\omega'}(t) - \fluxa{\omega'}{\omega}(t)  \right)\log \frac{\fluxa{\omega}{\omega'}(t) }{\fluxa{\omega'}{\omega}(t) } \right] \geq 0 \nonumber.
  \end{align} 
  \label{eq:splitting-continuous-time}%
\end{subequations}%
Equation \eqref{eq:transient-totent} is the transient version of \eqref{eq:total-ep-schnakenberg}, and hence we identify $\delta\totent\tind{t}$ with the transient generalization of the total (instantaneous) entropy variation.\footnote{
  Even without the thermodynamic considerations above, the expressions \eqref{eq:splitting-continuous-time} also arise naturally when treating non-equilibrium stochastic thermodynamics as a \emph{gauge theory} based on information-theoretical considerations \cite{Polettini2012}.
  More precisely, the gauge corresponds to the above-mentioned choice of a reference measure when defining entropies.
  Then, the expression for the (instantaneous) \emph{total} entropy production is the simplest non-trivial gauge invariant term associated with changes in entropy.  
}

From the second line of the equality, it is immediately clear that this quantity is positive for all $t\geq0$.
Positivity is also evident because $\sum_{\omega,\omega'}\left[ \p{t}{\omega}\tprob{\omega}{\omega'}\right] = \sum_{\omega,\omega'}\left[ \p{t}{\omega'}\tprob{\omega'}{\omega}\right]  < \infty$.
Then, the term $\delta\totent\tind{t}$ is a Kullback--Leibler divergence of two normalized distributions and hence always positive.

The right-hand side of Equation \eqref{eq:transient-medent} is the average of the motance $\mot$ over all jumps $\omega\to\omega'$ that appear at time $t$.
It involves the logarithmic ratio of forward and backward transitions, and thus provides an example of the relation \eqref{eq:medep-general}.
Consequently, one identifies $\delta\medent\tind{t}$  as the instantaneous variation of the entropy in the medium \cite{Seifert2005}.

\subsubsection{The time-discrete case}
In the last subsection we have introduced the instantaneous change of the system's and the medium's entropy for continuous-time Markov processes.
We will also need the expressions for the discrete-time case.
Using the discrete-time master equation \eqref{eq:master-equation}, we split the variation
\begin{subequations}
\begin{align}
  \Delta\sysent\tind{t} = \sysent\tind{t} -\sysent\tind{t-1}  \label{eq:defn-sysvar}
\end{align}
of the system entropy (\ref{eq:defn-sysent}) into $\Delta\sysent\tind{t} =  \Delta\totent\tind{t}- \Delta\medent\tind{t} $ using the expressions
\begin{align}
  \Delta\medent\tind{t} &:= \sum_{\omega,\omega'}\p{t-1}{\omega}\tprob{\omega}{\omega'}\log{\frac{\tprob{\omega}{\omega'}}{\tprob{\omega'}{\omega}}}\label{eq:defn-medent},\\
  \Delta\totent\tind{t} &:= \sum_{\omega,\omega'}\p{t-1}{\omega}\tprob{\omega}{\omega'}\log{\frac{\p{t-1}{\omega}\tprob{\omega}{\omega'}}{\p{t}{\omega'}\tprob{\omega'}{\omega}}}.\label{eq:defn-totent}
\end{align} 
  \label{eq:defn-ent}
\end{subequations}
In the continuous-time limit, the expressions \eqref{eq:defn-ent} yield their continuous-time analogues~\eqref{eq:splitting-continuous-time}.
Again, the motance of a transition determines the temporal variation of the entropy in the medium $\Delta\medent\tind{t}$. 
Note that we cannot write the variation $\Delta\totent\tind{t} $ of the total entropy as the sum over positive terms.
However, note that independently of time $t$ we have $\sum_{\omega,\omega'}\left[ \p{t}{\omega}\tprob{\omega}{\omega'}\right] = \sum_{\omega,\omega'}\left[ \p{t-1}{\omega'}\tprob{\omega'}{\omega}\right]=1$.
Hence, expression \eqref{eq:defn-totent} fulfils the properties of a Kullback-Leibler divergence and is thus always positive.

\subsection{Stochastic fluctuation relations}
\label{sec:stochastic-fr}
In this section we briefly discuss the notion of fluctuation relations (FR) in stochastic thermodynamics.
In their most common form \cite{Seifert2005}, the FR compares the probability \gls{symb:prob} of finding a certain value $a$ of the total entropy change $\delta\rlind{\tau}\totentrv\tind{t}$ in the interval $[t,t+\tau]$ with the probability of observing the negative value $-a$:
\begin{align}
  \frac{\prob[\delta\rlind{\tau}\totentrv\tind{t}=a]}{\prob[\delta\rlind{\tau}\totentrv\tind{t}=-a]} = \exp{a}.
  \label{eq:stochastic-fr}
\end{align}
First results in this direction are  Kurchan's~\cite{Kurchan1998} and Crooks'~\cite{Crooks1999} relations for Langevin  and Fokker--Planck equations.
Lebowitz and Spohn formulated a related result for the master equation approach \cite{Lebowitz+Spohn1999}.
A general framework was suggested by Maes \cite{Maes2004} and Seifert \cite{Seifert2005}.

Note that \eqref{eq:stochastic-fr} is indeed a \emph{detailed} rather than an \emph{integral} formulation of the second law \cite{Seifert2005}:
Rather than talking about averages, it is a statement about probabilities.
Because entropy (or heat) is an extensive quantity, for macroscopic systems and macroscopic time scales $\tau$ the mean value of the total entropy production becomes very large.
Hence, for macroscopic systems, Eq.~\eqref{eq:stochastic-fr} states that observing a negative value of the dissipation is not \emph{impossible}, but extremely \emph{improbable}:
In principle, heat can be extracted from a single reservoir to perform work on the system.
However, the FR states that the probability of such an event is extremely unlikely on macroscopic scales.

All of the fluctuation relations consider single realizations (so-called \emph{noise histories}) of the stochastic dynamics.
More precisely, the random variable $\delta\rlind{\tau}\totentrv\tind{t}$ for the total entropy production depends on a stochastic \emph{trajectory} $\traj \omega\rlind{\tau}$.
Consequently, the probability $\prob$ in equation \eqref{eq:stochastic-fr} is obtained by marginalizing the probability of noise-histories in an interval $[t,t+\tau]$.
In this thesis, we focus on the trajectory-dependent entropic random variables in the framework of time-discrete Markov chains.
For models in continuous time we refer the reader to the review article \cite{Seifert2012}.

In discrete time, the stochastic trajectory $\traj \omega\rlind{\tau}$ is a \emph{time series}
\begin{align}
  \traj \omega\rlind{\tau} = \left( \omega_0,\omega_1,\dots,\omega_\tau \right) \in \ospace^{\tau+1}.
  \label{eq:defn-traj}
\end{align}
For a Markov chain, the probability of seeing a finite sequence of states $\traj \omega\rlind{\tau}$ at time $t$ is given as
\begin{align}
  \prob\tind{t} \left[ \traj \omega\rlind{\tau} \right] = \p{t}{\omega_0}\prod_{t=1}^{\tau}\tprob{\omega_{t-1}}{\omega_{t}}.
  \label{eq:time-series-probability}
\end{align}
The probability of a time series of length one is the joint transition probability or probability flux \eqref{eq:flux-defn-nonstat}.

Let  $\obs\rlind{1}\tind{t_0}(\omega,\omega')$ be a (possibly time-dependent) function.
We define the \emph{jump average} as
\begin{align}
  \trav{\obs}\rlind{1}\tind{t_0} &:= \sum_{\omega,\omega'} \fluxa{\omega}{\omega'}(t_0) \obs\rlind{1}\tind{t_0}(\omega,\omega'),
  \label{eq:defn-jav}
\end{align}
where $ \fluxa{\omega}{\omega'}(t_0)$ denotes the flux at time $t_0$.

Because of the Markovian character of the dynamics, the expressions for the entropy changes depend only on the final transition $\omega_{\tau-1}\to\omega_\tau$ of $\traj \omega$.
They read
\begin{subequations}
\begin{align}
  \delta\sysentrv\tind{t}\left[ \traj \omega\rlind{\tau} \right] &:= \log{\frac{\p{t+\tau-1}{\omega_{\tau-1}}}{\p{t+\tau}{\omega_{\tau}}}}\label{eq:defn-sysentrv},\\
  \delta\medentrv\tind{t}\left[ \traj \omega\rlind{\tau} \right] &:=\log{\frac{\tprob{\omega_{\tau-1}}{\omega_{\tau}}}{\tprob{\omega_{\tau}}{\omega_{\tau-1}}}}\label{eq:defn-medentrv}, \\
  \delta\totentrv\tind{t}\left[ \traj \omega\rlind{\tau} \right] &:=\log{\frac{\p{t+\tau-1}{\omega_{\tau-1}}\tprob{\omega_{\tau-1}}{\omega_{\tau}}}{\p{t+\tau}{\omega_{\tau}}\tprob{\omega_{\tau}}{\omega_{\tau-1}}}}\label{eq:defn-totentrv}.
\end{align} 
  \label{eq:defn-entrv}
\end{subequations}

It is easy to see that for any of these quantities it holds that
\begin{align*}
  \sum_{\traj \omega\rlind{\tau}} \prob[\traj{\omega}\rlind{\tau}]\, \delta\!\gls{symb:entropy-rv}\tind{t}\left[ \traj \omega \rlind{\tau} \right] = \trav{\delta\entrv}\tind{t+\tau-1}\rlind{1} = \Delta \ent \tind{t+\tau}
\end{align*}
where \gls{symb:entropy-rv-var} and \gls{symb:ent-avg-var} stand for the expressions in Equations \eqref{eq:defn-entrv} and \eqref{eq:defn-ent}, respectively.

The above definitions can be generalized to the case of random variables that depend on the whole trajectory rather than only on its two last states.
In mathematics, such random variables are also known as $\tau$-chains.
We will use this more general notion in Chapter \ref{chap:information-st}.
Further, Appendix \ref{app:tau-chains} treats $\tau$-chains in a more formal way and proves some general results.

\section{Effective deterministic models for molecular dynamics}
\label{sec:md}

In the previous section, we have introduced stochastic processes as models for complex systems.
We have seen that it is not necessary to model every degree of freedom explicitly.
Rather, we can restrict our models to observable (collective) degrees of freedom.
The interaction of the system with the medium was modelled using both fluctuating microscopic forces and phenomenological macroscopic coefficients.
For the former we assumed stochastic white noise and arrived at a stochastic differential equation.

For the purposes of computer simulations it may be desirable to have a deterministic description instead.\footnote{
  After all, pseudo-random number generators as they are used in stochastic simulations are deterministic algorithms.
  }
In that case, one speaks of \emph{molecular dynamics} (MD) simulations.
The most common form of MD simply uses Hamiltonian dynamics.

Because Hamiltonian dynamics are consistent only with isolated systems, the ``environment'' of a non-isolated subsystems has to be modelled explicitly.
In that case, the large number of particles often render Hamiltonian dynamics unattractive for the purpose of simulations.
To deal with that problem, deterministic modifications to the equations of motion have been proposed.
Such equations are called \emph{thermostated} equations of motion, because the equations of motion contain the dynamics of an effective ``heat bath'' or \emph{thermostat}.

\subsection{Thermostated equations of motion}
\label{sec:thermostats}

In thermostated MD, one introduces \emph{auxiliary} degrees of freedom which are equipped with their own dynamics instead of stochastic noise.
Throughout this section, we follow the systematic approach to deterministic thermostats presented in Ref.~\cite{Samoletov_etal2007}.
We mostly focus on the thermostated equations that are used for equilibrium molecular dynamics (EMD).

The construction of EMD equations resembles the formulation of a Langevin equation:
One starts with a phenomenological equation of motion for the observable degrees of freedom under consideration.
In general, such equations include a mobility/drag term to model the dissipative effects of the environment.
Then, instead of adding a stochastic noise term, one chooses at least one of the following methods:
(i) Promoting the drag coefficient to a dynamical variable and specify an evolution rule for the latter or
(ii) adding a deterministic ``noise'' term which mimics the fluctuations caused by the environment.

The new dynamical variables introduced in that approach are the auxiliary degrees of freedom we mentioned above.
As in the stochastic case, the choice of these terms has to be consistent with the thermodynamic properties of the bath.
Again, this is achieved by demanding that the stationary distribution $\pden\tind{\infty}$ of the physical degrees of freedom becomes a Gibbs distribution $\pden\tsup{G}$ under equilibrium conditions.

\subsubsection{Nos\'e--Hoover thermostats}
As an example for the construction of thermostated equations of motion, we review Nos\'e--Hoover scheme.
Starting with Newton's equation for the macroscopic forces, we obtain a system of equations similar to \eqref{eq:langevin-overdamped}.
However, we do not add a noise term but rather promote the drag $\zeta$ to a dynamical variable $\widetilde \zeta(t)$ whose evolution is specified by a function~$g(\vec{q},\vec{p})$:
\begin{subequations}
\begin{align}
    \dot{\vec{q}} &= \frac{\vec p}{m},\\
    \dot{\vec{p}} &= -\nabla_{\vec q} V(\vec q) - \widetilde \zeta\frac{\vec p}{m},\\
    \dot{\widetilde \zeta}  &= g(\vec q,\vec p).
\end{align}
    \label{eq:NoseHoover}
  \end{subequations}
In order to find a thermodynamically consistent form of $g$, we refer to the steady state.
We know that in equilibrium \emph{equipartition} holds for the momenta, \ie 
\begin{equation}
  \eav{\sum \frac{\vec p^2}{m}}\tind{\infty}=d N T,
  \label{eq:Equipartition}
\end{equation}
where  $d$ is the dimension of physical space.
This motivates a choice of $g$ that accelerates the system if there is too little kinetic energy in the degrees of freedom and decelerates it otherwise.
Hence, one chooses
\begin{align}
  g(\vec q,\vec p) &= \frac{m}{Q}\left( \frac{\vec p^2}{m}-d N T \right) .
  \label{eq:NoseHooverG}
\end{align} 
where $Q=d N T \tmom^2$ is a constant related to the time scale $\tmom$ of the relaxation of the momenta.
This choice of $g$ leads to the following stationary distribution \cite{Samoletov_etal2007}:
\begin{equation}
  \pden^{\infty}\propto \exp\left( -\frac{1}{T}\left(V(\vec q)+ \frac{\vec p ^2}{2m} +F(\widetilde \zeta) \right) \right),
  \label{eq:NoseHooverStationary}
\end{equation}
where
\begin{align}
  F(\widetilde \zeta) &= \frac{Q}{2m^2}{\widetilde \zeta}^2 \label{eq:NoseHooverPhi}.
\end{align}

Momenta and coordinates are distributed according to a Maxwell--Boltzmann distribution, while the statistics of $\widetilde \zeta$ are determined by the ``potential'' $F(\widetilde \zeta)$.
The purely quadratic form of $F$ is generic for the following two reasons:
Firstly, it  leads to an expectation value $\eav{\widetilde \zeta}\tind{\infty} =0$ ensuring that the system is neither accelerated nor decelerated under equilibrium conditions.
Secondly, $\widetilde\zeta$ is used to model a force that is the outcome of many quasi-independent microscopic contributions.
Hence, the central limit theorem states that the distribution should be approximately Gaussian.
Note that $Q$ is the only free parameter.
The discussion of the Einstein FDR \eqref{eq:einstein-relation} relates the variance of the noise $2D$ to the phenomenological drag constant $\zeta$.
If $\zeta$ is given, this amounts to $Q \shouldbe d N T \frac{m^2}{\zeta^2}$.
Note that the drag defines a relaxation time scale $\tau_p := \frac{m}{\zeta}$.
Hence, without the reference to a phenomenological drag $\zeta$ we can understand $Q$ as defining the time scale $\tau_p$ via $Q=dNT\tau_p^2$.

Note that there are other motivations of the Nos\'e--Hoover equations of motion.
For instance, one can start with the stationary distribution \eqref{eq:NoseHooverStationary} and \emph{infer} $g=g(\vec q,\vec p,\tilde \zeta)$ by consistency.
For more details we refer the reader to Refs. \cite{Samoletov_etal2007} and \cite{Jepps+Rondoni2010}.

\subsubsection{Configurational thermostats}
Configurational thermostats are the deterministic analogue to the overdamped Langevin equation (\ref{eq:langevin-overdamped}).
It is assumed that the momenta have relaxed to their equilibrium values and the dynamics can be described by the equations
\begin{subequations} 
  \begin{align}
    \dot{\vec {q}} &= \widetilde \mu \,\nabla_{\vec q}V,
    \label{eq:OverdampedThermostat}\\
    \dot{\widetilde \mu} &= \frac{1}{Q_{\mu}}\sum\left[ \left( \nabla_{\vec q}V \right)^2- T (\nabla_{\vec q})^2 V \right].
    \label{eq:OverdampedG}
  \end{align}
  \label{eq:Overdamped}
\end{subequations}
Note that if one sets $\widetilde \mu = \widetilde \zeta^{-1}$, equation \eqref{eq:OverdampedThermostat} looks like the macroscopic part of \eqref{eq:langevin-overdamped}.
As above, the form of the dynamics of $\widetilde \mu$ is found by demanding a canonical form for the stationary distribution of the coordinates.
Whereas Eq.\,(\ref{eq:NoseHooverG}) ensured equipartition \eqref{eq:Equipartition}, Eq.\,\eqref{eq:OverdampedG} is consistent with Rugh's notion of a configurational temperature \cite{Rugh1997}.
As above, the constant $Q_{\mu}$ can be used to set a relaxation time scale $\tau_q$ for the positions $\vec q$.

The problem with Eqs.~\eqref{eq:Overdamped} is that the dynamics are not ergodic:
Mechanical equilibria $\nabla_{\vec q}V = 0$ act as attracting fixed points where the system comes to rest.
To restore ergodicity, further modifications of the equations of motion are required.
In one variant of  the so-called SDC scheme (after Samoletov, Dettmann and Chaplain \cite{Samoletov_etal2007}) sampling of phase space is enhanced by introducing another auxiliary dynamical variable $\vec \xi$.
This auxiliary variable is used to perturb the dynamics around mechanical equilibria, similar to what the stochastic noise term would do in the Langevin equation.
With the addition of $\vec \xi$ the equations of motion are
\begin{equation}
    \dot{\vec {q}} = \widetilde \mu \nabla_{\vec q}V + \vec \xi
  \label{eq:SDCthermostat}
\end{equation}
with the dynamics of $\widetilde \mu$ as in \eq{OverdampedG} and
\begin{equation}
  \dot{\vec \xi} = h(\vec \xi, \vec q).
\end{equation}
Consistency with the Boltzmann distribution requires that $h$ yields a dynamics that satisfies $\vec \xi \cdot \nabla_{\vec q} V = 0$.
There are essentially three different possibilities to satisfy this condition.
They correspond to differently constrained fluctuations around mechanical equilibria, which are described in detail in \cite{Samoletov_etal2007}.
Moreover, further generalizations can be found there and in Ref.~\cite{Samoletov_etal2010}.
The latter work focuses on a variant of the general scheme, the so-called Braga--Travis (BT) thermostat \cite{Braga+Travis2005}.
So far it is not known if or under which conditions the dynamics created by the SDC thermostating scheme are ergodic.
However, numerical simulations using deterministic SDC schemes indicate ergodicity at least in some variants \cite{Samoletov_etal2007}.

\subsubsection{Non-equilibrium molecular dynamics}
The schemes presented here are useful for thermostated equilibrium MD, \ie the modelling of non-isolated but closed systems.
However, there is also the need to model driven (open) non-equilibrium situations, which is the subject of \emph{non-equilibrium molecular dynamics} (NEMD).
To achieve non-equilibrium conditions one adds additional terms to the thermostated equations of motion.
For instance, one can couple the momentum equation to non-conservative forces, which constantly accelerates the system.
The thermostat then removes (\ie dissipates) the energy added in that way.
We will not go into the details of non-equilibrium molecular dynamics, but refer the reader to the literature, \eg Refs.\,\cite{Hoover1983,Evans+Morriss1990,Evans+Searles2002,Jepps+Rondoni2010}.

\subsection{Entropy and dissipation in deterministic dynamics}
Like for the case of stochastic dynamics, we would like to identify the entropy $\sent{\pden\tind{t}}$ of a density $\pden\tind{t}(x)$ with the system's entropy at a time $t$.
However, for deterministic dynamics of the form
\begin{align}
  \dot x = f(x)
  \label{eq:deterministic-dynamics}
\end{align}
there are certain problems with that interpretation.

Firstly, if we attempt to model systems in a thermodynamic environment using thermostated equations, the dynamical variables $x = (x\tsup{sys},\alpha)$ contain auxiliary degrees of freedom $\alpha$ in addition to the observable ones $x\tsup{sys}$.
Hence, the system entropy $\sysent\tind{t}$ should rather refer to the entropy of the marginalized ensemble $\pden\tind{t}\vert_{x\tsup{sys}}$.
However, we can only find the dynamical evolution of the joint distribution $\pden\tind{t}$.
For the EMD equations described above, the steady state density is a Gibbs distribution by construction.
However, we have no idea of its transient values.
For NEMD using modified EMD equations, one even loses such a physical interpretation for the steady-state distribution.

Secondly, even for Hamiltonian dynamics, where no auxiliary variables are present, the Shannon entropy of the ensemble $\pden$ shows the constant-entropy paradox \eqref{eq:const-entropy}.\footnote{ 
  We have mentioned that the constant-entropy paradox gets resolved if one considers coarse-grained degrees of freedom.
  A major part of this thesis is concerned with how this generalizes to non-Hamiltonian microscopic dynamics, \cf Chapters~\ref{chap:marksymdyn} and \ref{chap:information-st}.
}
Hence, there is no reason to expect that for more general dynamics such an interpretation would be useful.
We will see below that the situation actually gets much worse for non-equilibrium situations.

Although we cannot start with a definition of the system's entropy, one can still establish a connection to thermodynamics.
More precisely, in the remainder of this section we discuss the notion of dissipation, which we interpret as the entropy variation in the medium.
In particular, we consider its connection to the expansion and contraction of phase space volumes.

\subsubsection{Hamiltonian dynamics}
A first hint comes from Hamiltonian dynamics.
In the Boltzmann picture of statistical mechanics, thermodynamic entropy is related to the (logarithm of) accessible phase space volume $\Pi$:
\begin{align*}
  S\tsup{B} := \log \Pi
\end{align*}
The Gibbs-Liouville equation \eqref{eq:phase-space-expansion} ensures that phase space volume is conserved for Hamiltonian dynamics.
We have already mentioned that Hamilton's equation of motion are consistent with isolated systems.
Let us make this statement more precise:
The thermodynamic definition of an isolated system is that the entropy exchange with the outside world (\ie the dissipation) vanishes identically.
In statistical physics, this statement should hold for isolated systems independent of a particular point in time or any initial configuration.
In the discussion of \eq{phase-space-expansion} we have introduced the phase space expansion rate 
\begin{align*}
  \Lambda(x) := \vdiv f,
\end{align*}
which identically vanishes for Hamiltonian dynamics.
Thus, it has (at least some of) the desired properties of a quantity reflecting the entropy exchange with the environment.

\subsubsection{Thermostated dynamics}
Let us now look at the thermostated dynamics.
To be specific, consider the Nos\'e--Hoover equations \eqref{eq:NoseHoover}.
There we have 
\begin{align*}
  \Lambda(x) = \nabla_{\vec p}\cdot {\dot{\vec p}} =-\frac{\tilde \zeta}{m}.
\end{align*}
In general, this quantity takes both positive and negative values.
The same holds for the entropy (or heat) exchange of a closed system with its environment.
However, on average there should be no net exchange of heat with the environment when the system has reached equilibrium.

For thermostats that fall into the Nos\'e--Hoover (or the more general SDC) scheme outlined above, this is exactly the case.
Remember that we constructed the dynamics such that $\tilde\zeta$ has a Gaussian distribution around zero in the stationary state \eqref{eq:NoseHooverStationary}.
Thus, the steady-state mean of $\tilde \zeta$ (and hence the average phase space contraction rate) vanishes: 
\begin{align}
  \eav{\Lambda}\tind{\infty} = \int_\pspace\Lambda\pden\tind{\infty} \df x = 0.
  \label{eq:conservative-system}
\end{align}

\subsubsection{Non-equilibrium}
What about non-equilibrium situations?
Firstly, we mention that a generic deterministic dynamics \eqref{eq:deterministic-dynamics} does not necessarily admit an invariant density $\pden\tind{\infty}$.
Rather it might admit an invariant \emph{measure} $\gls{symb:probability-measure}\tind{\infty}$.
A detailed discussion of measure theory will be the subject of chapter \ref{chap:marksymdyn}.
Here, we just mention that the ensemble average $\eav{\Lambda}\tind{\infty}$ is still well-defined.

Sometimes, instead of ensemble averages one is interested in time averages.
For their definition, note that a differential equation $\partial_t x = f(x)$ formally defines a \emph{flow} $\Psi(\cdot,t) = \Psi\rlind{t}(\cdot)$.
It propagates an initial condition $x_0$ to the solution function $x_t = x(t)$ of the differential equation \eqref{eq:deterministic-dynamics}:
\begin{align*}
  \Psi\colon\pspace\times \timdom &\to \pspace,\\
  (x_0,t) &\mapsto x_t = \Psi\rlind{t}(x_0).
\end{align*}
Additionally, the flow obeys a semi-group structure, \ie $\Psi(t_2,\Psi(t_1,x)) = \Psi(t_1+t_2,x)$ for all $t_1,t_2$ in its time domain \gls{symb:time-domain}.
In the following we assume that the flow is invertible, \ie $\timdom = \reals$.
The time average $\bar\obs\tind{t}\rlind{\tau}$ of an observable $\obs(x)$ is defined as
\begin{align}
  \bar\obs\tind{t}\rlind{\tau}(x) = \frac{1}{\tau}\int_t^{t+\tau} \obs\left(\Psi\rlind{s}(x)\right) \df s.
  \label{eq:time-average}
\end{align}
%
We are interested in the time-averaged phase-space expansion rate:
\begin{align}
  \bar\Lambda\tind{t}\rlind{\tau}(x) := \frac{1}{\tau} \int_{t}^{t+\tau} \Lambda(\Psi\rlind{s}(x)) \df s.
  \label{eq:time-av-exp-rate}
\end{align}
Henceforth, we only consider situations where the limit
\begin{align}
  \bar\Lambda\tind{\infty} := \lim_{\tau \to \infty} \bar\Lambda\tind{t}\rlind{\tau}
  \label{eq:exp-time-av}
\end{align}
exists.
For autonomous systems that limit is independent of $t$.
Further, we are interested in globally attracting systems, where this limit is also independent of $x$.
By definition it is the sum of all Lyapunov exponents (for a detailed discussion of the latter \cf \cite{Guckenheimer+Holmes1983}).
For systems defined on compact phase spaces, this quantity is always non-positive, \ie $\bar\Lambda\tind{\infty}\leq 0$.
For a \emph{physical} dynamics the time average agrees with the ensemble average using a physical%
\footnote{
  A physical measure for an arbitrary dynamics is formally defined in Refs.~\cite{Young2002,Blank+Bunimovich2003}.
  Physicists usually call a dynamics ``ergodic'' when it is ``mixing'' in the sense that time and ensemble averages of physical observables agree.
  This intuitive notion agrees with the definition of a physical measure in mathematics.
  There, the term ``ergodic'' is already reserved for a slightly more general concept, \cf Section~\ref{sec:meas-top-ds}.
}
invariant measure $\invms$.
\begin{align*}
   \bar\Lambda\tind{\infty}  = \eav{\Lambda}\tind{\infty} \leq 0.
\end{align*}
We have already seen that the \emph{equality} $\eav{\Lambda}\tind{\infty}=0$ holds in the case of the equilibrium MD equations introduced above.
The generic case is
\begin{align}
   \bar\Lambda\tind{\infty}  = \eav{\Lambda}\tind{\infty} < 0.
   \label{eq:dissipative}
\end{align}
In particular, this holds for the equations of motions used for non-equilibrium molecular dynamics (NEMD).
With a reference to the standard literature \cite{Evans+Searles2002,Jepps+Rondoni2010}, we state that in that case $\eav{\Lambda}\tind{\infty}$ obtains the negative value of the thermodynamic dissipation rate $\Sigma$ divided by temperature $T$.
The latter is defined as
\begin{align}
  \Sigma = \sum_{\nu} J_\nu A_\nu
  \label{eq:dissipation}
\end{align}
where $J_\nu$ denotes the $\nu$th thermodynamic current which is driven by its corresponding \emph{conjugate thermodynamic force} or \emph{affinity} $A_\nu$, \cf Sec.~\ref{sec:discrete-st}.

The interpretation of $\Lambda(x)$ as the dissipation rate or entropy change in the environment is based on physical models for (NE)MD.
Regardless of any physical-thermodynamic interpretation, we characterize arbitrary deterministic dynamics by their (average) phase space expansion.
In the following we will refer to equations of motion $\dot x = f(x)$ that obey equations \eqref{eq:uniformly-conservative-system}, \eqref{eq:conservative-system} and \eqref{eq:dissipative} as \emph{uniformly conservative, conservative and dissipative systems}, respectively.

\subsection{Stroboscopic maps and time-discrete dynamics}
\label{sec:time-discrete}
Finally, we consider the case of discrete-time dynamical systems.
A discrete dynamical system can be obtained from a continuous flow $\gls{symb:flow}(x,t)$  as a \emph{stroboscopic map} $\gls{symb:strobo-map}(x) := \Psi(x,\Delta t)$ with a \emph{stroboscopic time interval} $\Delta t$.
Under the assumption that $\Psi(\cdot,t)$ is defined for all $t \in \gls{symb:reals}$, the map $\Phi(x)$ is invertible and maps $\pspace$ onto $\pspace$.
For $\tau \in \gls{symb:integers}$, we denote by  $\Phi\rlind{\tau}(x) := \Psi(x,\tau \Delta t)$.
Note that for positive and negative integers $\tau$, $\Phi\rlind{\tau}$ is the $\abs{\tau}$-fold iterate of $\Phi$ and $\Phi^{-1}$, respectively.

For continuous-time flows, we have classified systems by their phase space contraction rate $\bar\Lambda\tind{\infty}$.
We would like to do the same for discrete dynamics.
To that end consider the \emph{Jacobian determinant} $\jac{\tau}{x}$ of a $\tau$-times iterated map $\Phi$,
\begin{align}
  \jac{\tau}{x} := \abs{\det\Df \Phi\rlind{\tau}(x)},
  \label{eq:defn-jacobian}
\end{align}
where $\Df f(x)$ denotes the \emph{Jacobian matrix} of a differentiable function $f\colon \pspace \to \pspace$ at a point $x$.
If $\Phi(x)$ is a stroboscopic map obtained from an arbitrary dynamics \eqref{eq:deterministic-dynamics} with flow $\Psi(x,t)$ it holds that \cite{Ruelle1999}
\begin{align}
  \bar{\Lambda}\tind{\infty} = \int_\pspace\pden\tind{\infty} \vdiv f \df x = \int_\pspace\pden\tind{\infty} \log \ja\rlind{1} \df x.
  \label{eq:log-jac-phase}
\end{align}
If $\Phi$ is not obtained as a stroboscopic map, we lack the notion of the divergence of a vector field $\vdiv{f}$ which defines the dynamics.
However, the right hand side of \eq{log-jac-phase} is still well-defined.
The logarithm of the Jacobian determinant
\begin{align}
  \Lambda\rlind{1}\tind{t}(x):= \jac{1}{\Phi\rlind{t-1} s}
  \label{eq:phase-space-expansion-discrete}
\end{align}
describes the phase space expansion (in unit time) that a small volume around a point $x$ encounters at time $t$. 
Hence, it is natural to classify maps $\Phi$ with reference to $ \log \ja\rlind{1}$:
We say a map $\Phi$ is uniformly conservative, if $ \Lambda\rlind{1}\tind{t}$ vanishes identically.
Similarly, we call it conservative or dissipative if the steady-state average $\eav{ \Lambda\rlind{1}}\tind{\infty}$ vanishes or is negative, respectively.
Note that the log-Jacobian is also used in the definition of the Lyapunov exponents of discrete systems.

Finally, let us consider the variation of the Shannon entropy of the ensemble for a time-continuous dynamics in the time-interval $\Delta t$.
It is straightforward to realize that (\cf~\eq{pden-lambda} or \cite{Ruelle1999})
\begin{align}
  \Delta\tind{t}\rlind{1}H :=\left(\sent{\pden\tind{t}} - \sent{\pden\tind{t-1}}\right) =\Delta t \cdot \int_{\pspace}\pden\tind{t-1}\log \ja\rlind{1} \df x.
  \label{eq:change-differential-entropy}
\end{align}
Thus, in the steady state we have 
\begin{align}
  \Delta\tind{\infty}\rlind{1}H := \lim_{t\to\infty} \Delta\tind{t}\rlind{1}\sent{\pden\tind{t}} ={\Delta t} \int_{\pspace}\pden\tind{\infty}\log \ja\rlind{1} \df x ={\Delta t} \eav{\Lambda}\tind{\infty}.
  \label{eq:infinite}
\end{align}
Hence, $\Delta\tind{\infty}\rlind{1}H $ describes the asymptotic change of the ensemble entropy $\sent{\pden\tind{t}}$ in unit time.
Without loss of generality, from now on we fix the stroboscopic interval as the unit time, \ie $\Delta t=1.$

For conservative systems (like Hamiltonian or thermostated equilibrium dynamics) Eq.~\eqref{eq:infinite} evaluates to zero, \cf \eq{conservative-system}. 
For Hamiltonian dynamics, the entropy $\sent{\pden\tind{t}}$ retains its original value for \emph{all times}.
In contrast, for the EMD dynamics we have discussed above, the entropy of the ensemble $\sent{\pden\tind{t}}$ necessarily converges to the Gibbs entropy $\ent\tsup{G}$.
By definition, the latter is a (constrained) \emph{maximum entropy distribution}.

For dissipative non-equilibrium systems we have a completely different situation:
From \eq{infinite} we know that in the $t\to \infty$ limit the change per unit time of $\sent{\pden\tind{t}}$ is \emph{negative}.
Hence, $\lim_{t\to\infty} \sent{\pden\tind{t}} \to -\infty$.
Rather than having a maximum entropy distribution, the entropy of the ensemble is unbounded from below.
The reason for this is that the limit distribution is not a \emph{density} any more, but something which has a fractal distribution on a (generically) fractal support \cite{Ruelle1999}.
This is another reason, why outside of equilibrium one must not identify the entropy of the phase space ensemble with the system's entropy.
It is further at variance with the requirement that its value must be constant in any (non-equilibrium) steady state.

\subsection{Reversibility and deterministic fluctuation relations}
\label{sec:reversibility}
An important aspect of the Nos\'e--Hoover and many other thermostated (NE)MD models is their reversibility \cite{Jepps+Rondoni2010}.
For deterministic invertible dynamical systems, reversibility means that one can reverse the arrow of time by applying a time-reversal operator $\gls{symb:involution}\colon \pspace \to \pspace$ to its microstates.

For MD, the phase space $\pspace$ contains elements $x = (\vec q, \vec p, \alpha)$ which summarize coordinates $\vec q$, momenta $\vec p$ (which might be suppressed in underdamped dynamics) as well as the additional auxiliary variables
\begin{align}
  \alpha = (\widetilde \zeta, \vec \xi, \ldots ),
  \label{eq:AuxiliaryVariables}
\end{align}
which represent the degrees of freedom of the thermostat.

Let $\lambda$ denote the usual $d$-dimensional volume (Lebesgue measure) on $\reals^d$.
In the following, we are interested in measure-preserving time reversal involutions $\invo$.
Formally, we call a flow $\Phi\rlind{t} = \Psi(\cdot,t)$ (or an iterated map $\Phi\rlind{t} = \Phi\circ\cdots\circ\Phi$) \emph{reversible} if and only if there is a mapping $\invo$ such that
\begin{subequations}
\begin{align}
  \invo\circ\invo &= \text{id},  &\text{(involution)}&, \label{eq:defnInvolution}\\
  \lambda(\invo^{-1}A) &= \lambda(A), &\text{(measure-preserving)}& \label{eq:defnMeasurePreserving},\\
  \Phi^{(-t)} x &= \left(\invo \circ \Phi^{(t)}\circ \invo\right) x,~\forall t\in\reals~ (\text{or } \integers)  & \text{(time reversal)}&.
  \label{eq:defnReversibility}
\end{align}  
  \label{eq:reversibility}
\end{subequations}
In the present context of thermostated MD, the time-reversal operator that fulfils the above properties is given as  $\invo (\vec q, \vec p, \alpha) = (\vec q,-\vec p,-\alpha)$.
In Section \ref{sec:nmbm} we discuss a more abstract reversible dynamics in discrete time.

Reversibility with a measure-preserving involution is the central requirement for the existence of \emph{fluctuation relations} (FR) for deterministic dynamical systems.
Similar to the stochastic FR discussed in Section~\ref{sec:stochastic-fr}, deterministic FR are statements about the probability to observe a certain value of entropy production in an interval $[t,t+\tau]$.
Instead of a trajectory-dependent random variable for a stochastic processes, one considers the probability of observing a value $a$ for a phase space observable $\bar{\disfun}\tind{t}\rlind{\tau}$.
Again, the FR compares the probability of finding $\bar{\disfun}\tind{t}\rlind{\tau} = a$ with the probability of finding $-a$:
\begin{align}
  \frac{\prob[\bar{\disfun}\tind{t}\rlind{\tau} = a] }{\prob[\bar{\disfun}\tind{t}\rlind{\tau} = -a] } = \exp{a }.
  \label{eq:fluctuation-relation-deterministic}
\end{align}
The most famous examples of deterministic FR are the transient FR by Evans and Searles \cite{Evans+Searles2002} and the steady-state FR by Gallavotti and Cohen~\cite{Gallavotti+Cohen1995}.
A recent result ensures that deterministic FR follow generally from an abstract fluctuation theorem \cite{Wojtkowski2009}.
Like the stochastic FR \eqref{eq:stochastic-fr}, the deterministic formulation \eqref{eq:fluctuation-relation-deterministic} is a detailed probabilistic version of the second law of thermodynamics.

\subsubsection{The dissipation function}
To provide a more concrete example, we discuss the Evans--Searles dissipation function $\disfun\tind{t}\rlind{\tau}$ for reversible dynamics \eqref{eq:deterministic-dynamics}, \cf Refs.~\cite{Evans+Searles2002,Searles_etal2007}.
In order to construct this function, we define a logarithmic ratio of two probability densities $\pden,\pden'$ on phase space $\pspace$:
\begin{align}
  R[x;\pden,\pden'] := \log\frac{\pden(x)}{\pden'(x)}.
  \label{eq:R}
\end{align}
The average dissipation rate is given by the \emph{time-averaged dissipation function}
\begin{align}
  \bar{\disfun}\tind{t}\rlind{\tau}[x;\pden] := \frac {1}{\tau}\left( R[x;\pden,\Theta \pden] \right) - \bar\Lambda\tind{t}\rlind{\tau}(x),
  \label{eq:dissipation-function}
\end{align}
where $\Theta\pden := \pden \circ \invo \circ \Phi\rlind{\tau}$.
The first term  $R[x;\pden,\Theta \pden]$ compares a probability density $\pden$ at a point $x$ with its value at the conjugate point $\Theta x := \invo\Phi\rlind{\tau}x$.
The orbit of the conjugate point $\Theta x$ is the so-called \emph{conjugate} or \emph{anti-orbit} to the orbit of a point $x$.
Note that for $0\leq t \leq \tau$ we have $\Phi\rlind{t} x' = \invo\Phi\rlind{\tau-t)} x$.
Thus, the anti-orbit consists of the time-reversed points of the corresponding orbit, which are visited in reverse direction, \cf Fig.~\ref{fig:anti-orbit}.
The second term is just the negative value of the time-averaged phase-space expansion in an interval $[t,t+\tau]$, see \eq{time-av-exp-rate}.

Depending on the choices of $\pden$, one can obtain several of the above mentioned fluctuation relations.
The original Evans-Searles FR was obtained for $\bar{\disfun}_\tau := \bar{\disfun}\tind{0}\rlind{\tau}[x;\pden\tind{0}]$, where $\pden\tind{0}(x) = \pden\tind{0}(\invo x)$ is a time-reversal symmetric initial condition~\cite{Evans+Searles2002}:
\begin{align}
  \frac{\prob[\bar{\disfun}\tind{\tau} = a] }{\prob[\bar{\disfun}\tind{\tau} = -a] } = \exp(a\tau).
  \label{eq:evans-searlsfr}
\end{align}
In addition to that, Ref. \cite{Searles_etal2007} describes other fluctuation relations that can be obtained by choosing other probability distributions for $R[x;\pden,\pden']$.
\begin{figure}[t]
  \centering
  \includegraphics{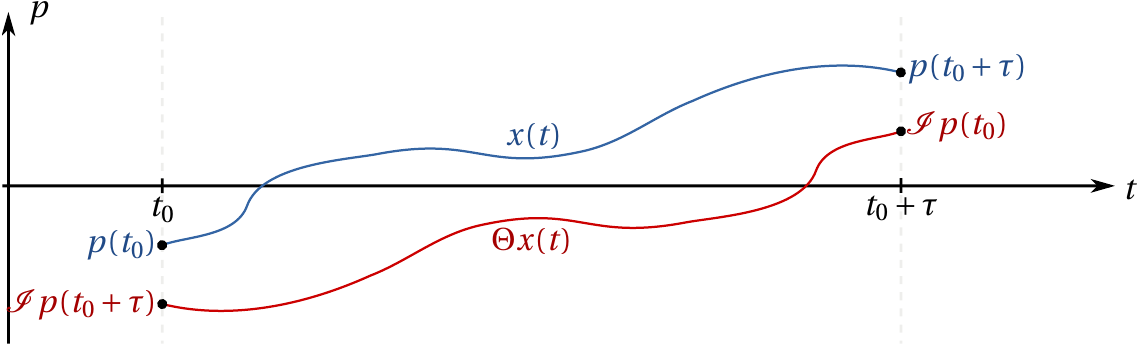}
  \caption{The anti-orbit $\Theta x(t)$ (red) to the orbit $x(t)$ (blue) in the time interval $[t_0,t_0+\tau]$.
  We only display the anti-symmetric (momentum) coordinate $p$.}
  \label{fig:anti-orbit}
\end{figure}

\section{Measurable dynamical systems}
\label{sec:measurable-ds}
In the previous section, we investigated how phase-space contraction for deterministic dynamics is interpreted as dissipation.
The main motivation for this connection to (thermodynamic) entropy came from the equations used in (NE)MD.
Independent from physical or thermodynamic interpretations, a purely information-theoretical connection between deterministic dynamical systems and entropy production has been worked out in the last fifty years.
Early investigations started with the works of Kolmogorov, who was interested in a way to characterize the \emph{complexity} of dynamical systems \cite{Kolmogorov1958}.
Together with his student Y.\,Sinai, the concept was further developed and has been summarized by Sinai himself in Ref.\,\cite{Sinai2009}.
Some similarities of that theory with thermodynamics have been discussed by D.\,Ruelle \cite{Ruelle2004}.
In the following, we only give a brief glimpse into the basic ideas.

\subsection{Mathematical prerequisites}
We  briefly review the basic concepts of topology and measure theory.
A detailed exposition is (freely) available in T.\,Tao's monograph \cite{Tao2011}.

Let $X$ be a set.
The set $\powset(X)$ containing all subsets of $X$ is called the \emph{power set} of $X$.
In the following, two kinds of subsets of  $\powset(X)$ will be important:
A \emph{$\sigma$-algebra} on a set $X$ is a family of subsets $\alg \subset \powset(X)$ which contains the \emph{empty set} $\emptyset$ and the entire set $X$ and is closed under the formation of countable unions and intersections.
For any family of sets $\fsets \subset \powset(X)$, the family $\sigma(\fsets)$ is the smallest $\sigma$-algebra that contains $\fsets$.
Elements of a $\sigma$-algebra are called \emph{measurable sets}.
A \emph{measurable space} $(X,\alg)$ consists of a set and a $\sigma$-algebra on that set.
We say that a map $\ms\colon\alg \to [0,\infty)$ is additive, if its value on the mutual union of a finite family $\fsets=\ifam{A_i}_i$ of disjoint subsets $A_i$ is equal to the sum $\sum_i\ms(A_i)$.
Further, we call the map $\sigma$-additive if the above statement also holds for countable families.
A \emph{measure} is a map $\ms\colon\alg \to [0,\infty)$  which is both additive and \emph{$\sigma$-additive}.
A \emph{probability measure} $\pms$ is a measure that obeys $\pms(X)=1$.

For any map $f \colon X_1 \to X_2$ the pre-image map $f^{-1}\colon \powset(X_2) \to \powset(X_1)$ is defined as
\begin{align*}
  f^{-1}(A_2) = \set{x \in X_1 \middle \vert f(x) \in A_2}.
\end{align*}
A \emph{measurable function} $f$ between two measurable spaces $(X_1,\alg_1)$ and $(X_2,\alg_2)$ is a function such that the pre-image of any measurable set is measurable, \ie $f^{-1}(A_2) \in \alg_1$, $\forall A_2 \in \alg_2$.
A random variable is a measurable function between a probability space $(X_1,\alg_1,\pms)$ and a measurable space $(X_2,\alg_2)$.
For a measurable function $f\colon (X_1,\alg_1)\to (X_2,\alg_2)$ we define the pushforward of a measure $\ms$ on $X_1$ as $(f)_*(\ms) := \ms\circ f^{-1}$, where $\circ$ denotes composition.

A family of subsets $\topo \subset \powset(X)$ containing  the \emph{empty set} \gls{symb:empty-set} and the entire set $X$, which is also closed under the formation of arbitrary unions and finite intersections is called a \emph{topology}.
The tuple $(X,\topo)$ is called a \emph{topological space} and elements $\pset \in \topo$ are called \emph{open sets}.
A mapping $f$ between two topological spaces $(X_1,\topo_1)$ and $(X_2,\topo_2)$ is called \emph{continuous}, if the pre-image of an open set is open, \ie $f^{-1}(\pset_2) \in \topo_1$, $\forall \pset_2 \in \topo_2$.
Often, we will assume that $(X,d)$ is a metric space with some metric $d$.
In that case, the open sets $\pset \in \topo$ are determined by the metric.
Similarly to the notion of a generated $\sigma$-algebra, the topology $\topo(\fsets)$ \emph{generated} by a family of sets $\fsets \subset \powset(X)$ is the smallest topology that contains $\fsets$.
The $\sigma$-algebra $\sigma(\topo)$ generated by the open sets is called the \emph{Borel $\sigma$-algebra}.

If not indicated differently, we assume that the phase space $\pspace$ is a metric space equipped with a topology $\ptopo$ and the corresponding Borel $\sigma$-algebra $\palg = \sigma(\ptopo)$.
A measure on such a space is called a \emph{Borel measure}.

Let $\pms$ and $\ms$ be measures for a measurable space $(\pspace,\palg)$.
We say that a measure $\pms$ is \emph{absolutely continuous} with respect to the measure $\ms$, if $\ms(\pset) =0 \Rightarrow \pms(\pset) =0$ for all $\pset \in \palg$ and write $\pms \ll \ms$.
In that case, $\pms$ has a \emph{density} $\den$ with respect to $\ms$, \ie
\begin{align*}
  \int_\pset g \df \pms = \int_\pset \den g \df \ms
\end{align*}
holds for all measurable functions $g$ and all sets $\pset \in \palg$.
The function $\den =: \frac{\df\pms}{\df\ms}$ is called the \emph{Radon--Nikodym derivative} of $\pms$ with respect to $\ms$.

If $\pms$ is a probability measure defined on the Borel sets and $\ms = \lms$ is the Lebesgue measure, the Radon--Nikodym derivative gives the \emph{probability density} $\pden := \frac{\df\pms}{\df\lms}$ of the probability measure $\pms$.

\subsection{Measurable and topological dynamical systems}
\label{sec:meas-top-ds}
In this subsection we consider \emph{$C^1$-diffeomorphisms}, which are differentiable maps with a differentiable inverse.\footnote{
We can relax that to \emph{almost-everywhere} differentiable maps with almost-everywhere differentiable inverse.
All that we need is that the log-Jacobian is well-defined almost everywhere.
For instance, a Lipshitz continuous functions with a Lipshitz-continuous inverse provides an example.}

Throughout the rest of this section, let $(\pspace,\palg)$ be a measurable space and $\Phi$ a measurable $C^1$-diffeomorphism.
The triple $(\pspace,\palg,\Phi)$ is called a \emph{measurable dynamical system} on \emph{phase space} $(\pspace,\palg)$.
Similarly, if $(\pspace,\topo)$ is a topological space and $\Phi$ is a continuous map, we call the triple $(\pspace,\topo,\Phi)$ a \emph{topological dynamical system}.

Usually,  we think of $\pspace$ as a separable metric space with $\palg$ denoting the Borel sets.
In that case, any measurable function is also continuous and a measurable dynamical system is also a topological one.
If a probability measure $\pms$ is given, we call $(\pspace,\palg,\Phi,\pms)$ a dynamical system with measure $\pms$.

Now we can formally define an \emph{invariant measure} $\invms$ as a (probability) measure that is a fixed point of the push-forward, \ie,
  \begin{align*}
    \Phi_*\invms \equiv \invms \circ \Phi^{-1} = \invms,
  \end{align*}
In that case we call $(\pspace,\palg,\Phi,\invms)$ a \emph{measure-preserving dynamical system}.  

A special kind of invariant measure is an \emph{ergodic} measure.
We call a set $\pset$ to be $\Phi$-invariant, if $\pset = \Phi^{-1}(\pset)$.
In mathematics, one says that a measure $\invms$ is ergodic with respect to $\Phi$ (or $\Phi$ is ergodic with respect to $\invms$) if for any invariant $\pset\in\palg$
\begin{align}
  \invms(\pset) = 0 \quad \text{or}\quad \invms(\pset) =1.
  \label{eq:defn-ergodic-measure}
\end{align}
For ergodic measures, ergodic theorems like the one of Birkhoff then ensure that properly defined asymptotic time-averages agree with averages taken with respect to $\invms$.
However, the mathematical definition of ergodicity does \emph{not} imply that the orbit of \emph{all} points densely covers the \emph{entire} phase space, as is sometimes assumed in physics.

The push-forward operator $\Phi_*$ defines an evolution rule for measures.
We will mostly be interested in the evolution of probability measures $\pms$ which are absolutely continuous with respect to a \emph{reference measure} $\ms$.
Most of the time  we choose $\ms = \lms$, where $\lms$ denotes the translation invariant (Lebesgue) measure on $(\pspace,\borel)$.
In that case, $\pms$ has a probability density $\pden\colon\pspace \to [0,\infty)$.

Because $\Phi$ is a measurable diffeomorphism, the $t$-times push-forwarded measure
\begin{align*}
  \pms\tind{t} := \Phi^t_*\pms
\end{align*}
is \emph{equivalent} to $\pms$, \ie $\pms$ and $\pms\tind{t}$ are mutually absolutely continuous.
One also says that $\Phi$ is \emph{non-singular with respect to $\pms$}.
In Refs.~\cite{Evans+Searles2002,Searles_etal2007} the authors call a density $\pden$ of $\pms$ to be \emph{ergodically consistent} with $\Phi$ if (i) $\Phi$ is non-singular with non-singular inverse and (ii) time-reversal can be represented by a measure-preserving involution \eqref{eq:reversibility}.
A non-singular $\Phi$ yields a sequence of measures $\set{\pms\tind{t}}$ with densities $\set{\pden\tind{t}}$ that evolve according to
\begin{align}
  \pden\tind{t}= \frac{\pden\tind{0}\circ \Phi^{-t}}{\ja\rlind{t}\circ \Phi^{-t}},
  \label{eq:density-evolution}
\end{align}
where $\ja\rlind{t}$ is the Jacobian determinant \eqref{eq:defn-jacobian}.

This provides the following perspective on observables and their averages:
In the present setting, an observable $\obs\colon \pspace \to \reals$ is a measurable function that assigns real values to points in phase space.
Usually, we interpret an observable as the outcome of a measurement procedure.
The (Lebesgue) integral over the full phase space is the \emph{ensemble average} of an observable $\obs$:
\begin{align}
  \eav{\obs}\tind{t} := \int_\pspace \obs \df{\pms\tind{t}} = \int_\pspace \obs \pden\tind{t} \df x.
    \label{eq:ensemble-average}
\end{align}

The properties of the Jacobian
\begin{subequations}
  \begin{align}
\ja\rlind{t} &= \frac{1}{\ja\rlind{-t}\circ \Phi^t}\label{eq:inverse-jacobian},\\
     &= \prod_{k=1}^t\left( \ja\circ\Phi^{t-k} \right) \equiv\prod_{k=1}^t\left( \ja\circ\Phi^{k} \right)  \label{eq:iterated-jacobian},
  \end{align} 
  \label{eq:jacobian-rules}
\end{subequations}
together with the usual transformation rules of the integral, yield for the ensemble average at time $t$:
  \begin{align}
    \eav{\obs}\tind{t} = \int_\pspace \left( \obs\circ\Phi^t \right)\pden\tind{0}\df x = \int_{\pspace}\obs \pden\tind{t}\df x.
    \label{eq:integral-over-pushforward density}
  \end{align}
  Hence, there are two ways to interpret the average $\eav{\obs}\tind{t}$:
In the first, one can think of a time-dependent observable $\obs\tind{t} := \obs \circ \Phi^t$ which is averaged with the initial density $\pden\tind{0}$.
In the second one, the observable does not depend on time whereas $\pden\tind{t}$ evolves according to  \eq{density-evolution}.
A similar duality exists in quantum mechanics:
The first interpretation is called ``Heisenberg picture'', whereas the latter is named after Schr\"odinger.

\subsection{Topological and measure-theoretic entropy}
\label{sec:topent-ksent}
In this subsection, we define entropy for both topological and measurable dynamical systems.
Although these expressions are called ``entropies'', they rather resemble asymptotic time-averages.
Hence, they are better understood as rates of change of entropy than as entropy itself.
This is a consequence of the fact that these entropies were designed to quantify the (asymptotic) \emph{complexity} of the dynamics, rather than that of an ensemble at a specific point in time.

The \emph{topological entropy} is the appropriate notion of the complexity of a topological dynamical system $(\pspace,\topo,\Phi)$.
It provides an upper bound to the \emph{metric entropy} defined for the measurable dynamical system $(\pspace,\sigma(\topo),\Phi)$.
Here, we only give a brief overview.
For a more rigorous discussion and explicit proofs we refer the reader to chapters 4--6 of Ref.~\cite{Jost2006}.

Both the definition of the topological and the metric entropy start with the notion of a (minimal) cover of phase space.
The elements of that cover can be understood as an approximate description of the position of a point $x\in\pspace$.
For instance, they may correspond to a certain value of an observable representing a measurement on the system.\footnote{
  We will make that notion more precise in Chapter~\ref{chap:marksymdyn}.
}
If we measure the system initially and after each of $\tau$ iterations, we obtain a time series $\traj \omega\rlind{\tau}$.
The topological entropy specifies the exponential growth rate of the number of distinguishable time series $\traj \omega\rlind{\tau}$ with $\tau$.

Hence, in a sense it specifies the (rate of) additional \emph{information} we obtain about the topological structure of the flow when we observe longer and longer time series.
The measure-theoretic entropy additionally takes a (non-uniform) probability for time series into account.

\subsubsection{Covers and partitions}
Henceforth it will be useful to consider \emph{indexed families of sets} $\fsets=\ifam{\gls{symb:cell}}_{\omega \in \alphabet}$.
An element $\cell_\omega \subset \pspace$ is indexed by an entry $\omega$, which is an element of the index set $\alphabet$.
If we do not care about the indexing, we consider $\fsets\subset\powset(\pspace)$ as a subset of the power set.

A finite (indexed) family $\fsets = \ifam{\cell_\omega}_{\omega\in\alphabet}\subset\powset\left( \pspace \right)$ is called a \emph{cover} of $\pspace$ if $\bigcup_{\omega}\cell_\omega =\pspace$.
For a topological space, a cover $\fsets \subset \topo$ consisting of open sets is called an \emph{open} or \emph{topological cover}.
A cover $\fsets$ is called \emph{minimal} if it does not contain any true subset $\fsets' \subsetneq \fsets$ which already is a cover.
In the following, we will always assume that $\fsets$ is minimal.
A cover consisting of disjoint sets is called a \emph{partition}.
Note that for a connected phase space $\pspace$, a partition can never be an open cover.\footnote{
  Later we will define so-called topological partitions, which are neither partitions nor covers, but turn out to be a useful generalizations.
  Topological partitions have been introduced by R.\,Adler \cite{Adler1998}.}

For two families of subsets $\fsets$ and $\fsets'$ we define their \emph{mutual refinement} or \emph{join} as the set of all mutual intersections of their elements:
\begin{align}
  \fsets \vee \fsets' := \set{\cell \cap \cell' \,\vert\, \cell\in \fsets, \cell' \in \fsets'}
  \label{eq:refinement}
\end{align}
Note that the join of two partitions again is a partition.
Further, the join of two open covers is an open cover, because a topology is closed under finite intersections.

The pre-image of an indexed family of subsets $\fsets = \ifam{\cell_\omega}_{\omega \in \alphabet}$ under the map $\Phi$ is defined as
\begin{align}
  \Phi^{-1} (\fsets) := \set{\Phi^{-1} \cell_\omega\,\vert\, \omega \in \alphabet}
  \label{eq:pre-image-fsets}
\end{align}
Hence, if $\Phi$ is continuous, the pre-image of a topological cover again is a topological cover.
Similarly, if $\Phi$ is measurable, the pre-image of a family of measurable subsets again is a family of measurable subsets.

We define the \emph{topological entropy of a minimal cover} $\fsets$ as the logarithm of the number of its elements:
\begin{align*}
  \Topent(\fsets) := \log\abs{\fsets}
\end{align*}
The dynamics creates a sequence of refined covers $\ifam{\bigvee_{n=1}^t \Phi^{-t}(\fsets)}_{t\in\naturals}$.
For a topological dynamical system $(\pspace,\topo,\Phi)$ and a cover $\fsets$ we define the \emph{topological entropy} of the asymptotic dynamical refinement as
\begin{align*}
  \Topent(\fsets,\Phi) := \lim_{t \to \infty}\frac 1 t \Topent\left(\bigvee_{n=1}^t \Phi^{-t} (\fsets)\right).
\end{align*}
As we have mentioned above, this quantity is rather an asymptotic entropy \emph{rate} than an entropy.
The \emph{topological entropy of a topological dynamical system} $(\pspace,\topo,\Phi)$ is the maximum of that quantity over all minimal covers:
\begin{align}
  \topent(\Phi) := \sup_{\fsets\text{ is a minimal cover}}\Topent(\fsets,\Phi).
  \label{eq:topological-entropy}
\end{align} 

The definition of the measure-theoretic \emph{Kolmogorov--Sinai} entropy (KS-entropy) proceeds in an analogue way, if we replace ``cover'' by ``partition'' and ``continuous'' by ``measurable''.
To that end let $(\pspace,\alg,\invms,\Phi)$ be a measure-preserving dynamical system and $\gls{symb:partition}=\ifam{\cell_\omega}_{\omega \in \alphabet}$ be a finite partition.
The measure-theoretic entropy of the partition is
\begin{align*}
  \KSent_{\invms}(\parti) = -\sum_{\cell \in \parti}\invms(\cell)\log\left(\invms(\cell)\right).
\end{align*}
With that, the \emph{metric entropy of $(\pspace,\alg,\invms,\Phi)$ with respect to $\parti$} reads
\begin{align*}
  \KSent_{\invms}(\parti,\Phi) := \lim_{t \to \infty}\frac 1 t \KSent_{\invms}\left(\bigvee_{n=1}^t \Phi^{-t} \parti\right).
\end{align*}
Finally, the \emph{Kolmogorov--Sinai entropy} of $(\pspace,\Phi,\invms,\Phi)$ is obtained as the supremum over all partitions:
\begin{align}
  \ksent_{\invms}(\Phi) := \sup_{\parti \text{ is a partition}}\KSent_{\invms}(\parti,\Phi).
  \label{eq:KS-entropy}
\end{align}  
Generally each invariant measure $\invms$ yields another value of the Kolmogorov--Sinai entropy.
If these values have a maximum, the associated measure is called the \emph{measure of maximum entropy}.
The following variational principle states that the topological entropy is an upper bound for the Kolmogorov--Sinai entropy:

\begin{theorem}[Variational principle \cite{Dinaburg1971,Goodman1971,Goodwyn1971}]
  \begin{align}
    \topent(\Phi) = \sup_{\invms \text{ is $\Phi$-invariant}}\ksent_{\invms}(\Phi)
    \label{eq:variational-principle-entropy}
  \end{align}
  \label{theo:variational-principle-entropy}
\end{theorem}


Finally we state a result that relates the Kolmogorov--Sinai entropy to phase-space contraction:
\begin{theorem}[Pesin's formula \cite{Hasselblatt+Pesin2008}]
  Let $\bar{\Lambda}\tind{\infty}^+(x)$ be the sum of all positive Lyapunov exponents associated with a point $x$ for a map $\Phi$, counted according to their multiplicity.
  Let $\invms$ be a $\Phi$-invariant probability measure.
  \emph{Pesin's formula} states that
  \begin{align}
    \ksent_{\invms} = \int_\pspace \bar{\Lambda}\tind{\infty}^+\df{\invms}.
    \label{eq:pesins-formula}
  \end{align}
  \label{theo:pesins-formula}
\end{theorem}

Note that Pesin's formula \eq{pesins-formula} only talks about the \emph{positive} part of the Lyapunov spectrum.
The reason for that is that the entropies defined above distinguish the \emph{direction of time}:
The refinements of $\parti$ are constructed using the \emph{pre}-images rather than the images of partition elements.
If $\Phi$ is invertible with measurable (or continuous) inverse,  we can define similar quantities for the reversed dynamical systems $(\pspace,\Phi^{-1})$.
Then, Pesin's formula yields that the sum of the KS-entropy for dynamical system and its reverse agree with the sum of \emph{all} Lyapunov exponents, and hence with the average phase-space contraction in the steady state, \cf Ref.~\cite{Gaspard2004}.

\section{Summary}
In the present chapter, we have reviewed different notions of entropy together with the contexts they appear in. 
We started with the definition of entropy in classical thermodynamics for macroscopic heat engines.
After that, we presented the notion of entropy as information, as it arises in a mathematical treatment of the elements of communication.
Even much earlier, Gibbs introduced a similar probabilistic notion of (equilibrium) entropy in the context of his statistical ensembles.
This idea together with the assumption of an underlying microscopic Hamiltonian dynamics are the key assumptions of classical statistical physics.
Consistency between different macroscopic thermodynamic constraints is achieved in the thermodynamic limit, \ie the case where the particle number becomes very large.

Modern statistical physics aim to generalize the classical concept.
More precisely, it drops certain classical assumptions and aims to define appropriate generalizations of entropy and dissipation.
For instance, the thermodynamic limit does not apply when one deals with small systems, where noise plays a major role.
We reviewed modern model paradigms, using either stochastic or deterministic, but non-Hamiltonian dynamics.
Finally, we briefly gave an overview of the abstract mathematical treatment of the latter, \ie the ergodic theory of measurable dynamical systems.

Throughout the chapter we stressed the importance of a distinction between a ``system'' and its surrounding ``medium'' by means of operational accessibility.
Or, as Oliver Penrose puts it in his work on a deductive treatment of the foundations of statistical mechanics~\cite{Penrose1970}:
\begin{quote}
  ``This limitation on our powers of observation is an essential part of statistical mechanics; without it the theory would be no more than a branch of ordinary mechanics.''
\end{quote}
In his work, Penrose carefully reviews the assumptions made by (classical) statistical mechanics.
The key assumptions are those of \emph{causality of the microscopic dynamics} and the \emph{Markovian postulate} for the mesoscopic dynamics.
The microscopic causality manifests itself in the assumptions of a \emph{deterministic} evolution.
Penrose focuses on Hamiltonian mechanics as the underlying deterministic evolution, though he is well aware that this is only another (good) approximation.
The Markovian postulate characterizes the nature of the statistics of the time series obtained by consecutive measurements on a system:
They have limited memory in the sense that the probability of measuring a future state must only depend on the current state and not on the history of past measurements.

While Penrose used these postulates in his deductive treatment of classical statistical mechanics, the present thesis looks at them in the light of more modern paradigms.
The success of (non-equilibrium) molecular dynamics simulations with non-Hamiltonian equations of motion encourages a deeper look at more \emph{general} microscopic dynamics.
Further, the abstract mathematical concept of entropy in ergodic theory is most interesting for non-Hamiltonian, dissipative dynamics.
Thus, Chapters~\ref{chap:marksymdyn} and \ref{chap:information-st} investigate the behaviour of observable time series produced by reversible deterministic, but not necessarily Hamiltonian dynamics.

Within that framework, Chapter \ref{chap:marksymdyn} investigates the implications of the Markovian postulate on observable states and microscopic ensembles.
Chapter~\ref{chap:information-st} outlines an abstract mathematical framework for entropy and entropy production based on information theory.
Although Chapter~\ref{chap:information-st} is largely independent of the Markovian postulate, we explicitly give an analytically tractable example where the Markovian postulate holds exactly.
Further, we establish a connection to the modern theory of stochastic thermodynamics, which is similarly based on the Markovian postulate.

Stochastic thermodynamics is generally understood as a paradigm for small systems in (non-equilibrium) thermodynamic environments.
For such systems, noise is not negligible and may even play a functional role.
Consequently, the second half of this thesis, Chapters \ref{chap:cycles} and \ref{chap:fluctuations}, is concerned with the quantification  of fluctuations in Markovian stochastic thermodynamics.

In the final Chapter \ref{chap:discussion}, we put our results in the context of the bird's-eye perspective given in the present introductory chapter.
At that point, the author hopes that the reader will have a good idea of the notions of entropy and dissipation in modern statistical physics;
both in general terms and in the particular framework of stochastic thermodynamics.
To come back to the initial quote:
The present chapter should be understood as the common basis for the discussion we attempt in the final chapter.

\clearpage

  \chapter{Markovian symbolic dynamics}
  \label{chap:marksymdyn}
  \begin{fquote}[O.~Penrose][Foundations of Statistical Mechanics][1970]
  The crucial [postulate in idealized models of real physical systems] is expressing the assumption, that the successive observational states of a physical system form a Markov chain.
  This is a strong assumption, [\ldots], but even so, it has been adopted here because it provides the simplest precise formulation of a hypothesis that appears to underlie all applications of probability theory in physics.
\end{fquote}

\section*{What is this about?}

In the discussion of the previous chapter we have already mentioned the two crucial assumptions of classical statistical mechanics as identified by O.~Penrose.
The first one was the assumption that the ``phase-space density at any time is completely determined by what happened to the system before that time and is unaffected  by what will happen to the system in the future''~\cite{Penrose1970}.
The second one is the Markovian postulate, which we chose as the initial quote for this chapter.

In his considerations, Penrose thought of Hamiltonian mechanics as the underlying microscopic deterministic evolution.
From that perspective, he presented a deductive treatment of the foundations of classical statistical mechanics.
As he mentions himself (\cf the initial quote), the Markovian postulate is crucial for the success of such a treatment.

In the present chapter, rather than adopting it, we establish requirements under which the Markovian postulate holds rigorously.
Stochasticity in the observed time series arises from the sensitive dependence on initial conditions for such deterministic-chaotic systems.
In our framework, a measurement outcome is given by the value of an observable $\partmap$ on an underlying microscopic phase space $\pspace$.
Usually, the microscopic evolution rule $\Psi\colon x_0\mapsto x_t$ is described by non-linear deterministic equations $\dot x = f(x)$.
The (NE)MD equations from section \ref{sec:md} provided physical examples for such equations.
For simplicity, in this chapter we will restrict ourselves to iterated invertible (stroboscopic) maps $\Phi$.

We investigate the implications of the Markovian postulate for measurement observables $\partmap$ and microscopic ensembles (measures) $\pms$.
The rigorous treatment demands a larger amount of formality than  previous and subsequent chapters.
However, we hope that the chapter is self-contained enough, such that it can be read without any further reference.
This chapter is supposed to appear as a paper in the \emph{Journal of Statistical Physics} \cite{Altaner2014}.

It is structured as follows:
We start by formally introducing our mathematical framework in Section~\ref{sec:symbolic-stochastic-dynamics}. 
In particular, we describe how a measurable dynamical system on phase space together with an observable gives rise to a measurable symbolic dynamical system on a shift-invariant subset of all infinite time series, a so-called subshift.
In Section~\ref{sec:partitions} we review the notion of partitions created by observables.
So-called Markov partitions generate a topology on the space of possible time series, which provides the necessary topological backbone of a Markov chain.
In Section~\ref{sec:markov-measures} we equip these spaces with measures that make the symbolic dynamics a Markov chain.
In the final Section~\ref{sec:mms-discussion} we discuss our findings in the context of earlier results in ergodic theory.
Further, we propose operational interpretations of the abstract results in the context of an experimenter taking (idealized) measurements.

%
%
%
%
\section{Symbolic stochastic dynamics}
\label{sec:symbolic-stochastic-dynamics}
\begin{figure}[ht]
  \centering
  \includegraphics{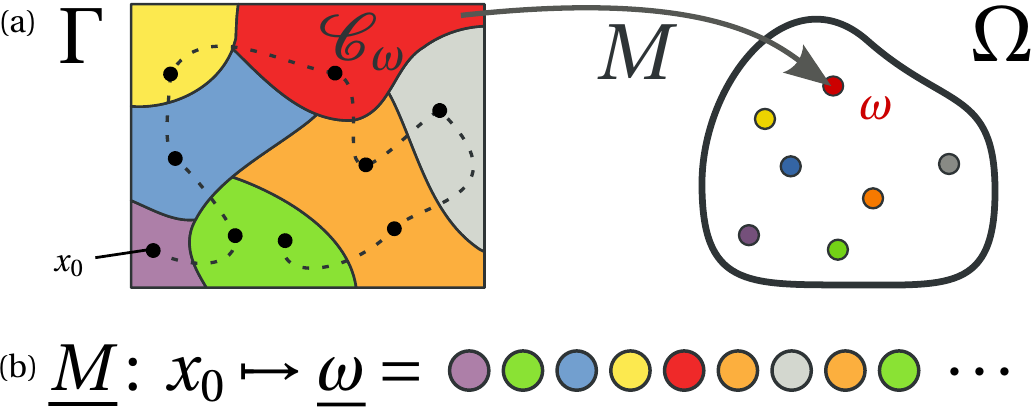}
  \caption{a) All ``microstates'' $x\in\cell_\omega$ are mapped to the ``measurement result'' $\omega$ by an observable $\partmap$.
    Consequently, a cell $\cell_\omega$ is the pre-image $\partmap^{-1}(\omega)$ of $\omega$ for the observable.
    b) A microscopic (discrete) orbit $x(t)$ starting at at $x(0)=x_0$ produces a (discrete) time series $\traj \omega=\traj\partmap(x_0)$ of subsequent measurement results.
  }
  \label{fig:partmap}
\end{figure}

We start by introducing the basic mathematical notions needed for the formal treatment of the following situation, \cf Fig.~\ref{fig:partmap}:
Consider an observable $\partmap\colon \pspace \to \alphabet$, which describes the possible outcomes of a measurement on a physical system.
In reality, any measurement apparatus only has a finite resolution.
Hence, the number  $\abs{\alphabet}$ of \emph{mesoscopic} measurement results $\omega \in \alphabet$ is smaller than the number of \emph{microscopic states} $x \in \pspace$.
In particular, we are interested in the situation where $N:=\abs{\alphabet}$ is finite, but may be large.
This assumption of finiteness will simplify the mathematical treatment considerably.\footnote{
Moreover, if one takes the finiteness of any numerical representation of data on a piece of paper or in the storage of a computer seriously, this assumption is also physically justified.}
Now consider an experimenter who records subsequent values of the observable $\partmap$ in a time series $\traj \omega\rlind{\tau} = (\omega_0,\omega_1,\cdots,\omega_\tau)$.
Upon repeating the experiment, she is careful to prepare the system each time in the same initial state.
Nonetheless, she observes a different time series of successive measurement results every time.
The reason for this phenomenon lies in the finite resolution of her experimental apparatus:
She can only \emph{prepare} a mesoscopic initial condition, \ie one that is experimentally accessible to her.
However, she has no chance to ensure that the system is at a certain microscopic configuration.

The mathematical subjects we need in order to formalize this situation are \emph{measure theory} and the derived concept of a \emph{stochastic process}.
In accordance with the causality hypothesis, the evolution of the system in phase space is described as an abstract \emph{dynamical system} (\cf Section \ref{sec:measurable-ds}).
The mesoscopic or \emph{coarse-grained} description on the level of time series is the subject of \emph{symbolic dynamics}.
Putting everything together we arrive at the central Definition~\ref{def:observed-sp} of a \emph{stochastic process of observed time series} which are generated by a dynamical system and an observable.

\subsection{Stochastic processes}
In Section~\ref{sec:measurable-ds} we have already introduced the necessary definitions and notation for measurable dynamical systems.
Further, we have given an informal discussion of stochastic processes in Section~\ref{sec:stochastic-models}.
The rigorous treatment in the present chapter requires a much higher degree of formality.
In the following, we revisit the theory of Markov chains on finite state spaces.
For a detailed review of continuous-time stochastic processes on arbitrary state spaces  we refer the reader to Ref.~\cite{Capasso+Bakstein2012}.

\begin{definition}[Stochastic process]
  \label{def:stochastic-process}
  Let $(\spc,\palg,\pms)$ be a probability space.
  Further, let $\ifam{f\tind{t}}_{t \in \timdom}$ be a family of random variables $f_t\colon (\spc,\palg) \to (\ospace,\salg)$ indexed by $t\in\timdom$.
  We call $\ifam{f\tind{t}}_{t \in \timdom}$ a \emph{stochastic process}, $(\spc,\palg,\pms)$ its \emph{underlying probability space} and $(\ospace,\salg)$ its \emph{state space}.
  Further, for $x\in\spc$ the family $\traj f(x) := \ifam{f_t(x)}_{t \in \timdom}$ is called the \emph{trajectory} of $x$.  
\end{definition}

In the above definition, the index set $\timdom$ was arbitrary.
In many cases it is understood as a \emph{time domain} and thus assumed to be a totally ordered set with a (semi-)group structure like $\reals$, $\integers$ or $\naturals$.
In the following, $\timdom$ will always denote $\integers$ or $\naturals$, \ie we are interested in \emph{discrete-time stochastic processes}.
Similarly, the state space $\ospace$ can be either discrete or continuous.
In the discrete case, we call the process a \emph{stochastic jump process}.
Henceforth, we often consider infinite tuples $\traj \omega := \ifam{\omega_t}_{t\in\timdom} \in \ospace^\timdom$ and their finite projections $\traj \omega\rlind{\tau} \in \ospace^{\tau+1}$, which we will refer to as infinite and finite \emph{time series}, respectively.

In the present context of stochastic process, time series appear as the values of the trajectory map $\traj f(\cdot)$ for a given point $x \in X$.
Yet, the space of infinite time-series $\ospace^\timdom$ is interesting on its own.
Formally, it is a \emph{product space}.
Let us recall some useful definitions about the structure (in particular the topology and the $\sigma$-algebra) of product spaces:

\begin{definition}[Product spaces, projection operator, cylinders]
  Let $\indset$ and $\ospace$ be arbitrary sets.
  The Cartesian product of $\ospace$ over \indset, $\prod_{i \in \indset} \ospace$, is the set containing all generalized tuples $\ifam{\omega_i}_{i \in \indset}$ such that $\omega_i \in \ospace$.
  The short form $\ospace^\indset := \prod_{i \in \indset} \ospace$ yields a more compact notation.
  Moreover, we write $\ospace^\indset \equiv \ospace^N$ with $N \in \naturals\setminus{0}$, if either $\abs{\indset} = N$ and we do not care about the structure of the index set $I$, or if $I = \set{1,2,\dots,N}$.
  
  For $\indset'\subset \indset$, the \emph{projection map} $\proj_{\indset'\leftarrow I}\colon \ospace^{\indset} \to \ospace^{\indset'}$ singles out the entries at indices $i \in \indset' \subset \indset$.
  If $\indset'$ is finite, we say $\proj_{\indset'\leftarrow I}$ is \emph{finite}.
  If $\indset' = \set{i}$, we write $\proj_i := \proj_{\set{i}\leftarrow I}$ for the projector on the $i$th component.
  
  If $(\ospace,\topo)$ is a topological space, the pre-image $\proj_{i}^{-1}(\sset)$ of an open-set $\sset \in \topo$ is called a \emph{cylinder} with base in $\ospace$.
\end{definition}

The \emph{product topology} is the topology generated by all cylinders, \ie it is the smallest topology such that any finite projection is continuous.
Similarly, the \emph{product $\sigma$-algebra} is defined to be the smallest $\sigma$-algebra such that every finite projection is measurable.

In the present work we consider \emph{only} the case of finite state spaces with $\abs{\ospace} = N<\infty$.
Then, it is natural to choose both the $\sigma$-algebra and topology on the factor set $\ospace$ to be the power set, \ie $\topo = \salg = \powset(\ospace)$.
Thus, the product $\sigma$-algebra, by definition, is the $\sigma$-algebra generated by the cylinders.
However, note that the product topology and the product $\sigma$ algebra are distinct if $\timdom$ is an infinite set.

Now that we have the notion of a measurable structure on the space of time series, we can define the \emph{probability of a trajectory}:
\begin{definition}[Probabilities of trajectories]
  \label{def:prob-traj}
  Let  $\traj f := \ifam{f\tind{t}}_{t \in \timdom}$ be a discrete time jump process on a finite state space $(\ospace,\powset(\ospace))$ with underlying probability space $(X,\palg,\pms)$, \cf Def.~\ref{def:stochastic-process}.
 Further, let $\sset \in \salg^\timdom$ be an element of the product $\sigma$-algebra on $\ospace^\timdom$.
 Then, the probability $\prob(\sset)$ is defined as the push-forward measure with respect to the trajectory, \ie
 \begin{align*}
   \prob(\sset) := (\traj f^*\pms)(\sset) \equiv \pms(\traj f^{-1}(\sset)).
 \end{align*}
\end{definition}

\begin{remark}
  The Daniell--Kolmogorov theorem, (\cf Appendix~\ref{app:construction-mms}), ascertains that a  consistent choice of the values of the measure $\prob$ fully determines the stochastic process.
  Then, a prescription of the probabilities $\prob(\sset)$ of events $\sset \in \salg^\timdom$ is sufficient, \ie there is no need to explicitly refer to the underlying probability space.
  In contrast, in the present work we are explicit:
  The underlying probability space is the phase space of the deterministic microscopic dynamics $\Phi$ of a physical system together with an ensemble measure.
  The trajectory $\traj f$ corresponds to the recording of (infinite) time series.
  We will formally define such an \emph{stochastic process of observed time series} in Def.~\ref{def:observed-sp}.
\end{remark}

Usually, one is interested in the probability of so-called \emph{cylinder sets}, which are finite intersections of cylinders.
Such sets define stochastic \emph{events} by specifying which \emph{elementary event} $\omega_t\in\ospace$ may or may not occur at a given position in time $t \in \timdom$.
In particular, we are interested in the probabilities 
of a special sort of cylinder sets:
\begin{definition}
  \label{def:forward-cylinder-subshift}
  Let $\traj \omega\rlind{\tau}=(\omega_0,\omega_1,\cdots,\omega_\tau) \in \ospace^{\tau+1}$ be a finite time series of length $\tau+1$ and $t \in \timdom$.
  We define the \emph{$t$-shifted forward cylinder} for $\traj\omega\rlind{\tau}$ as
  \begin{align*}
    \cyl\tind{t}[\traj \omega\rlind{\tau} ]&:= \bigcap_{k=0}^\tau \proj_{t+k}^{-1}\set{\omega_k}\\
    &\equiv \set{\traj \nu \in \ospace^\timdom \,\middle\vert\, \nu_{t+k} = \omega_k, \, 0\leq k \leq \tau}.
 \end{align*}
 The measure of such a cylinder, $\prob\tind{t}[\traj \omega \rlind{\tau}] = \prob(\cyl\tind{t}[\traj \omega\rlind{\tau}])$, is called the \emph{probability} for the finite time series $\traj\omega\rlind{\tau}$ to \emph{occur} at time $t$.  
\end{definition}

Henceforth, we use the following convention with respect to temporal indices:
A \emph{super}script $(\tau)$ denotes the \emph{run length} of finite sequences.
A \emph{sub}script $t$ characterizes a \emph{distinct point} in time.
If both indices appear, we usually refer to a finite sequence that starts at time $t$ and extends to time $t+\tau$.

In the rest of this work, we will focus our attention on the arguably simplest (non-trivial) kind of stochastic process, that is a \emph{discrete time Markovian jump process} or \emph{Markov chain} on a finite state space.
We say that $\bvec p = \ifam{p_\omega}_{\omega \in \alphabet}$ is a \emph{stochastic vector} if it obeys $0\leq p_\omega \leq 1$ for all $\omega \in \alphabet$ and it sums to unity, \ie $\sum_{\omega\in \alphabet} p_\omega = 1$.
Further, a stochastic matrix $\tmat = \ifam{\tprob{\omega}{\omega'}}_{\omega,\omega' \in \alphabet}$ obeys $0\leq \tprob{\omega}{\omega'} \leq 1$ for all $\omega,\omega' \in \alphabet$ and each row is normalized, \ie $\sum_{\omega'}\tprob{\omega}{\omega'}=1$.

\begin{definition}[Homogeneous Markov chain]
  \label{def:markov-chain}
  Let $\traj\omega^{(\tau -1)} \in \alphabet^\tau$ and $\omega_\tau \in \alphabet$ be arbitrary.
  Define $\traj\omega^{(\tau)} := (\traj\omega^{(\tau-1)},\omega_\tau ) \in \alphabet^{\tau+1}$.
  A jump process is called a \emph{homogeneous Markov chain} if there exists a stochastic matrix $\tmat$ with entries $\tprob{\omega}{\omega'}$, such that
  \begin{align}
    \prob\tind{t}[\traj\omega^{(\tau)} ] =  \prob\tind{t}[\traj\omega^{(\tau-1)} ] \cdot\tprob{\omega_{\tau-1}}{\omega_\tau}, \forall t \in \timdom
    \label{eq:markov-property}
  \end{align}
  The above relation is called the \emph{Markov property}.
  The numbers $\tprob{\omega}{\omega'}$ are called the \emph{transition probabilities} of the Markov chain.
  The stochastic matrix $\tmat$ is called the \emph{transition matrix} of the process.
\end{definition}

The Markov property is a very strong property of a stochastic process.
In particular, it allows for a recursive construction of time-series probabilities from a stochastic vector $\bvec p\tind{t}$ with elements $\p{t}{\omega}$:
\begin{align*}
  \prob\tind{t}[\traj \omega^{(\tau)}] = \p{t}{\omega_0} \prod_{k=1}^\tau \tprob{\omega_{k-1}}{\omega_{k}}.
\end{align*}
Summing over all possible finite time series of a given length $\tau$ which end in a state $\omega_{t+\tau}$ yields the connection between the elements of $\bvec p^{(t)}$ and $\bvec p^{(t+\tau)}$:
\begin{align*}
  \p{t+\tau}{\omega_{t+\tau}} = \tsum{\omega}{T} \left[\p{t}{\omega_t}\prod_{k=t+1}^{t+\tau} \tprob{\omega_{k-1}}{\omega_k}\right]
\end{align*}
where $T:=\set{t, t+1, \cdots,t+\tau-1 } \subset\timdom$ and the sum symbol $\tsum{\omega}{T}$ is an abbreviation for the multi-sum
  \begin{align*}
    \tsum{\omega}{T}\left[\, \cdot\, \right]:= \prod_{t_i \in T}\left[ \sum_{\omega_{t_i}\in\ospace} \left[ \,\cdot\, \right] \right] := \sum_{\omega_{t_1}\in\alphabet}\quad\sum_{\omega_{t_2}\in\alphabet}\cdots\sum_{\omega_{t_k}\in\alphabet}\left[\,\cdot\,\right].
  \end{align*}
  The special case for time series $\traj \omega\rlind{1} = (\omega,\omega')$ of length $\tau = 1$ is called the \emph{master equation}, \cf Sec.~\ref{sec:discrete-st}: 
\begin{align}
  \p{t+1}{\omega'} = \sum_{\omega} \p{t}{\omega} \tprob{\omega}{\omega'} .  \label{eq:master-equation}
\end{align}

Markov processes are \emph{memoryless processes}, in the sense that the probability of the next state only depends on the current state.
Hence, at any time the process has completely ``forgotten'' its past.

Let us summarize the present subsection:
We have introduced a formal way to treat sequences of random variables $\traj f = \ifam{f_t}_{t\in\timdom}$ as a stochastic process, without specifying the nature of $f_t$ or the underlying probability space $X$.
On the other hand, the setting outlined at the beginning of this section, already suggests a certain interpretation:
The underlying probability space is the phase space of some physical dynamics, \ie $X=\pspace$, and $f_t$ should correspond to subsequent measurements.
Such dynamical systems will be the subject of the next subsection.

\subsection{The stochastic process of observed time series}
For the remainder of this chapter, we focus on \emph{dynamical systems in discrete time}, which are characterized by a map $\Phi$ from phase space $\pspace$ onto itself.
We further assume that the system is \emph{autonomous}, \ie the map $\Phi$ is independent of time $t$.
Physically, we understand it as a \emph{stroboscopic} map obtained from some physical microscopic dynamics $\Psi$ evolving in continuous time, \cf Section \ref{sec:time-discrete}.

We restate the definition of a measurable dynamical system from Section \ref{sec:measurable-ds}:
\begin{definition}[Measurable dynamical system]
  Let $(\pspace,\palg)$ be a measurable space and let $\Phi\colon(\pspace,\palg) \to (\pspace,\palg)$ be a measurable map.
  We call $(\pspace,\palg,\Phi)$ a \emph{measurable dynamical system}.
  If a probability measure $\pms$ is given, we call $(\pspace,\palg,\Phi,\pms)$ a dynamical system with measure $\pms$.
\end{definition}
In statistical physics, $\mu$ is also called a microscopic \emph{statistical ensemble}.
Then, it is understood to represent the (distribution of) configurations of a large number of copies of the system.

An observable $\partmap\colon (\pspace,\palg) \to (\ospace,\salg)$ is a measurable map from phase space $\pspace$ to the space of observations $\ospace$.
As mentioned above, we assume that $\ospace$ is of finite cardinality $N$ and choose $\salg = \powset(\ospace)$ as the $\sigma$-algebra on $\ospace$.
With these ingredients we formalize the process of taking measurements at equidistant time steps as a stochastic process:
\begin{definition}[Stochastic process of observed time series]
  Let $(\pspace,\palg,\Phi,\pms)$ be a dynamical system with measure $\pms$ and $\partmap\colon (\pspace,\palg) \to (\ospace,\salg)$ a measurable function.
  The sequence
  \begin{align}
    \mtraj &:= \ifam{\partmap\tind{t}}_{t\in\timdom}\nonumber\\
      &:= \ifam{\partmap \circ \Phi^t}_{t \in \timdom}.
    \label{eq:trajectory-sym}
  \end{align}
  is called the \emph{stochastic process of observed time series}.
    \label{def:observed-sp}
\end{definition}
Note that $\mtraj$ is only well-defined for $\timdom = \integers$ if $\Phi$ is invertible.
Henceforth, we will assume this to be the case.

The image of phase space under the trajectory of the process, $\subshift_\partmap^\timdom := \mtraj(\pspace) \subset \ospace^\timdom$ contains all the infinite time series that an observable $\partmap$ can produce.
Further, we will prove that this set is invariant under shifting the whole sequence one step to the left, possibly dropping the zeroth element if $\timdom = \naturals$.
In general, sets $\subshift^\timdom \subset \alphabet^\timdom$ that obey such a dynamic shift-invariance are called \emph{subshifts}.
The formal treatment of the topology and the dynamics of subshifts is the goal of \emph{symbolic dynamics}. 

\subsection{Symbolic dynamics}
Symbolic dynamics is the study of infinite symbol sequences produced by a finite \emph{alphabet} $\alphabet$.
Hence, it is closely related to the study of \emph{messages} in information theory, \cf Section \ref{sec:info-theory}.
In the present case, the alphabet $\alphabet$ is the co-domain of the observable $\partmap$.
However, for now we will forget about $\partmap$ and the underlying dynamics.
Rather, we first characterize certain subsets $\subshift^\timdom \subset \alphabet^\timdom$ which are invariant under the action of a \emph{shift map}:
\begin{definition}[Shift, alphabet, shift map]
  Let $\alphabet$ be a finite set.
  The sets $\alphabet^\integers$ and $\alphabet^\naturals$ are called the \emph{full shift} and \emph{forward shift} over $\alphabet$, respectively.
  The set $\alphabet$ is referred to as an \emph{alphabet} and its elements are called \emph{symbols}.\\  
  A finite string $\traj\omega^{(\tau)}=\left( \omega_t \right)_{t \in \{0,1,\cdots,\tau\}}\in \alphabet^{\tau+1}$ consisting of $\tau+1$ symbols is called a \emph{block}.\footnote{
  Above we used the expressions ``finite'' and ``infinite time series'' for what in the framework of symbolic dynamics is called a block and a symbol sequence, respectively.
  We make no difference in their meaning, and distinguish them just by the mathematical framework they appear in.
  }\\  
  The \emph{shift map} $\shiftmap$ acts on (bi-)infinite symbol sequences $\traj \omega = \left( \omega_t \right)_{t \in \timdom} \in \alphabet^\timdom$ in shifting the whole sequence by one step to the left: 
  \begin{align}
    \shiftmap\colon\alphabet^\timdom &\to \alphabet^\timdom,\nonumber\\
    \ifam{\omega_t}_{t\in\timdom} &\mapsto \ifam{ \omega_{t+1}}_{t \in \timdom}.
    \label{eq:shift-map}
  \end{align}
\end{definition}
A \emph{subshift} is a shift-invariant subset of the full shift:
\begin{definition}[Subshift]
  Let $\subshift^\timdom \subset \alphabet^\timdom$ be a subset of the full or forward shift, respectively.
  If $\subshift^\timdom$ is \emph{shift-invariant}, \ie if $\shiftmap \subshift^\timdom = \subshift^\timdom$, we call $\subshift^\timdom$ a \emph{subshift}.
  The elements $\traj \omega \in \subshift^\timdom$ are called \emph{allowed} or \emph{admissible} sequences.
\end{definition}

By definition, the set of all cylinders is invariant under the action of the shift map $\shiftmap$.
Hence, the $\sigma$-algebra generated by the cylinders is shift-invariant by construction.
Thus, $\shiftmap$ is measurable and specifies a measurable dynamical system on $\subshift^\timdom$:
\begin{definition}[Symbolic dynamics]
  Let $\alphabet^\timdom$ be a shift space, $\shiftmap$ the shift map and  $\subshift^\timdom$ a subshift.
  Further, let $\salg_\subshift^\timdom$ be the restriction of the product $\sigma$-algebra on $\alphabet^\timdom$ to $\subshift^\timdom$.
  The dynamical system $(\subshift^\timdom,\salg_\subshift^\timdom,\shiftmap)$ is called a \emph{symbolic dynamical system} or just short a \emph{symbolic dynamics}.
\end{definition}

For the rest of this subsection, we are concerned with special kinds of subshifts.
Consider a finite set $F$ of blocks, possibly of different lengths.
We will refer to $\traj \omega\rlind{\tau}$ as a \emph{forbidden block} of length $\tau+1$.
For finite $F$ it is obvious that the set
\begin{align*}
  \subshift = \set{ \traj \omega \in \ospace^\timdom \,\vert\, \traj \omega \text{ does not contain any block $\traj\omega\rlind{\tau} \in F$ as a sub string  }}
\end{align*}
is a subshift.
More precisely, we call $\subshift$ a $m$-step subshift of finite type (SFT), where $m+1$ denotes the maximal length of the blocks in $F$.
For a much deeper treatment of this and related subjects, we refer to Refs.~\cite{Weiss1973,Williams1973}.
In addition, Ref.~\cite{Bai-Lin1989} contains a presentation which may be more accessible to physicists.

In the remainder of this work, we will only consider $1$-step SFTs.
For finite alphabets $\alphabet$, this assumption can be made without any loss of generality:
For an $m$-step SFT one simply considers the finite set $\alphabet^m$ as the alphabet of a larger shift space $(\alphabet^m)^\timdom$.
Henceforth we use ``SFT'' as a synonym for ``$1$-step SFT''.

A SFT on an alphabet $\alphabet$ with cardinality $N$ can be characterized by a simple matrix:
\begin{definition}[Subshift of finite type]
  Let $\subshift^\timdom$ be a subshift and let $\adjm$ be a $N\times N$-matrix with entries $a^\omega_{\omega'}\in\set{0,1}$.
  The set
  \begin{align}
    \subshift^\timdom := \set{\traj \omega = \ifam{\omega_t}_{\in \timdom}\,\vert\,a^{\omega_t}_{\omega_{t+1}} = 1 }
    \label{eq:sft-definition}
  \end{align}
  is called a (1-step) subshift of finite type (SFT).
  Then, the matrix $\adjm$ is called the adjacency matrix of $\subshift^\timdom$.
%
\end{definition}

The reader might have noticed the similarity of Equation~\eqref{eq:sft-definition} with the Markov property (\ref{eq:markov-property}):
Both state that the possibility to see a block (respectively finite time series) $\traj \omega\rlind{\tau+1} := \left( \traj \omega \rlind{\tau}, \omega_{\tau+1} \right)$ only depends on the last symbol $\omega_{\tau}$ of $\traj\omega\rlind{\tau}$.
Hence, both equations characterize a memoryless process where the future only depends on the present and not on the past.
The Markov property can be understood as specifying the \emph{probability} rather than (or more precisely: in addition to) the \emph{possibility} of the appearance of sequences.
It is clear that the topological structure of a subshift $\subshift^\timdom=\traj f(  \spc)$ created by a Markovian stochastic process $\traj f$ is a SFT.
This is the reason why SFTs are also known in the literature as \emph{topological Markov shifts} \cite{Adler_etal1977}, \emph{topological Markov chains} \cite{Krieger1980} or \emph{intrinsic Markov chains} \cite{Parry1964}.
If a Markov chain has a transition matrix $\tmat$, the adjacency matrix of the corresponding SFT obeys $\adjm = \sgn(\tmat)$.
In that case we say that the transition matrix $\tmat$ is \emph{compatible} with the adjacency matrix $\adjm$.

Now let us come back and draw a connection to the stochastic process of observed time series $\traj \omega \in \subshift_M^\timdom$:
\begin{proposition}
  \label{theo:stochastic-symbolic-dynamics-subshift}
  Let $\mtraj := \ifam{\partmap\tind{t}}_{t\in\timdom}= \ifam{\partmap \circ \Phi^t}_{t \in \timdom}$ be the trajectory of a stochastic process of observed time series.
  Further let $\subshift_\partmap^\timdom := \mtraj(\pspace)$.
  Then, the following statements are true:
  \begin{itemize}
    \item  The shift map $\shiftmap$ on $\subshift_{\partmap}^\timdom$ plays the role of $\Phi$ on $\pspace$, \ie
  \begin{align}
    \mtraj \circ \Phi = \shiftmap \circ \mtraj. \label{eq:symbolic-dynamics}
  \end{align}
    \item The set of all possible infinite time series, $\subshift_{\partmap}^\timdom$, is a subshift.
  \end{itemize}
\end{proposition}
\begin{proof}
  For the first statement observe that for $x\in\pspace$ we have
  \begin{align*}
    \left( \mtraj \circ \Phi \right)(x) &= \left( \partmap \Phi^{t}\Phi x \right)_{t\in\timdom} = \left( \partmap \Phi^{t+1} x \right)_{t\in\timdom} = \shiftmap\left( \partmap \Phi^{t} x \right)_{t\in\timdom} 
   =  \left( \shiftmap \circ \mtraj\right)(x)
  \end{align*}
  and hence $\shiftmap \subshift_{\partmap}^{\timdom} = \shiftmap \mtraj \pspace = \mtraj \Phi \pspace$.
  Then, because $\Phi\pspace = \pspace$ we have $\shiftmap \subshift_{\partmap}^{\timdom} = \subshift_{\partmap}^{\timdom}$ and thus the second statement follows.
\end{proof}

In the next section, we investigate how the choice of the observable $\partmap$ influences the properties of the stochastic process of measured time series, \cf definition \ref{def:observed-sp}.
Henceforth, $\subshift_\partmap^\timdom := \mtraj(\pspace)$ will contain the possible sequences generated by such a process.

\section{Observables and partitions on phase space}
\label{sec:partitions}
Equipped with the notions of an SFT and that of the stochastic process of observed time series, we can ask the following question:
For a given dynamics $\Phi$, what are necessary conditions on the observable $\partmap$ such that the subshift $\subshift^\timdom_\partmap$ is of finite type?

Unfortunately, it turns out that this is not the case in general.
Even for the most simple dynamics, there are certain sequences $\traj \omega$ missing from $\subshift^\timdom_\partmap$, which would be needed to make it an SFT.
In a sense, the subshift $\subshift^\timdom_\partmap$ is ``too small'':
Every $x\in\pspace$ has \emph{exactly one} symbolic representation,  due to the fact that $\partmap$ is a function.
In order to make $\subshift^\timdom_\partmap$ an SFT --- and therefore a subshift with an easy topological structure --- we need to admit \emph{multiple} symbolic sequences $\traj \omega$ for points $x\in\pspace$.

In Ref.~\cite{Adler1998}, R.\,Adler discusses exactly this problem, which is also well-known from the decimal encoding of the real numbers.
Consider the unit interval $\pspace =[0,1)$.
Points $x \in \pspace$ have decimal representations $0.\traj\omega$ where $\traj \omega \in  \set{0,1,2,\cdots,9}^{\naturals}$ is an semi-infinite string of digits.
Obviously, the set $\set{0,1,2,\cdots,9}^{\naturals}$ is an SFT, but the encoding of the unit interval by infinite symbols is not unique:
The real number $\frac{1}{2}$ has the two equivalent decimal expansions $\frac{1}{2} = 0.4\bar{9} \equiv 0.5\bar{0}$.

\begin{figure}[ht]
  \begin{center}
    \includegraphics{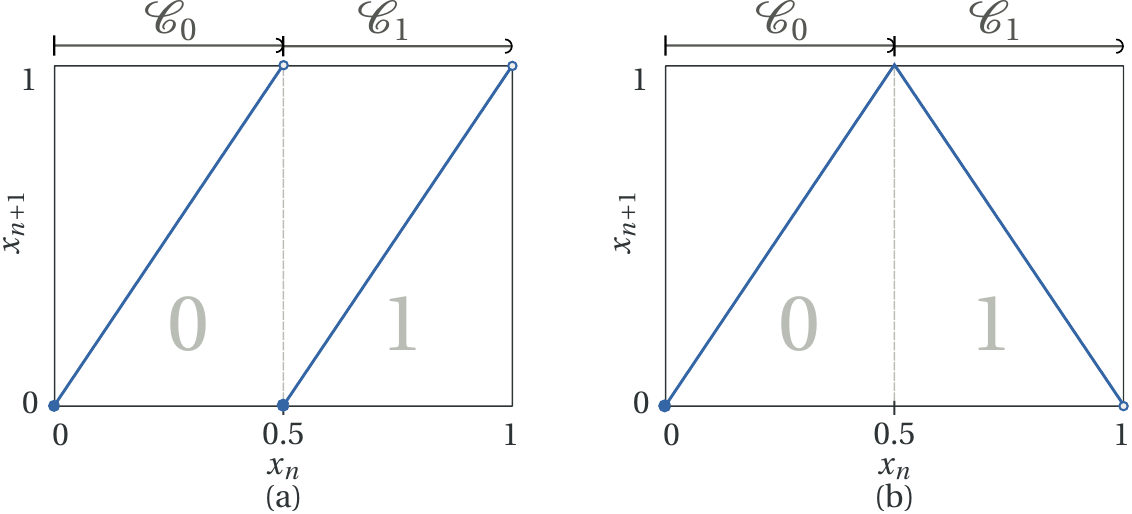}
   \end{center}
  \caption{
    (a) The graph of the Bernoulli map $\Phi\colon x\mapsto (2x \mod 1)$  on the unit interval $[0,1)$.
    The partition $\parti = (\cell_{0},\cell_{1}) = ([0,0.5),[0.5,1))$ for the observable $\partmap(x) = \chi_{[0.5,1)}(x)$ creates a subshift $\subshift^\naturals_\partmap$.
    However, this is not a subshift of finite type as the sequence $\traj\omega = (0,1,1,\cdots)$ is not part of $\subshift_\partmap^\naturals = \mtraj(\pspace)$.
    (b) The tent map $\Phi\colon x \mapsto 2\min{\set{x,1-x}}$ continuously  maps the unit interval $\pspace = [0,1)$ to itself.
      It is topologically conjugate to the map shown in (a).   
  }
  \label{fig:mult2-tent}
\end{figure}

Let us formulate this observation in the language of generated time series.
To that end, consider the \emph{binary} encoding of the unit interval $\pspace = [0,1)$.
As a dynamics $\Phi$, we consider ``multiplication by two modulo one'', also known as the \emph{Bernoulli map}, which is depicted in Figure~\ref{fig:mult2-tent}a).
The observable $\partmap$ is defined to yield $0$ if $0\leq x < \frac{1}{2}$ and $1$ otherwise.
It partitions $\pspace$ into the intervals $\cell_0 = [0,\frac{1}{2})$ and $\cell_1 = [\frac{1}{2},1)$.
By definition, $\mtraj\colon \pspace \to \subshift^\naturals_\partmap$ is surjective.
One can easily verify that the function $\mtraj^{-1}\colon (\omega_t)_{t\in\naturals} \mapsto \sum_{t\in\naturals} \omega_t 2^{-(t+1)}$ is its inverse and thus $\mtraj$ is a one-to-one map.
However, observe that both $\mtraj^{-1}(1,0,0,\dots) = \mtraj^{-1}(0,1,1,\ldots) = 0.5$.
Because $\mtraj$ is bijective, one of these infinite sequences (in the present case, the second one) is not an element of $\subshift^\naturals_\partmap$ and thus the latter cannot be an SFT.
Unlike one may expect, the origin of that problem is \emph{not} the discontinuity at $x = \frac{1}{2}$.
The same is true for the point $x=\frac{1}{2}$ in the continuous tent map shown in Fig.~\ref{fig:mult2-tent}b, which is topologically conjugate to the Bernoulli map.

As one might already guess from these examples, the issue arises for points $x\in \pspace$ that lie (or are mapped to) the boundary of partition elements.
The solution as presented in Ref.~\cite{Adler1998} is to define the factor map $\factormap\colon (\omega_t)_{t\in\naturals} \mapsto \sum_{t\in\naturals} \omega_t 2^{-(t+1)}$ for \emph{all} sequences in the SFT $\subshift^\naturals = \set{0,1}^\naturals$.
In this \emph{extension} of $\partmap^{-1}$ on $\subshift^\naturals_\partmap$ to $\factormap$ on $\subshift^\naturals\supset\subshift^\naturals_\partmap$ we give up bijectivity for a nicer topology.
This is the general idea behind \emph{topological partitions}, which we review in Section~\ref{sec:topological-partitions}.

%
%
%

In the following, it is instructive to look at the structure on phase space generated by $\partmap$ as opposed to $\partmap$ itself.
In Section~\ref{sec:topent-ksent} we have already defined the concept of a \emph{partition} $\ifam{\cell_\omega}_{\omega \in \alphabet}$ of $\pspace$ as a subset of $\powset(\pspace)$ whose elements are disjoint and whose union is $\pspace$.

Now consider the set $\cell_\omega:= \partmap^{-1}(\set{\omega})$ which is the pre-image of a singleton subset $\set{\omega}$.
The \emph{partition induced by $\partmap$} is defined as $\parti := \ifam{\cell_\omega}_{\omega \in \alphabet}$. 
We will refer to its elements $\cell_\omega$ as the \emph{cells} induced by $\partmap$.
Note, that from any (finite) partition indexed by elements $\omega \in \ospace$,  one can construct a map $\partmap\colon\pspace \to \ospace$ defined by $\partmap(x) := 
\omega \text{ if } x \in \cell_\omega$.
%

Henceforth, we will slightly abuse the notation and drop the set delimiters when we refer to singletons and write, for instance, $\cell_\omega = \partmap^{-1}(\omega)$.
Further, we will use the expressions ``observation'', ``state'' and ``symbol'' synonymously to refer to $\omega \in \alphabet$, depending on whether we want to emphasize its role as the value of an observable, the state of a stochastic process or the symbol of an alphabet, respectively.

\subsection{Generating partitions}
In this subsection we are concerned with so-called \emph{generating partitions}.
Rather than embracing the topological definition of a generating partition (\cf Ref.~\cite{Adler1998}), we use Sinai's measure-theoretic definition of such partitions \cite{Sinai2009}:
Informally speaking, the elements $\cell_\omega$ of a generating partition $\parti$ and their (pre-)images under iterations of the dynamics $\Phi$ generate the measurable structure on phase space.

We need the following elementary result of set theory:
\begin{lemma}
  \label{theo:sets}
    Let $f\colon X_1 \to X_2$ be a map and $\fsets_1 \subset \powset(X_1)$ and $\fsets_2 \subset \powset(X_2)$.
    Then it holds that
   \begin{align*}
      f^{-1}\left(\bigcup_{\pset\in\fsets_2}\pset\right) &= \bigcup_{\pset \in\fsets_2}f^{-1}(\pset), &  f^{-1}\left(\bigcap_{\pset\in\fsets_2}\pset\right) &= \bigcap_{\pset \in\fsets_2}f^{-1}(\pset),\\
      f\left(\bigcup_{\pset\in\fsets_1}\pset\right) &= \bigcup_{\pset \in\fsets_1}f(\pset), & f\left(\bigcap_{\pset\in\fsets_1}\pset\right) &\subseteq \bigcap_{\pset \in\fsets_1}f(\pset).
  \end{align*}
  Further, $f^{-1}\left(\sigma(\fsets_2) \right) = \sigma\left( f^{-1}(\fsets_2) \right)$ and  $f\left(\sigma(\fsets_1) \right) \subseteq \sigma\left( f(\fsets_1) \right)$.
\end{lemma}
\begin{proof}
  The first part of the claim is (trivially) verified using the definition of the image and pre-image of a point.
  The second claim then follows from the definition of the generated $\sigma$-algebra.
\end{proof}

Now consider the set $\cyls\tsup{a}=\set{\proj_{t\leftarrow\timdom}^{-1}(\omega) \,\vert\, \omega \in \alphabet, t \in \timdom}\subset\alphabet^\timdom$ consisting of the cylinders whose base (\ie the image of the projection) is a singleton $\set{\omega}$.
Because the base of an element $\cyl\tsup{a} \in \cyls\tsup{a}$ cannot be subdivided into non-empty sets, we call them the \emph{atomic cylinders}.
All (open) subsets of $\alphabet$ are countable (in fact finite) unions of singletons $\set{\omega}$, because $\alphabet$ is finite.
Taking the pre-image of a set commutes with set-theoretic operations (\cf Lemma~\ref{theo:sets}).
This implies that the atomic cylinders are a large enough set to generate the product $\sigma$-algebra $\salg^\timdom$ on $\alphabet^\timdom$, \ie $\sigma(\cyls\tsup{a}) = \salg^\timdom$.
For the rest of this subsection, we take $\timdom = \integers$, and sequences $\traj \omega \in \subshift^\timdom$ are bi-infinite.

We denote the set $\pcyls\tsup{a} := \bigcup_{t \in \integers} \Phi^{-t} \parti$, which contains all pre-images and images of a partition $\parti = \partmap^{-1}\ospace$ under the (iterated) action of $\Phi$, as the \emph{atomic cells}.
Using Eq.~(\ref{eq:symbolic-dynamics}), we find that $\mtraj$ applied to any atomic cell $\cell\tsup{a} \in \pcyls\tsup{a}$ uniquely identifies an atomic cylinder:
\begin{align*}
  \mtraj \Phi^t \cell_\omega = s^t \mtraj \cell_\omega = s^t \proj_0^{-1}\set{\omega} = \proj_{-t}^{-1}\set{\omega} \in \cyls\tsup{a}.
\end{align*}
Thus, taking the image $\mtraj$ of $\pcyls\tsup{a}$ is a bijection onto $\cyls\tsup{a}$, \ie $\pcyls\tsup{a} = \mtraj^{-1}\cyls\tsup{a}$ and $\mtraj\pcyls\tsup{a} = \cyls\tsup{a}$.
For a \emph{generating partition}, the atomic cells generate the measurable structure on phase space:
\begin{definition}[Generating partition]
  Let $\left( \pspace,\palg,\Phi \right)$ be a measurable dynamical system and $\parti$ a partition of $\pspace$.
  Let $\pcyls\tsup{a} := \bigcup_{t \in \integers} \Phi^{-t} \parti $ denote the set of atomic cells.

  A partition $\parti$ is called \emph{generating} for $\left( \pspace,\palg,\Phi \right)$, if $\sigma(\pcyls\tsup{a})=\palg$, \ie if the $\sigma$-algebra generated by the atomic cells $\pcyls\tsup{a}$ agrees with the $\sigma$-algebra for the dynamical system.
\label{def:generating-partition}
\end{definition}

\begin{figure}[th]
  \centering
  \includegraphics{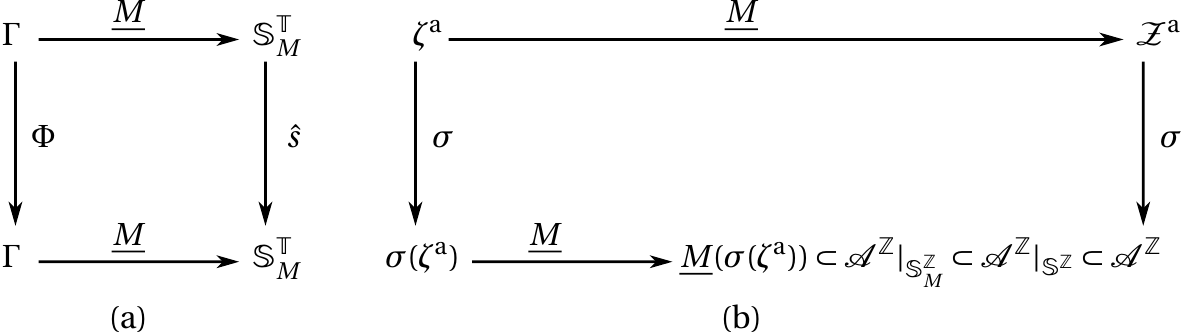}
  \caption{
    (a) A commuting diagram illustrating $\shiftmap\circ\mtraj = \mtraj \circ \Phi$.
    If $\Phi$ is invertible and $\timdom = \integers$, one can reverse the vertical arrows.
    If $\partmap$ is obtained from a generating partition w.r.t.~the Borel sets, also the horizontal arrows can be reversed.
    (b) The relation between the atomic cylinders and the atomic cells and the corresponding generated $\sigma$-algebras.
    Every measurable subset of $\subshift^\integers_\partmap$ is also measurable in the restriction $\salg^\integers\vert_{\subshift^\integers}$ of the product $\sigma$-algebra $\salg^\integers$.
  }
  \label{fig:generating}
\end{figure}
%
The next lemma states that the images of measurable sets on phase space are measurable, \cf~Fig.~\ref{fig:generating}b:
\begin{lemma}
  \label{theo:measurable}
  Let $\parti$ be a generating partition for the measurable dynamical system $\left(\pspace,\palg,\Phi\right)$.
  Let $\mtraj = \ifam{\partmap \circ \Phi^t}_{t \in \integers}$ denote the stochastic process of observed time series.
  Let $\salg^\integers_\partmap:=\salg^\integers\vert_{\subshift^\integers_\partmap}$ be the symbol product $\sigma$-algebra restricted to $\subshift^\integers_\partmap$.
  Then, $\palg$ is the pre-image of $\salg^\integers_\partmap$ for $\mtraj$, \ie
  \begin{align*}
    \mtraj^{-1}\left( \salg^\integers_\partmap \right) = \palg
  \end{align*}
  and the image of any measurable set $\pset \in \palg$ is measurable in $\salg^\integers_\partmap$, \ie
  \begin{align*}
    \mtraj\left( \palg \right) \subseteq \salg^\integers_\partmap.
  \end{align*}
\end{lemma}
\begin{proof}
  For a generating partition, by definition we have $\sigma(\pcyls\tsup{a}) \equiv \palg$.
  We already established that the symbolic product $\sigma$-algebra $\salg^\integers$ is generated by the atomic cylinders.
  Hence $\salg^\integers_\partmap := \sigma(\cyls\tsup{a})\vert_{\subshift_\partmap^\integers}$.
  The proof of the first statement is straightforward:
  \begin{align*}
    \mtraj^{-1}\left( \sigma(\cyls\tsup{a})\vert_{\subshift_\partmap^\integers}\right) &=\left( \mtraj^{-1} \sigma(\cyls\tsup{a})\vert_{ \mtraj^{-1}\subshift_\partmap^\integers}\right) \\
  &= \sigma( \mtraj^{-1} \cyls\tsup{a})\vert_{ \pspace} = \sigma( \mtraj^{-1} \cyls\tsup{a})\\
  &  = \sigma(\pcyls\tsup{a})\equiv\palg,
  \end{align*}
  where to get to the second line we used that by definition $ \mtraj^{-1} \subshift_\partmap^\integers \equiv \mtraj^{-1}\mtraj\pspace = \pspace$ and Lemma~\ref{theo:sets}.
  Also by Lemma~\ref{theo:sets}, observe that $\mtraj\left( \sigma(\pcyls\tsup{a})\right) \subseteq \sigma(\mtraj\pcyls\tsup{a}) = \sigma(\cyls\tsup{a})$.
  Further, for $\pset \in \sigma(\pcyls\tsup{a})$ we have (by definition of $\mtraj$) that $\mtraj\pset \subset \subshift_\partmap^\integers$ and thus $\mtraj \sigma(\pcyls\tsup{a})\subset \powset(\subshift_\partmap^\integers)$.
  Hence, we have established that $\mtraj\left( \sigma(\pcyls\tsup{a})\right) \subseteq \sigma(\cyls\tsup{a})\cap \powset(\subshift_\partmap^\integers) \equiv \salg^\integers_\partmap $ and thus proved the second statement.
\end{proof}

Note that the notion of a generating partition is always defined with respect to the $\sigma$-algebra $\palg$ on phase space.
If $\palg$ is the family of Borel sets, we have the following corollary:
\begin{corollary}
  \label{theo:faithful}
  If $\parti$ is a generating partition for $(\pspace,\palg,\Phi)$, where $\palg$ denotes the Borel sets, $\mtraj\colon\pspace \to \subshift^\integers_\partmap$ is an invertible map with measurable inverse.
\end{corollary}

\begin{proof}
  Let $\fsets_\pspace$ and $\fsets_\partmap$ contain the singleton subsets of $\pspace$ and $\subshift^\integers_\partmap$, respectively.
  Then, to establish invertibility of $\mtraj$ we need to show bijectivity of $\mtraj\colon \fsets_\pspace \to \fsets_\partmap$.
  Surjectivity is trivial, because $\subshift^\integers_\partmap$ is defined to be the image of $\pspace$ under $\mtraj$.
  For injectivity, we show that the pre-image of $\set{\omega}$ under $\mtraj$ is a singleton for any $\omega \in \subshift^\integers_\partmap$.
  We will prove the result by contradiction.
  To that end let $x \in \pspace$.
  By surjectivity, there is an element $\omega \in \subshift^\integers_\partmap$, such that $\mtraj(x) = \omega$.
  Now suppose that $\mtraj^{-1}\set{\omega}$ is not a singleton.
  Then, the first statement of Lemma \ref{theo:measurable} ensures that the singleton $\set{x}$ cannot be an element of $\palg$.
  However, this is a contradiction to the assumption that $\palg$ is the Borel $\sigma$-algebra.
  Hence, $\mtraj$ is injective and surjective and thus a bijection with inverse $\mtraj^{-1}\colon \subshift^\integers_\partmap \to \pspace$.
  The fact that this inverse is measurable is a just the second statement of  lemma \ref{theo:measurable}.
\end{proof}

Under the assumptions of the above lemma, \ie if $\mtraj$ is invertible, one says that it provides a \emph{faithful encoding} of the dynamical system on phase space as a symbolic dynamics.
Then, any measure on $\subshift^\integers_\partmap$ can be pushed forward by $\mtraj^{-1}$ to a Borel measure on phase space.
In particular, we can \emph{construct} certain measures in a much easier way referring to $\subshift$ and $\shiftmap$ instead of using the dynamics $\Phi$.

\subsection{Topological and Markov partitions}
\label{sec:topological-partitions}

The ultimate goal of this chapter is to find measures on $\pspace$, such that the stochastic process of observed time series obeys Markovian statistics.
A necessary requirement for this is that the space of allowed time series forms an SFT.
However, we have seen that (generating) partitions induced by observables generally do not yield subshifts $\subshift^{\integers}_\partmap$ which are of finite type.

From the simple chaotic maps described at the beginning of this section (\cf Fig~\ref{fig:mult2-tent}), we have identified the problem that arise from elements that lie on the boundary of the partition induced by $\partmap$.
In this subsection, we review how this problem is fixed by slightly enlarging $\subshift^{\integers}_\partmap$. 

From the measure-theoretic point of view, boundary points are negligible for sets $\cell_\omega$ with a Riemann-integrable characteristic function $\chi_{\omega}(x)$.
Thus, R.\,Adler suggested the use of \emph{topological partitions} rather than usual partitions \cite{Adler1998}.
Topological partitions consist of open sets, whose closures cover the phase space:
\begin{definition}[Topological partition]
  Let $(\pspace,d)$ be a metric space with topology $\topo$ and let $\parti\tsub{top} =\ifam{\cell_\omega}_{\omega \in \alphabet} \subset \topo$ be a finite collection of open sets.
  We call $\parti$ a \emph{topological partition} if
  \begin{align*}
    (i) &\quad\cell_\omega \cap \cell_{\omega'} = \emptyset, \,\omega \neq \omega',\\
    (ii)&\quad\pspace = \bigcup_{\omega \in \alphabet} \bar{\cell_\omega},
  \end{align*}
  where the overbar denotes closure of sets.
\end{definition}

Thus, from a partition $\parti =\ifam{\cell_\omega}_{\omega \in \alphabet} = \partmap^{-1}(\alphabet)$ of a metric phase space $(\pspace,d)$ we obtain a topological partition $\parti\tsub{top} = \mathring{\parti} := \big(\mathring{\cell_\omega}\big)_{\omega \in \alphabet}$.
Like a usual partition, a topological partition on a dynamical system creates a subshift $\subshift^\integers$.
In order to construct this subshift, it is useful to define the following sets:
\begin{definition}[Bi-infinitely refined cell]
  \label{def:infinite-cell}
  Let $\parti\tsub{top} = \ifam{\cell_{\omega}}_{\omega \in \ospace}$ be a topological partition.
  For $\traj \omega \in \ospace^\integers$ we call the set 
  \begin{align}
    \label{eq:infinite-cell}
    \cell[\traj \omega] :=  \bigcap_{T=0}^{\infty} \bar{\bigcap_{t=-T}^{t=T}\Phi^{-t}\cell_{\omega_t}},
  \end{align}
  the \emph{bi-infinitely refined cell} for $\traj \omega$.
\end{definition}

Because $ \bigcap_{T=0}^{\infty} \bar{\bigcap_{t=-T}^{t=T}\Phi^{-t}\cell_{\omega_t}} \subset  \bigcap_{t=-\infty}^{t=\infty}\Phi^{-t}\bar{\cell_{\omega_t}}$ the set $\cell[\traj \omega]$ contains only points such that $\Phi^t x \in \bar{\cell_{\omega_t}}$ \cite{Adler1998}.%
\footnote{In general, $\cell[\traj\omega]\neq \bigcap_{t=-\infty}^{t=\infty}\Phi^{-t}\bar{\cell_{\omega_t}}$.
However, the missing points are (pre-)images of boundary points and do not matter measure theoretically.
  Again, we refer to Ref.~\cite{Adler1998} for the details.}
Assume for now that the topological partition $\parti\tsub{top} = \ifam{\mathring{\cell}_\omega}_{\omega \in \ospace}$ is obtained from taking the interior of elements of a proper partition $\parti=\ifam{{\cell}_\omega}_{\omega \in \ospace}$ of phase space.
Then, the orbit of any point $x \in\cell[\traj \omega]$ visits the \emph{closures} $\bar{\mathring{\cell}}_{\omega_t}=\bar{\cell_{\omega_t}}$ of the partition elements recorded by the time series $\traj \omega$, \ie it is at least infinitesimally close to an orbit whose iterates $x_t=\Phi^t x$ are in $\cell_{\omega_t}$, where $\omega_t$ is specified by the time series $\traj \omega$.

With that, we define the set of symbol sequences:
\begin{definition}[Symbol sequences obtained from a topological partition]
  \label{def:subshift-topological}
  Let $\parti\tsub{top}$ be a topological partition and $\cell[\traj\omega]$ the set defined in equation (\ref{eq:infinite-cell}).
  Further, let $(\pspace,\Phi)$ be a dynamical system.
  Then, we define the \emph{set of $(\parti\tsub{top},\Phi)$-sequences} $\subshift^\integers \subset \alphabet^\integers$ as
  \begin{align}
  \label{eq:defn-subshift-topological}
  \subshift^\integers := \set{\traj \omega =\ifam{\omega_t}_{t\in\integers} \,\middle\vert\, \cell[\traj\omega] \neq \emptyset}
\end{align}
\end{definition}
Thus this set consists of time series $\traj \omega$, such that there are orbits of points $x\in\pspace$ which are at any point in (or at least infinitely close to) the partition elements labelled by the indexed elements of $\traj \omega$.
Adler proved that $\subshift^\integers$ defined in equation (\ref{eq:defn-subshift-topological}) is indeed a subshift \cite{Adler1998}.
Further, because $\cell_\omega \subseteq \bar{\mathring{\cell_\omega}}$ it contains the subshift $\subshift^\integers_\partmap\subset\subshift^\integers$.
Hence, the corresponding extended $\sigma$-algebra $\salg^\integers := \alphabet^\integers\vert_{\subshift^\integers}$ contains $\salg^\integers_\partmap$.
In the following, when talking about the stochastic symbolic dynamics of observed time series, we will always mean the measurable dynamical system $(\subshift^\integers,\salg^\integers,\shiftmap)$.
Note, that the first statement of Lemma~\ref{theo:measurable} also holds true for the slightly extended measurable symbolic dynamics:
If $\parti$ is generating, the image of any measurable set $\pset \in\palg$ is measurable in $\salg^\integers \supset\salg^\integers_\partmap$.
We will make use of that fact in Section~\ref{sec:phase-space}.

\begin{remark}
  In Ref.~\cite{Adler1998} the notion of a generating partition is purely topological.
  However, if $\parti$ is generating with respect to the Borel $\sigma$-algebra on a metric space in the sense of Def.~\ref{def:generating-partition}, then the topological partition $\parti\tsub{top}=\mathring\parti$ is generating in the topological sense.
  For completeness, we mention that in analogy to the example presented in Figure~\ref{fig:mult2-tent}, a factor map $\factormap\colon \subshift^\integers \to \pspace$ can be constructed.
  It maps the time series $\traj \omega \in \subshift^\integers$ to the unique element $x \in \cell[\traj \omega]$ and commutes with the dynamics, \ie $\Phi \circ \factormap = \factormap \circ \shiftmap$, \cf Ref.~\cite{Adler1998}.
\end{remark}

Now we can finally define a Markov partition:
\begin{definition}[Markov partition]
  \label{def:markov-partition}
  Let $\parti$ be a partition, $\mathring{\parti}$ the topological partition containing the interiors of sets in $\parti$, and $\subshift^\integers \subset \alphabet^\integers$ be the subshift generated by $\mathring{\parti}$ via Definition~\ref{def:subshift-topological}.
  
  Then, if $\subshift^\integers$ (and hence also its projection to positive times, $\subshift^\naturals$) is a subshift of finite type, we call $\parti$ (or $\mathring{\parti}$) a (topological) \emph{Markov partition}.
\end{definition}

Although our definition is equivalent to Adler's \cite{Adler1998}, it is different from previous definitions~\cite{Sinai1968,Bowen1970}.
In those references, a Markov partition is also always a topological generator, \ie a partition that generates the Borel $\sigma$-algebra.
For the remainder of the present chapter, we will work under the assumptions of Definition~\ref{def:markov-partition} with a Markov partition $\parti$.
However, we do not demand that $\parti$ is generating --- neither in the topological, nor in the measure-theoretic sense.

Let us summarize the results of this section.
For a dynamical system $(\pspace,\palg,\Phi)$ we have introduced the notion of a generating partition.
If we have such a generating partition, the images of measurable sets $\pset \in \palg$ under $\mtraj$ are measurable sets in the subshift $\subshift^\integers_\partmap$.
From any partition, we can obtain a topological partition consisting of the interiors of its elements.
Via Def.~\ref{def:subshift-topological} a topological partition defines a subshift $\subshift^\integers \supset \subshift^\integers_\partmap$, which contains some additional sequences corresponding to orbits that move along the boundaries of partition elements.
In the next section, we will assume that we have a Markov partition.
Thus the dynamical system $(\subshift^\integers,\salg^\integers,\shiftmap)$ is a measurable dynamical system on a SFT.
Then, the associated (topological) partition on phase space allows for a \emph{stochastic} symbolic dynamics with Markovian statistics for its time series.

\section{Markov measures}
\label{sec:markov-measures}
In this section, we define probability measures $\mms$ on SFTs $\subshift^\timdom$, such that the probabilities of $t$-shifted forward cylinders, $\prob\tind{t}[\traj \omega \rlind{\tau}] := \mms(\cyl\tind{t}[\traj \omega\rlind{\tau}])$ obey the Markov property (\ref{eq:markov-property}) for $t\in\timdom$.
We will see that there is a non-trivial difference between the case of $\timdom = \naturals$ and $\timdom = \integers$ when it comes to transient, \ie not shift-invariant measures.

\subsection{Markov measures for semi-infinite sequences}
In the present subsection, we are only interested in the forward dynamics, \ie we set $\timdom = \naturals$ and consider semi-infinite time series.
Hence, we always consider the $\sigma$-algebra $\salg^\naturals$ on $\alphabet^\naturals$.
\begin{definition}[Markov measure]
  \label{def:one-sided-mms}
  Let $\subshift^\naturals \subset \alphabet^\naturals$ be an SFT with adjacency matrix $\adjm$, $\tau \in \naturals$ and $\cyl\tind{0}[\traj \omega\rlind{\tau}]$ be a forward cylinder.
  Further, let $\tmat = \ifam{\tprob{\omega}{\omega'}}_{\omega,\omega' \in \alphabet}$ be a stochastic matrix compatible with $\adjm$ and $\bvec p_0 = \ifam{\p{0}{\omega}}_{\omega \in \alphabet}$ a stochastic vector.
  A measure $\fmms=\fmms(\tmat,\bvec p_0)$ on $(\alphabet^\naturals,\salg^\naturals)$ obeying
  \begin{align*}
    \prob\tind{0}[\traj \omega \rlind{\tau}] \equiv \fmms(\cyl\tind{0}[\traj \omega\rlind{\tau}]) = \p{0}{\omega_0}\prod_{k=1}^{\tau}\tprob{\omega_{k-1}}{\omega_k}
  \end{align*}
  is called a (one-sided) \emph{Markov measure}.
\end{definition}
In Appendix~\ref{app:construction-mms} we use the Daniell--Kolmogorov extension theorem to prove that this measure exists and is unique for any given choice of $\tmat$ and $\bvec p_0$.
Note that the forward arrow on $\fmms$ is not a vector arrow.
Rather, it indicates that the Markov measure is a ``forward measure'', \ie it is defined for the \emph{semi-infinite} sequences extending in forward time.
Further, in Appendix~\ref{app:construction-mms} we proof the following property of the Markov measure:
\begin{proposition}
  \label{theo:pushforward}
  Let $\fmms\tind{0} \equiv \fmms(\tmat, \bvec p_0)$ be a Markov measure.
  Then, for $t \in \naturals$
  \begin{align*}
    \fmms\tind{t} := s^t_*\fmms\tind{0} = \fmms(\tmat,\bvec p_t)
  \end{align*}
  where $\bvec p\tind{t} = \bvec p\tind{0} \tmat^t$.
\end{proposition}

Thus, the Markov measure $\fmms\tind{\infty} = \fmms(\tmat,\bvec p\tind{\infty})$ is stationary if and only if the stochastic vector $\bvec p\tind{\infty}$ is a left eigenvector of $\tmat$.
If the SFT $\subshift^\naturals$ is \emph{irreducible} (\ie if its adjacency matrix $\adjm$ is that of a strongly connected graph), $\bvec p\tind{\infty}$ and thus $\fmms\tind{\infty}$ are uniquely determined by a compatible transition matrix $\tmat$.
If the SFT is additionally aperiodic,  $\lim_{t \to\infty}(\fmms\tind{t}) = \fmms\tind{\infty}(\tmat,p_{\infty})$ exists and is the same for \emph{any} (transient) Markov measure.
Under these constraints, $\fmms\tind{\infty}$ was first introduced by W.~Parry \cite{Parry1964}.
He further showed that there is a unique stationary Markov measure $\fmms\tind{\infty}\tsup{P}$, which maximizes the Kolmogorov--Sinai entropy \eqref{eq:KS-entropy} of the measurable dynamical system $(\subshift^\naturals,\salg^\naturals,\shiftmap)$.
This measure has become known as the \emph{Parry measure}. 

\subsection{Markov measures for bi-infinite sequences}
A stationary Markov measure can be extended to a stationary Markov measure on $\subshift^\integers$ by slightly changing definition \ref{def:one-sided-mms}:
\begin{definition}[Stationary Markov measure on the bi-infinite sequences]
  \label{def:stationary-mms}
  Let $\subshift^\integers \subset \alphabet^\integers$ be an SFT with an adjacency matrix $\adjm$.
  For $t \in \integers$ let $\cyl\tind{t}[\traj \omega\rlind{\tau}]$ be a $t$-shifted forward cylinder.
  Further, let $\tmat = \ifam{\tprob{\omega}{\omega'}}_{\omega,\omega' \in \alphabet}$ be a stochastic matrix compatible with $\adjm$ and $\bvec p\tind{\infty} = \ifam{\p{\infty}{\omega}}_{\omega \in \alphabet}$ a stochastic left eigenvector of $\tmat$.
  A measure $\imms=\imms(\tmat,\bvec p\tind{\infty})$ obeying
  \begin{align*}
    \prob\tind{t}[\traj \omega \rlind{\tau}] \equiv \imms(\cyl\tind{t}[\traj \omega\rlind{\tau}]) = \p{\infty}{\omega_0}\prod_{k=1}^{\tau}\tprob{\omega_{k-1}}{\omega_k}
  \end{align*}
  is called a \emph{stationary Markov measure for the bi-infinite shift}.
\end{definition}
Again, this measure exists, is unique and shift-invariant (\cf appendix \ref{app:construction-mms}).
If $\subshift^\integers$ is irreducible, $\fmms\tind{\infty}$ is uniquely determined by $\tmat$.

Apart from stationary Markov measures, there are other \emph{non}-stationary extensions to the bi-infinite sequences.
Maybe surprisingly, it turns out that any non-stationary extension to the full shift violates the Markov property (\ref{eq:markov-property}) for $t$ smaller than some finite $t_0 \in \integers$.
This can be easily understood from the master equation (\ref{eq:master-equation}), which is a consequence of the Markov property. 
It states that the probability vectors $\bvec p := \bvec p\tind{t}$ and $\bvec p' := \bvec p\tind{t+1}$ must obey $\bvec p' =\bvec p \tmat $ independently of $t$.
Clearly, this condition is fulfilled for the stationary Markov measure, \ie when $\bvec p = \bvec p'$ is a left unity-eigenvector of $\tmat$.

It is instructive to attempt a construction of a non-stationary Markov measure for the full shift in the same spirit as the one-sided Markov measure.
Then, we can use the master equation to see where and why such a measure fails to fulfil the Markov property for times lower than some finite time $t_0$:
Generically, a stochastic matrix $\tmat$ is invertible.
Hence, $\tmat^{t}$  and thus $\bvec p\tind{t} := \bvec p \tmat^t$ is well-defined for all $t\in\integers$ for any (initial) stochastic vector $\bvec p\tind{0}$.
At first glance, this might seem like a reasonable ansatz.
However, in general the inverse of a stochastic matrix is itself not a stochastic matrix,\footnote{
This is only the case if $\tmat$ is both orthogonal and double stochastic.
Then, the transposed matrix $\tmat\tsup{T}$ is its inverse.}
with the consequence that $\bvec p\tind{t}$ for $t<0$ is not a stochastic vector any more.
Generally, $\tmat^{-1}$ is not even positive semi-definite.
Thus, $\bvec p\tind{t}$ obtains negative entries for some $t< t_0\leq 0$.
The same problem persists also for singular stochastic matrices, given that solutions to the master equation can be found at all.

This leads to an important insight:
The \emph{only} stationary measure on the full subshift that obeys the master equation (\ref{eq:master-equation}) for all times is a stationary Markov measure.
We will provide a physical interpretation of this statement in Section~\ref{sec:operational}.

Let us finally discuss another kind of extension of the Markov measure to the full shift, which we will refer to as the \emph{two-sided Markov measure}.
Such measures can be used to formalize the notion of a stochastic backward process.
A backward process specifies the dynamics one would observe if time were to run backward after the initialization at the initial time $t_0 = 0$.
The notion of a backward process is crucial for a general formulation of stochastic fluctuation relations \cite{Seifert2012}.
In order to define a backward process, we give the following definition:
\begin{definition}
  Let $\traj \omega\rlind{\tau}=(\omega_0,\omega_1,\cdots,\omega_\tau) \in \ospace^{\tau+1}$ be a finite time series of length $\tau+1$ and $t \in \timdom$.
  We define the \emph{$t$-shifted backward cylinder} for $\traj\omega\rlind{\tau}$ as
  \begin{align}
    \back\cyl\tind{t}[\traj \omega\rlind{\tau} ]&:= \bigcap_{k=0}^\tau \proj_{t-k}^{-1}\set{\omega_k}\\
    &\equiv \set{\traj \nu \in \ospace^\timdom \,\vert\, \nu_{t-k} = \omega_k, \, 0\leq k \leq \tau}.
   \label{eq:backward-cylinder-subshift}
 \end{align}
 The measure of such a cylinder, $\back\prob\tind{t}[\traj \omega \rlind{\tau}] = \prob(\back\cyl\tind{t}[\traj \omega\rlind{\tau}])$, is called the \emph{probability} for the finite time series $\traj\omega\rlind{\tau}$ to \emph{occur in reverse direction} at time $t$.
\end{definition}
Note that, by definition, we have $\cyl\tind{t}[\traj \omega\rlind{\tau} ] = \back\cyl\tind{t+\tau}[\revop \traj \omega\rlind{\tau} ]$ where $\revop\colon\alphabet^{\tau+1}\mapsto \alphabet^{\tau+1}$ is the reversal operator mapping $\ifam{\omega_k }_{0\leq k \leq\tau}\mapsto \ifam{\omega_{\tau-k} }_{0\leq k \leq\tau}$.

Using the backward cylinders, we define:
\begin{definition}[Two-sided non-stationary Markov measure]
  \label{def:two-sided-mms}
  Let $\subshift^\integers \subset \alphabet^\integers$ be an SFT with adjacency matrix $\adjm$.
  Let $\tau \in \naturals$ and let  $\cyl\tind{0}[\traj \omega\rlind{\tau}]$ and $\back\cyl\tind{0}[\traj \omega\rlind{\tau}]$ be forward and backward cylinders, respectively.
  Let $\tmat = \ifam{\tprob{\omega}{\omega'}}_{\omega,\omega' \in \alphabet}$ and $\back \tmat= \ifam{\btprob{\omega}{\omega'}}_{\omega,\omega' \in \alphabet}$ be stochastic matrices compatible with $\adjm$ and $\adjm^\mathrm{T}$, respectively.
  Let $\bvec p\tind{0} = \ifam{\p{0}{\omega}}_{\omega \in \alphabet}$ be a stochastic vector.
  A measure $\fbmms=\fbmms(\tmat,\back\tmat,\bvec p\tind{0})$ obeying
  \begin{align*}
    \prob\tind{0}[\traj \omega \rlind{\tau}] \equiv \fbmms(\cyl\tind{0}[\traj \omega\rlind{\tau}]) =
      \p{0}{\omega_0}\prod_{k=1}^{\tau}\tprob{\omega_{k-1}}{\omega_k}\\
    \back \prob\tind{0}[\revop \traj \omega \rlind{\tau}] \equiv \fbmms(\back\cyl\tind{0}[\revop\traj \omega\rlind{\tau}]) =
      \p{0}{\omega_0}\prod_{k=1}^{\tau}\btprob{\omega_{k}}{\omega_{k-1}}
  \end{align*}
  and whenever $\cyl[\traj \omega\rlind{\tau},\revop\traj \nu\rlind{\tau}] := \cyl\tind{0}[\traj \omega\rlind{\tau}]\cap   \back\cyl\tind{0}[\revop\traj \nu\rlind{\tau}] \neq \emptyset$:
  \begin{align*}
    \fbmms\left(\cyl[\traj \omega\rlind{\tau},\revop\traj \nu\rlind{\tau}]\right)=      \p{0}{\omega_0}  \prod_{k=1}^{\tau}\left[\tprob{\omega_{k-1}}{\omega_k}\right]               \prod_{k=1}^{\tau}\left[\btprob{\nu_{k}}{\nu_{k-1}}\right]
  \end{align*}
  is called a \emph{two-sided non-stationary Markov measure for the bi-infinite shift}.
\end{definition}
In Appendix~\ref{app:construction-mms} we proof that this measure exists and is unique.
The following proposition allows for a more intuitive definition of the two-sided Markov measure:
\begin{proposition}
  Let $\subshift^\integers$ be a subshift and $\subshift^+ = \proj_{\naturals \leftarrow \integers}\subshift^\integers \subset \alphabet^\naturals$ its restriction to the forward sequences.
  For a measurable set $\sset \in \salg^\integers$ define the projections $\sset^+ := \proj_{\naturals \leftarrow \integers} \sset$ and  $\sset^- := \proj_{-\naturals \leftarrow \integers} \sset$ onto the forward and backward trajectories, respectively.
  Then,
  \begin{align*}
    \fbmms(\sset) = \frac{\fmms^{_+}(\sset^{+})\,\fmms^{_-}(\revop\sset^{-})}{\fmms^{_+}(\sset^{0})}.
  \end{align*}
  where  $\fbmms = \fbmms(\tmat,\back\tmat,\bvec p\tind{0})$ is a two-sided Markov measure and the measures $\fmms^{_+} = \fmms(\bvec p \tind{0},\tmat)$ and $\fmms^{_-} = \fmms(\bvec p\tind{0},\back\tmat)$ are Markov measures on $\subshift^+$ and $\subshift^- = \revop \subshift^+$, respectively.
  \label{theo:fbmms-alternative}
\end{proposition}

The stationary Markov measure and the two-sided Markov measure are just two examples of an extension to the full shift.
Many other non-stationary extensions of one-sided Markov measures are possible.
However, for the purpose of this work we only consider the stationary and the two-sided Markov measure on the full shift.
In the next subsection, we will finally connect Markov measures on $\subshift^\integers$ to a dynamical system on phase space.

\subsection{(Natural) Markov measures on phase space}
\label{sec:phase-space}
%

For the remainder of this subsection, let $\parti = \partmap^{-1}\alphabet$ be a generating Markov partition for $(\pspace,\palg,\Phi)$.
Then, Lemma~\ref{theo:measurable} ensures that the image of any measurable set $\pset \in \palg$ is measurable in $\salg^\integers$.
Thus, any (Markov) measure $\mms$ on $(\alphabet^\integers,\salg^\integers)$ yields a measure $\mms_\pspace := \mms \circ \mtraj$ on $(\pspace,\palg)$.
More precisely, we consider the push-forward measures $\mtraj^{-1}_*\fmms\tind{\infty}$ and $\mtraj^{-1}_*\fbmms$ of the stationary and the two-sided Markov measure, respectively:
\begin{definition}[Markov measure on phase space]
  Let $(\pspace,\palg,\Phi)$ be a measurable dynamical system.
  Further, let $(\subshift^\integers,\salg^\integers,\shiftmap)$ be the symbolic dynamical system obtained from a generating Markov partition $\parti = \partmap^{-1}(\ospace)$.
  Let $\fmms(\tmat,\bvec p\tind{\infty})$ and $\fbmms(\tmat, \back\tmat, \bvec p\tind{0})$ be stationary and two-sided Markov measures on $\subshift$.
  Then, we call
  \begin{align*}
    \fmms^\pspace(\tmat,\bvec p\tind{\infty}) &:= \mtraj^{-1}_*\fmms(\tmat,\bvec p\tind{\infty})\text{ and}\\
    \fbmms^\pspace(\tmat, \back\tmat, \bvec p\tind{0}) &:= \mtraj^{-1}_*\fbmms(\tmat, \back\tmat, \bvec p\tind{0})
  \end{align*}
  \emph{stationary and two-sided Markov measures on phase space}, respectively.  
  \label{def:ps-mms}
\end{definition}
We are mostly interested in Borel measures on phase space.
In that case, the topological and measure-theoretic notion of a generating partition agree.%
\footnote{Note that in Def.~\ref{def:ps-mms} we do not require that the partition generated the Borel sets.}

As of now, we have not required anything else of $\tmat$ other than that it is compatible with the adjacency matrix $\adjm$ of $\subshift^\integers$.
In the case where $\subshift^\integers$ is obtained from a phase-space dynamics, $\adjm$ only contains information about the topological structure of the latter.
However, we would like to identify a natural transition matrix in order to capture the natural behaviour one would expect for real experiments.
To that end, we first need to establish what we mean by a ``natural'' behaviour for a physical dynamical system.
In fact, the concept of a \emph{natural measure} already exists and does express exactly what we are looking for \cite{Young2002,Blank+Bunimovich2003}:

\begin{definition}[Natural measure]
 A probability measure $\pms_\Phi$ is called \emph{natural} (with respect to $\Phi$ and the reference measure $\ms$) if there exists an open subset $U \in \topo$ such that for any absolutely continuous measure $\pms\ll\ms$ with support $U_\pms \subset U$ we have:
  \begin{align*}
    \frac{1}{\tau}\sum_{t=0}^\tau (\Phi^t)_*\pms \stackrel{\tau\to\infty}{\rightarrow}\pms_\Phi .
  \end{align*}
\end{definition}
In other words, $\pms_\Phi$ is a stable fixed point of the push-forward operator $\Phi_*$.
It can be constructed from the action of $\Phi_*$ on absolutely continuous measures.
The  open subset $U$ is called the \emph{basin of attraction} of $\pms_\Phi$:
All measures with support in $U$ are ``attracted'' towards $\pms_\Phi$ as time passes.
In the following, we take the Lebesgue measure as a reference, \ie $\ms=\lms$.
Hence, the above definition means that all measures which admit a probability density in $U$ ``average'' to $\pms_\Phi$. 

Let us now assume that $U=\pspace$, \ie the measure has a basin of attraction that is the whole phase space.
Further, we assume that the measure is ergodic.
Under these assumptions, we define the \emph{natural transition matrix}:
\begin{definition}[Natural transition matrix]
  \label{def:natural-tmat}
  Let $(\pspace,\palg,\Phi)$ be a measurable dynamical system with a Markov partition $\parti=\ifam{\cell_\omega}_{\omega \in \alphabet}$.
  Let $\mu_\Phi$ be an ergodic natural measure with basin of attraction $U = \pspace$.
  Then, the matrix $\ptmat = \ptmat(\pms_\Phi,\parti)$ with entries
  \begin{align*}
    \ptprob{\omega}{\omega'} := \frac{\pms_\Phi(\Phi^{-1}\cell_{\omega'} \cap \cell_{\omega}) }{\pms_\Phi(\cell_\omega)}
  \end{align*}
  is called the \emph{natural transition matrix}.
\end{definition}
Ergodicity of $\mu_\Phi$ together with the fact that $\Phi$ is onto, ensures that $\ptmat$ is also ergodic and hence irreducible and aperiodic.
Thus, there exists a unique \emph{natural stationary measure} $\fmms_\pspace := \mtraj^{-1}_*\fmms(\ptmat)$.

In Section~\ref{sec:nmbm}, we encounter a special natural Markov measure on phase space defined for invertible dynamics $\Phi$.
Recalling Proposition~\ref{theo:stochastic-symbolic-dynamics-subshift}, denote by $\back\mtraj:= \left(\partmap\circ\Phi^{-t}  \right)_{t\in\timdom}$ the trajectory map of the stochastic symbolic dynamics generated by $\Phi^{-1}$.
Further, the (topological) partition associated to $\partmap$ defines the subshift $\back\subshift \supset \back\subshift_{\partmap} := \back\mtraj(\pspace)$.
Note that the latter contains all the sequences of the subshift $\subshift\supset\subshift_\partmap = \partmap(\pspace)$  in reverse order, \ie $\back \subshift = \revop(\subshift)$.
In particular, if $\subshift$ is an SFT with adjacency matrix $\adjm$, $\revop \subshift$ is an SFT characterized by $\adjm^\mathrm{T}$.

Hence, if $\parti$ is a (generating) Markov partition for $\Phi$ it is also one for the inverse $\Phi^{-1}$.
Thus, if the natural transition matrix $\ptmat$ for $\Phi$ exists, so does the natural transition matrix $\back \ptmat$ defined using the natural measure $\pms_{\Phi^{-1}}$.
This enables the definition of the \emph{natural two-sided Markov measure} for invertible $\Phi$:
\begin{definition}
  \label{def:natural-two-sided-mms}
  Let $\ptmat$ and $\back \ptmat$ be the natural transition matrices with respect to $\Phi$ and $\Phi^{-1}$, respectively.
  Then, the measure $\fbmms(\ptmat,\back \ptmat,\bvec p_0)$ and its phase space analogue, $\fbmms^\pspace(\ptmat,\back \ptmat, \bvec p_0)$ are called the \emph{natural two-sided measures with the initial condition $\bvec p_0$} on the symbolic and phase space dynamics, respectively.
\end{definition}
The natural two-sided Markov measure has already been used implicitly in the literature on so-called \emph{multibaker maps} \cite{Gaspard2005,Vollmer_etal1997,Vollmer_etal1998,Breymann_etal1998,Tel+Vollmer2000,Vollmer2002,Colangeli_etal2011}.
In Chapter~\ref{chap:information-st} we discuss the role of the two-sided natural measure for \emph{network multibaker maps}, which are the most general variant of multibaker maps.
Additionally, we discuss the (stochastic) thermodynamic properties of and the notion of entropy for such maps in a general information-theoretic framework.

\section{Discussion}
\label{sec:mms-discussion}
After the technical considerations above, we put our findings in the context of previous mathematical and physical results.
Firstly, we revisit the concept of natural measures $\pms_\Phi$, which we used to define the natural transition matrix.
In particular, we discuss their relation to the natural stationary Markov measure $\fmms\tind{\infty}(\ptmat,\bvec p\tind{\infty})$.

After that, we relate our abstract results to the \emph{operational} framework outlined at the beginning of Section~\ref{sec:symbolic-stochastic-dynamics}.
There, we see how our abstract mathematical results obtain natural physical interpretations.

\subsection{Connection to ergodic theory}
\label{sec:discussion-math}
An ergodic natural measure with full support on the whole phase space is the attractor of measures that are absolutely continuous with respect to the Lebesgue measure.
Then, any probability measure with a density will converge to the natural measure.
Further, the natural transition matrix is well-defined, \cf Def.~\ref{def:natural-tmat}.
Note that generically the natural measure is \emph{not} absolutely continuous.

For the remainder of this subsection, we stick with the above assumptions.
Then, the Markov chain defined by the natural transition matrix is both ergodic and aperiodic with unique stationary distribution $\bvec p\tind{\infty}$.
Hence, any extensions of a one-sided natural Markov measure eventually converges to the unique stationary Markov measure $\fmms\tind{\infty}(\tmat):=\fmms\tind{\infty}(\tmat,\bvec p\tind{\infty})$.

Let us now assume that the partition $\parti$, which we use to define the natural transition matrix, generates the Borel $\sigma$-algebra.
Further, let $\mms$ be an extension of a (non-stationary) natural one-sided Markov measure $\mms(\ptmat,\bvec p)$.
In that case, $\mms^\pspace:= \mms\circ\mtraj$ is a Borel measure on phase space.
Two questions naturally arise:
\begin{itemize}
  \item What is the relation of the natural measure $\pms_\Phi$ and
    \begin{align*}
      \mms^\pspace\tind{\infty} := \lim_{t\to\infty} (\Phi^t)_*\mms^\pspace,
    \end{align*}
    given the latter exists?
  \item Under which conditions is a non-stationary measure $\mms^\pspace$ absolutely continuous with respect to the Lebesgue measure?
\end{itemize}

Let us first assume a positive answer to the second question.
In that case, $\mms^\pspace$ necessarily converges to the natural measure under push-forward by $\Phi$ and we have $\mms^\pspace\tind{\infty}=\pms_\Phi$.
Moreover, $\mms$ converges to the unique natural stationary measure $\fmms\tind{\infty}(\ptmat)$ defined on the whole shift, \ie for $\timdom =\integers$.
Because $\mtraj \circ \Phi = \shiftmap \circ \mtraj$, we find that the natural stationary measure on phase space actually \emph{agrees} with the natural measure, \ie
\begin{align}
  \fmms^\pspace\tind{\infty}(\ptmat) := \fmms\tind{\infty}(\ptmat)\circ\mtraj =  \mms\tind{\infty}^\pspace = \pms_\Phi. 
  \label{eq:natural-measure-agrees}
\end{align}

However, this fact crucially depends on the fact that $\mms^\pspace$ was absolutely continuous in the first place.
Remember that we demanded that $\mms^\pspace$ is a pull-back to phase space of an (arbitrary) extension of a natural one-sided measure $\fmms(\ptmat)$.
The latter is defined on the symbolic $\sigma$-algebra $\salg^\naturals$ for the forward sequences.
This $\sigma$-algebra uniquely defines the restricted $\sigma$-algebra $\palg^\naturals = \mtraj^{-1}(\salg^\naturals)$.

The most refined sets in $\palg^{\naturals}$ are the pre-images of semi-infinite sequences $\traj\omega \in \alphabet^\naturals$ under $\mtraj$.
That is, they are sets containing points with the same \emph{symbolic future}.
Such sets lie on the same (local) stable manifold.
Specifying a measure on $\palg^\naturals$ can thus be understood at defining a \emph{marginalized} measure, where the local stable manifold is integrated out.
What remains is the topological structure that governs the \emph{future} of points, which are also known as the (local) unstable manifolds.
One also says that a measure on $\palg^\naturals$ defines a \emph{conditional measure on the unstable manifolds} \cite{Young2002}.

The stable and unstable manifolds always intersect transversally.
In a sense, they define a curvilinear transversal coordinate system \cite{Adler1998}.
Extending a one-sided Markov measure to the bi-infinite sequences thus means specifying a structure along the stable directions.
Or to put it differently:
Specifying the \emph{past} of a system defines a density along the stable directions, specifying the \emph{future} does the same along the unstable directions.

Let us now choose the measure $\mms$ as an extension of $\fmms(\ptmat)$ such that along the stable direction we have a density with respect to the (restricted) Borel measure.
Then, the question whether $\mms$ converges to the natural measure can be reformulated:
Is the measure $\fmms^\pspace(\ptmat)$ (restricted to $\palg^\naturals$) absolutely continuous with respect to $\lms$ (restricted to $\palg^\naturals$)?
Or equivalently:
Does the measure $\mms^\pspace$, constructed as an extension of a \emph{natural} one-sided Markov measure, have absolutely continuous conditional measures on the unstable manifolds?

A positive answer to that question for \emph{invariant} measures defines (a generalization) of a so-called SRB measure \cite{Young2002}.
The latter were discovered by Sinai, Ruelle and Bowen in 1970s in the context of Anosov systems and Axiom-A flows \cite{Sinai1972,Bowen+Ruelle1975}.
The natural measure has been proposed to be the proper generalization of the SRB measure in a general context \cite{Young2002,Blank+Bunimovich2003}.

These facts suggest that it is useful to think of natural measures as stationary Markov measures defined for a generating Markov partition.
In fact, the SRB measure can be defined using generating Markov partitions for hyperbolic systems~\cite{Sinai1972,Gallavotti+Cohen1995}.
Hence, it seems natural to ask whether under the present assumptions, the invariant natural measure on phase space $\mms\tind{\infty}$ \emph{agrees} with the natural measure $\pms_\Phi$.

What we know is that both measures (by definition) agree on sets of the form $\pcyl\tind{t}[(\omega,\omega')] \equiv \pcyl\tind{t}[\omega] \cap \pcyl\tind{t+1}[(\omega')]$.
However, as the family of sets
\begin{align*}
  \set{\pcyl\tind{t}[(\omega,\omega')] \,\middle\vert\,t \in \integers,\, \omega,\omega'\in\alphabet}
\end{align*}
is not closed under intersections, this does not imply that the measures are the same, \cf Lemma~\ref{theo:pi-system}.

Further, it is possible to construct ergodic invariant measures with full support on $\pspace$ which are not (pull-backs) of Markov measures.
The easiest example is provided by considering the pull-back of a stationary $m$-step Markov measure \cite{Bowen+Chazottes1975}.
The latter is a Markov measure constructed for an alphabet $\alphabet^m$, where the symbols are allowed blocks of $m$ symbols that may appear in some 1-step SFT.

Although the author did not find a rigorous proof during the preparation of this thesis, he believes that under the present assumptions  $\mms\tind{\infty}$ agrees with $\pms_\Phi$.
If this is in fact true, we would have an alternative approach to the natural measure.
That said, we continue with a more physics-motivated discussion of our results.

\subsection{Operational interpretation}
\label{sec:operational}
In this subsection, we discuss our findings in the light of the experimental situation outlined at the beginning of Section~\ref{sec:symbolic-stochastic-dynamics}.
In particular we consider the situation where an experimenter records time series which appear to obey Markovian statistics.
As we have seen in Section \ref{sec:stochastic-models}, such experimental situations are often modelled as Markovian stochastic processes.
In the remainder of this section we discuss the implicit assumptions on the microscopic dynamics, the microscopic ensemble and the nature of the observable in the light of our mathematical treatment above.

Before we continue, let us review the assumptions of our idealized framework of the measurement process:
The map $\Phi$ is understood as a stroboscopic map obtained from the real dynamics which proceeds in continuous time.
To ensure that an autonomous (\ie time-independent) dynamics $\Phi$ is a good approximation of the real situation, we require a huge temporal precision on the (stroboscopic) measurement intervals.
Further, the measurement needs to be either perfectly reproducible (because then the disturbance of the system by the measurement apparatus is the same at each iterated measurement) or completely interaction free.

Obviously, neither of these conditions will ever be rigorously fulfilled.
However, we are still able to have these requirements fulfilled in a \emph{gedankenexperiment}, and look for non-trivial interpretations.

\subsubsection{Observables, partitions and the measurement process }
We start by discussing the nature of observables and the corresponding partitions of phase space.
For the moment suppose that we know the microscopic dynamics $\Phi$ and we also have a theoretical model (\ie a well-defined observable $\partmap$) for a real measurement apparatus.
Generically, such an observable will not induce a Borel-generating Markov partition on phase space.
Just by considering the resolutions which are available in typical experiments, we would expect the partition to be far too coarse to generate the Borel-sets.
However, no one can ever say if the statistics of experimentally observed time series \emph{really} are Markovian.
The only thing we might be able to say is that they \emph{appear} memoryless within the finite time span of a given experiment.

\paragraph{Generating partitions and measurability}
Let us continue with the notion of ``measurability'', both in the abstract mathematical and the operational sense.
By the former we mean the concept of a $\sigma$-algebra on phase space, whereas by the latter we mean the experimental observation of time series, \ie the recording of subsequent (elementary) measurement results $\omega$.
Generating partitions (in the sense of the measure-theory) provide the connection between these two notions.
The advantage of the measure-theoretic Definition~\ref{def:generating-partition} over the topological one is that it holds for arbitrary $\sigma$-algebras $\palg$ on phase space.
In contrast to the topological definition, we do not require $\palg$ to be as refined as the Borel sets.
All that is needed is that $\palg$ agrees with the $\sigma$-algebra $\sigma(\pcyls\tsup{a})$ generated by the atomic cells $\pcyls\tsup{a}$.

To appreciate the operational interpretation of $\sigma(\pcyls\tsup{a})$, note that an atomic cell $\cell_t[(\omega)]$ contains exactly the points in phase space, which yield the measurement result $\omega$ at time $t$.
The generated $\sigma$-algebra $\sigma(\pcyls\tsup{a})$ thus contains countable unions and intersections of these ``elementary events''.
Measurable sets $\sset \in \sigma(\pcyls\tsup{a})$ thus describe events which can be formulated as (countable) Boolean statements about these elementary events.
For instance, $\sigma(\pcyls\tsup{a})$ contains events like ``We observe the same measurement result $\omega$ for each time in the discrete interval $\ifam{t_0,t_0+1,\cdots,t_0+\tau}$'' or ``Neither result $\omega$ nor $\omega'$ do occur at time $t=t_0$''.
Also statements about infinite time series like ``The measurement result $\omega$ never appears after some time $t=t_0$'' are possible.

If a partition is generating, then by definition $\sigma(\pcyls\tsup{a})=\palg$.
This means that the ``usual'' descriptions of operationally accessible measurement events (like the examples given above) agree with the mathematically measurable sets on phase space.
Or to put it differently:
If a partition is generating, then all events that can be described by Boolean predicates are measurable in the mathematical sense.
In fact, it is hard to describe an event that is not an element of the $\sigma$-algebra generated by the elementary measurement results.

If we understand measurability in that operational sense, it seems sensible to \emph{choose}  $\palg:=\sigma(\pcyls\tsup{a})$ as the measurable structure on phase space.
Then, the mathematical and operational notions of measurability coincides and \emph{every} partition induced by \emph{any} observable is generating.

However, there is a caveat to this interpretation:
Usually we have an initial time $t=0$ where we prepare the system and then record  only for $t\geq0$.
Hence, to be precise we must consider the partition generated by the atomic cells in forward time only.
It is worthwhile to mention the following consequence of a theorem by Kolmogorov--Sinai for invertible $\Phi$:
The mere existence of a one-sided generator for a dynamical system $(\pspace,\palg,\Phi)$ implies that its dynamical notion of entropy (\cf Sec.~\ref{sec:topent-ksent}) vanish identically.

Finally, consider the case where one is fortunate enough to have a high-resolution measurement apparatus whose phase space partition generates the Borel $\sigma$-algebra.
In that case, single points in phase space are represented by elements in the $\sigma$-algebra.
Unfortunately, these events are statements about infinite time series and hence not operationally accessible:
Their recording might surpass the time scale of a typical graduate student \ldots

\paragraph{Markov partitions and local equilibrium}
Next, let us discuss the Markovian postulate.
We have already seen that this implies a topological constraint on the structure that the observable induces on phase space:
The corresponding partition must be a Markov partition.
We stress again, that we do not require that this partition generates the Borel sets, \ie its elements do not need to be ``small''.

Hence, on first sight, this requirement seems not too restrictive.\footnote{
For instance, it would allow the trivial partition obtained by a constant observable $\partmap(x) = 1,\,\forall x$.}
However, let us come back to the example of the Bernoulli map, Fig.~\ref{fig:mult2-tent}a).
If we just slightly misplace the partition such that $\cell_0 = [\epsilon,\frac{1}{2})$ the sequence $\traj \omega = (0,0,\cdots)$ (which formerly belonged to the fixed point at $x=0$) would not be allowed.
We cannot expect that we are fortunate enough to obtain such an alignment in reality.

However, we observe that Markovianity holds (at least for all practical purposes) for many single-molecule experiments \cite{Seifert2011}.
As we have discussed in Section~\ref{sec:stochastic-models}, the loss of memory  is a necessary consequence of the physical assumption of \emph{local equilibrium}.
More precisely, it relies on the notion of an (infinite) \emph{separation of time scales}:
We assume that within  the stroboscopic measurement interval, the ensemble constrained to an observable (``local'') state (approximately) relaxes to a stationary\footnote{
  There is also the corresponding notion of a ``local steady state'', \cf \cite{Hatano+Sasa2001}, which might be more appropriate here.}
(``equilibrium'') distribution.
If there exists a stationary distribution which attracts most initial conditions, it necessarily loses the information of the latter.

Obviously, a system with a smooth evolution in continuous time never relaxes to such a stationary distribution in \emph{finite} times.
However, it might already ``mix'' points in phase space to a sufficient degree, such that for all practical purposes a stationary distribution is a valid approximation.
A form of this ``smearing out'' of  points over the elements of the partition also manifests in the \emph{$n$-fold intersection property} used by Adler for his (equivalent) definition of Markov partitions \cite{Adler1998}.

Finally, it is interesting to ask \emph{why} Markovian statistics are so commonly found in experiments.
In the final Chapter in Section~\ref{sec:ubiquity} we revisit the Markovian postulate as a  kind of ``anthropic principle'' imposed by the \emph{scientific method}~\cite{Popper2002}.

\subsubsection{The arrow of time and stationarity on two levels}
In the present subsection, we are concerned with reversibility and the \emph{arrow of time} as it appears on the two levels of description.
We start with the invertible microscopic evolution on phase space, where we demanded \emph{determinism}.
On that level, we have a symmetry with respect to the direction of time.
In principle, by applying a time-reversal operator (\cf Sec.~\ref{sec:reversibility}), one can always find the original ensemble at time $t_0$ from an evolved microscopic ensemble at time $t_0+\tau$.
In that sense, an \emph{arbitrary} (normalized) initial ensemble at time $t_0$ plays the same role as the push-forwarded ensemble at time $t_0 + \tau$.

In a popular article on the \emph{thermodynamic} arrow of time,  J.\,Lebowitz points out that the irreversibility on the observational level originates from particular choices of initial conditions \cite{Lebowitz1993}.
Here we extend this discussion to the Markovian postulate.
First of all, note that a Markovian dynamics breaks time-symmetry in a peculiar way:
Take an \emph{arbitrary} stochastic vector $\bvec p\tind{t_0}$ as an initial condition.
Because the transition matrix is stochastic, we know that any subsequent distribution $\bvec p\tind{t_0+\tau} = \bvec p\tind{t_0} \tmat^\tau$ is equally stochastic.
Further, as stochastic matrices are generically invertible, we can obtain $\bvec p\tind{t_0}$ from $\bvec p\tind{t_0+\tau}$.
However, if we use the inverse of $\tmat$ to obtain a vector $\bvec p\tind{t}$ for any $t<t_0$, we cannot be sure if this vector is still stochastic.
Moreover, one can always find a time $t_1 \leq t_0$ such that $\bvec p_t$ is not stochastic for $t<t_1$ --- unless $p\tind{t_0}=p\tind{\infty}$ is already a left unity eigenvector to $\tmat$.
Hence, unlike in the microscopic case, $t_0$ distinguishes a particular point in time where we know that $\bvec p\tind{t}$ is stochastic for $t \in [t,\infty)$.
Without loss of generality, in the remainder of this chapter we choose $t_0=0$ and refer to it as the \emph{present}.

The necessity of a ``present'', \ie a distinguished initial point in time, for generic Markovian dynamics manifests in the fact that there is no non-stationary measure that yields Markovian statistics for all $t\in\integers$.
Operationally, the initial point $t=0$ is distinguished by the \emph{preparation procedure}.
Hence, we interpret the mathematical statement in an operational way:
The manipulations on a system during the experimental preparation of a certain observable state are a non-Markovian process.

We also want to stress the distinction between microscopic and observational steady states.
First of all, it is clear that the former implies the latter.
However, we can only \emph{observe} the statistics of time series.
\emph{Physically}, the microscopic states should not play a role, though they might be very useful as tools in the mathematical description \cite{Ruelle1999,Ruelle2004}.
Moreover, we have no chance to prepare them even with the best experimental equipment, because in general they concentrate finite probability on infinitesimal regions in phase space.
Equally, as they are obtained only in the $t\to\infty$ limit, we can never wait long enough for such a state to evolve naturally.

Finally, let us stress that it is not needed to have microscopic stationarity for observational stationarity:
\emph{Any} non-stationary extension of a one-sided stationary Markov measure to the full shift yields observational stationarity.
Moreover, if the natural measure is an SRB measure (\ie it has absolutely continuous density along the unstable directions), we can initialize \emph{non-stationary} absolutely continuous measures yielding \emph{stationary} observational statistics.
In Section~\ref{sec:foundations-st} we discuss this result in the light of the foundations of the stochastic thermodynamics of Markov chains, \cf the introductory remarks in~Section~\ref{sec:discrete-st}.

\section{Summary}
In the present chapter, we have outlined a gedankenexperiment where an experimenter performs perfectly reproducible successive measurements on a physical system.
In many experimental situations, such observations seem to obey Markovian Statistics.
The main goal of this chapter was the rigorous discussion of the consistency of two common assumptions:
Firstly, the assumption of microscopic causality or determinism implies that the underlying microscopic dynamics is prescribed by a deterministic (measurable) dynamical system \cite{Penrose1970}.
Secondly, the Markovian postulate for the observed process is central for the foundations of (classical) statistical mechanics \cite{Penrose1970} as well as for modern stochastic thermodynamics \cite{Seifert2012}, \cf Sec.~\ref{sec:stochastic-models}.

In summary, the constraints imposed by the Markovian postulate are two-fold:
The partition of phase space induced by the observable must be a Markov partition in the sense of Definition~\ref{def:markov-partition}.
However, it does not need be a particularly fine partition.
More precisely, we do not require that the partition generates the Borel $\sigma$-algebra.

For observables which induce a Markov partition, we identified measures on phase space which yield Markovian statistics for the observed time series after a preparation at $t=0$.
We found that there is a large class of measures which fulfil this requirement.
Measures within that class differ in terms of their past, \ie the statistics of events happening at times $t<0$.
Moreover, we have discussed how ``measurability'' in the mathematical and the operational sense agree, if we accept $\sigma$-algebras on phase space that are coarser than the Borel sets.
In that case, one can still find Borel measures on phase space that yield Markovian statistics.
It is only important that their restriction on the $\sigma$-algebra generated by the atomic cells agrees with the values of the Markov measure.
Hence, in addition to the ambiguity with respect to the past, we have an ambiguity with respect to the ``fine structure'' within measurable events.
Consequently, this leads to a large class of ensembles that obey the Markovian postulate.

In spite of that, we must admit that observables corresponding to real measurements on physical systems do not induce Markov partitions in the mathematical sense.
However, Markovian statistics seem to be readily observed --- at least \emph{for all practical purposes}.
We discussed this fact in the context of the assumption of local equilibrium.
Further we have hinted at a ``Markovian anthropic principle'', which will be discussed in more detail in Section~\ref{sec:ubiquity}.

In the next chapter, we consider the same idealized framework of taking measurements as in the present chapter.
However, we will not assume that the coarse-grained observations behave in a Markovian way.
Rather, the subject of the next chapter is an information-theoretical analysis of our idealized framework.
More precisely, we are going to compare the uncertainty we have about the system after iterated measurements with the information contained in its actual microscopic configuration.
Moreover, we present an analytically tractable model dynamics with an observable that induces a generating Markov partition.
Thus, the next chapter will also provide concrete examples for the results obtained in the present chapter.
%
%
%
%
%

  \chapter{An information-theoretical approach to stochastic thermodynamics}
  \label{chap:information-st}
  \begin{fquote}[E.\,T.~Jaynes][A Backward Look on The Future][1993]
  [\ldots] it was necessary to think of probability theory as extended logic, because then probability distributions are justified in terms of their demonstrable information content, rather than their [\ldots]  frequency connections.
\end{fquote}

\section*{What is this about?}

In the previous Chapter~\ref{chap:marksymdyn}, we started our discussion of the \emph{microscopic} foundations of ST.
More precisely, we were interested in the following question:
When does a microscopic-deterministic dynamics yield Markovian statistics for the time series of coarse-grained observables?

In the present chapter, rather than investigating the microscopic foundation of the Markovian postulate, we discuss the connection of microscopic and mesoscopic notions of entropy.
In Chapter~\ref{chap:entropy} we have already reviewed the identification of entropy and entropy production in statistical mechanics.
There, we advocated the distinction of system and medium by their \emph{observability} in experiments, \cf Section~\ref{sec:stat-phys}.
We motivated the notions of entropy used in stochastic thermodynamics in Section~\ref{sec:discrete-st}.
In addition, Section~\ref{sec:md} reviewed the identification of \emph{dissipation} in molecular dynamics simulations with the phase space contraction prescribed by a microscopic model dynamics $\Phi$.

Here, we present an approach where these identifications emerge naturally.
In our reasoning we apply some of Jaynes' ideas, as summarized in the initial quote \cite{Jaynes1993}.
Instead of using phenomenological thermodynamic notions of entropy, we base our reasoning on information theory as a theory of statistical inference.

Often, the ensembles of statistical mechanics are interpreted in the so-called \emph{frequentist} picture, where the phase space density is interpreted as characterizing the statistics of trials obtained by measurements.
However, microscopic configurations (\ie points in a system's phase space) are never operationally accessible.
Hence, we follow Jaynes instead and interpret microscopic ensembles as our best estimate of the real microscopic situation --- given any prior knowledge about the dynamics.

The present chapter is structured as follows:
We begin by using the deterministic microscopic framework presented in the previous chapter in order to formalize the measurement process.
In this endeavour, we rely on Jaynes' approach to statistical physics as a theory of inference.
Consequently, we introduce two different phase-space ensembles, which reflect the information available on the microscopic and the mesoscopic levels of description, respectively:
On the mesoscopic level, we obtain a \emph{coarse-grained ensemble} $\cgden\tind{t}$ based on the statistics of mesoscopic observations, that gives rise to a coarse-grained, \ie inferred, entropy $\cgent\tind{t}$.
In contrast, if we have knowledge about the microscopic dynamics $\Phi$, we can also calculate the ``real'', \emph{fine-grained} evolution $\fgden\tind{t}$ of an initial ensemble and thus the \emph{fine-grained} entropy $\fgent\tind{t}$.
In addition, the comparison of these ensembles by the means of a Kullback--Leibler divergence yields a time-dependent relative entropy $\relent\tind{t}$.

The entropies obtained in this way are averages over the entire phase space.
Considering the phase space cylinders introduced in the previous chapter,  we obtain more \emph{detailed} notions of entropy and entropy variation.
Like the phase space cylinders, these detailed entropies are associated with observable time-series $\traj\omega\rlind{\tau}$.
Eventually, they will take the role of the entropic $\tau$-chains, which have been introduced in the context of stochastic thermodynamics (ST) in Section~\ref{sec:discrete-st}.

In order to have a clear information-theoretic interpretation of our results, we motivate the definition of four \emph{fundamental $\tau$-chains}.
We see that the detailed versions of $\fgent\tind{t}$, $\cgent\tind{t}$ and $\relent\tind{t}$ can be obtained as linear combinations of these fundamental $\tau$-chains.
In addition, we discuss how they yield \emph{thermodynamic} $\tau$-chains for the variation of the entropy in a system and its medium.
Under the assumption of a physical, \ie reversible microscopic dynamics we finally arrive at a central result of this work:
A conjecture regarding the generic microscopic foundations of ST, which involves the  concept of the natural two-sided Markov measure introduced in the previous chapter.

Another central aspect of the present chapter is the introduction of network multibaker maps (NMBM).
We use them as abstract model dynamics, which are both analytically tractable and versatile in mimicking more complicated situations.
In particular, NMBM can be tuned to exhibit the hallmarks of the physical microscopic dynamics used in molecular dynamics simulations:
They can be made time-reversible with a measure-preserving involution.
Further, one can tune them to be uniformly conservative, conservative or dissipative, where the latter case is the generic one.
Consequently, we use NMBM to exemplify the ideas presented in the current and the previous chapter.

\section{A general information-theoretic framework}
\label{sec:info-framework-gen}

\subsection{The measurement process revisited}
\label{sec:measurement-revisited}
Let us return to the experimental situation described in the previous chapter:
A scientist performs measurements on a system by recording the output $\omega_t\in\ospace$ of a measurement apparatus at equidistant times $t$ in a time-series $\traj \omega\rlind{\tau}$.
Upon repeating the experiment, she samples the probabilities $\prob_{t_0}[\traj\omega\rlind{\tau}]$ for the measured time-series between time $t_0$ and $t_0+\tau$.

At the initial time $t_0 = 0$ of preparation, the system finds itself in a certain microscopic state $x\in\pspace$.
However, the scientist never has access to this microscopic information.
What she knows is the initial observable state $\omega_0$.

Hence, her initial measurements specify the initial \emph{visible} or \emph{observable} ensemble.
For instance, she may ensure a specific initial condition $\omega\tsub{init}$ by dropping all trials where $\omega_0 \neq \omega\tsub{init}$.
Alternatively, the histogram of the initial measurements provides (a frequentist approach to) an initial distribution $\bvec p\tind{0} = \ifam{\p{0}{\omega}}_{\omega \in \ospace}$.

In addition to the distribution $\bvec p_0$ for observable states, we require \emph{microscopic ensembles} to specify a distribution on the phase space $\pspace$ of the underlying microscopic dynamics.
In contrast to the observable ensemble, the microscopic ensemble cannot be sampled in a frequentist way.
From Jaynes' point of view, the microscopic ensemble formally expresses our expectation about the microscopic distribution based on whatever information is available \cite{Jaynes1957}.
Or differently put:
It has to be \emph{inferred} in a way consistent with our knowledge of physics and mathematics.

A distribution that maximizes the (differential) Shannon entropy with respect to a set of constraints (which formalize prior knowledge) is the \emph{least biased} or \emph{maximally non-committal} prior \cite{Jaynes2003}.
In thermodynamics, the Gibbs distribution \eqref{eq:gibbs-entropy} provides an example:
It is the least biased prior with respect to the available \emph{macroscopic} thermodynamic information.

In that light, let us review the (mesoscopic) information available at the initial time $t=0$.
On the one hand, we have the coarse-grained information specified by the initial observable ensemble $\bvec p\tind{0}$.
This yields the constraint
\begin{align}
  \int \chi_\omega \pden_0 \df x \shouldbe \p{0}{\omega},
  \label{eq:compatibility}
\end{align}
for the initial microscopic density $\pden_0$, where $\chi_\omega$ is the indicator function for $\cell_\omega$.
Recall that $\cell_\omega$ is the set of phase space points yielding the measurement result $\omega$.

On the other hand, there might be additional information (or assumptions) regarding the observable \emph{dynamics} or the thermodynamic interpretation of individual measurement results.
In statistical mechanics, a common assumption is that of local equilibrium (LE), \cf Secs.~\ref{sec:langevin} and \ref{sec:operational}.
In its most common form, LE assumes that the (marginalized) distribution $\pden_0(x\,\vert\,\omega)=: \pden_\omega\colon\cell_\omega\to\reals$ of microstates $x \in \cell_\omega$ is a Gibbsian, \ie a maximum-entropy (MaxEnt) distribution --- compatible with the physical interpretation of the measurement result $\omega$.

After a measurement at time $t=1$, we obtain a new observable distribution $\bvec p_1$.
Using the principles described above, we get an updated, inferred distribution $\cgden\tind{0}$.
However, this inferred distribution is usually not the same as distribution $\fgden\tind{1}$, which is obtained by the microscopic dynamics acting on $\pden\tind{0}$.
Figure~\ref{fig:densities-fg-cg} illustrates this situation.

Consistency requires that the sampled observable probabilities $\bvec p\tind{t}$ obey
\begin{align}
  \int_{\cell_\omega}\cgden\tind{t} \df x \shouldbe \p{t}{\omega} \shouldbe \int_{\cell_\omega} \fgden\tind{t} \df x.
  \label{eq:compatibility-other}
\end{align}
The first equality sign expresses consistency between the coarse-grained ensemble and measurements.
The second equality expresses the fact that the microscopic theory is \emph{physically valid}:
If its predictions do not agree with our measurements, the theory should better be discarded.
Consequently, \eq{compatibility-other} ensures that the microscopic model $\Phi$ has been obtained via the \emph{scientific method} \cite{Popper2002}.

In the present thesis, we do not report on (numerical) experiments on particular systems.
Consequently, we cannot sample the statistics of time-series  in order to obtain the observable distribution $\bvec p\tind{t}$ at different moments in time.
Instead, we consider some microscopic (not explicitly specified) model dynamics $\Phi$, which we assume to be a good physical theory.
Hence, equation~\ref{eq:compatibility-other} \emph{defines} the observable ensemble $\bvec p\tind{t}$ at time $t$.

In the previous Chapter we have looked for conditions on $\Phi$ and $\pden_0$ such that $\bvec p\tind{t}$ evolves according to Markovian statistics.
Here, we initially drop this assumption and do not impose any further requirements on the microscopic dynamics $\Phi$ or the measurement observable $\partmap$.
Instead, we discuss how entropies quantify our knowledge of the physics of a system, without invoking thermodynamic arguments.
Only later we return to dynamics which yield Markovian observable statistics --- and conjecture how the framework outlined so far may provide a microscopic foundation of Markovian ST.

\begin{figure}[th]
  \centering
  \includegraphics{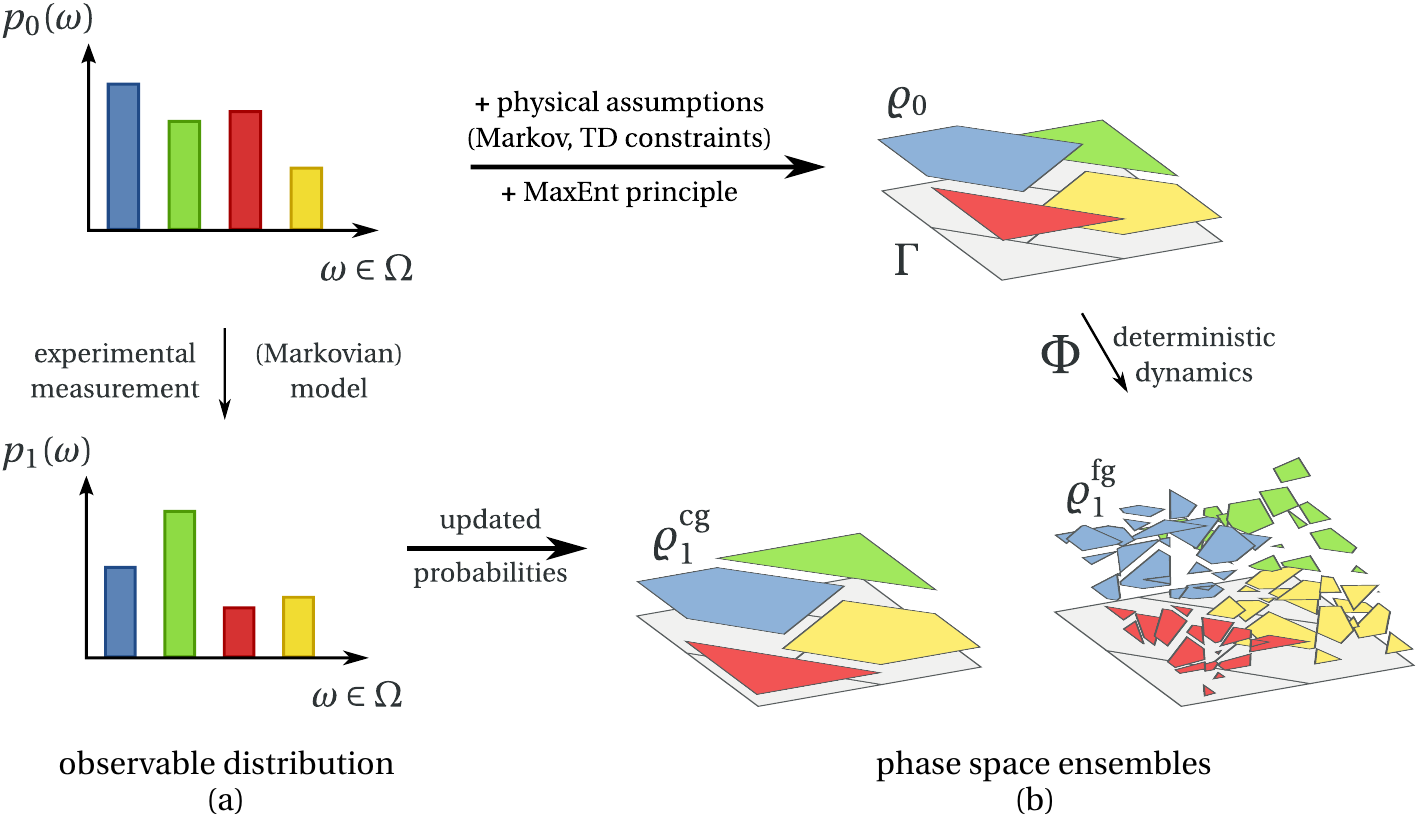}
  \caption{
    Inferring microscopic distributions from mesoscopic and macroscopic constraints.
    (a)~Top: The mesoscopic information comes in the form of an observable distribution $p_t(\omega)$.
    As observable are experimentally accessible, we can prepare them as a mesoscopic initial condition $\bvec p_0$.
    Bottom: At a subsequent time step, $\bvec p_1$ is obtained from a new measurement or the prediction of a (Markovian) mesoscopic model.
    (b)~Top: From a mesoscopic initial condition, the microscopic initial density $\pden\tind{0}$ is inferred.
    The microscopic density is a maximum-entropy distribution that respects both i) the observable distribution and ii) any additional (dynamical, macroscopic or thermodynamic) constraints.
    Bottom left: At a subsequent time step, we infer the coarse-grained density $\cgden\tind{t}$ again by a maximum-entropy principle.
    Bottom right: In contrast, the initial density is propagated by the microscopic deterministic dynamics $\Phi$ to yield the fine-grained density $\fgden\tind{1}$, which shows a more complicated structure.
}
  \label{fig:densities-fg-cg}
\end{figure}

\subsection{Fine- and coarse-grained entropy}
\label{sec:two-level-ent}
We start with the formal definitions of the fine- and coarse-grained ensembles motivated above.
In Chapter~\ref{chap:entropy} we defined the (differential) entropy $\sent{\pden}$ with respect to a density $\pden$ rather than using a (probability) measure $\pms$.
Hence, we have implicitly chosen a \emph{reference measure} $\ms$ and the density $\pden = \frac{\df \pms}{\df \ms}$ amounts to the Radon--Nikodym derivative of $\pms$ with respect to $\ms$.
In addition, we discussed in Section~\ref{sec:phase-space} how a reference measure $\ms$ is needed for the definition of a natural measure.

Throughout the present chapter, we choose the Borel sets $\palg$ as a $\sigma$-algebra and use the translation invariant (Lebesgue) measure $\lms =\ms $ as a reference.
Further, we assume that the phase-space dynamics $\Phi$ is invertible and non-singular with respect to $\lms$.
This ensures that the iterated Jacobian determinant $\jac{\tau}{x}$ of $\Phi$ is well-defined for all $\tau \in \integers$ and almost all $x\in\pspace$.

%
%

The \emph{fine-grained measure at time $t$} is nothing else than the image (pushforward) of an initial probability measure $\pms_0$: 
\begin{align}
  \fgms\tind{t} := (\Phi^t)_*\pms\tind{0} \equiv \pms\tind{0} \circ \Phi^{-t}.
\end{align}
Its density is determined by Equation~\eqref{eq:density-evolution}, \ie
\begin{align}
  \fgden\tind{t}(x):= \frac{\df{\fgms\tind{t}}}{\df \ms}= \frac{\fgden\tind{0}(\Phi^{-t}(x))}{\jac t {\Phi^{-t}x} }.
    \label{eq:evolvedDensity}
\end{align}

In the following, we need to condition the fine-grained density on \emph{phase space cylinders} $\pcyla{t}{\traj\omega\rlind{\tau}}\subset\pspace$.
A phase space cylinder contains the initial conditions $x$, such that the (finite) time-series $\traj\omega\rlind{\tau}$ occurs at time $t$ in the trajectory $\mtraj(x)$.
Recall that $\pcyla{t}{\traj\omega\rlind{\tau}}$ is the pre-image of the symbolic cylinder $\cyla{t}{\traj\omega\rlind{\tau}}$ from Definition~\ref{def:forward-cylinder-subshift}:
\begin{align}
  \pcyla{t}{\traj\omega\rlind{\tau}}:= \mtraj^{-1}\left(   \cyla{t}{\traj\omega\rlind{\tau}}   \right).
  \label{eq:phase-space-cylinder}
\end{align}
For a probability measure $\pms$ on $(\pspace,\palg)$, we define the \emph{conditioned measure}
  \begin{align}
    \left.\pms\tind{t}\right\vert_{\traj\omega\rlind{\tau}}(\pset) := \frac{\pms\left(\Phi^{-t}\pset\cap\pcyla {t}{\omega\rlind{\tau}}\right) }{\pms\left(\pcyla{t}{\omega\rlind{\tau}}\right)} \equiv \frac{\pms\tind{t}\left(\pset\cap\pcyla {0}{\omega\rlind{\tau}}\right)  }{\proba{t}{\omega}{\tau}}
  \end{align}
  with density $    \left.\pden\tind{t}\right\vert_{\traj\omega\rlind{\tau}}(x) := \frac{\df{ \pms\tind{t}\vert_{\traj\omega\rlind{\tau}}}}{\df \ms}$.

For real systems, we know neither $\Phi$ nor $\pden_0$, so the fine-grained distribution $\fgms\tind{t}$ cannot be inferred.
All that we know are individual values of the measurement observable $\partmap$.
From performing measurements on a large number of system, we can estimate the distribution $\bvec p\tind{t} = \ifam{\p{t}{\omega}}_{\omega \in \ospace}$.

In Section \ref{sec:partitions} we discussed the partition $\parti = (\cell_\omega)_{\omega \in \ospace} = \partmap^{-1}(\ospace)$ induced by the observable $\partmap$.
Denote by $\icms_\omega := \pms_0\vert_{\cell_\omega}$ the (unknown) \emph{initial measure conditioned on cell $\cell_\omega$} and by $\icden_\omega$ its density.
The usual assumption for $\icden_\omega$ ensures that it is a Gibbs distribution $\pden\tsup{G}$, \ie  a maximum-entropy distribution, \cf \eq{gibbs-entropy}.
From the point of information theory, constrained maximum entropy distributions are the least biased estimates which are compatible with the given prior information expressed by the constraints \cite{Jaynes1957}.

In the thermodynamic context, this prior information is physical and usually amounts to a certain knowledge about the macroscopic state of the system, \eg its volume or its temperature.
For more abstract considerations without the reference to physics, such constraints arise from (known) model symmetries.
In the mathematical framework introduced in the previous chapter, a measurement result $\omega$ provides us with incomplete information about the system's position in phase space:
We know that its microstate $x$ must belong to the partition element $\cell_\omega$.

In addition to any other prior knowledge we have, this induces an additional constraint on the microscopic measure:
The measure capturing this information must be supported on $\cell_\omega$ only.
We call the maximum-entropy measure $\priorms_\omega$  supported on cell $\cell_\omega$ the \emph{prior measure} for state $\omega$.
The entropy associated to its density $\priorden_\omega$ is called the \emph{assumed internal} entropy of state $\omega$.
Note that the assumed entropy of a state is independent of time.
It is therefore independent of the ``real'' internal entropy $\sent{\fgden\tind{t}\vert_{\cell_\omega}}$, which we obtain by constraining $\fgden\tind{t}$ to $\cell_\omega$.

Combining the information of the measurement at time $t$ with the prior leads to the  \emph{coarse-grained measure} $\cgms\tind{t}$, \cf Fig.~\ref{fig:densities-fg-cg}.
It is fully defined by its density
\begin{align}
  \label{eq:cgms}
  \cgden(x):= \sum_{\omega \in \ospace} \chi_{\omega}(x) \p{t}{\omega} \priorden_\omega(x).
\end{align}
The probability of an observable state is $\p{t}{\omega} = \fgms(\cell_\omega)$.
Because the prior density $\priorden_\omega$ is normalized on $\cell_\omega$, $\cgden$ automatically fulfils \eq{compatibility-other}.

The fine- and coarse-grained measures give rise to the \emph{fine- and coarse-grained entropies}
\begin{align}
  \fgent\tind{t}:= \sent{\fgden\tind{t}} \text{ and }
  \cgent\tind{t}:= \sent{\cgden\tind{t}},
\end{align}
where $\sent{\pden}$ denoting the differential Shannon entropy \eqref{eq:differential-entropy}.

Given the family of maximum-entropy priors $\ifam{\priorden_\omega}_{\omega \in \ospace}$ for each of the cells, the microscopic density fully specifies the coarse-grained density.
The additional information is expressed by the Kullback--Leibler divergence of $\cgden$ from $\fgden$.
In the following, we refer to it as the \emph{relative entropy}
\begin{align}
  \label{eq:relent}
  \relent\tind{t} := \dkl\left[ \fgden\tind{t} \middle\Vert \cgden\tind{t} \right] \geq0.
\end{align}

\subsection{The fundamental and derived entropic $\tau$-chains}
\label{sec:entropic-tau-chains}
In Section \ref{sec:discrete-st}, we have introduced observables $\obs[\traj \omega\rlind{\tau}]$ that depend on (parts of) the random trajectory $\traj\omega\rlind{\tau}$ generated by a noise history in a stochastic process.
Moreover, we focused on the interpretation of the quantities $\sysentrv\tind{t}[\traj \omega\rlind{\tau}]$ and $\medentrv\tind{t}[\traj \omega\rlind{\tau}]$ as the entropy variations in the system and its medium, respectively.
The expressions yielding their definition were motivated by the fact that their (time-series) averages amount to the variation of the entropies $\Delta\tind{t+\tau}\sysent$ and  $\Delta\tind{t+\tau}\medent$, respectively.

In the present section, we aim to achieve the same for the (variations of the) entropies $\fgent\tind{t}$, $\cgent\tind{t}$ and $\relent\tind{t}$.
More precisely, we define time-series dependent observables $\fgentrv\tind{t}$, $\cgentrv\tind{t}$ and $\relentrv\tind{t}$ yielding $\fgent\tind{t}$, $\cgent\tind{t}$ and $\relent\tind{t}$ as their averages.

In order to achieve this goal, it is convenient to introduce some notation, which is properly formalized in Appendix~\ref{app:tau-chains}.
A \emph{$\tau$-chain} $\obs\rlind{\tau}\tind{t}$ is a function
\begin{align*}
  \obs\rlind{\tau}\tind{t} \colon \ospace^{\tau+1}\times\timdom &\to \reals,\\
  (\traj \omega\rlind{\tau},t_0) &\mapsto \obs\tind{t_0}[\traj\omega\rlind{\tau}].
\end{align*}
Note that for a more concise notation, we drop the temporal index on $\obs$ when talking about its value $\obs\tind{t_0}[\traj\omega\rlind{\tau}]$, as the run length $\tau$ is specified explicitly by the symbol $\traj \omega\rlind{\tau}$.

The run-length index $(\tau)$ is important for the definition of the \emph{canonical sequence} $\ifam{\obs\rlind{\tau}\tind{t}}_{\tau \in \timdom}$ of $\tau$-chains which are obtained from a (time-dependent) state observable $\obs\tind{t}(\omega)\colon\alphabet\times\timdom \to \reals$.
The elements $\obs\rlind{\tau}\tind{t}$ of a canonical sequence evaluate $\obs\tind{t}$ at the state of the system at time $t+\tau$:
\begin{align}
  \obs\rlind{\tau}\tind{t}[\traj \omega\rlind{\tau}] := \obs\tind{t+\tau}(\omega_\tau).
  \label{eq:canonical-sequence}
\end{align}
In the following, we will encounter certain $\tau$-chains which are obtained as canonical sequences of state observables (\ie zero-chains).
Consequently, in our wording we synonymously refer to $\obs\tind{t}(\omega)$ and its canonical $\tau$-chain.

Before we discuss the $\tau$-chains associated to the fine-grained, coarse-grained and relative entropy, we introduce four \emph{fundamental} entropic $\tau$-chains.
We call them \emph{fundamental}, because they have a clear information-theoretic interpretation in our framework for taking measurements on an underlying  microscopic-deterministic system.
From the fundamental $\tau$-chains, we construct the \emph{derived} $\tau$-chains $\fgentrv\tind{t}$, $\cgentrv\tind{t}$ and $\relentrv\tind{t}$. 

The first fundamental $\tau$-chain is the  \emph{visible self-information}:
\begin{align}
  \visentrv\tind{t}(\omega) := -\log \p{t}{\omega}.
  \label{eq:visentrv}
\end{align}
It quantifies the \emph{uncertainty} or \emph{surprisal} of finding the state $\omega$ at time $t$, if we know the distribution $\bvec p \tind{t}$.
We call it visible, because $\bvec p \tind{t}$ is experimentally accessible through measurements.%
\footnote{At the moment we refrain from calling it the system's entropy, although this would be the correct interpretation.}

Further, we have time-independent \emph{assumed internal entropy} of a state $\omega$ associated to the prior distribution $\priorden_\omega$: 
\begin{align}
  \priorentrv(\omega) :=\sent{\priorden_\omega}= -\int_{\cell_{\omega}}\priorden_{\omega} \log \priorden_{\omega} \df x.
  \label{eq:priorentrv}
\end{align}

Another fundamental quantity is the so-called \emph{cross entropy} between the prior $\priorden_\omega$ and the real fine-grained entropy on cell $\cell_\omega$:
\begin{align}
  \crossentrv\tind{t} (\omega):= -\int_{\cell_{\omega}}\fgden\tind{t}\vert_{\cell_{\omega}}\log\left( \priorden_{\omega} \right)\df x.
  \label{eq:crossentrv}
\end{align}
It quantifies  the mismatch of the (constrained) real microscopic cell and the assumed prior on that cell.

The three quantities introduced so far are state variables.
As just mentioned, we identify them with with their canonical (sequence of) $\tau$-chains $(\visentrv\tind{t})\rlind{\tau}$, $(\priorentrv)\rlind{\tau}$ and $(\crossentrv\tind{t})\rlind{\tau}$.

The fourth fundamental observable is already defined as a $\tau$-chain. 
It quantifies the expansion of a phase space cylinder $\pcyla{0}{\traj\omega\rlind{\tau}}$, \ie the (average) phase space expansion experienced by the points that give rise to a time series $\traj\omega\rlind{\tau}$ in the interval $[t_0,t_0+\tau]$:
\begin{align}
  \contentrv\tind{t_0}\rlind{\tau}[\traj \omega\rlind{\tau}] :=    \int_{\cell\tind{0}[\traj\omega\rlind{\tau}]}\left.\fgden\tind{t_0}\right\vert_{\traj\omega\rlind{\tau}}  \,\log (\ja \rlind{\tau}) \df x. 
    \label{eq:contentrv}
\end{align}

\subsubsection{Averages}

The probability of observing a time-series $\traj\omega\rlind{\tau}$ is defined as the initial measure of the associated phase space cylinder, \cf Definition~\ref{def:forward-cylinder-subshift}:
\begin{align*}
  \prob\tind{t_0}\left[\traj \omega\rlind{\tau}\right] :=  \fgms\tind{t_0}\left(\cell\tind{0}\left[\traj \omega\rlind{\tau}\right]\right).
\end{align*}
The average of a $\tau$-chain is defined as (\cf \eqref{eq:ts-average}) 
\begin{align*}
  \trava{\obs\rlind{\tau}}{t_0}{\tau} := \sum_{\traj \omega\rlind{\tau} \in \alphabet^{\tau+1}} \prob\tind{t_0}\left[\traj \omega\rlind{\tau}\right] \obs\tind{t_0}\left[\traj \omega\rlind{\tau}\right].
\end{align*}
Lemma \ref{theo:trav-to-eav} ensures that the members of a canonical sequence obey 
\begin{align}
  \trava{\obs\rlind{\tau}}{t_0}{\tau} = \eav{\obs}\tind{t_0+\tau}.
  \label{eq:trav-to-eav-canonical}
\end{align}
\begin{subequations}
  Hence, the assumed entropy and the cross entropy have time-series averages that agree with the state averages at time $t = t_0+\tau$ 
  \begin{align}
    \trava{\priorentrv}{t_0}{\tau} &= \eava{\priorentrv}{t_0+\tau} \equiv \sum_\omega  \left[\p{t_0+\tau}{\omega} \priorentrv(\omega)\right]\label{eq:priorent-av}\\
    \trava{\crossentrv}{t_0}{\tau} &= \eava{\crossentrv}{t_0+\tau} \equiv \sum_\omega  \left[\p{t_0+\tau}{\omega} \crossentrv\tind{t_0+\tau}(\omega)\right]\label{eq:crossent-rv}
  \end{align}  
Similarly, the time-series average of the visible self-information is the Shannon entropy of the visible ensemble at time $t = t_0+\tau$:
  \begin{align}
  \trava{\visentrv}{t_0}{\tau} & =\sum_\omega\left[ \p{t_0+\tau}{\omega}\log\p{t_0+\tau}{\omega}\right] \equiv \sent{ \bvec p\tind{t_0+\tau}} .
  \label{eq:visent-av}
  \end{align}
  Finally, the time-series average of $\contentrv\tind{t_0}$ yields the average total phase space expansion between $t_0$ and $t_0+\tau$, calculate with the microscopic density $\fgden_{t_0}$:
  \begin{align}
    \trav{\contentrv}\rlind{\tau}\tind{t_0} &=  \int_\pspace\fgden\tind{t_0} \log \left( \ja\rlind{\tau} \right)\df x= \Lambda\rlind{\tau}\tind{t_0}
    \label{eq:content-av}
  \end{align}%
  \label{eq:fundamental-avs}%
\end{subequations}%
From the four fundamental sequences of $\tau$-chains, we define the \emph{$\tau$-chains for the coarse-grained, fine-grained} and \emph{relative entropy}
\begin{subequations}
  \begin{align}
    \left(\cgentrv\tind{t}\right)\rlind{\tau} &:= \left(\visentrv\tind{t}\right)\rlind{\tau} + \left(\priorentrv\tind{t}\right)\rlind{\tau},\label{eq:cgent-rv}\\
    \left(\fgentrv\tind{t}\right)\rlind{\tau} &:= \contentrv\tind{t}\rlind{\tau} + \fgent\tind{t}\label{eq:fgent-rv} \hspace{5em}\text{ and }\\
    \left(\relentrv\tind{t}\right)\rlind{\tau} &:= \left(\visentrv\tind{t}\right)\rlind{\tau}- \left(\fgentrv\tind{t}\right)\rlind{\tau}  + \left(\crossentrv\tind{t}\right)\rlind{\tau}\label{eq:relent-rv},
  \end{align}%
  \label{eq:ent-rvs}%
\end{subequations}%
respectively.
In order to justify their names, we look at their respective time-series averages, which obey 
\begin{subequations}
\begin{align}
  \trava{\cgentrv}{t_0}{\tau} &= \cgent\tind{t_0+\tau},\label{eq:cgent-av}\\ 
  \trava{\fgentrv}{t_0}{\tau} &= \fgent\tind{t_0+\tau},\label{eq:fgent-av}\\ 
  \trava{\relentrv}{t_0}{\tau} &= \relent\tind{t_0+\tau} \label{eq:relent-av}.
\end{align}
\label{eq:entrv-averages}%
\end{subequations}%
Equations \eqref{eq:entrv-averages} are obtained as the result of the following calculations.
We start with \eq{cgent-rv}:
   \begin{align*}
     \cgent\tind{t} &=-\int_{\pspace}\left( \sum_\omega \iverson{x\in\cell_\omega}p_\omega\rlind{t}\priorden_\omega \right) \log \left(  \sum_\omega \iverson{x\in\cell_\omega}p_\omega\rlind{t}\priorden_\omega \right)\df x\\
    &=\sum_{\omega}\int_{\cell_\omega}p_{\omega}\rlind{t}\priorden_{\omega}(x)\left(-\log p_{\omega}\rlind{t} -\log \priorden_{\omega}(x) \right)\df x\\
    &=\sum_{\omega}\left[p_{\omega}\rlind{t}\left(-\log p_{\omega}\rlind{t}\right)\right] + \sum_{\omega}\left[p_{\omega}\rlind{t}\left(-\int_{\cell_\omega} \priorden_{\omega}(x)\log\priorden_{\omega}(x) \df x\right)\right]\\
    &= \eav{\visentrv\tind{t}(\omega) + \priorentrv(\omega)}\tind{t}
  \end{align*}
Finally, equation (\ref{eq:trav-to-eav-canonical}) yields the claim.
As mentioned in Section~\ref{sec:time-discrete}, an easy calculation shows
  \begin{align}
    \fgent\tind{t_0+ \tau} &= -\int_\pspace \fgden\tind{t_0 + \tau}(x) \log \fgden\tind{t_0 + \tau}(x) \df x= - \int \frac{\fgden\tind{t_0}(\Phi^{-\tau}(x))}{\jac {\tau}{\Phi^{-\tau}(x)}}\log \frac{\fgden\tind{t_0}(\Phi^{-\tau}(x))}{\jac {\tau}{\Phi^{-\tau}(x)}} \df x\nonumber\\
    &= - \int_\pspace \fgden\tind{t_0} \log \frac{\fgden\tind{t_0}(x)}{\jac {\tau}{x}} \df x=- \int_\pspace \fgden\tind{t_0} \log {\fgden\tind{t_0}(x)}\df x + \int_\pspace \fgden\tind{t_0} \log {\jac {\tau}{x}} \df x\nonumber \\
    &= \fgent\tind{t_0} +\Lambda\rlind{\tau}\tind{t_0} = \trava{\fgent\tind{t} + \contentrv\tind{t}\rlind{\tau}}{t_0}{\tau} \label{eq:pden-lambda}.
  \end{align}
In the first line we used Equation~(\ref{eq:density-evolution}).
To obtain the second line we used the transformation theorem for integrals.
Finally, we used \eq{content-av} in the last line and thus prove \eq{fgent-av}.
For \eq{relent-av} we use the averages of the fundamental sequences of observables \eqref{eq:fundamental-avs}:
\begin{alignat*}{3}
    \trava{\relentrv}{t_0}{\tau} 
    = & \trava{\visentrv}{t_0}{\tau}
    &\,&{}-\trava{\fgentrv}{t_0}{\tau} 
    &\,&{}+ \trava{\crossentrv}{t_0}{\tau} \\
    = & -\sum_\omega\left[ \p{t_0+\tau}{\omega} \log \p{t_0+\tau}{\omega}\right]  
    &\,&{}- \fgent\tind{t_0 +\tau}
    &\,&{}+ \crossent\tind{t_0+\tau}.
  \end{alignat*}
Now, we split the phase space integrals $\int_\pspace$ into the integrals over the partition elements $\sum_\omega \int_{\cell_\omega}$ to obtain
  \begin{align*}
    \trav{\relentrv}\tind{t_0}\rlind{\tau   }
     &=  \sum_\omega& \Bigg[ &\int_{\cell_\omega} \fgden\tind{t_0+\tau}     
       \bigg(-\log \p{t_0+\tau}{\omega} + \log  \fgden\tind{t_0 +\tau} - \log \priorden_\omega  \bigg) \df {\ms} \Bigg] 
\\
     &=  \sum_\omega& \Bigg[ &\int_{\cell_\omega} \fgden\tind{t_0+\tau} 
     \log\left( \frac{\fgden\tind{t_0 +\tau}}{ \p{t_0+\tau}{\omega} \priorden_\omega}\right) \df {\ms} \Bigg] %
\\
      &= & -&\int_{\pspace} \fgden\tind{t_0+\tau}     \log \left(\frac{\sum_\omega\chi_\omega \p{t_0+\tau}{\omega} \priorden_\omega} {\fgden\tind{t_0 +\tau}}\right)  \df {\ms}%
\\
     &= & &\dkl\left[ \fgden\tind{t_0+\tau} \Vert \cgden\tind{t_0+\tau}\right] \\
     &=  && \relent\tind{t_0 + \tau}   \end{align*}
which proofs \eq{relent-av}. 

Let us summarize this subsection.
We established $\tau$-chains for a set of fundamental information-theoretic quantities which appear naturally within our framework of the measurement process.
From those (time-series dependent) fundamental expressions we constructed the three derived $\tau$-chains $\cgentrv$, $\fgentrv$ and $\relentrv$.
We showed that an average over time-series running from time $t_0$ to $t_0+\tau$  amounts to the value of the coarse-grained, fine-grained and relative entropy at the final time, respectively.
In the next section, we calculate their temporal variations.

\subsection{Temporal variations}
In Section~\ref{sec:discrete-st} we have introduced time-series dependent expressions $\delta\sysentrv$ and $\delta\medentrv$ in the context of stochastic thermodynamics.
Their respective averages yield the temporal variations $\Delta\sysent$ and $\Delta\medent$.
Now we do the same for the changes of the fine-grained, coarse-grained and relative entropy.

Let $\ifam{\obs\rlind{\tau}}_{\tau \in \naturals}$ be a sequence of $\tau$-chains, which does \emph{not} need to be canonical, \cf ~\eqref{eq:canonical-sequence}.
For a given finite time-series $\traj\omega\rlind{\tau}=(\omega_0,\omega_1,\cdots,\omega_\tau)$, the elements of the corresponding \emph{sequence of the variations} $\ifam{\delta\rlind{\tau} \obs}_{\tau \in \naturals^+}$ are defined as
\begin{align}
  \delta\rlind{\tau}\obs\tind{t_0}[\traj\omega\rlind{\tau}] :=  \obs\rlind{\tau}\tind{t_0}\left[ (\omega_0,\omega_1,\cdots\omega_{\tau-1},\omega_\tau) \right] - \obs\rlind{\tau-1}\tind{t_0}\left[ (\omega_0,\omega_1,\cdots\omega_{\tau-1}) \right].
  \label{eq:defn-sequence-of-variations}
\end{align}
The fundamental variation $\tau$-chains associated to the fundamental chains $\visentrv$, $\priorentrv$, $\crossentrv$ and $\contentrv$ read:
\begin{subequations}
\begin{align}
  \delta\rlind{\tau}\visentrv\tind{t}[\traj \omega\rlind{\tau}] &= \log\frac{\p{t+\tau-1}{\omega_{\tau -1}}}{\p{t+\tau}{\omega_{\tau}}},
  \label{eq:visvarrv}\\
  \delta\rlind{\tau}\priorentrv[\traj \omega\rlind{\tau}] &=\priorentrv(\omega_\tau) - \priorentrv(\omega_{\tau-1}),
  \label{eq:priorvarrv}\\
  \delta\rlind{\tau}\crossentrv[\traj \omega\rlind{\tau}] &=\crossentrv\tind{t+\tau}(\omega_\tau) - \crossentrv\tind{t+\tau-1}(\omega_{\tau-1}),
  \label{eq:crossvarrv}\\
  \delta\contentrv\tind{t_0}\rlind{\tau}[\traj \omega\rlind{\tau}] &=    \int_{\cell[\traj\omega\rlind{\tau}]} \left.\fgden\tind{t_0}\right\vert_{\traj\omega\rlind{\tau}} \,\log\left( \ja\rlind{1}\circ\Phi\rlind{t-1}\right) \df x, 
    \label{eq:contvarrv}
\end{align}  %
  \label{eq:fundamental-varrvs}%
\end{subequations}
Thus, the variations of the derived quantities \eqref{eq:ent-rvs} amount to:
\begin{subequations}
  \begin{align}
    \delta\rlind{\tau}\cgentrv\tind{t} &\equiv \delta\rlind{\tau}\visentrv\tind{t} + \delta\rlind{\tau}\priorentrv\tind{t} =  \log\frac{\p{t+\tau-1}{\omega_{\tau -1}}}{\p{t+\tau}{\omega_{\tau}}} +\priorentrv(\omega_\tau) - \priorentrv(\omega_{\tau-1}) ,\label{eq:cgvarrv}\\
    \delta\rlind{\tau}\fgentrv\tind{t} &\equiv \delta\rlind{\tau} \contentrv\tind{t} =   \int_{\cell[\traj\omega\rlind{\tau}]} \left.\fgden\tind{t_0}\right\vert_{\traj\omega\rlind{\tau}} \,\log\left( \ja\rlind{1}\circ\Phi\rlind{t-1}\right) \df x \label{eq:fgvarrv} \hspace{5em}\text{ and }\\
    \delta\rlind{\tau}\relentrv\tind{t}&\equiv \delta\rlind{\tau}\visentrv\tind{t}  - \delta\rlind{\tau}\fgentrv\tind{t} + \delta\rlind{\tau}\crossentrv\tind{t} \label{eq:relvarrv},
  \end{align} %
  \label{eq:var-rvs}%
\end{subequations}
Lemma~$\ref{theo:variations}$ in Appendix~$\ref{app:tau-chains}$ ensures that the time-series average $\trav{\delta \entrv}\rlind{\tau}\tind{t}$ of $\delta \rlind{\tau}\entrv$ equals the temporal variations $\Delta\tind{t+\tau}S =\ent\tind{t+\tau} - \ent\tind{t+\tau-1} $ of the time-dependent average $S_t := \trav{\entrv}\rlind{\tau}\tind{t}$.
Hence, the time-series averages of the $\tau$-chains \eqref{eq:var-rvs} yield the variations of the corresponding entropies:
\begin{subequations}
  \begin{align}
  \cgent\tind{t_0+\tau} - \cgent\tind{t_0+\tau -1} &= \Delta\cgent\tind{t_0+\tau} = \trava{\delta \cgentrv}{t_0}{\tau},\\
  \fgent\tind{t_0+\tau} - \fgent\tind{t_0+\tau -1} &= \Delta\fgent\tind{t_0+\tau} = \trava{\delta \fgentrv}{t_0}{\tau},\\
  \relent\tind{t_0+\tau} - \relent\tind{t_0+\tau -1} &= \Delta\relent\tind{t_0+\tau} = \trava{\delta \relentrv}{t_0}{\tau}.
\end{align}%
  \label{eq:variance-average}%
\end{subequations}

Finally, let us discuss these averages in more detail:
The variation of the coarse-grained entropy
\begin{align}
  \Delta\cgent\tind{t_0+\tau} &= \Delta\visent\tind{t_0+\tau} + \Delta\priorent\tind{t_0+\tau} \nonumber \\
  &= \sum_{\omega,\omega'}\left[  \prob\tind{t_0+\tau-1}[(\omega,\omega')]\left( \log \frac{\p{t_0+\tau-1}{\omega}}{\p{t_0+\tau}{\omega'}} +\priorentrv(\omega' ) - \priorentrv(\omega) \right) \right]
  \label{eq:var-cg-ent}
\end{align}
consists of the change of the visible and the assumed internal entropy.
The former is obtained as the average over the well known logarithmic ratio of the ensemble probabilities before and after the transition.
The latter is just the difference of the assumed internal entropies of the respective cells.

We have already encountered the variation of the fine-grained entropy at several points in this thesis:
\begin{align}
  \Delta\fgent\tind{t_0+\tau} &= \trava{\delta \contentrv}{t_0}{\tau} = \Lambda\tind{t_0+\tau-1}\rlind{1}= \int_\pspace\fgden\tind{t_0+\tau-1}\,  \log (\ja\rlind{1})\df x.
  \label{eq:var-fg-ent}
\end{align}
It has the same value as the averaged phase-space expansion rate and can be calculated from the logarithm of the Jacobian determinant.

Finally, consider the variation of the relative entropy:
\begin{align}
  \Delta \relent\tind{t} &= \int_{\pspace}\fgden\tind{t}\log\left( \frac{\fgden\tind{t}}{\cgden\tind{t}} \right) - \fgden\tind{t-1} \log\left( \frac{\fgden\tind{t-1}}{\cgden\tind{t-1}} \right)\df x\nonumber\\
   &= \int_{\pspace}\fgden\tind{t-1}\log\left( \frac{\fgden\tind{t}\circ \Phi}{\cgden\tind{t}\circ \Phi} \right) - \fgden\tind{t-1} \log\left( \frac{\fgden\tind{t-1}}{\cgden\tind{t-1}} \right)\df x\nonumber\\
   &= \int_{\pspace}\fgden\tind{t-1}\log\left( \frac{\fgden\tind{t}\circ \Phi}{\fgden\tind{t-1}}\frac{\cgden\tind{t-1}}{\cgden\tind{t}\circ \Phi}  \right) \df x\nonumber\\
   &= \int_{\pspace}\fgden\tind{t-1}\log\left( \frac{1}{\jac{1}{x}}\frac{\cgden\tind{t-1}(x)}{\cgden\tind{t}\left(\Phi(x)\right)}  \right) \df x \label{eq:tentative-diss-fun}.
\end{align}
Note that though $\relent\tind{t}$ is a Kullback--Leibler divergence and hence always positive, this is not generally true for the variation $\Delta\relent\tind{t}$.
This is easily seen from the following example:
Consider that at some point in time $t>0$ we have $\fgden\tind{t} = \cgden\tind{t}$.
In that case, in general $\fgden\tind{t-1} \neq \cgden\tind{t-1}$ and hence $0=\relent\tind{t}<\relent\tind{t-1}$.
Hence, the positivity of $\Delta\relent\tind{t}$ crucially depends on the microscopic initial condition $\pden\tind{0}$.
We will discuss the issue of positivity in detail in Section~\ref{sec:info-st-discussion}.
Before we do so, however, let us illustrate the results of the present section using a concrete example.

\section{Network multibaker maps}
\label{sec:nmbm}

In this section, we exemplify the information-theoretical framework presented in Section~\ref{sec:info-framework-gen} on a concrete model.
As our chaotic microscopic dynamics, we use a variant of so-called multibaker maps, which were originally introduced by Hopf \cite{Hopf1948}.
Gaspard stressed their role as a generic example of a strongly mixing hyperbolic map \cite{Gaspard2005}.
In analogy to models of transport theory, Vollmer and co-workers introduced reversible multibaker maps \cite{Vollmer_etal1998,Vollmer2002}.
Being reversible, they share many of the features of the (NE)MD models discussed in Section \ref{sec:md}.

A multibaker map consists of a (countable) number of rectangular \emph{cells}.
The dynamics maps rectangular subsets of each cell to rectangular subsets in \emph{adjacent} cells.
Historically, multibaker maps are arranged on a regular one-dimensional lattice with either open or periodic boundary conditions.
Several variants of this linearly arranged multibaker maps exist in the literature \cite{Gaspard2005,Vollmer2002,Colangeli_etal2011}.

Here, we generalize this setting to more complex phase-space topologies.
More precisely, we extend the neighbouring relations between the cells of a multibaker maps to arbitrary networks of states.
In spite of being more general, these \emph{network multibaker maps} (NMBM) still admit an explicit analytical treatment of the evolution of their phase-space densities.
For multibaker maps on linear chains, such calculations have been the subject of earlier work \cite{Vollmer_etal1997,Vollmer2002}.

Generalizing the approach followed in Ref.~\cite{Rondoni_etal2000}, we focus on the statistical behaviour of certain sets of microscopic orbits rather than on global averages.
More precisely, we consider the evolution of the microscopic trajectories that belong to a phase space cylinder $\pcyla{t_0}{\traj\omega\rlind{\tau}}$. 
We will see how the expressions used in Markovian ST emerge naturally as a result.

\subsection{Formulation of the model}

The definition of the model starts with an arbitrary network specified by a directed graph $\graph = (\verts,\edges)$ on $N$ vertices $i \in \verts := \{1,2,\ldots,N\}$, and edges $e\in\edges \subset \verts\times\verts$.
To keep notation at bay, we assume that the graph is simple, \ie that there is at most one edge connecting vertex $i$ to vertex $j$.
However, note that the prescription of a NMBM trivially extends to graphs with more than one edge between two given vertices.

Now denote by $\verts_i := \set{j\in\verts: (i,j)\in \edges}$ the set of vertices that are connected to a state $i$ and by $\abs{\verts_i}$ the \emph{degree} of vertex $i$.
The microscopic phase space consists of rectangular \emph{cells} $\cell_i := [0,1]\times\vcell_i\cdot[0,1]$ with area (\ie Lebesgue measure) $\vcell_i$ for each vertex $i$.
The overall \emph{phase space} $\pspace$ of the system is the disjoint union $\pspace := \bigsqcup_{i=1}^N \cell_i$.
Hence, a point $x \in \pspace$ is a triple $x = (x_1,x_2,i)$ where $(x_1,x_2) \in \cell_i$.
\begin{figure}[t]
  \begin{center}
     \includegraphics{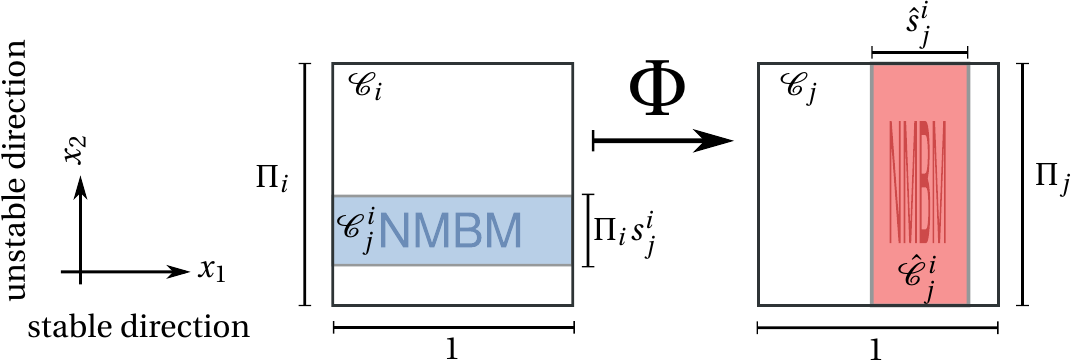}
  \end{center}
  \caption{Dynamics of the network multibaker map.
    For each of the $\abs{\verts_i}$ cells adjacent to the cell indexed by $i$, a \emph{horizontal} strip (blue) $\cell^i_j \subset \cell_i$ of relative height $s^i_j$ is affine-linearly mapped to a vertical strip (red) $\hat \cell^i_j \subset \cell_j$ of width $\hat s^i_j$.
    The horizontal coordinate $x_1$ specifies the position along the unstable direction, \ie the direction where $\Phi$ is expanding.
    Similarly, phase space is contracted along the stable, vertical direction $x_2$. 
    The deformation of a strip is made visible by the distortion of the letters ``NMBM''.
  }
  \label{fig:multibaker-defn}
\end{figure}

A NMBM, $\Phi\colon\pspace \to \pspace$, deterministically maps phase-space points $x \in \cell_i$ to adjacent cells $\cell_j$.
It is specified geometrically (\cf Fig.\,\ref{fig:multibaker-defn}) by dividing each cell $\cell_i$ into $\abs{\verts_i}$ horizontal strips of finite relative height $s^i_j>0$ and an offset $b^i_j = \sum_{k<j}s^i_k$:
\begin{equation}
  \cell^i_j :=
  \begin{cases}
  [0,1]\times \vcell_i\cdot[b^i_j,b^i_j+s^i_j)  & i \in \verts, j \in \verts_i,\\
  \emptyset & \text{else}.
  \end{cases}
  \label{eq:defnCell}
\end{equation}
The dynamics $\Phi$ maps each horizontal strip, $\cell^i_j \subset\cell_i$, to a \emph{vertical} strip, $\hat \cell^i_j:=\Phi\cell^i_j \subset \cell_j$
defined as
\begin{equation}
  \hat\cell^i_j :=
  [\hat b^i_j,\hat b^i_j+\hat s^i_j)\times \vcell_j\cdot[0,1].
  \label{eq:defnCellImage}
\end{equation}
The number $\hat s^i_j\in[0,1]$ denotes the relative width of the vertical strips in cell $\cell_j$ and thus fulfil $\sum_i \hat s^i_j =1$.
The \emph{offsets} read $\hat b^i_j := \sum_{k<i}\hat s^k_j$.
To obtain an analytically tractable model, we choose $\Phi$ to act in an affine-linear way on the cells.
This means that points in any strip are mapped such that the horizontal direction is contracted uniformly by a factor $\hat s^i_j<1$ whereas the vertical direction is expanded by a factor $(s^i_j)^{-1}>1$.
Formally,
\begin{align}
  \Phi\colon \pspace &\to \pspace\nonumber\\
  (x_1,x_2,i) &\mapsto \left(\hat{b}^i_j + \hat{s}^i_jx_1,\,\frac{\vcell_j}{ s^i_j}\left(\frac{x_2}{\vcell_i} - b^i_j\right),j \right)  \label{eq:nmbm-analytical}
\end{align}
for $b^i_j<\frac{x_2}{\vcell_i}\leq b^i_{j+1}$.
Hence, a NMBM is fully defined by the numbers $\vcell_i$, $s^i_j$ and $\hat s^i_j$.
Note that the matrix $A$ with entries $a_{ij} = \operatorname{sgn}(s^i_j) = \operatorname{sgn}(\hat s ^i_j)$ is the adjacency matrix of the graph~$\graph$.

\subsection{Reversibility and further constraints}

By imposing further constraints on the numbers $s^i_j$, $\hat s^i_j$ and $\vcell_i$, one can implement additional features into the dynamics.
More precisely, one can abstractly realize time-reversible dynamics similar to the models used in (NE)MD, \cf Section~\ref{sec:md}.
For the remainder of this work, let us make the following assumption for the graph $\graph$:
(i) $\graph$ is connected, \ie that its adjacency matrix is irreducible.
(ii) $\graph$ is dynamically reversible, \ie the presence of a directed edge $e:=(i,j) \in \edges$ implies that $- e := (j,i) \in \edges$, too.
(iii) Each vertex is connected to itself, \ie\,$(i,i) \in \edges,\,\forall i\in\verts$.

Then, the dynamics can be made reversible with a measure-preserving involution~$\invo$.
It has been shown \cite{Tel+Vollmer2000,Vollmer2002} that a necessary requirement for the existence of such an involution is the symmetry
\begin{align}
  \hat s^i_j = s^j_i,\,\forall i,j.
  \label{eq:ReversibilityRelativeVolumes}
\end{align}
Then, the map
\begin{align}  
  \invo \colon \pspace &\to \pspace\nonumber\\
  (x_1,x_2,i) &\mapsto (1-\vcell_i^{-1}x_2, \vcell_i(1-x_1),i),\,\label{eq:multibakerInvolution} 
\end{align}
fulfils the conditions (\ref{eq:reversibility}), \ie it is a measure-preserving ($\det{(\Df\invo)} = 1$) involution that facilitates time-reversal.
Moreover, it acts locally on the cells, \ie points in a cell $\cell_i$ stay there under the action of $\invo$.
The consequences of this symmetry are depicted in Figure~\ref{fig:multibaker-rev}.

\begin{figure}[t]
  \begin{center}
     \includegraphics{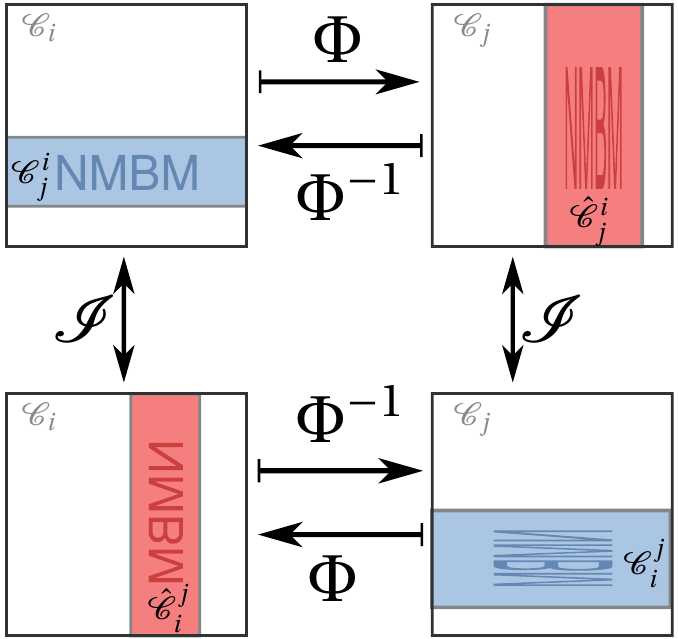}
  \end{center}
  \caption{
    A reversible NMBM obeying \eq{ReversibilityRelativeVolumes} features a volume preserving involution $\invo$, \cf\,Eq.\,\eqref{eq:multibakerInvolution}, which interchanges horizontal and vertical stripes in one cell, \ie\, \mbox{$\invo \hat\cell^i_j = \cell^j_i$}.
  The dynamics $\Phi$ is only volume preserving, if \eq{volume-preserving} holds as well.
  Again, the letters ``NMBM'' reflect how the dynamics and the involution acts on individual points $x \in \pspace$.
  }
  \label{fig:multibaker-rev}
\end{figure}

For a generic choice of the parameters, the dynamics is dissipative.
This fact is not influenced by the assumption of reversibility as prescribed by Equation~\eqref{eq:ReversibilityRelativeVolumes}.
However, by imposing further constraints we are able to mimic conservative and uniformly conservative dynamics, which mimic isolated systems and other non-driven systems, respectively.

From the geometric specification of the multibaker map or from (\ref{eq:nmbm-analytical}) it is clear that
\begin{align}
  \jac{1}{x} = \frac{\abs{\hat\cell^i_j}}{\abs{\cell^i_j}} = \frac{\vcell_j\hat s^i_j}{\vcell_i s^i_j} \text{ for }x \in \cell^i_j.
  \label{eq:jacobian}
\end{align}
This means, a reversible multibaker map is uniformly conservative if and only if
\begin{align}
  \label{eq:volume-preserving}
  \vcell_i s^i_j \equiv \vcell_j s^j_i.
\end{align}
In order to check whether a NMBM is non-uniformly conservative, which is a weaker property, we need to know the steady-state distribution.
We will see in Section~\ref{sec:entropic-chains} that reversible NMBM are conservative if and only if the relative volumes $s^i_j$ fulfil the Kolmogorov cycle criterion \eqref{eq:kolmogorov-crit}, \ie
\begin{align}
  \prod_{t = 1}^{\tau} s^{i_{t-1}}_{i_{t}} = \prod_{t = 1}^{\tau} s^{i_{t}}_{i_{t-1}}.
  \label{eq:nmbm-conservative}
\end{align}
for any sequence of $\tau+1$ states obeying $i_0 = i_\tau$.

In summary, NMBM are very versatile and can be used to reproduce the key features of general reversible dynamics.
This is not limited to the case of reversible dynamics with a measure-preserving involution $\invo$.
For instance, Gaspard and co-workers also introduced non-reversible uniformly conservative maps of the unit square to itself \cite{Gaspard+Wang1993,Gaspard+Dorfman1995}.
We just mention that the most general form of NMBM also contains these maps as a special case.

\subsection{NMBM observables, priors and initial conditions}
\label{sec:nmbm-obs}
In the previous section we demonstrated the versatility of NMBMs to serve as a toy model for more realistic dynamics.
Now, we choose an observable $\partmap\colon\pspace \to \ospace$ that associates observations (measurement results) $\omega \in \ospace$ to points $x\in\pspace$ in phase space.
In Chapter~\ref{chap:marksymdyn} we have discussed in detail, how an observable $\partmap$ induces a phase space partition $\parti = \partmap^{-1}(\ospace)$.
The elements $\cell_\omega$ of that  partition contain the pre-images of elements in $\ospace$.

For NMBM it is natural to choose $\ospace = \verts$ and define $\partmap(x) := i$ for $x \in\cell_i$.
Then, the induced partition $\parti = \ifam{\cell_\omega}_{\omega \in \ospace}$ agrees with the partition $\parti = \left( \cell_i \right)_{i\in\verts}$ we used to define the NMBM.
For \emph{reversible} network multibaker maps, this partition also has an important symmetry:
The involution $\invo$ factorizes on the partition elements (\cf Fig.\,\ref{fig:multibaker-rev}), \ie
\begin{equation}
  \invo \cell_\omega = \cell_\omega,\, \forall \omega \in \ospace.
  \label{eq:cellSymmetry}
\end{equation}
In that case we say that the partition $\parti = \ifam{\cell_\omega}_{\omega \in \ospace}$ is \emph{absolutely $\invo$-invariant}.\footnote{
This is a stronger condition than \emph{$\invo$-invariance} $\invo \parti = \parti$, \ie $\forall \cell \in \parti\colon \exists \cell'=\invo \cell \in \parti$.}

In the discussion of the previous chapter (Sec.~\ref{sec:operational}) we have discussed properties of partition induced by \emph{real} measurements on real (\ie physical) systems.
From that point of view, the assumption of an absolutely $\invo$-invariant partition seems rather restrictive.
However, we have also seen that the thermostated equations of motion used in (NE)MD are reversible with a measure-preserving involution (\cf Sec.~\ref{sec:thermostats} and Ref.~\cite{Jepps+Rondoni2010}).

In that case, the measure-preserving time-reversal involution is given by flipping momenta and (possibly) auxiliary variables.
Note that then \emph{any} (MD-)observable which depends \emph{only} on configurational degrees of freedom obeys the symmetry (\ref{eq:cellSymmetry}) on the induced cells.
That is, any partition constructed in such a way is absolutely $\invo$-invariant.
This factorization of the involution over the partition is the reason why such a dynamics \emph{always} satisfies a fluctuation theorem for the phase space expansion of phase space cylinders \cite{Wojtkowski2009}.
For reversible multibaker maps on a one-dimensional lattice this was first realized in Ref.~\cite{Rondoni_etal2000}.

Now let us discuss the priors we use in the specification of the initial (and coarse-grained) measure.
For our abstract dynamics we have no special physical model in mind.
Hence, there are no further constraints reflecting any additional knowledge.
The compatible maximum-entropy distributions $\priorden_\omega$ are uniform on each cell:
\begin{align}
  \priorden_\omega = \vcell_\omega^{-1}.
  \label{eq:uniform-cell-prior}
\end{align}
Thus, the assumed entropy for each cell can be obtained in a ``Boltzmannian'' way as the logarithm of the associated phase-space volume:
\begin{align}
  \priorentrv(\omega) = \log{\vcell_\omega}.
  \label{eq:nmbm-priorent}
\end{align}

In order to calculate the other fundamental $\tau$-chains \eqref{eq:visentrv}--\eqref{eq:crossentrv}, we need to specify the initial condition $\pden\tind{0}$.
To be consistent with the arguments brought forward in Section~\ref{sec:info-framework-gen},  we take the coarse-grained density as our initial ensemble, \ie
\begin{align}
  \fgden\tind{0}=  \pden\tind{0} = \cgden\tind{0} = \sum_{\omega \in \ospace} \left[\chi_{\omega}(x) \p{t}{\omega} \vcell_\omega^{-1}\right].
  \label{eq:NMBM-initialization}
\end{align}

\subsection{Evolution of the densities}
Let us recall Definition~\ref{eq:trajectory-sym} of the trajectory $\mtraj\rlind{\tau} \colon\pspace \to \ospace^{\tau+1}$.
It maps an initial condition $x$ to its observable time-series of length $\tau$.
The latter is obtained from measurements on the iterations of $x$ produced by the successive action of the map $\Phi$:
\begin{align*}
  \mtraj^{(\tau)} = \left( \partmap \circ \Phi^t \right)_{t\in\{0,1,\dots,\tau\}} 
\end{align*}
Then, for a point $x \in \pspace$ with time-series $\traj\omega\rlind{\tau} = \mtraj\rlind{\tau} x$, equations (\ref{eq:evolvedDensity}) and (\ref{eq:NMBM-initialization})  together with (\ref{eq:jacobian}) yield the fine-grained density:
\begin{align}
  \fgden\tind{t}(\Phi^{t}x) &= \pden_0 (x) \left(\prod_{t=1}^\tau \jac{1}{\Phi^t (x)}\right)^{-1}\nonumber\\
  &= \frac{\p{0}{\omega_0}}{\vcell_{\omega_\tau}} \prod_{t=1}^\tau\left( \frac{s^{\omega_{t-1}}_{\omega_{t}}}{s^{\omega_{t}}_{\omega_{t-1}}} \right).
  \label{eq:fgden-evolution}
\end{align}
The phase space cylinder $\pcyla{0}{\traj\omega \rlind{\tau}}:= (\mtraj\rlind{\tau})^{-1}\set{\traj \omega\rlind{\tau}}$ contains all the points $x\in \pspace$ that give rise to the time-series $\traj \omega\rlind{\tau}$ starting at $t=0$.
Its $\tau$-fold iterated image is the image of the (negatively shifted) forward cylinder $\cyla{-\tau}{\traj\omega\rlind{\tau}}$
\begin{align}
  \pcyla{-\tau}{\traj\omega \rlind{\tau}} : = \Phi\rlind{\tau}\left(\pcyla{0}{\traj\omega \rlind{\tau}}\right) = \mtraj^{-1}\left( \cyla{-\tau}{\traj\omega\rlind{\tau}} \right).
  \label{eq:iterated-pcyl}
\end{align}
Note that it can be defined recursively as 
\begin{align}
  \pcyla{-\tau}{\traj\omega \rlind{\tau}} \equiv \Phi\left(\pcyla{-{\tau-1}}{\traj\omega \rlind{\tau-1}} \cap \cell^{\omega_{k-1}}_{\omega_{k}}  \right),  \label{eq:defnTrajectoryVolumeElementRecursive}
\end{align}
where $\pcyla{0}{ (\omega_0)} =\cell_{\omega_0}$.
%

Consider the volume (Lebesgue-measure) of the phase space cylinder $\pcyla{-\tau}{\traj\omega \rlind{\tau}}$.
One can easily verify that for any $t\leq \tau \in \naturals$, $\pcyla{-t}{\traj\omega \rlind{t}}$ is a \emph{vertical} strip.
Intersecting it with the horizontal strip $\cell^{\omega_{k-1}}_{\omega_{k}}$ reduces its volume by a factor \mbox{$s^{\omega_{k-1}}_{\omega_{k}}<1$}.
Through contraction and expansion in horizontal and vertical directions, respectively, $\Phi$ changes the volume by another factor $\hat s^{\omega_{k-1}}_{\omega_{k}}/s^{\omega_{k-1}}_{\omega_{k}}$. 
Hence,  $\lms\left( \pcyla{-t}{\traj \omega\rlind{t}}\right)  = \hat s^{\omega_{t-1}}_{\omega_{t}}\lambda\left( \pcyla{-t}{\traj \omega\rlind{t-1}} \right)$ and after iteration to $t=\tau$,
\begin{align}
  \lms\left( \pcyla{-\tau}{\traj \omega\rlind{\tau}}\right)=\vcell_{\omega_\tau}\prod_{t=1}^{\tau}\hat s^{\omega_{t-1}}_{\omega_t}=\vcell_{\omega_\tau}\prod_{t=1}^{\tau} s^{\omega_{t}}_{\omega_{t-1}},
  \label{eq:VolumeTrajectoryElementFinal}
\end{align}
where for the last equality we used Eq.\,\eqref{eq:ReversibilityRelativeVolumes}.
The probability of observing a time-series $\traj \omega\rlind{\tau}$ at starting at time $t=0$ is
\begin{align}
  \prob_0\left[ \traj\omega\rlind{\tau} \right] &= \int_{\pcyla{0}{\traj\omega \rlind{\tau}} } \fgden_t \df x
  = \int_{\pcyla{-\tau}{\traj\omega \rlind{\tau}} } \fgden(\Phi^t x) \df x\nonumber  \\
  &= \left( \lms\left( \pcyla{-\tau}{\traj \omega\rlind{\tau}}\right)\right)\cdot\left( \fgden\tind{t}(\Phi^{t}x)\right) \nonumber\\
  &=\left(\vcell_{\omega_\tau}\prod_{t=1}^{\tau} s^{\omega_{t}}_{\omega_{t-1}}\right)\cdot\left( \frac{\p{0}{\omega_0}}{\vcell_{\omega_\tau}} \prod_{t=1}^\tau\left( \frac{s^{\omega_{t-1}}_{\omega_{t}}}{s^{\omega_{t}}_{\omega_{t-1}}} \right)\right)\nonumber \\
  &= \p{0}{\omega_0} \prod_{t=1}^{\tau} s^{\omega_{t-1}}_{\omega_{t}}. \label{eq:probability-evolution}
\end{align}
Note that in the second line we used that the fine-grained density $\fgden(\Phi^tx)$ is the same for all $x\in\pcyla{-\tau}{\traj \omega\rlind{\tau}}$.

Equation~\eqref{eq:probability-evolution} reveals an interesting fact:
The probabilities evolve according to a Markov chain for a transition matrix $\tmat$ with elements $\tprob{\omega}{\omega'} = s^{\omega}_{\omega'}$.
This means that our assumed initial distribution must be the extension of a one-sided Markov measure with transition matrix $\tmat$.
We will come back to that observation in Section \ref{sec:foundations-st}.

\subsection{The entropic $\tau$-chains and their variations}
\label{sec:entropic-chains}
Now we have all the ingredients to write down the fundamental and derived entropic $\tau$-chains.
For brevity, we only state the variations.
They are obtained by applying Eqs.~\eqref{eq:fundamental-varrvs} to expressions \eqref{eq:visentrv}--\eqref{eq:crossentrv}, which are evaluated using the expressions \eqref{eq:jacobian}, \eqref{eq:nmbm-priorent} and \eqref{eq:fgden-evolution}:
\begin{subequations}
  \begin{align}
    \delta\rlind{\tau}\visentrv\tind{t} &= \log\frac{\p{t+\tau-1}{\omega_{\tau-1}}}{\p{t+\tau}{\omega_{\tau}}},
    \label{eq:visvarrv-nmbm}\\
    \delta\rlind{\tau}\priorentrv\tind{t} = \delta\rlind{\tau}\crossentrv\tind{t} &=   \log\frac{\vcell_{\omega_{\tau}}}{\vcell_{\omega_{\tau-1}}},
    \label{eq:priorvarrv-nmbm}\\
    \delta\rlind{\tau}\contentrv\tind{t}   &=   \log\frac{\vcell_{\omega_{\tau}}s^{\omega_{\tau}}_{\omega_{\tau-1}}}{\vcell_{\omega_{\tau-1}}s^{\omega_{\tau-1}}_{\omega_{\tau}}}.
    \label{eq:contvarrv-nmbm}  \end{align}
  \label{eq:fundvarrv-nmbm}%
\end{subequations}%
From the expressions for the fundamental variations we obtain the variations of the derived quantities as the linear combinations \eqref{eq:var-rvs}:
\begin{subequations}
  \begin{align}
    \delta\rlind{\tau}\cgentrv\tind{t} &= \log\frac{\p{t+\tau-1}{\omega_{\tau-1}}}{\p{t+\tau}{\omega_{\tau}}} + \log\frac{\vcell_{\omega_{\tau}}}{\vcell_{\omega_{\tau-1}}},
    \label{eq:fgvarrv-nmbm}\\
    \delta\rlind{\tau}\fgentrv\tind{t}   &=   \log\frac{s^{\omega_{\tau}}_{\omega_{\tau-1}}}{s^{\omega_{\tau-1}}_{\omega_{\tau}}}+ \log\frac{\vcell_{\omega_{\tau}}}{\vcell_{\omega_{\tau-1}}},\\
    \delta\rlind{\tau}\relentrv\tind{t}   &=   \log\frac{\p{t+\tau-1}{\omega_{\tau}}s^{\omega_{\tau-1}}_{\omega_{\tau}}}{\p{t+\tau}{\omega_{\tau}}s^{\omega_{\tau}}_{\omega_{\tau-1}}}.
  \end{align}
  \label{eq:varrv-nmbm}%
\end{subequations}%

Equation \eqref{eq:probability-evolution} tells us that the relative strip volumes $s^\omega_{\omega'}$ are the transition probabilities of the Markovian process that generates the observed time series.
In order to connect the expressions obtained here with the ones used in Markovian ST (\cf Section~\ref{sec:discrete-st}), we define:
\begin{subequations}
  \begin{align}
    \delta\rlind{\tau}\sysentrv\tind{t} &=  \log\frac{\p{t+\tau-1}{\omega_{\tau-1}}}{\p{t+\tau}{\omega_{\tau}}} = \delta\rlind{\tau}\visentrv\tind{t} \equiv \delta\rlind{\tau}\cgentrv\tind{t} - \delta\rlind{\tau}\priorentrv\tind{t},
    \label{eq:sysvarrv-nmbm}\\
    \delta\rlind{\tau}\medentrv\tind{t} &=  \log\frac{s^{\omega_{\tau-1}}_{\omega_\tau}}{s^{\omega_\tau}_{\omega_{\tau-1}}} = -\delta\rlind{\tau}\contentrv\tind{t} + \delta\rlind{\tau}\priorentrv\tind{t}, 
    \label{eq:medvarrv-nmbm}\\
    \delta\rlind{\tau}\totentrv\tind{t} &=  \log\frac{\fluxa{\omega_{\tau-1}}{\omega_{\tau}}(t+\tau-1)}{\fluxa{\omega_{\tau}}{\omega_{\tau-1}}(t+\tau)}  = \delta\rlind{\tau}\visentrv\tind{t} -\delta\rlind{\tau}\contentrv\tind{t} +\delta\rlind{\tau}\crossentrv\tind{t} .
    \label{eq:totvarrv-nmbm}
  \end{align}
  \label{eq:tdvarrv-nmbm}
\end{subequations}
This is a remarkable result:
We have reproduced the $\tau$-chains used in ST within an general information-theoretical framework for deterministic dynamics.
Admittedly, the identifications made in Eqs.~\eqref{eq:tdvarrv-nmbm} were made in order to be consistent with Markovian ST.
Consequently, one wonders how much of this result is truly general and how much is due to the special structure of NMBM as a model dynamics.

Regardless of the generality of this result, it yields the proof of \eq{nmbm-conservative}:
For NMBM, the average phase space contraction in the steady state is equivalent to the average entropy change in the medium identified in ST.
In the latter framework, the Kolmogorov criterion~\eqref{eq:kolmogorov-crit} ensures reversibility and hence detailed balance.
This in turn ensures that the average phase-space contraction in the steady state vanishes, \ie that the NMBM is conservative.

In the following section, we discuss how to generalize this results to arbitrary microscopic dynamics, which we consider physical.
Amongst other things, we motivate Eqs.~\eqref{eq:tdvarrv-nmbm} \emph{without} referring to ST.
Rather, we base our argument on the identification of dissipation with phase-space contraction, as is common in thermostated MD \cite{Evans+Searles2002,Searles_etal2007,Jepps+Rondoni2010}.

\section{Discussion}
\label{sec:info-st-discussion}
Let us now discuss how the information-theoretical framework outlined in the present chapter may serve as a microscopic foundation for ST.
First, we give a motivation of Equations~\eqref{eq:tdvarrv-nmbm} without an explicit reference to ST.
After that, we focus on the expression for the total entropy production as a relative entropy.
In order to connect the present results to the those of Chapter~\ref{chap:marksymdyn}, we discuss how the natural two-sided measure appears naturally for NMBM.
Finally, we comment on the influence of the reference measure on our results.

\subsection{Consistent identification of system and medium entropy}
\label{sec:stentrv-from-info}
In the introduction to the present chapter we emphasized that the information-theoretic formalism introduced here is independent of ST.
However, we have motivated the definitions~\eqref{eq:tdvarrv-nmbm} to ensure consistency with Markovian ST.
In contrast, now we motivate them from the point using the general perspective on the distinction of system and medium presented in Section~\ref{sec:entropy-statphys}.
In addition, we make use of the arguments in Section~\ref{sec:md}, where we identified thermodynamic dissipation with phase space contraction.

To be consistent with the notion of a system's entropy as the entropy of a coarse-grained, experimentally accessible and thus \emph{visible} ensemble, we set:
\begin{align}
  (\sysentrv)\rlind{\tau}\tind{t}[\traj\omega\rlind{\tau}] &:= (\visentrv)\rlind{\tau}\tind{t}[\traj\omega\rlind{\tau}]
  \nonumber\\
  &\equiv -\log \p{t+\tau}{\omega_\tau}.
  \label{eq:sysentrv-general}
\end{align}

The entropy of the medium is more complicated.
Firstly it should take into account the integrated dissipation into the medium, $\dissentrv$.
This term accounts for the macroscopic irreversibility characterized by the calometrically accessible heat flow from the system to the medium.
In correspondence to the (NE)MD models discussed in Section \ref{sec:md} we thus set:
\begin{align}
  (\dissentrv)\rlind{\tau}\tind{t}[\traj\omega\rlind{\tau}] &:= -\contentrv\rlind{\tau}\tind{t}[\traj\omega\rlind{\tau}].
  \label{eq:dissentrv-general}
\end{align}
Moreover, the medium entropy should contain a hidden contribution $\hiddentrv$, which arises due to our (subjective) assumptions of the distribution on the cells.
As we mentioned in Section~\ref{sec:info-theory}, the cross-entropy \eqref{eq:crossentrv} quantifies this mismatch, which we interpret as one contribution to the hidden entropy.
Additionally, we have a second contribution that amounts to the entropy $\sent{\fgden\tind{t}}=\fgent_{t}$ of the microscopic ensemble at time $t_0$.
The difference of these two terms defines the \emph{hidden entropy}: 
\begin{align}
  (\hiddentrv)\rlind{\tau}\tind{t}[\traj\omega\rlind{\tau}] &:= (\crossentrv)\rlind{\tau}\tind{t}[\traj\omega\rlind{\tau}] - \fgent\tind{t}.
  \label{eq:hiddentrv-general}
\end{align}
Hence, the medium entropy consists of the contributions
\begin{align}
  (\medentrv)\rlind{\tau}\tind{t}[\traj\omega\rlind{\tau}] &:= (\dissentrv)\rlind{\tau}\tind{t}[\traj\omega\rlind{\tau}] +(\hiddentrv)\rlind{\tau}\tind{t}[\traj\omega\rlind{\tau}] \nonumber \\
  &\equiv -\int_{\cell_0[\traj\omega\rlind{\tau}]}\left.\fgden\tind{t}\right\vert_{\traj\omega\rlind{\tau}} \,\log (\ja \rlind{\tau})  \df x + \crossentrv(\omega_\tau) - \fgent\tind{t}.
  \label{eq:medentrv-general}
\end{align}
In the context of Markovian ST, Seifert pointed out that such a hidden contribution follows from the assumption of local equilibrium \cite{Seifert2011}.
In contrast, our argument explicitly establishes the role of the deterministic dynamics and the real microscopic distribution.

The total entropy is the sum of the contributions assigned to the medium and the system:
\begin{align}
  (\totentrv)\rlind{\tau}\tind{t}[\traj\omega\rlind{\tau}] &:= (\sysentrv)\rlind{\tau}\tind{t}[\traj\omega\rlind{\tau}] + (\medentrv)\rlind{\tau}\tind{t}[\traj\omega\rlind{\tau}]\nonumber\\  
  &\stackrel{\eqref{eq:medentrv-general}}{\equiv}  (\visentrv)\rlind{\tau}\tind{t}[\traj\omega\rlind{\tau}] -\contentrv\rlind{\tau}\tind{t}[\traj\omega\rlind{\tau}]
 +(\crossentrv)\rlind{\tau}\tind{t}[\traj\omega\rlind{\tau}] - \fgent\tind{t_0}\nonumber\\
  &\stackrel{\eqref{eq:fgent-rv}}{\equiv}  (\visentrv)\rlind{\tau}\tind{t}[\traj\omega\rlind{\tau}]  +(\crossentrv)\rlind{\tau}\tind{t}[\traj\omega\rlind{\tau}]  -(\fgentrv)\rlind{\tau}\tind{t}[\traj\omega\rlind{\tau}]\nonumber\\
  &\stackrel{\eqref{eq:relent-rv}}{\equiv} (\relentrv)\rlind{\tau}\tind{t}[\traj\omega\rlind{\tau}].
  \label{eq:totentrv-general}
\end{align}
This is an important result:
It identifies the total entropy as a relative entropy between the real (fine-grained) density and our assumed coarse-grained density.
Note that the latter is ``subjective'' in the sense that it depends on the maximum-entropy priors $\pden_\omega$.

The total entropy is a Kullback--Leibler divergence and thus always positive.
Moreover, we see that the value of the fine-grained entropy $\fgent\tind{t}$ at time $t$, which we identified as a part of the hidden entropy production, cancels in the corresponding variation $\delta\rlind{\tau}\totentrv\tind{t}$.
In addition, the NMBM dynamics presented in Section~\ref{sec:nmbm} provides a consistency check:
For this analytically tractable dynamics, the identifications \eqref{eq:tdvarrv-nmbm}, which we have just motivated generally, agree with the expressions used in ST.

\subsection{Positivity of the variation of the total entropy}
In the previous Subsection~\ref{sec:stentrv-from-info}, we motivated the identification of the total entropy with the relative entropy. 
Let us discuss this fact in the light of the second law of thermodynamics.
In that perspective, we should have that $\Delta\rlind{\tau}\relent\tind{t_0} \geq 0$ for all $t_0,\tau >0$.
The present author conjectures that the following (non-rigorous) argument can be transformed into a proper proof.

Without loss of generality we choose $t=0$ as the initial time where the system is prepared. 
That point in time is special, because in the framework presented above it represents the point in time where the coarse-grained density $\cgden\tind{0}$ agrees with the fine-grained density $\fgden\tind{0}$.
The initial density was chosen as a distribution with maximum entropy.
By definition, a maximum entropy (MaxEnt) distribution is the least biased distribution compatible with a set of constraints formalizing prior knowledge about the system \cite{Jaynes2003}.
Hence, we understand a MaxEnt distribution as the one with the least amount of ``intrinsic structure'' among all compatible distributions.
In the course of the dynamics (unless they are uniformly conservative), phase space is contracted and expanded.
Thus, additional structure is introduced into each cell.
Therefore, in the course of time the difference of the real distribution within a cell $\fgden\tind{t}\vert_{\cell_\omega}$ and the maximum entropy prior becomes more and more pronounced.

This difference of one probability distribution with respect to another is quantified by the Kullback--Leibler divergence.
Hence, the relative entropy and thus the total entropy should always increase.
Consequently, a second law stating that  $\Delta\rlind{\tau}\relent\tind{t_0} \geq 0$ should also hold in this case.

Note that this holds for the dynamically reversible NMBM we have discussed above.
They reproduce the expressions known from ST and we have already discussed in Section~\ref{sec:discrete-st}.
As such the total entropy production is a KL-divergence and thus always positive, \cf Equation~\eqref{eq:transient-totent}.


\subsection{Foundations of Markovian stochastic thermodynamics}
\label{sec:foundations-st}
%

After having motivated the identification of system and medium entropies in the last previous in general terms, we return to Markovian dynamics.
For NMBM, we have seen that the expressions \eqref{eq:sysentrv-general}, \eqref{eq:medentrv-general} and \eqref{eq:totentrv-general} reproduce the expressions known from ST, \cf Eqs.~\eqref{eq:tdvarrv-nmbm}.
Although NMBM constitute an abstract model without a physical justification, they intuitively represent the dynamics of more general (hyperbolic) dynamics.
In particular, we have demonstrated how  NMBM can be made reversible and further tuned to show features of conservative and dissipative systems.

In the light of Chapter~\ref{chap:marksymdyn}, let us review the properties of NMBMs that yield Equations \eqref{eq:fundvarrv-nmbm}:
Firstly, a necessary requirement for the Markovian evolution of time-series is that the observable defining the coarse-grained states $\omega \in \ospace$ induces a Markov partition $\parti$.\footnote{Note that the natural partition for NMBM is also \emph{generating}, but this is not needed for the argument.}\\
Secondly, the fact that $\parti$ is absolutely $\invo$-invariant with a measure-preserving time-reversal involution $\invo$ ensures the validity of Eq.~\eqref{eq:cellSymmetry}.
For NMBM it yields the relation $s^i_j = \hat s^j_i$, which relates the relative weights of images and pre-images of the strips.
Thirdly, even if the topological requirements for a Markovian evolution of the observed time-series is fulfilled, we still need the right initial conditions.

Let us formulate this conditions in a more general context.
We have already discussed the issue of Markov partitions for real, \ie physical systems in Section~\ref{sec:operational}.
For now, assume that this assumption is fulfilled at least for all practical purposes.

Regarding the existence of a measure-preserving time-reversal involution $\invo$ and the existence of an absolutely $\invo$-invariant partition, recall the discussion of thermostated equations of motions in Section~\ref{sec:thermostats}.
We have seen that \emph{any} observable that depends only on the coordinates yields a partition that obeys $\invo \cell_\omega = \cell_\omega$.
Hence, absolute $\invo$-invariance might be in fact a generic symmetry of partitions induced by physical observables on physical microscopic dynamics.
Additionally, recent work points out that care has to be taken when applying the framework of ST to systems where momenta of particles are treated explicitly~\cite{Kawaguchi+Nakayama2013}.

The most subtle, but arguably most important question regards the choice of initial conditions.
In Chapter~\ref{chap:marksymdyn} we have discussed Markov measures on phase space.
These are measures, such that (forward) time-series yield Markovian statistics.
The natural Markov measure used the notion of a natural measure on phase space.
The next paragraph comments on the significance of the natural two-sided measure introduced in Definition~\ref{def:natural-two-sided-mms}.

\paragraph{Significance of the two-sided natural measure for reversible dynamics}
The natural two-sided measure is constructed using the natural transition matrices $\ptmat$ and $\back\ptmat$ obtained from the natural measure $\pms_\Phi$ for $\Phi$ and the natural measure $\pms_{\Phi^{-1}}$ for the inverse dynamics $\Phi^{-1}$.
Suppose the natural measure for both $\Phi$ and $\Phi^{-1}$ is an SRB measure, \ie it has absolutely continuous densities along the unstable manifolds.
The unstable manifolds of $\Phi^{-1}$ are the stable manifolds of $\Phi$ and vice versa.

For a dynamics featuring a measure-preserving time-reversal involution $\invo$ which acts locally on the partition elements, this yields a geometric interpretation of the natural measure.
First note that every measure preserving time-reversal involution $\invo$ ensures that $\back \pms := \pms \circ \invo$ is absolutely continuous, if $\pms$ is absolutely continuous.
Hence, by the definition of the natural measure and the time-reversal symmetry, we have
\begin{align*}
  \pms_\Phi \circ \invo &= \lim_{\tau\to\infty}\left(\frac 1 \tau \sum_{t=1}^\tau \pms \circ \Phi^{-1}\right) \circ \invo 
  = \lim_{\tau\to\infty}\frac 1 \tau \sum_{t=1}^\tau  \pms \circ \invo \circ \Phi
  = \lim_{\tau\to\infty}\frac 1 \tau \sum_{t=1}^\tau  \back \pms \circ \Phi\\
  &= \pms_{\Phi^{-1}}.
\end{align*}
This means that applying $\invo$ allows us to switch between $\pms_{\Phi^{-1}}$ and $\pms_\Phi$.
If $\parti$ is absolutely $\invo$-invariant, this relation factorizes on the partition elements, \ie it holds also if we constrain the natural measures to any cell $\cell_\omega$.
Within each cell, a natural Markov measure conditioned on the forward cylinders (\ie the unstable manifolds) has to be proportional to the natural measure on this cell.
The natural two-sided measure is a special way to specify the (transversal) density along the stable manifolds of $\Phi$, \ie the \emph{un}stable manifolds of $\Phi^{-1}$:
On the latter, it obtains the density of the natural measure $\pms_{\Phi^{-1}} =\pms_\Phi \circ \invo$.
Hence, for an absolutely $\invo$-invariant partition the natural two-sided measure constitutes an $\invo$-invariant initial condition, \cf the discussion of the dissipation function in Section~\ref{sec:reversibility}.


For a piecewise linear dynamics like NMBM, the natural measure $\pms_\Phi$ conditioned on the unstable manifolds is piecewise constant \cite{Blank+Bunimovich2003}.
With the above symmetry, the natural measure corresponds to the uniform initialization on each cell.
Thus, for NMBM the coarse-grained measure obtained from the maximum entropy distributions on each cell \emph{is} a two-sided natural Markov measure.

Let us return to Definition~\ref{def:natural-tmat} of the natural transition matrix.
Given an absolutely $\invo$-invariant partition obeying \eq{cellSymmetry}, we see that
\begin{align*}
    \ptprob{\omega}{\omega'} &\equiv \frac{\pms_\Phi(\Phi^{-1}\cell_{\omega'} \cap \cell_{\omega}) }{\pms_\Phi(\cell_\omega)}  = \frac{\pms_\Phi(\cell_{\omega'} \cap \Phi \cell_{\omega}) }{\pms_\Phi(\cell_\omega)} \\
    &=  \frac{\pms_{\Phi^{-1}}(\invo \cell_{\omega'} \cap \invo \Phi \cell_{\omega}) }{\pms_{\Phi^{-1}}(\invo \cell_\omega)} =  \frac{\pms_{\Phi^{-1}}(\cell_{\omega'} \cap \Phi^{-1} \cell_{\omega}) }{\pms_{\Phi^{-1}}(\cell_\omega)} \equiv \back{q}^{\omega}_{\omega'}.
\end{align*}
The numbers $ \back{q}^{\omega}_{\omega'}$ are the entries of the natural transition matrix for the backward process $\back \ptmat$.
Because the adjacency matrix of the backward process is the transpose of the adjacency matrix of the forward process, we have
\begin{align}
  \adjm = \sgn{\ptprob{\omega}{\omega'}} = \sgn(\back{q}^{\omega}_{\omega'}) =  \adjm^T.
  \label{eq:dynamical-reversibilty}
\end{align}
Hence, an absolutely invariant partition yields a \emph{dynamically} reversible stochastic process.
For a NMBM, the natural measure $\pms_{\Phi^{-1}}$ for the inverse dynamics is constant along the $x_2$ direction.
Hence, the relative \emph{volumes} of vertical strips $\hat s^{\omega}_{\omega'}$ correspond to the entries of the natural matrix of the backward process $\back q^{\omega'}_{\omega}$.
Then, the symmetry \eqref{eq:ReversibilityRelativeVolumes} expresses the fact that $s^{\omega}_{\omega'} = q^{\omega}_{\omega'} = \back{q}^{\omega}_{\omega'} = \hat{s}^{\omega'}_{\omega}$.

\paragraph{A general conjecture}

In Chapter \ref{chap:entropy} we have pointed out that the medium entropy relates to the irreversibility of the system's transitions.
For a jump process, it generically compares a forward jump $\omega \to \omega'$ to the corresponding backward jump $\omega'\to\omega$ in an appropriately chosen backward process.

Hence, for the entropy change in the medium it seems natural to consider the logarithmic ratio:
\begin{align*}
  \medentrv \sim \log\frac{ \ptprob{\omega}{\omega'}}{\back{q}^{\omega'}_{\omega} }.
\end{align*}
Indeed, for reversible NMBM this just yields the desired expression, \ie $\medentrv \sim \log\frac{ s^{\omega}_{\omega'}}{s^{\omega'}_{\omega} }.$
We formally conjecture that this should also hold generally:
\begin{conjecture}[Foundations of dynamically reversible, Markovian ST]
  \label{conj:foundations}
  Let $(\pspace,\borel,\Phi)$ be a dynamical system with a measure-preserving time-reversal involution $\invo$.
  Let $\parti$ be an absolutely $\invo$-invariant Markov partition.
  Under the assumption that we take the two-sided natural Markov measures as initial conditions and as priors, we postulate that:
  \begin{align}
    \delta\rlind{\tau}\medentrv\tind{t}[\traj\omega\rlind{\tau}] &:= -\delta\rlind{\tau}\contentrv\tind{t}[\traj\omega\rlind{\tau}] +\delta\rlind{\tau}\priorentrv\tind{t}[\traj\omega\rlind{\tau}]\\
    &= \log\frac{ \ptprob{\omega}{\omega'}}{\ptprob{\omega'}{\omega} },
    \label{eq:conjecture}
  \end{align}
  where $\ptprob{\omega}{\omega'}$ are the elements of the natural transition matrix for $\Phi$ and $\parti$.  
\end{conjecture}

If proven true, this conjecture provides a proper deterministic foundation of stochastic thermodynamics.
It further enables the systematic study of the relaxation of certain assumptions.
The obvious generalization is to consider $\invo$-invariant partitions which are not absolutely $\invo$-invariant.
This gives rise to a stochastic process which is not dynamically reversible.

In Secs.~\ref{sec:nmbm-obs} and \ref{sec:reversibility} we have mentioned Wojtkowski's abstract fluctuation theorem for the phase space contraction \cite{Wojtkowski2009}.
For NMBM, the latter is connected to the entropy production identified in ST and thus reproduces the stochastic fluctuation relations.
Actually, Wojtkowski's theorem is formulated for a more general situation where the backward microscopic dynamics $\Psi$ does not need to be the inverse process $\Phi^{-1}$.
All that is needed that $\Phi$ and $\invo\circ\Psi\circ\invo$ are related by a measure-preserving involution $\invo$.
For instance, this situation is found when the dynamics is driven out of equilibrium by magnetic rather than electrical fields.
From the perspective of ST, the natural measure induced by $\Psi$ then yields the transition rates that define the appropriate backward process.
We expect this to provide a dynamical picture of the stochastic master fluctuation relations, which has been formulated by Seifert in Ref.\,\cite{Seifert2012}.
A more detailed discussion on that subject can be found in Section~\ref{sec:master-fr}.

However, there is a caveat regarding the maximum entropy priors:
For NMBM, the uniform distribution on each cell is both a maximum-entropy prior as well as the natural two-sided measure on phase space.
This fact has its origin in the linearity of NMBM.
For any abstract dynamics, which is non-linear, the natural two-sided measure has a more complicated structure, \ie it is \emph{not} uniform on the cells.
In the beginning of the present chapter, we argued that the prior measure needs to maximize entropy with respect all the \emph{additional knowledge} we have about the dynamics.

Throughout this work we emphasized that \emph{physical} microscopic dynamics --- like Hamiltonian dynamics or (NE)MD equation --- exhibit certain features:
They allow for a measure-preserving involution.
Given a measurement observable, that is invariant under time-reversal, this involution factorizes on the cells of the induced partition.
Stochastic thermodynamics additionally assumes a Markovian evolution of the observable states.
In that regard, the natural two-sided Markov measure appears as the natural candidate for a maximum-entropy measure that reflects these additional constraints.

\subsection{Influence of the reference measure}
\label{sec:ref-meas}
Finally, we comment on the influence of the reference measure in our considerations.
In a recent work, Polettini investigates how the choice of a reference measure formalizes another aspect of subjectivity.
This subjectivity is reflected in the fact that probability densities $\pden$ and thus entropies  $\sent{\pden}$ depend on the choice of reference \cite{Polettini2012,Polettini+Vedovato2013}.
In addition, the value of the differential entropy change upon coordinate transformations $T: \pspace \to \pspace$.

Polettini argues that the choice of a reference measure needs to be understood as expressing some prior assumption about the system.
The Lebesgue measure thereby reflects the \emph{microcanonical} prior, where all states are assumed to have equal a-priori likelihood to appear.\footnote{
  Ref.\,\cite{Polettini+Vedovato2013} beautifully explains this fact using the different perspective that parents and their children have regarding the ``disorder'' they perceive in the children's playing room.}
In his interpretation, the choice of prior constitutes a gauge transformation.
We will come back to that interpretation in the following Chapter~\ref{chap:cycles}.
In ST, the expressions associated to the entropy (changes) in the system and the medium change under the action of the gauge transformation, \ie the choice of reference \cite{Polettini2012}.
However, the expression associated with the change of the \emph{total entropy} is gauge-invariant.

In the present framework, this statement is equally true if we identify the total entropy with the relative entropy, \cf \eq{relent}, \ie a Kullback--Leibler divergence.
As such it has the property to be \emph{invariant} under the choice of the reference measure or coordinate transformations.
Hence, also Polettini's considerations point to the validity of the identifications made here.

\section{Summary}
In this chapter, we have proposed an information-theoretic framework to constitute the microscopic foundations of stochastic thermodynamics.
We started by formalizing the process of taking repeated measurements in the spirit of Jaynes' view of statistical physics as a theory of statistical inference.
In this framework, we used a deterministic microscopic dynamics to motivate fundamental information-theoretic $\tau$-chains $\visentrv$, $\priorentrv$, $\crossentrv$ and $\contentrv$.
These $\tau$-chains are observables that depend on finite time-series $\traj \omega\rlind{\tau}$ of length~$\tau$.

From the fundamental $\tau$-chains we were able to construct the derived chains $\cgentrv$, $\fgentrv$ and $\relentrv$ as linear combinations.
The derived chains average to the entropy of the inferred coarse-grained ensemble, the real microscopic ensemble and their relative entropy, respectively.
We further motivated the expressions $\sysentrv$, $\medentrv$ and $\totentrv$, which correspond to the entropy of the system, the medium and the sum of both, respectively.
In that step, we used the connection of dissipation and phase space contraction in physical microscopic models.

In order to exemplify our considerations, we introduced network multibaker maps as a versatile and generic, yet analytically tractable model systems.
We showed how for these maps our abstract thoughts are consistent with stochastic thermodynamics and how the entropic concept of the latter emerge naturally.
Further, we used network multibaker maps to motivate the importance of the natural two-sided Markov measure, which was introduced in Chapter~\ref{chap:marksymdyn}.

This in turn lead to Conjecture~\ref{conj:foundations} which formalizes the microscopic foundations of ST in a general information-theoretical framework.
At this point, we have concluded the first part of the thesis regarding the \emph{microscopic} foundations.
In the following two Chapters~\ref{chap:cycles} and \ref{chap:fluctuations} we are concerned with the mathematical structure of Markovian ST on finite state spaces.

Hence, the discussion there is independent of the hypothesis of a deterministic phase space dynamics we have assumed in the previous and present Chapters~\ref{chap:marksymdyn} and \ref{chap:information-st}.
However, we will continue using notions of entropy and entropy production as consistency criteria.
Interestingly, (an extension of) the gauge invariance discussed in Section~\ref{sec:ref-meas} will play a central role.


  \chapter{The structure of Markov jump processes}
  \label{chap:cycles}
  \begin{fquote}[G.~Kirchhoff][Ueber die Aufl\"osung der Gleichungen, auf welche man bei der Untersuchung der linearen Vertheilung galvanischer Str\"ome gef\"uhrt wird] [1847 ]
  Ich will jetzt beweisen, da\ss~die Aufl\"osung der Gleichungen [auf welche man bei der Untersuchung der linearen Vertheilung galvanischer Str\"ome gef\"uhrt wird], sich allgemein angeben lassen.
\end{fquote}

\section*{What is this about?}
In the present chapter we are concerned with the algebraic and topological structure of Markovian dynamics on finite networks.
In Section~\ref{sec:discrete-st} we have seen how the network of states for a model of ST can be represented as an (undirected) graph.
Graphs are further commonly used to represent electrical circuits.

The introductory quote of Kirchhoff is from his work on the distribution of electrical currents in circuits build out of resistors and batteries.
An attempted translation reads:
\begin{quote}
  I am going to prove that the solution of the equations [encountered in the study of the distribution of galvanic currents] can be stated in a general way.
\end{quote}
The results proved by Kirchhoff are commonly known as Kirchhoff's current and voltage law, respectively.

Nowadays these results have been generalized by mathematicians as the \emph{matrix-tree theorem} of graph theory \cite{Tutte1998}.
As such, they are algebraic-topological results that hold for arbitrary graphs --- independent of the physical or mathematical problem they represent.
Here, we discuss their relevance for Markov processes on finite state spaces.
Consequently, they are going to be useful for the understanding of the structure of models used in ST.

We start the presentation in Kirchhoff's original setting, \ie the study of electrical networks.
The electrical currents running in an electrical circuit have their analogue in the  probability currents of a stationary Markov process.
Building up on this idea, the first result of the present chapter is a complete analogy between electrical circuits built from resistors and batteries with dynamically reversible Markov processes.
Moreover, this analogy carries over to ST.

After that we investigate the algebraic structure of the \emph{physical observables} corresponding to measurable currents in ST.
A result is the generalization of the so-called Schnakenberg decomposition \cite{Schnakenberg1976} to arbitrary physical observables \cite{Altaner_etal2012}.

The classical Schnakenberg decomposition is a result formulated for the steady-state ensemble averages of physical observables.
Here, we generalize this result further to statements about \emph{fluctuations} of physical observables away from their expectation value.
The main tool in our analysis will be the theory of large deviations for Markov processes.
A central object of that theory is the \emph{rate function} \gls{symb:rate-function} for physical observes $\obs$, which is a statement about their entire spectrum of fluctuations.

The main result of this treatment can be summarized as follows:
Firstly, we provide a fully analytical approach to the fluctuation spectrum of arbitrary physical observables, which does \emph{not} involve the solution of a complicated eigenvalue problem.
Secondly, we show that the fluctuation spectrum of \emph{any} physical observable is determined by the fluctuation spectrum of the probability currents on a special set of edges.

Parts of the results presented in the present chapter have already been published by the author in Ref.~\cite{Altaner_etal2012}.
The main theorems have been obtained in the context of the Artur Wachtel's M.Sc.~thesis~\cite{Wachtel2013}, which was jointly supervised by J\"urgen Vollmer and the author of the present thesis.
The proofs of the main results can be found there and in an unpublished manuscript by these authors~\cite{Wachtel_etal2014}.

\section{Kirchhoff's laws and an electrical analogy}
\label{sec:electric}
In the previous Chapters~\ref{chap:marksymdyn} and \ref{chap:information-st} we have discussed the deterministic microscopic origins of (Markovian) stochastic processes.
Here, we complement this perspective with a discussion of the structure of a stochastic network.
More precisely, we focus on (non-equilibrium) steady states of continuous time Markovian processes.
Let us briefly review the set-up for this chapter:

Similar to Section \ref{sec:discrete-st}, let $\omega \in \ospace = \set{1,2,\cdots,N}$ denote the elements of a finite state space.
The probability distribution $\bvec p\tind{t}$ evolves according to the continuous-time master equation \eqref{eq:master-continuous}
\begin{align*}
  \partial_t \bvec p\tind{t} = \bvec p\tind{t} \tmat.
\end{align*}
For continuous-time dynamics, the matrix $\tmat$ is a \emph{rate} matrix, rather than a stochastic matrix.
Its off-diagonal elements $\tprob{\omega}{\omega'},~\omega\neq\omega'$ are the time-independent transition rates.
The diagonal element $\tprob{\omega}{\omega} =-\sum_{\omega'} \tprob{\omega}{\omega'}=-\eav{\tau_\omega}^{-1}$ amounts to the negative of the escape rate from state $\omega$.
The inverse of the escape rate is the average \emph{staying time}:
For a Markov process, the staying time $\tau_\omega$ for a state $\omega$ obeys an exponential distribution $\prob[\tau_\omega = t] \propto \exp\left(-\frac{t}{\eav{\tau_\omega}}\right)$.
Similar to the previous chapters, we allow only one type of transition between any two states.
A generalization to multiple types of transitions is possible and has been discussed in Refs.~\cite{Andrieux+Gaspard2007,Esposito+vdBroeck2010,Faggionato+dPietro2011,Esposito2012}.
As we have discussed above, the reversibility of microscopic physical laws motivates dynamical reversibility, \ie $\tprob{\omega}{\omega'}>0\Leftrightarrow \tprob{\omega'}{\omega}>0$.

We visualize the network of states as a graph $\graph=(\verts,\edges)$ with vertices $v \in \verts$ and edges $\edge \in \edges$.
In the present context of Markovian processes, we identify $\verts \sim \ospace$ and draw an edge $\edge = (\omega,\omega')$ wherever $\tprob{\omega}{\omega'}>0$.
At time $t$ the system is in state $\omega$ with a probability $\p{t}{\omega}$.
If the network is connected (\cf Section~\ref{sec:discrete-st}), there exists a unique (invariant) steady-state distribution $\bvec p\tind{\infty}$ \cite{Feller1968}.

\subsection{Steady states and Kirchhoff's current law}
\label{sec:kirchhoff}
Henceforth, we focus on the steady state of a Markov process on a finite set of states $\omega \in \ospace$.
With the invariant distribution $\bvec p\tind{\infty}$, the Master equation \eqref{eq:master-continuous}  in matrix form reads
\begin{align}
  0 = \bvec p\tind{\infty}\tmat.
  \label{eq:master-equation-ss-matrix}
\end{align}
Note that for the time-continuous case the invariant density is a left eigenvector of the rate matrix $\tmat$ for the eigenvalue \emph{zero}.

It is instructive to discuss this equation also by looking at the individual components of the vector equation~\eqref{eq:master-equation-ss-matrix}.
With the definition of the steady state probability fluxes
\begin{align*}
  \fluxa{\omega}{\omega'} := \p{\infty}{\omega} \tprob{\omega}{\omega'},\quad~\omega\neq \omega',
\end{align*}
and the steady state currents
\begin{align*}
  \curra{\omega}{\omega'}:= \fluxa{\omega}{\omega'}- \fluxa{\omega'}{\omega},
\end{align*}
the master equation \eqref{eq:master-equation-ss-matrix} in component form reads
\begin{equation}
  \sum_{\omega'} \curra{\omega}{\omega'} = \sum_{\omega'} \left[ \fluxa{\omega}{\omega'} -\fluxa{\omega'}{\omega} \right]=0,~\forall \omega.
  \label{eq:master-equation-ss}
\end{equation}
Because of probability conservation, the net current $\sum_{\omega}\curra{\omega}{\omega'}$ arriving at each vertex $\omega'$ in the graph must vanish.
Equivalently, the \emph{influx} $\sum_{\omega'} \fluxa{\omega'}{\omega}$ into a state $\omega$ equals its \emph{outflux}  $\sum_{\omega'} \fluxa{\omega}{\omega'}$.

By definition, an \emph{equilibrium system} fulfils detailed balance.
Then, we have that $\curra{\omega}{\omega'}=0$ for all $\omega,\omega' \in \ospace$, \ie each individual term in the above sum in Eq.~\eqref{eq:master-equation-ss} vanishes.
In contrast, for a non-equilibrium steady state (NESS), there are at least some edges $(\omega,\omega')$ that support non-vanishing currents $\curra{\omega}{\omega'}$.

For currents flowing in electrical networks, Eq.~\eqref{eq:master-equation-ss} is known as \emph{Kirchhoff's current law} \cite{Kirchhoff1847}.
It has its origin in the conservation of electrical charge.
In the next subsection we will elaborate further on the analogy between electrical and stochastic networks.

\subsection{Kirchhoff's second law and an electrical analogy}
\label{sec:electric-analogy}
\begin{figure}[ht]
  \centering
  \includegraphics[scale=.8]{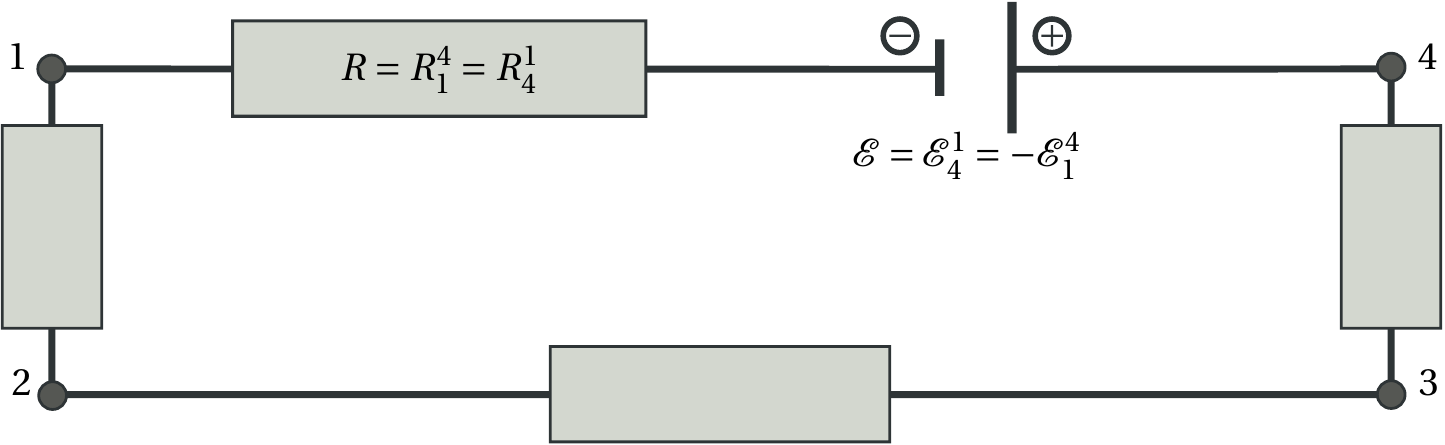}
  \caption{An electrical network.
    Each wire connecting two vertices $\omega$ and $\omega'$ has a positive resistance $\resa{\omega}{\omega'}=\resa{\omega'}{\omega}$.
    In addition, a battery-like element with electromotance $\mot = \mota{1}{4} = -\mota{4}{1}$ between vertices $1$ and $4$ results in a current $\curr = \curra{1}{4}$ with the same sign as $\mot$.%
  }
  \label{fig:electrical}
\end{figure}%
For our electrical analogy, let us review Kirchhoff's second law.
This law is also known as the \emph{mesh rule} or \emph{voltage law} and relates the voltage drops $V$ due to electrical resistance to the electromotive force $\mot$ of a battery.

Consider an electrical circuit consisting of battery-like elements and resistors like the one shown in Fig.~\ref{fig:electrical}.
We label the vertices by $\omega \in \ospace$.
Then, a wire connecting two vertices is identified by an edge $(\omega,\omega')$.
A battery-like element on an edge $(\omega,\omega')$ is characterized by its \emph{electromotive force} or \emph{electromotance} $\mota{\omega}{\omega'}$.
It can be either positive or negative, where the sign characterizes the direction of the flow of the mobile charges.
Consequently, the matrix containing the electromotances for all wires is anti-symmetric, \ie $\mota{\omega}{\omega'} = - \mota{\omega'}{\omega}$.
In contrast, the \emph{Ohmian resistances} $\resa{\omega}{\omega'} = \resa{\omega'}{\omega}$ are positive and symmetric.
If no current is flowing between two vertices $\omega$ and $\omega'$, a battery with electromotance $\mota{\omega}{\omega'}$ creates a voltage difference $\Delta \pota{\omega}{\omega'}= -\mota{\omega}{\omega'} $.
The voltage $\pot_\omega$ is measured with respect to an arbitrary reference potential $\pot_0\equiv0$.
If a current $\curra{\omega}{\omega'}$ is flowing over a resistor $\resa{\omega}{\omega'}$, the \emph{absolute value} of the difference to the reference potential $\Delta \pota{\omega}{\omega'}$ changes by the amount $-\vola{\omega}{\omega'} = \resa{\omega}{\omega'}\curra{\omega}{\omega'}$:
\begin{align}
  \Delta \pota{\omega}{\omega'}=-\mota{\omega}{\omega'} - \vola{\omega}{\omega'} = -\mota{\omega}{\omega'} +  \resa{\omega}{\omega'}\curra{\omega}{\omega'}.
  \label{eq:kirchhoff-equation}
\end{align}
Note that the motance has the same sign as the current it produces and hence we have  $\sgn(\mota{\omega}{\omega'})= \sgn(\resa{\omega}{\omega'}\curra{\omega}{\omega'})\equiv  -\sgn(\vola{\omega}{\omega'})$.
The dissipated power $\powa{\omega}{\omega'}$ along edge $(\omega,\omega')$ equals the product of the current times the negative of the voltage drop $\vola{\omega}{\omega'} $:
\begin{align}
  \powa{\omega}{\omega'} = -\vola{\omega}{\omega'} \curra{\omega}{\omega'} =\mota{\omega}{\omega'} \curra{\omega}{\omega'} \geq0.
  \label{eq:power}
\end{align}
The above mentioned \emph{voltage law} then states that the difference of reference potential $\pot_\omega$ along any cycle $\alpha\rlind{\tau} = (\omega_0,\cdots,\omega_{\tau-1},\omega_{\tau})$ in the network is conserved:
\begin{align}
  0 &= \sum_{i=1}^{\tau} \Delta \pota{\omega_{i-1}}{\omega_{i}}.
  \label{eq:kirchhoff-second-law}
\end{align}
An equivalent statement is that the sum off all voltage drops $\vola{\omega_{i-1}}{\omega_{i}}$ equals the sum of the electromotances $\mota{\omega_{i-1}}{\omega_{i}}$ along any cycle $\alpha\rlind{\tau}$:
\begin{align}
   \sum_{i=1}^{\tau} \mota{\omega_{i-1}}{\omega_{i}} = \sum_{i=1}^{\tau} \vola{\omega_{i-1}}{\omega_{i}}.
  \label{eq:kirchhoff-second-law-alt}
\end{align}
\begin{table}[t]
  \caption{Electrical and thermodynamic analogies, \cf~Ref.~\cite{Altaner_etal2012}. FED denotes free energy differences as in Hill's theory, \cite{Hill1977}.
 Note that the analogy is based in the steady state values of the fluxes and currents.}
 \centering
 \begin{tabular}{|c|c|cc|}
 \hline {\bf symbol}&%
  {\bf definition}&%
  {\bf thermodynamic}&%
  {\bf electric}\\
  \hline \hline \(\pot_\omega\) &%
    \( -\log \p{\infty}{\omega}\) &%
    {\small state variable}&%
    {\small potential}\\
  \hline \(\Delta \pota{\omega}{\omega'}\) &%
  \( \log [\p{\infty}\omega/\p{\infty}{\omega'}]\)&%
  {\small difference of a state variable}&%
  {\small potential difference}\\
  \hline \(\curra{\omega}{\omega'}\) &%
  \(\fluxa{\omega}{\omega'} - \fluxa{\omega'}{\omega}\) &%
  \multicolumn{2}{c|}{\small current} \\
  \hline \(\affa{\omega}{\omega'}\) &%
  \(\log[\fluxa{\omega}{\omega'} / \fluxa{\omega'}{\omega}]\) &%
   {\small affinity, \it {gross FED} }&%
  --- \\
  \hline \(\mota{\omega}{\omega'}\) &%
  \(\log[\tprob{\omega}{\omega'}/ \tprob{\omega'}{\omega}]\) &%
   {\small motance, \it basic FED}&%
  {\small electromotance}\\
  \hline \(\vola{\omega}{\omega'}\) &%
  \( -\mota{\omega}{\omega'}\)&%
  {\small negative motance}&%
  {\small voltage drop}\\
  \hline \(\resa{\omega}{\omega'}\) &%
  \( \vola{\omega}{\omega'}/\curra{\omega}{\omega'}\) &%
  ---&%
  {\small resistance}\\
  \hline \(\powa{\omega}{\omega'}\) &
  \(-\vola{\omega}{\omega'}\curra{\omega}{\omega'}\)
  &%
  {\small ---}&%
  {\small dissipated power}\\
  \hline \(\pow\) &
  \(\frac 1 2 \sum_{\omega,\omega'}\powa{\omega}{\omega'}\)
  &%
  {\small med/tot entropy production}&%
  {\small tot.~dissipated power}\\
  \hline
 \end{tabular}
 \label{tab:analogies}
\end{table}%
Let us now come back to the case of Markov processes.
In Section \ref{sec:discrete-st} we have introduced the \emph{affinity} $\affa{\omega}{\omega'}$ and the motance $\mota{\omega}{\omega'}$ of an edge $(\omega,\omega')$.
We saw how they appear in Schnakenberg's network theory \cite{Schnakenberg1976} and discussed their relation to the (change of) entropy in ST.
In the context of a steady-state Markov process, it is possible to define quantities on edges and vertices that fulfil expressions that are the analogues of Kirchhoff's laws.
We define the following quantities for any vertex $\omega$ and any edge $(\omega,\omega')$ where the steady-state current $\curra{\omega}{\omega'}$ does not vanish:
\begin{subequations}
  \begin{align}
    \pot_\omega &:= - \log {\p{\infty}{\omega}},\\
    \Delta \pota{\omega}{\omega'} &:= \log\frac{\p{\infty}{\omega}}{\p{\infty}{\omega'}},\\
    \vola{\omega}{\omega'} &:= - {\mota{\omega}{\omega'}} = - \log\frac{\tprob{\omega}{\omega'}}{\tprob{\omega'}{\omega}},\\
    \resa{\omega}{\omega'} &:= \frac{\vola{\omega}{\omega'}}{\curra{\omega}{\omega'}}.
  \end{align}
  \label{eq:electric-definitions}
\end{subequations}
The definitions above fulfil the properties of their electrical counter-parts:
The ``resistance'' matrix $\res$ is positive and symmetric, whereas the motance matrix $\mot$ as well as the current matrix $\curr$ are anti-symmetric.
By definition they fulfil the relations \eqref{eq:kirchhoff-equation}, \eqref{eq:kirchhoff-second-law} and \eqref{eq:kirchhoff-second-law-alt}.\footnote{This even holds for the transient case, if we substitute $\bvec p\tind{\infty}$ by $\bvec p \tind{t}$.}
With a factor $\frac{1}{2}$ due to the double-counting of edges, the dissipated power \eqref{eq:power} of an electrical network in the steady state amounts to
\begin{align}
  \pow &:= \frac{1}{2}\sum_{\omega,\omega'}\powa{\omega}{\omega'} =-\frac{1}{2}\sum_{\omega,\omega'}\curra{\omega}{\omega'}\vola{\omega}{\omega'} \nonumber\\
  &= \sum_{\omega,\omega'}\fluxa{\omega}{\omega'} \log\frac{\tprob{\omega}{\omega'}}{\tprob{\omega'}{\omega}} 
  \label{eq:total-diss-power}
\end{align}
This expression agrees with the medium entropy production for a stationary Markov process.
Note that because we are considering a steady state, the entropy change in the system vanishes and \eq{total-diss-power} similarly agrees with the \emph{total} entropy production.

Different electrical analogies have been presented in the literature and are suitable for different purposes (see e.g. \cite{Zia+Schmittmann2007,Feller1968}).
In Table~\ref{tab:analogies} we summarize our analogy, which has first appeared in a slightly less general form in Ref.~\cite{Altaner_etal2012}.
In addition, we give the \emph{thermodynamic} interpretations as discussed in Section~\ref{sec:discrete-st} and in Hill's work \cite{Hill1977}.

\section{Cycles and trees as the fundamental building blocks of networks}
In the introduction to this chapter, we mentioned that Kirchhoff's results are a special case of more general \emph{matrix-tree theorems} \cite{Tutte1998}.
In the present section, we provide an abstract algebraic framework for treating networks represented as graphs.

Let us recall the notation introduced in Section~\ref{sec:discrete-st}.
A \emph{directed graph} $\graph\tsup{d}=(\verts,\diredges)$ consists of vertices $v\in\verts$ and edges $e=(v,v') \in \diredges$.
For an \emph{undirected graph} $\graph\tsup{u} = (\verts,\udiredges)$ the edges $e = \set{v,v'} \in \udiredges$ are not ordered.
Henceforth, we will only consider bi-directional graphs, where the presence of an edge $(v,v')\in \diredges$ implies that $(v',v) \in \diredges$, too.

In contrast to the presentation in Section~\ref{sec:discrete-st}, we use the symbol $v_i \in\verts$ rather than $\omega$ to refer to a vertex.
In doing so, we stress the generality of the present algebraic discussion:
It is completely independent from the interpretation of the graph as a stochastic network.
Similarly, we choose an (arbitrary) enumeration for the edges $\edge_m \in \edges$.

\subsection{Anti-symmetric observables on the edges}
\label{sec:physical-obs}
As a motivation for what follows, let us introduce the notion of observables associated to edges.
The quantities defined in Table \ref{tab:analogies} can be understood as real-valued functions $\obs$ on the set of directed edges:
\begin{align}
  \obs\colon\diredges &\to \reals,\nonumber\\
  e = (v,v') &\mapsto \obs(e) = \obs^{v}_{v'}.
  \label{eq:edge-function}
\end{align}
More abstractly, they can be interpreted as (time-independent) \emph{one-chains}, \cf Chapter~\ref{chap:information-st} and Appendix~\ref{app:tau-chains}.

In particular, we are interested in \emph{anti-symmetric observables}.
For a graph which corresponds to a \emph{dynamically reversible} Markov process representing a mesoscopic description of a physical system, such observables have a physical interpretation.
To appreciate this fact, recall our motivation for a dynamically reversible graph in Section~\ref{sec:discrete-st}:
The states $\omega$ represent the values of some physical observable which depends only on the symmetric coordinates.
Physical currents are the instantaneous changes of measurable quantities.
Hence, they have to be anti-symmetric with respect to time-reversal.
This implies that the value associated to an observable transition $\omega\to\omega'$ must take the negative value if time were to run backward.
Thus, in the following we will refer to anti-symmetric observables defined on the edges as \emph{physical observables}.

With the definition of the \emph{negative} edge $-e:=(v',v)$ for any directed edge $e=(v,v')$, anti-symmetric observables obey
\begin{align}
  \obs(e) = \obs^{v}_{v'} = -\obs^{v'}_{v} = \obs(-e).
  \label{eq:anti-symmetry}
\end{align}
The definition of the negative (\ie inverse) element of a physical observable $\obs$ allows for an abstract treatment in the framework of linear algebra.

The \emph{averages} of such quantities have the interpretation of physical currents per unit time.
Examples include the current running through a wire in an electrical network, the heat flow in a thermodynamic system, reaction rate of chemical compounds or the velocity of a molecular motor.

However, we can also understand them as the increments of a counting process \cite{Harris+Schuetz2007}.
A jump along an edge $e=(v,v')$ increases a physical observable by an value $\obs(e)$.
Upon a jump along the reverse edge $-e$, the variable is decreased by the same amount.
The corresponding examples then are the transported charge, the motance interpreted as a basic free energy difference, the change in the number of certain chemical molecules and the distance of a mechanical step, respectively.

\subsection{Algebraic graph theory}
\label{sec:cycles}
\begin{figure}[th]
  \centering
  \includegraphics[scale=1]{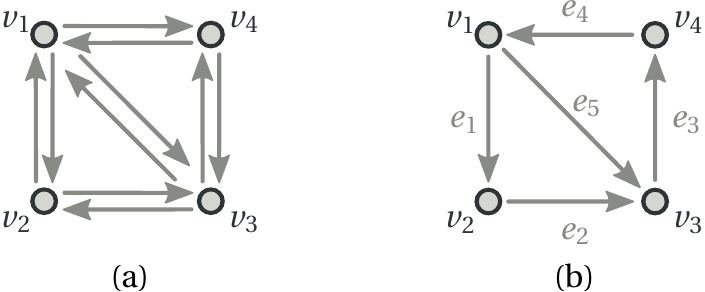}
  \caption{The relation between bi-directional and oriented graphs.
    a) In a bi-directional graph \(\left( \verts, \diredges \right)\) of a dynamically reversible Markovian jump process, we have that $e \in \diredges$ implies that also $-e \in \diredges$.
    b) An oriented graph \(\left( \verts, \oredges \right)\) is obtained by arbitrarily choosing one of the directed edges between any two vertices $v$ and $v'$.
  }
  \label{fig:graphs}
\end{figure}

Let us formally define the space of all anti-symmetric observables
\begin{align}
  \currents := \set{\obs\colon \diredges \to \reals \,\middle\vert\, \obs \text{ is anti-symmetric}}.
  \label{eq:defn-currentspace}
\end{align}
Anti-symmetric observables are completely defined by their values  on a set of \emph{oriented} edges $\oredges$ obtained by arbitrarily picking an ordered pair $(v,v')$ for each unordered pair $\set{v,v'} \in \udiredges$.
For the dynamically reversible (\ie bi-directional) graphs we are interested in, we have $\diredges = \oredges \cup -\oredges$.
The space of anti-symmetric observables $\currents$ is isomorphic to the space of functions defined on the oriented edges $\reals^{\oredges}$.
Consequently, we identify them with each other in what follows.

The notion of the negative $-e$ of an edge $e$ allows us to treat $\currents$ as a linear (vector) space.
If we identify an edge $e \in \oredges$ with its indicator function
\begin{align*}
  e\colon\oredges &\to \reals,\\
  e' &\mapsto \delta_{e,e'} = 
  \begin{cases}
    1 & e=e',\\
    0 & e\neq e',
  \end{cases}
\end{align*}
we can use the set of all oriented edges $\oredges$ as a basis for the space of anti-symmetric variables $\currents$.
An element $\obs\in\currents$ can hence be written as
\begin{align}
  \obs = \sum_{e\tsub{o} \in \oredges}\obs(e\tsub{o})e\tsub{o} = \sum_{m=1}^M \obs_m e_m,
  \label{eq:current-edge-decomposition}
\end{align}
where $M := \abs{\oredges}$ is the number of oriented edges and $\ifam{e_m}_{1\leq m\leq M}$ assigns some arbitrary enumeration to the oriented edges.
This allows for the definition of a bilinear \emph{scalar product} for any two anti-symmetric observables $\obs,\psi\in\currents$:
\begin{align}
  \sprod{\obs}{\psi} := \sum_{m=1}^M \left[\obs_m \psi_m\right].
  \label{eq:edge-sp}
\end{align}
Note that the basis given by $\oredges$ is orthonormal with respect to that scalar product.

In analogy to our algebraic treatment of the edges, we do the same for the vertices.
Again, we identify a vertex $v \in \verts$ with its indicator function $v\colon \verts \to \reals; v' \mapsto \delta_{v,v'}$.
Consequently, the space of all real functions on the vertices, $\potentials := \reals^\verts$ is identified with the linear space spanned by the $v\in\verts$.
Similarly, the scalar product for $\potentials$ treats the vertices as an orthonormal basis.

Any linear operator between linear spaces is fully defined by its action on the respective basis elements.\footnote{ The matrix representation of operators in quantum mechanics is a well-known example.}
Hence, the so-called \emph{boundary} operator
\begin{align}
  \del\colon\oredges&\to\potentials\nonumber\\
  (v,v') &\mapsto v - v'
  \label{eq:boundary}
\end{align}
linearly extends to an operator $\del\colon\currents \to \potentials$.
It is dual (with respect to the natural scalar products on $\currents$ and $\potentials$) to the \emph{co-boundary} operator $\del^*\colon\potentials \to \currents$.
For illustrative purposes, in Figure~\ref{fig:boundary-coboundary} we display the action of the boundary and the co-boundary operator on edges and vertices, respectively.

\label{sec:cycles-cocycles}
\begin{figure}[htb]
  \centering
  \includegraphics[scale=1]{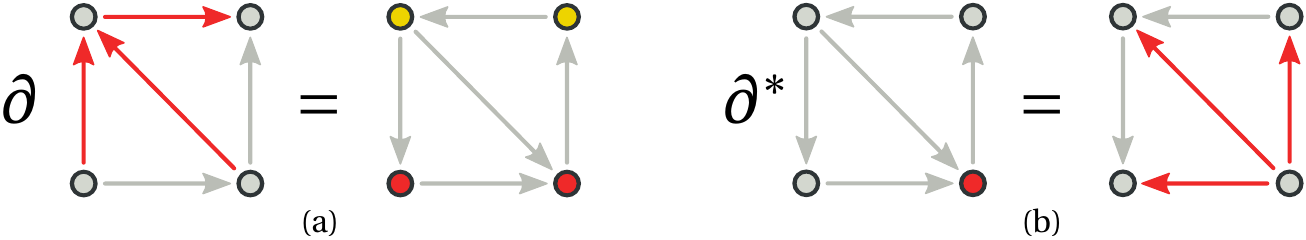}
  \caption{
  The action of the boundary and the co-boundary exemplified on the graph of Figure~\ref{fig:graphs}.
  a) The boundary operator acts on the directed edges marked in red, whose linear combination is an element of the space $\currents$.
  b) The co-boundary acts on vertex space $\verts$.
  An edge or vertex with weight \(0\) is marked grey, a weight of \(1\) is indicated red, while a weight of \(-1\) is yellow.
  Note that a negative weight on an edge is equivalent to a positive weight with reverse orientation.
  }
  \label{fig:boundary-coboundary}
\end{figure}

The definition of the boundary operator allows us to express Kirchhoff's current law and thus the stationarity condition for master equations \eqref{eq:master-equation-ss} as
\begin{align*}
  \del \curr = 0.
\end{align*}
It can be understood as the discrete analogue of the divergence of a vector field.
The \emph{cycle space} $\cycles$ contains the \emph{divergence-free} currents:
\begin{align}
  \cycles := \ker{\del} \subset \currents,
  \label{eq:cycle-space}
\end{align}
where ``$\ker\del$'' denotes the kernel (null-space) of the linear boundary operator $\del$.

The image of the dual co-boundary operator,
\begin{align}
  \cocycles := \img{\del^*}\subset\currents
  \label{eq:cocycles}
\end{align}
is called the \emph{co-cycle space} and can be interpreted as the set of \emph{gradient fields}.
We will discuss the relation to vector calculus and field theory in more detail in Section~\ref{sec:field-theory}.

Duality of $\del$ and $\del^*$ ensures that the cycles and co-cycles form an orthogonal decomposition of the space of anti-symmetric observables $\currents$, \ie $\cocycles = \cycles^{\perp}$.
Thus we can decompose the space of anti-symmetric observables as
\begin{align*}
  \cycles \oplus \cocycles = \cycles \oplus \cycles^{\perp} = \currents.
\end{align*}
Consequently, if we pick arbitrary elements $z\in\cycles$ and $y \in \cocycles$ of the cycle and co-cycle space, respectively, we have $\sprod{z}{y}=0$.

\subsection{Trees and chords}
\label{sec:graph-topology}

\begin{figure}[htb]
  \centering
  \includegraphics[scale=1]{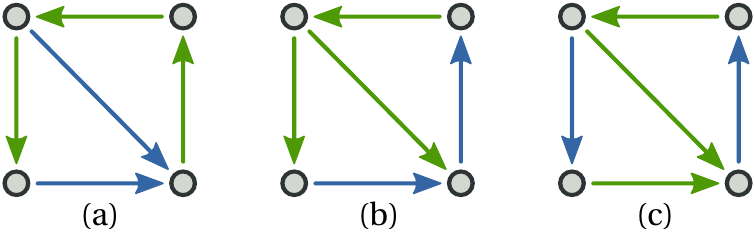}
  \caption{
  Three different spanning trees for the graph given in Figure~\ref{fig:graphs}.
  The edges \(\tredges\) of the different trees are marked green, while the chords \(\chords\) are depicted blue.
  Every other spanning tree of the graph, up to symmetries, looks like one of the depicted ones.
  In the following we will mainly consider the spanning tree \ref{fig:trees}a.
  }
  \label{fig:trees}
\end{figure}

In the previous subsection, we have defined cycles as elements of the kernel of the boundary operator.
This agrees with the picture of a cycle as ``something that has no boundary''.
However, the usual intuition of a cycle is that of a closed path $\alpha\rlind{\tau} = (\omega_0,\omega_1,\cdots,\omega_\tau)$ with $\omega_0 = \omega_\tau$.
In Section~\ref{sec:discrete-st} we have mentioned \emph{fundamental cycles} and that they play an important role in Schnakenberg's network theory for master equations \cite{Schnakenberg1976}.
In this subsection, we will make clear what we mean by a fundamental cycle and how this notion fits into the general algebraic framework.

Fundamental cycles naturally emerge from the topology of the graph.
To make that statement precise, we consider \emph{subgraphs} $\graph'\subset\graph$ of a graph $\graph$.
We say that a graph $\graph' = (\verts',\edges')$ is  a subgraph of $\graph = (\verts,\edges)$ if $\verts' \subset \verts$ and $\edges'\subset\edges$. 

A \emph{circuit} is a connected graph $(\verts,\oredges)$ where the number of edges and vertices is the same, \ie $\abs{\oredges} = \abs{\verts}$.
Consequently, every vertex has exactly two neighbours.
Intuitively, a circuit can be understood as the graph of a discretised circle whose edges are not necessarily all pointing in the same direction.

A \emph{tree} is a connected graph $(\verts,\tredges)$ that contains no circuits as subgraphs.
It necessarily satisfies $\abs{\tredges} = \abs{\verts}-1$.
Every connected graph $\graph = (\verts,\oredges)$ has a tree $(\verts,\tredges)$ as a subgraph which contains all vertices $\verts$ of $\graph$.
Such a tree is called a  \emph{spanning tree} for the graph $\graph$.

For a given graph which itself is not a tree, the choice of the spanning tree is not unique.
Figure~\ref{fig:trees} shows different spanning trees of the oriented graph shown in Figure~\ref{fig:graphs}b.
The set of edges of $\graph$ that do not belong to a given tree, are called its \emph{chords} $\eta \in \chords = \edges\setminus\tredges$.
Adding a chord $\eta$ to a spanning tree creates the subgraph $(\verts,\tredges\cup\set{\eta})$ which contains exactly one circuit.
By aligning the edges on that circuit to point in the same direction as $\eta$, we obtain the \emph{fundamental cycle} $\zeta_\eta$.
The fundamental cycle is a subgraph of the bidirectional graph $\graph = (\verts,\diredges)$.

For the example of the spanning trees depicted in Figure~\ref{fig:trees}, the fundamental cycles are shown in Figure~\ref{fig:cycles}.
A fundamental cycle has no boundaries and is thus an element of the cycle space.
Note that  all fundamental cycles are linearly independent (as they do not share any chords) and their number is $\abs{\chords} = \abs{\oredges} - \abs{\verts} +1$.
This number is called the \emph{first Betti} or \emph{cyclomatic} number and is a topological constant.
It coincides with the dimension of the cycle space $\cycles$, \cf Ref.~\cite{Tutte1998}.
Hence, the set of fundamental cycles provide a basis for $\cycles$.

\begin{figure}[h]
  \centering
  \includegraphics[scale=1]{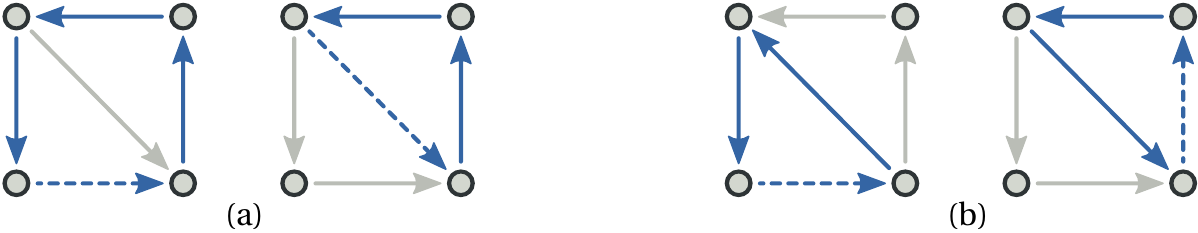}
  \caption{Fundamental cycles of the spanning trees in figure~\ref{fig:trees}: every chord \(\eta\in\chords\) (dashed here, blue in figure~\ref{fig:trees}) generates a fundamental cycle, also marked in blue here.
  The grey edges and vertices are not part of the fundamental cycles.
  Note that both spanning trees in figure~\ref{fig:trees}b and c generate the exact same fundamental cycles, but with different chords.
  In contrast, the spanning tree in figure~\ref{fig:trees}a shares only one fundamental cycle with the spanning trees b and c.
  }
  \label{fig:cycles}
\end{figure}

Consequently, every abstract cycle $z \in \cycles$ can be written as a linear combination of fundamental cycles 
\begin{align}
 z = \sum_{\eta \in \chords} \zeta_\eta z(\eta),
  \label{eq:fundamental-cycle-decomposition}
\end{align}
where $z(\eta)$ is the value of $z$ on chord $\eta$.
The relation \eqref{eq:fundamental-cycle-decomposition} yields Schnakenberg's \emph{cycle decomposition} for the steady state probability current $\curr$ \cite{Schnakenberg1976}:
\begin{align}
  J = \sum_{\eta \in \chords} \zeta_\eta J(\eta).
  \label{eq:cycle-decomposition-current}
\end{align}
It implies that the steady state current $\curr$ is completely defined by its values $\curr(\eta)$ on the chords.
Equation \eqref{eq:fundamental-cycle-decomposition} is the appropriate generalization.

The orthogonal decomposition of $\currents$ into abstract cycles and co-cycles implies that
\begin{align}
  \sprod{z}{y}=0,\quad\forall z\in \cycles, y\in \cocycles.
  \label{eq:second-Kirchhoff-law-abstract}
\end{align}
Using the intuition provided by the fundamental cycles, this equation is a generalization of Kirchhoff's second law.
In order to appreciate this fact, note that any cycle $z$ is a linear combination of fundamental co-cycles and thus
\begin{align}
  0=\sprod{z}{y}=\sum_{\eta \in\chords} z(\eta)\sprod{\zeta(\eta)}{y}, \quad\forall z\in \cycles, y\in \cocycles.
  \label{eq:second-Kirchhoff-law-explained}
\end{align}
Then, any term appearing in the sum is a discrete integral around the edges of the cycle $\zeta(\eta)$, \ie $\sprod{\zeta(\eta)}{y} = \sum_{\edge \in \zeta(\eta)}y(\eta)$.
Thus, expression~\eqref{eq:second-Kirchhoff-law-abstract} is just a weighted sum of observables ``integrated'' around (independent) cycles in the graph.

\begin{figure}[h]
  \centering
  \includegraphics[scale=1]{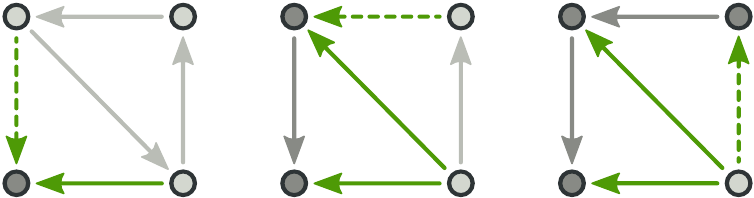}
  \caption{Fundamental co-cycles of the spanning tree in figure~\ref{fig:trees}a: every edge \(\tau\in\tredges\) (dashed here, green in figure~\ref{fig:trees}) generates a fundamental co-cycle, also marked in green here.
  The grey edges and vertices are not part of the fundamental co-cycles.
  }
  \label{fig:cocycles}
\end{figure}

A similar, but slightly less intuitive approach allows us to construct a basis of the co-cycle space $\cocycles$.
The removal of an edge $\tau \in \tredges$ from a spanning tree yields a disconnected subgraph consisting of exactly two connected components.
Note that these components can be as small as a single vertex without any edges.
Within the set of edges $\chords\cup\set{\tau}$ there is a subset of edges containing the edge $\tau$ together with all other edges that connect these two components.
This subset can be understood as the \emph{boundary} between the two components.
Reorienting the boundary edges to align with $\tau$ yields the \emph{fundamental co-cycle} corresponding to edge $\tau$.
Figure~\ref{fig:cocycles} illustrates this construction.

\section{Quantification of fluctuations of physical observables}
\label{sec:large-deviations}
The algebraic structure of physical observables on finite networks is reflected by the stochastic properties of Markov processes.
In particular, it allows us to quantify the fluctuations of physical observables in an analytic way.
In what follows, a fluctuation means the behaviour of a system away from its average.

Mathematically, the fluctuations of a random variable $X$ are contained in its probability distribution $\pden^X(x)$.
For distributions which are strongly peaked around a certain value, its \emph{mean} (or \emph{expectation})
\begin{align}
  \mean (X):=\eav{x} = \int x \pden^X(x) \df x
  \label{eq:expectation-value}
\end{align}
and its \emph{variance}
\begin{align}
  \var (X) := \eav{\left(x - \mathbb E(X)\right)^2} = \int \left(x - \mathbb E(X)\right)^2 \pden^X(x) \df x
  \label{eq:mean}
\end{align}
are a good characterization of the distribution.
The mean defines a typical value and the variance (or its square root, the so-called \emph{standard deviation}) characterize the fluctuations around that value.
However, for distributions that have non-negligible probability in the \emph{tails} of the distribution, one requires more information to capture their atypical behaviour.

\subsection{Cumulants of random variables}
The mean and the variance of a distribution are the first two \emph{cumulants} of a distribution.
The general definition for an \(\reals^d\)-valued random variable \(\bvec{X}=\ifam{X_i}_{i\in\set{1,2,\cdots,d}}\) proceeds via the cumulant-generating function (CGF)
\begin{align}
  g_{\bvec{X}}\colon \reals^d&\to\reals,\nonumber\\
  \bvec q &\mapsto \log \langle \exp\left( \bvec{q}\cdot\bvec{X} \right)\rangle.
  \label{eq:cgf}
\end{align}
The \emph{joint cumulants} are obtained as the partial derivatives of the CGF with respect to the components of $\bvec q$ evaluated at $\bvec q = 0$:
\begin{align}
  \kappa(X_{i_1},X_{i_2}, \dots,X_{i_\ell}) := \frac{\del}{\del q_{i_1}}\frac{\del}{\del q_{i_2}}\dots\frac{\del}{\del q_{i_\ell}} g_{\bvec{X}}(\bvec{0})\,,
\end{align}
We call the (countable) set of all cumulants the \emph{fluctuation spectrum} of a distribution.
If the components $X_i$ of $\bvec X$ are independent random variables, the mixed joint cumulants vanish.
Moreover, the cumulants are multi-linear in their arguments.
We emphasize that the CGF and thus the fluctuation spectrum contains \emph{all} the information of the original distribution.

The cumulant-generating function \(g_{\bvec{X}}(\bvec{q})\) is (non-strictly) convex and satisfies \(g_{\bvec{X}}(\bvec{0})=0\) for all \(\bvec{X}\).
If $\bvec X$ is scalar (\ie \(d=1\)), one defines the $\ell$th cumulant 
\begin{align}
  \kappa(\underbrace{X, \dots, X}_{\ell\text{ times}}) := \kappa_\ell(X)\,.
  \label{eq:ho-cumulant}
\end{align}


\subsection{Asymptotic properties in stochastic processes}
In Chapter~\ref{chap:marksymdyn} we have formally defined a stochastic process as a sequence of random variables $\ifam{X_t}_{t\in\timdom}$.
Here, we consider the case of sequences of random vectors $\ifam{\bvec X_t}_{t\in\timdom}$ and $\timdom = \reals^+$ or $\timdom =\naturals$.
We are interested in the distribution $\pden\rlind{\tau}_{\bvec X}$ of the time-average
\begin{align*}
  \bar{\bvec X}\rlind{\tau} := \frac{1}{\tau}\int_{t=0}^\tau \bvec X_t \df t,
\end{align*}
where for $\timdom = \naturals$ the integral turns into a sum.
More precisely, we are interested in its asymptotic properties for $\tau \to \infty$.

If the individual random vectors $\bvec X_t=\bvec X$ are independent and identically distributed (i.i.d.) with mean $\mean(X) = \mu$, the \emph{strong law of large numbers} ensures that 
\begin{align*}
  \lim_{\tau \to \infty} \bar{\bvec X}\rlind{\tau} \xrightarrow{\assure} \mu.
\end{align*}
This means that fluctuations around the mean value asymptotically vanish.
In that case, the \emph{central limit theorem} (CLT) makes this statement more precise, as it specifies an asymptotic distribution.
If $\sigma^2 = \var(X)$ denotes the variance of $\bvec X$, one formulation of the CLT states that
\begin{align}
  \lim_{\tau \to \infty} \sqrt{\tau}\left( \bar{\bvec X}\rlind{\tau} - \mu \right)\xrightarrow{\text{in distribution}} \mathcal N(0,\sigma^2),
  \label{eq:clt}
\end{align}
where $\mathcal N(\mu,\sigma^2)$ is a \emph{normally distributed} random variable with mean $\mu$ and variance $\sigma^2$.
A consequence of the CLT is the square-root scaling behaviour of the variance $\var\left( \bar{\bvec X}\rlind{\tau}\right)$ with $\tau$.

Large-deviations theory (LDT) generalizes scaling statements about the distribution of sequences of random variables.
A comprehensive introduction to the topic can be found in Ref.\,\cite{Touchette2011}.
The standard textbook on the subject and its relation to (classical) statistical physics is the one by Ellis, \cf Ref.\,\cite{Ellis2005}.
The relation is discussed in even more detail in the review article \cite{Touchette2009}.

Consider a sequence of random vectors $\ifam{ \bar{\bvec X}\rlind{\tau}}_{t \in \timdom}$.
We say that this sequence obeys a \emph{Large deviation principle} if the limit
\begin{align}
  I_{\bvec X}(\bvec{x}) \coloneqq - \lim_{\tau\to\infty}\frac{1}{\tau}\log\left[ \pden\rlind{\tau}_{\bvec{X}}(\bvec{x})\right]\,. \label{eq:large-deviation-principle}
\end{align}
exists (at least in an open neighbourhood of $\reals^d$ around the origin).
The function \(I_{\bvec X}(\bvec{x})\) is called \emph{rate function}.
If a large deviations principle holds,  the probability density $\pden_{\bvec{X}}\rlind{\tau}$ obeys the scaling
\begin{align}
  \pden_{\bvec{X}}\rlind{\tau}(\bvec{x})\propto \exp\left[ -\tau\,I_{\bvec{X}}(\bvec{x})+R(\tau,\bvec{x}) \right],
  \label{eq:ldt-pden-approx}
\end{align}
where $R(\tau,\bvec{x}) \sim o(\tau)$ scales sub-linearly with $\tau$.
Hence, for large $\tau$ we can approximately write $\pden_{\bvec{X}}\rlind{\tau}(\bvec{x})\propto\exp\left[-\tau\,I_{\bvec{X}}(\bvec{x}) \right]$.

For the case of i.i.d.~random variables, we have see that asymptotically the variance, \ie its second cumulant vanishes.
In that case,the CLT \eqref{eq:clt} implies the same for all higher cumulants.
In fact, this is generally true for random variables that obey a large deviation principle.

Hence, the fluctuation spectrum for such random variables is not useful in the asymptotic limit $\tau \to \infty$.
However, the scaling relation \eqref{eq:large-deviation-principle} suggest the definition of the \emph{scaled cumulant-generating function} (SCGF):
\begin{align}
  \lambda_{\bvec{X}}(\bvec{q}) := \lim_{\tau\to\infty}\frac{1}{\tau} g_{\bar{\bvec{X}}\rlind{\tau}}(\tau\bvec{q}) = \lim_{\tau\to\infty}\frac{1}{\tau}\log \left\langle \exp\left[\tau\,\bvec{q} \cdot \bar{\bvec{X}}\rlind{\tau}\right]\right\rangle\,. \label{eq:def-scgf}
\end{align}
where again $\bvec{q}\in\reals^d$.
Like the normal cumulant-generating function, the SCGF is convex and \(\lambda_{\bvec{X}}(\bvec{0})=0\) for every family of random vectors.
The cumulant-generating function $g_{\bvec{X}}(\bvec{q})$ of a random variable \(\bvec{X}\) is always a smooth function.
However, the scaled limit $\lambda_{\bvec{X}}(\bvec{q})$ is no necessarily differentiable any more.

If  \(\lambda_{\bvec{X}}(\bvec{q})\) is differentiable on \(\reals^d\), the G\"{a}rtner--Ellis Theorem then assures that the sequence $\ifam{\bar{\bvec{X}}\rlind{\tau}}_{\tau \in \timdom}$ satisfies a Large Deviation Principle \cite{Ellis2005}.
Moreover, its rate function 
\begin{align}
  I_{\bvec X}(\bvec{x}) = \bvec{x}\cdot \bvec{q}(\bvec{x}) - \lambda_{\bvec{X}}(\bvec{q}(\bvec{x}))  \label{eq:gaertner-ellis}
\end{align}
 is obtained as the Legendre transform of $\lambda$.
Note that the functional dependence $\bvec{q}(\bvec{x})$ is given by inverting the relation $\bvec{x}=\nabla \lambda_{\bvec{X}}(\bvec{q})$.

In that case, we define the \emph{scaled fluctuation spectrum} as the set of \emph{scaled cumulants}:
\begin{align}
  c(X_{i_1},X_{i_2}, \dots,X_{i_\ell}) \coloneqq  \frac{\del}{\del q_{i_1}}\frac{\del}{\del q_{i_2}}\dots\frac{\del}{\del q_{i_\ell}} \lambda_{\bvec X}(\bvec{0}) \,.
\end{align}

The scaled cumulants directly inherit multi-linearity from the cumulants.
From the definition in equation~(\ref{eq:def-scgf}) we immediately infer the scaling of the cumulants to be  
\begin{align}
  c(X_{i_1},X_{i_2}, \dots,X_{i_\ell}) = \lim_{\tau\to\infty} \tau^{\ell-1} \kappa( X_{i_1}\rlind{\tau},X_{i_2}\rlind{\tau}, \dots,X_{i_\ell}\rlind{\tau})\,. \label{eq:cumulant-scaling}
\end{align}

So far we dealt with general sequences of random vectors $\bar{\bvec X}\rlind{\tau}$.
In the next section we will explicitly consider the sequence of time-averages of increasing length.
If these time-averages are obtained for physical observables in a Markov process, we have analytical access to the SCGF.
Henceforth, we are always interested in the characterization of the asymptotic fluctuations.
Hence, we omit the word ``scaled'' and refer simply to the \emph{fluctuation spectrum} of an observable.

\subsection{Large deviation theory of Markovian jump processes}
%
In this section we consider the time-averages of the physical observables for Markov processes in continuous time.
In Section~\ref{sec:physical-obs} we introduced the notion of a physical observable $\obs\in\currents$ as an anti-symmetric variable defined on the edges of a graph:
\begin{align*}
  \obs\colon\edges&\to\reals\\
  (\omega,\omega')&\mapsto \obs^{\omega}_{\omega'}
\end{align*}
For an ergodic Markov chain on a finite state space we define the time-average as
\begin{align}
  \bar{\obs}\rlind{\tau} := \frac{1}{\tau} \sum_{n=1}^{N(\tau)} \obs^{\omega_{n-1}}_{\omega_n}.
  \label{eq:time-average-ldt}
\end{align}
The sum is over the states $(\omega_n)_{n \in \set{0,1,\cdots n(\tau)}}$ that occur in the trajectory $\traj \omega\rlind{\tau}$ for the interval $[0,\tau]$ .
For dynamics in continuous time the number of jumps $n(\tau)$ is itself a random variable.
The distribution of these time averages always satisfies a Large deviation principle \cite{Touchette2011}.
Moreover, the scaled cumulant-generating function \gls{symb:scgf} is differentiable and can be calculated as an eigenvalue of an appropriately defined matrix.

To be concrete, we consider the multi-variate case where $\bvec{\obs}=(\obs_{1}, \obs_{2}, \dots, \obs_{d})\in\currents^d$ is a $d$-tuple of anti-symmetric observables.
Its component-wise time average \eqref{eq:time-average-ldt} is a family of random vectors $\bar{\bvec{\obs}}\rlind{\tau}$ taking values in \(\reals^d\).
The multivariate SCGF \(\lambda_{\bvec{\obs}}(\bvec{q})\) is given by the unique dominant eigenvalue of the \emph{tilted matrix} \(\tmat_{\bvec{\obs}}(\bvec{q})\) with components \cite{Touchette2011}
\begin{align}
  \left(\tmat_{\bvec{\obs}}(\bvec{q})\right)^i_j
  := \tprob{i}{j} \exp\left( \bvec{q} \cdot\bvec{\obs}^i_j \right)
  = \tprob{i}{j} \exp\left(\sum_{\ell=1}^d q_\ell \obs_{\ell}(e^i_j) \right)\,.
  \label{eq:skewed-generator}
\end{align}
The expression \(\obs_{\ell}(e^i_j)\) stands for the value of the $\ell$th random variable in $\bvec \obs$ on the edge $e^i_j$, along which transitions occur at rate \(\tprob{i}{j}\).

The scaled cumulants are obtained as the partial derivatives of the dominant eigenvalue.
Generally, the analytic calculation of the dominant eigenvalue of a big matrix is difficult.
At this point one usually applies numerical algorithms to find (approximations) to $\lambda_{\bvec \obs}(\bvec q)$ \cite{Touchette2011}.
Thus, in general it is difficult to find the large-deviation function $I_{\bvec{\obs}}(\bvec x)$.

The key result of the present chapter is an easy method for the determination of the fluctuation spectrum of \emph{any} physical observable $\bvec \obs$ --- without explicitly referring to $I_{\bvec \obs}(\bvec x)$.
The crucial point is to realize that all of the relevant information is already contained in the coefficients $a_i(\bvec q)$ of the characteristic polynomial
\begin{align}
  \chi_{\tmat_{\bvec{\obs}}(\bvec{q})}(\lambda) := \sum_{i=0}^{N} a_{i}(\bvec{q})\,\lambda^{i}  := \det\left( \lambda\mathbb{1} -\tmat_{\bvec{\obs}}(\bvec{q})  \right)\,.
\end{align}
The roots of the characteristic polynomial are the eigenvalues \(\lambda\) of the tilted matrix \(\tmat_{\bvec \obs}(\bvec{q})\).
The characteristic equation
\begin{align*}
  \chi_{\tmat_{\bvec{\obs}}(\bvec{q})}(\lambda) \shouldbe 0
  \label{eq:characteristic}
\end{align*}
contains all the information about the largest eigenvalue \(\lambda_{\bvec{\obs}}(\bvec{q})\)  of the tilted matrix.
Since it is the cumulant-generating function, it fulfils \(\lambda_{\bvec{\obs}}(\bvec{0})=0\).
This relation uniquely determines the \emph{solution branch} of \eq{characteristic} which gives the $\bvec q$-dependence of \(\lambda_{\bvec{\obs}}(\bvec{q})\).
Thus, we can use the Implicit Function Theorem to extract the scaled cumulants iteratively from the coefficients \(a_i(\bvec{q})\) of $\chi_{\tmat_{\bvec{\obs}}(\bvec{q})}$.
Their \(\bvec q\)-dependence is directly analytically accessible from the determinant formula for matrices.

The most important case is that of a single current-like observable, where both \(\bar{\obs}\rlind{\tau}\) and \(q\) are scalar.
Its first two scaled cumulants read
\begin{subequations}
\begin{align}
  c_1(\obs) &=-\frac{a_{0}'}{a_{1}}\,, \label{eq:magic-formula-a}\\
     c_2(\obs) &=\frac{-2\, a_2 c_1^2(f)-2\, c_1(f) a_1'-a_0''}{a_1}
             =2\frac{a_1' a_0'}{a_1^2} -2\frac{a_2\left(a_0'\right)^2}{a_1^3} - \frac{a_0''}{a_1}\,,
\end{align}
  \label{eq:scaled-cumulants-example}
\end{subequations}
where all of the \(a_i\) and their derivatives must be evaluated at \(q=0\).
Higher cumulants can be calculated iteratively.
For a more detailed account on the analytical calculation of the fluctuation spectrum we refer to the M.Sc. thesis~\cite{Wachtel2013} and the pre-print~\cite{Wachtel_etal2014}.

Let us stress the following fact:
The only ingredients needed for the calculation of the asymptotic fluctuation spectrum are the transition matrix \(\tmat\) and the observable \(\obs\).
At no point we explicitly need the (steady-state) distribution $\bvec p\tind{\infty}$ of the Markov chain.
In the previous chapter, we have seen that the latter is completely determined by the currents on a set of fundamental chords.
Moreover, we have seen how a physical observable can be decomposed into fundamental cycles.
The next subsection deals with an important consequence of this fact for the fluctuation spectra of time-averages $\bar{\obs}\rlind{\tau}$ taken for physical observables $\obs \in\currents$.

\subsection{Cycles and fluctuations of physical observables}
\label{sec:cycle-ldt}
In this subsection, we clarify how the topological structure of the Markovian jump process influences the fluctuation spectrum of scaled cumulants.
The results in this section have mainly been developed by A.\,Wachtel.
We state them without proof and refer the reader to Ref.~\cite{Wachtel_etal2014}.

In Section~\ref{sec:cycles-cocycles} of the previous chapter we have introduced algebraic concept of a cycle $z\in\cycles$.
Remember that we identified the space of anti-symmetric observables $\currents$ with the vector space $\reals^\oredges$, where $\oredges$ was a set of oriented edges of a bi-directional graph.
In that abstract sense, a cycle was identified with an element of the subspace $\cycles\subset\currents$ defined as the kernel of the boundary operator $\del$.
The orthogonal complement of $\cycles$ in $\currents$ is the co-cycle space $\cocycles$.
This orthogonal decomposition of $\currents$ guarantees that any anti-symmetric observable can be written as a unique sum \(\obs=z+y\) of a cycle \(z\in \cycles\) and a co-cycle \(y \in \cocycles\).

Because the (scaled) cumulants are multi-linear, we can calculate the fluctuation spectrum of $\obs$ from the fluctuation spectrum of \(z\) and \(y\).
By ergodicity of the Markov chain, we have $\lim_{\tau \to \infty} \obs\rlind{\tau} \xrightarrow{\assure} \eav{\obs}\tind{\infty}$.
As the first scaled cumulant agrees with the first cumulant of $\obs$, it can be written as the scalar product of $\obs$ with the steady state current $\curr$:
\begin{align*}
  \kappa_1(\obs) = c_1(\obs) = \mean(\obs) = \eav{\obs}\tind{\infty} = \sprod{\obs}{\curr}.
\end{align*}
We already know that for a co-cycle $y \in \cocycles$, Kirchhoff's second law holds, \ie
\begin{align}
  \kappa_1(y) = \sprod{y}{\curr} = 0
  \label{eq:second-law-first-cumulant}
\end{align}
The following proposition thus generalizes Kirchhoff's second law to the entire fluctuation spectrum:
\begin{proposition}
  Let \(y\in \cocycles\) be a co-cycle.
  Then its scaled cumulant-generating function \(\lambda_y(q)\equiv 0\) vanishes.
  Thus, all of its scaled cumulants vanish.
\end{proposition}
Andrieux and Gaspard~\cite{Andrieux+Gaspard2007} proved a special case of this result.
Similar considerations can be found in Ref.\,\cite{Faggionato+dPietro2011}.
The general proof has the following corollary~\mbox{\cite{Wachtel2013,Wachtel_etal2014}}:
  \begin{corollary}
    \label{theo:cycles-matter}
  The scaled cumulant-generating function \(\lambda_\obs(q)\) of a current-like observable \(\obs\in\currents\) satisfies \(\lambda_\obs(q)=\lambda_z(q)\) where \(z\in \cycles\) is the unique cycle part of \(\obs\).
\end{corollary}
In other words, the SCGF and thus the rate function of all the observables in the subspace \(\obs+\cocycles\subset \currents\) agree with that of both \(\obs\) and its cycle part \(z\).
Regarding the asymptotic fluctuations, this can be understood as a form of gauge invariance.
Consequently, for calculating \(\lambda_\obs\) we can use any representative within the class \(\obs+\cocycles\).
\begin{figure}[thb]
  \centering
  \includegraphics[scale=1.3]{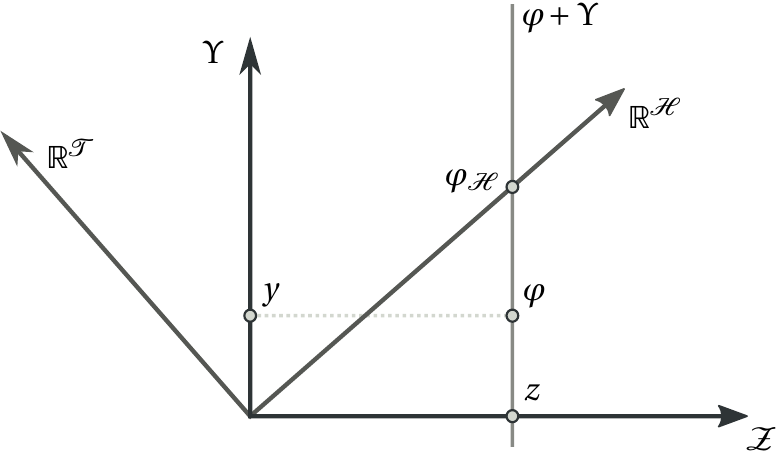}
  \caption{Geometrical interpretation of the chord representation: Projecting \(\obs\) in parallel to \(\cocycles\) onto the chord space \(\reals^\chords\) yields \(\obs_\chords\).
    Fore more worked examples and additional theory, \cf the pre-print~\cite{Polettini2014}.
  }
  \label{fig:projection}
\end{figure}

Although we have seen that both  \(\obs\) and its cyclic part \(z\) have the same fluctuation spectra, neither choice is necessarily the most convenient one for the purpose of analytical calculations.
A good candidate in that respect is the \emph{chord representation} of an observable~$\obs$,
\begin{align}
  \obs_{\chords}\coloneqq \sum_{\eta\in\chords} \langle \obs, \zeta_\eta\rangle\, \eta,
  \label{eq:chord-representation}
\end{align}
where \(\zeta_\eta\in \cycles\) is the fundamental cycle corresponding to the chord \(\eta\in\chords\).
The numbers  $\sprod{ \obs}{ \zeta_\eta}$ appearing in the chord representation are obtained as the components of oblique projections of $\obs$ onto the space $\reals^\chords$ spanned by the chords.
Oblique projections, are not necessarily, but rather project parallel to a linear subspace as depicted in Figure~\ref{fig:projection}.
In our case, the oblique projection is parallel to \(\cocycles\), \cf also reference~\cite{Polettini2014}.

The chord representation \(\obs_\chords\) of $\obs$ is the unique element in the intersection \(\reals^\chords\cap \obs + \cocycles\).
By Corollary~\ref{theo:cycles-matter}, the scaled cumulant-generating functions of \(\obs\) and \(\obs_\chords\) agree.
Moreover, the chord representation \(\obs_\chords\) is supported on at most $b = \abs{\chords}=\abs{\edges}-\abs{\verts}+1$ edges, where $\obs$ generically takes non-trivial values on all edges.

This in turn may reduce the effort to calculate the scaled cumulants:
Denote the component of the oblique projection onto $\eta$ by \(\obs_\eta\coloneqq\langle \obs, \zeta_\eta\rangle\).
We summarize them in the $b$-dimensional vector \(\bvec \Phi = \left( \obs_{\eta_1}, \obs_{\eta_2}, \dots, \obs_{\eta_b} \right)\transpose\), where $b = \dim\cycles$ is also the dimension of the cycle space.
Writing also \( \bvec H = \left( \eta_1, \eta_2, \dots, \eta_b \right)\transpose\), the chord representation can be thought of as the matrix product \(\obs_\chords=\bvec \Phi\transpose \bvec H\).
Note that this matrix product formally looks like a scalar product $\sprod{\cdot}{\cdot}$, but its value is an element of $\reals^\chords\cap \obs+\cocycles$.

The scaled cumulant-generating functions satisfy \(\lambda_\obs(q)=\lambda_{\obs_{\chords}}(q)=\lambda_{\bvec H}(\bvec{\Phi}q)\).
Consequently, in order to determine the scaled cumulants \(c_1(\obs), c_2(\obs), \dots, c_\ell(\obs)\) of \(\obs\), it is sufficient to calculate the vector \(\bvec \Phi\) and the joint scaled cumulants \(c(\eta_{i_1}, \dots, \eta_{i_\ell})\) of the chords up to the order \(\ell\).
Due to multi-linearity, the scaled cumulants of \(\obs\) read
\begin{align}
  c_\ell(\obs) &= \left( \prod_{i=1}^\ell \sum_{\eta_i\in\chords} \obs_{\eta_i}\right) c(\eta_1, \eta_2, \dots,\eta_\ell)\label{eq:gen-schnakenberg}
\end{align}
such that
\begin{subequations}
\begin{align} 
  c_1(\obs) &= \sum_{\eta\in\chords}\obs_{\eta}\, c_1(\eta)=\sum_{\eta\in\chords} \obs_\eta\, J(\eta)\,,\label{eq:schnakenberg}\\
  c_2(\obs) &= \sum_{\eta_1\in\chords}\sum_{\eta_2\in\chords} \obs_{\eta_1} \obs_{\eta_2}\, c(\eta_1, \eta_2)\,.
\end{align}
  \label{eq:schnakenberg-c1c2}
\end{subequations}
For the first cumulant --- the expectation value --- the representation in equation~(\ref{eq:schnakenberg}) is called the \emph{Schnakenberg decomposition} \cite{Schnakenberg1976}. 
Equation~(\ref{eq:gen-schnakenberg}) generalizes this decomposition to the entire fluctuation spectrum.

\section{Discussion}
In the discussion of the present chapter, we focus on three aspects.
Firstly, we comment on other cycle decompositions that have been used in the literature, and how they relate to the Schnakenberg decomposition \eqref{eq:cycle-decomposition-current}.
Secondly, we give an intuition about the role of the boundary and co-boundary operator.
Finally, we review the significance of our main results for ST.

\subsection{Different cycle decompositions and the flux-cycle transform}
\label{sec:flux-cycle-transform}
The decomposition of the steady-state current $\curra{\omega}{\omega'}$ into its fundamental cycles using \eq{fundamental-cycle-decomposition} is just one way of a cycle decomposition.
Different variants have been proposed in the literature \cite{Schnakenberg1976,Kalpazidou1993, Kalpazidou2006,Jiang_etal2004}.

\paragraph{Decompositions using fundamental cycles}
The Schnakenberg decomposition of the current \eqref{eq:cycle-decomposition-current} is a special case of the general decomposition using fundamental cycles.
It has been used by several authors in the context of Markov processes \cite{Andrieux+Gaspard2007,Faggionato+dPietro2011}.
Algorithms have been proposed for the efficient construction of fundamental cycles in (large) random graphs \cite{Paton1969}.
A recent preprint~\cite{Polettini2014} reviews the relevance of (oblique) projections in the context of ST.
It also discusses the duality relation between the boundary and co-boundary operators $\del$ and $\del^*$ in more detail. 
An explicit method for the construction of the steady state currents is reviewed in Ref.~\cite{Zia+Schmittmann2007}.

In summary, we can say that the ideas pioneered by Kirchhoff \cite{Kirchhoff1847} are still relevant nowadays in both abstract mathematics and theoretical physics.
In mathematics, the result is usually known under the name of the matrix-tree or matrix-forest theorem \cite{Tutte1998}.
The basis for the Schnakenberg decomposition which uses the notion of a set of fundamental cycles was laid by Hill \cite{Hill1977}.
In addition, Hill has also considered another set of cycles which we briefly review in the next paragraph.

\paragraph{Hill's stochastic cycles}
Kalpazidou \cite{Kalpazidou2006} and Jiang \textit{et.~al.}~\cite{Jiang_etal2004} have reviewed the significance of cycles for Markov processes.
In particular, they present several methods for a decomposition of the network of states using cycles.
These authors distinguish between stochastic and deterministic decomposition algorithms.

The extended set of cycles used by Hill  includes all distinct closed paths formed by a random walk on the network of states \cite{Hill1977}.
This allows for the treatment of networks with absorbing states.
Hill emphasized that for biological processes, the knowledge of which cycles are completed how many times before an absorbing state is reached is important.
As an example consider a model for an enzyme which includes the process of the enzyme's degradation as an absorbing state.
The number of completed cycles than quantifies the biological activity during an enzyme's life time.
Moreover, the stochastic dynamics of cycle completion can be formalized as an equivalent Markov process on a transformed state space which has the cycles as its vertices, \cf Refs.~\cite{Jiang_etal2004,Kalpazidou2006}.

\paragraph{Flux cycles and the cycle transformation}
In Ref.~\cite{Altaner_etal2012} we introduced another cycle decomposition, complementary to both Hill's and Schnakenberg's cycles.
Rather than being based on the steady-state currents $\curr$, it is based on the steady state fluxes $\flux$.
Consequently, we call it the \emph{flux-cycle decomposition}.
The number of flux cycles in this decomposition is generally larger than the number of fundamental cycles (\ie the cyclomatic number) but smaller than the number of all distinct closed paths.

Similarly to the latter approach, the cycles in the flux-cycle decomposition for a graph $\graph$ can be viewed as the vertices of a transformed graph $\graph'$.
The dynamics on the new graph $\graph'$ are those of an equilibrium system, \ie a system where detailed balance holds and no currents are flowing.
A set of flux cycles can be constructed using the algorithm published in Ref.~\cite{Altaner_etal2012}, where also consider an analogy to networks of public transport.

In Ref.~\cite{Altaner_etal2012} we also discuss how changes of weights assigned by the algorithm correspond to the change of the dominant paths in a network.
We illustrate this fact using the example of a periodic \emph{totally asymmetric exclusion process} (TASEP) with two particles on four sites.
The change in the decomposition amounts to the change of the cycles that dominate the steady state upon changing an external driving.
For the TASEP example, it results in the change of ``gait'' observed in the movement of the two particles.
In the context of biological systems the notion of dominant cycles has been discussed in Ref.~\cite{Hill1979}.

\subsection{Analogy to field theory}
\label{sec:field-theory}
Let us discuss the role of the boundary  and co-boundary operators $\del\colon\currents \to \potentials$ and $\del^*\colon\potentials \to \currents$ in physical terms.
They are maps between the \emph{edge space} $\currents\sim\reals^\oredges$ and the \emph{vertex space} $\potentials \sim \reals^\verts$.
In an analogy to continuous field theory, these spaces correspond to the notion of \emph{vector} and \emph{scalar} fields.

In a dynamic field theory like hydrodynamics, one distinguishes between currents of conserved and non-conserved quantities.
Let $\vec z$ denote the current of a conserved quantity.
Then, the continuity equation for the steady state reads
\begin{align}
  \nabla\cdot\vec z = 0.
  \label{eq:continuity-general}
\end{align}
If we interpret the boundary operator $\del$ as a discrete divergence, the discrete analogue of Eq.~\eqref{eq:continuity-general} is the definition of the elements of the cycle space Eq.~\eqref{eq:cycle-space}.
Hence, the cycle space contains the discrete analogue of \emph{currents of conserved quantities}, which amount to \emph{divergence-free fields} in a steady state.

Similarly, the co-boundary operator is a discrete version of the vector gradient.
In field theory, the gradient acts on scalar fields $U$ to produce irrotational \emph{gradient fields}
\begin{align}
  \vec y := \nabla U.
  \label{eq:gradient-field-general}
\end{align}
Again, the above equation is the continuous version of the definition of the  \emph{co-cycle space} in Eq.~\eqref{eq:cocycles}.
This is why co-cycles take the role of potential differences. 
Kirchhoff's second law \eqref{eq:kirchhoff-second-law} expresses this fact for the difference $\Delta U^{\omega}_\omega$ of the reference voltage $U_\omega=-\log \p{\infty}{\omega}$ associated to a vertex.

Finally, the duality of $\del$ and $\del^*$ ensures that the cycles and co-cycles form an orthogonal decomposition of the space of anti-symmetric observables $\currents$.
Consequently, we can write $\obs = z+y$ with $z\in \cycles$ and $y \in \cocycles$ for any $\obs\in \currents$.
This decomposition is the discrete analogue of the Helmholtz decomposition for vector fields in $\reals^3$, which can be found in any standard textbook on fluid mechanics, \eg Ref.~\cite{Acheson1990}.

Figure~\ref{fig:trinity} illustrates these analogies.
It also motivates the interpretation of co-cycles as \emph{tidal currents} which was introduced by Polettini~\cite{Polettini2014}.
From that point of view, fluxes can be characterized by their cyclic contribution as well as their tendency to ``flood'' across a boundary, \cf Fig.~\ref{fig:trinity}c).
Note that generically a current $\obs\in\currents$ has both tidal and cyclic contributions.
Yet, only the latter are important for the (asymptotic) characterization of fluctuations in non-equilibrium steady states.

\begin{figure}[ht]
  \centering
  \includegraphics{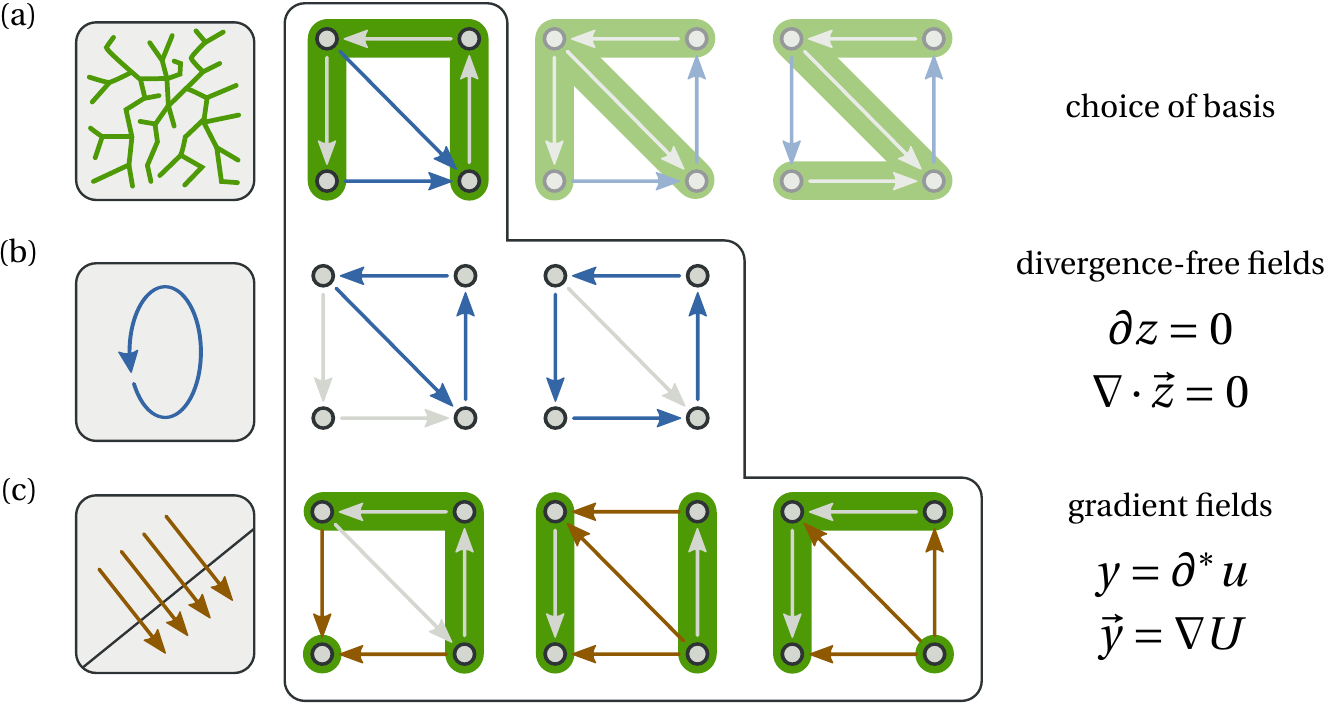}
  \caption{Trees, cycles and co-cycles for discrete graphs.
    a) A spanning tree (green) determines the basis of physical, \ie anti-symmetric, observables.
    The corresponding chords are depicted in blue.
    Alternative choices of a spanning tree (semi-transparent) yield different chords and a different set of basis vectors.
    b) The fundamental cycles constitute a basis of the cycle space.
    In the analogy of vector calculus, cycles $z \in \cycles$ are divergence-free vector fields expressing the stationary currents of conserved quantities.
    c) The co-cycles $y\in\cocycles$ can be visualized as tidal currents across a boundary (brown).
    They correspond to gradient fields in vector calculus.
  }
  \label{fig:trinity}
\end{figure}

\subsection{Relevance for stochastic thermodynamics}
\label{sec:relevance}
Finally, we discuss the relevance of our results for ST.
More precisely, we discuss how they provide useful tools for the study of dynamically reversible Markov processes in general.
Consequently, here we do not refer to any particular \emph{model} or a particular (physical) observable.
We will come back to these considerations in much more detail in the following Chapter~\ref{chap:fluctuations}.
There, we apply our general results to models of molecular motors in non-equilibrium environments.

Before we start the discussion let us emphasize once more that our methods are completely \emph{analytical}.
They can be implemented in any computer algebra system to yield \emph{exact results}.

\subsubsection{Access to the steady-state distribution}
None of the results presented in the present chapter requires access to the steady state distribution of a Markov process.
Usually, the steady state distribution $\bvec p\tind{\infty}$ is obtained as the left eigenvalue of $\tmat$ corresponding to the zero eigenvalue.
As the eigenvalue is know, the determination of $\bvec p\tind{\infty}$ can be obtained by solving a set of linear equations.
The fact that the stationary distribution is accessible using combinatorial aspects of graphs was already known to Hill~\cite{Hill1977}.
A nice presentation of the method can be found in Ref.~\cite{Zia+Schmittmann2007}.

For the purpose of \emph{symbolic} calculations in computer algebra systems, there is another elegant way to obtain the stationary distribution:
Consider the random variable characterizing the faction of time the system stays in state $v_k$.
Its SCGF is obtained as the dominant eigenvalue of the matrix $\mathbb W_{v_k}(q)$ with entries 
\begin{align}
  \left(\mathbb W_{v_k}(q)\right)^i_j := \tprob{i}{j}\exp\left( q\delta\!_{jk} \right).
  \label{eq:staying-ldp}
\end{align}
The first cumulant of the staying time distribution characterizes the average occupancy of state $v_k$ in the limit of large times.
For an ergodic system, it thus amounts to the steady state probability $p\tind{\infty}(k)$.
This result hold in general, \ie not only for dynamically reversible Markov processes \cite{Touchette2011}.

The essence of our approach applies here as well: 
We do not need to calculate the largest eigenvalue of the tilted matrix.
Rather all necessary information is contained in the characteristic polynomial $\chi_k(\lambda,q)$ of $\mathbb W_{v_k}$.
Equation~\eqref{eq:magic-formula-a} then reduces to
\begin{align}
  p\tind{\infty}(k) = c_1(v_k) = -\frac{a_0'}{a_1} \equiv -\left.\frac{\frac{\del \chi_k}{\del q}}{\frac{\del \chi_k}{\del \lambda}}\right\vert_{\lambda=q=0}.
  \label{eq:magic-equation}
\end{align}

\subsubsection{Benefits of the chord representation}
The chord representation $\obs_\chords$ of an observable $\obs$ takes only non-vanishing values on the chords.
That means, it is supported by at most $\abs{\chords}=\abs{\oredges} -\abs{\verts} +1$ edges, which is the number of independent cycles.
For any observable supported on more than $\abs{\chords}$ edges, the chord representation might be more suitable for the purpose of calculation.
The value of $\obs_\chords$ on a chord $\eta$ is $\sprod{\obs}{\zeta_\eta}$, which is simply the integral of the observable along the fundamental cycle $\zeta_\eta$.

Moreover, another simplification for the purpose of calculation can be obtained by a suitable choice of spanning tree.
Assume we are only interested in the fluctuation spectrum of certain observables, of which we know that they are conserved along some particular cycles of the network.
In that case, it is convenient to choose a spanning tree such that these cycles are elements of the basis of fundamental cycles.
This choice of basis ensures that in the chord representation of the observables the corresponding chords do not show up.
In turn, this reduces the number of terms appearing in the general expression~\eqref{eq:schnakenberg} for $\ell$th joint scaled cumulant.

\section{Summary}

In the present chapter we discussed the algebraic and topological structure of Markovian dynamics on finite networks.
As a motivation, we considered Kirchhoff's laws for electrical circuits and gave a complete analogy between the latter and the steady state of Markov processes.
On the abstract level, Kirchhoff's laws are formulated using an abstract boundary operator $\del$ and its dual $\del^*$.
These two latter operators can be understood as the discrete version of the vector divergence and gradient for vector and scalar fields, respectively.

Physical observables $\obs\in\currents$ are the analogue of the vector fields representing currents in field theory.
In field theory, steady state currents of conserved quantities are divergence free.
In analogy to the Helmholtz composition from fluid mechanics, one can always write $\obs = z+y$, where $z\in \cycles$ is its divergence-free contribution.
The contribution $y \in \cocycles = \cycles^\perp$ is called the co-cyclic contribution to $\obs$.

The main part of the present chapter discussed the quantification of fluctuations of physical observables.
We found that the fluctuation spectrum, and equivalently the rate function $I_\obs$ of a physical observable $\obs \in \currents$ only depends on its cyclic part $z$.
Hence, with respect to their fluctuations, two observables $\obs=z+y$ and $\obs'=z+y'$ are equivalent, if their cyclic contribution $z$ agrees.
Consequently, we can choose $\obs \in z + \cocycles$ such that it is most convenient for the purpose of calculation.
In that context, we present a generalization of the Schnakenberg decomposition for arbitrary scaled cumulants of arbitrary observables.

A corollary follows from the multi-linearity of cumulants.
It states that in order to obtain the fluctuation spectrum of an arbitrary physical observable $\obs$, it is enough to know the fluctuation spectrum of the currents $\curr$ on a subset of edges.

The results in this chapter are both analytic and constructive.
Rather than having to solve an eigenvalue problem to obtain the fluctuation spectrum, it suffices to differentiate the coefficients of a polynomial.
Our method thus neither requires advanced combinatorics nor the solution of any linear equation or an eigenvalue problem.

In the following Chapter~\ref{chap:fluctuations}, we demonstrate the power of our methods on models of ST.
More precisely, we study several models for the molecular motor kinesin.
Our methods allow us to analytically obtain a phase diagram for the operation modes of the motor in dependence of its chemo-mechanical driving forces.
Moreover, our methods allow a much more detailed look at the structure of the phase diagram than what was possible before.

  \chapter{Modelling molecular motors}
  \label{chap:fluctuations}
  \begin{fquote}[E.~Schr\"odinger][What is Life?][1944]
  [Can] a physicist who [has learnt] the statistical foundation of his science [and] begins to think about organisms and about the way they behave [make] any relevant contributions?
[\ldots]
It will turn out that he can.
\end{fquote}

\section*{What is this about?}
In small systems, fluctuations are relevant and have recently attracted thorough attention from scientists in various fields.
The molecular machinery of life, \ie the biological macromolecules responsible for any metabolic activity in living cells has arguably been at the centre of this research.
\emph{In vivo}, these systems are always found in a chemical environment far from equilibrium provided by the cell's cytosol.
In recent years, scientists have also performed \emph{in vitro} experiments on these systems, thus opening the door for a quantitative treatment.

The initial quote by Schr\"odinger is about the role of theoretical physics for the study of living systems.
He affirms that statistical physics can yield valuable contributions to the study of the arguably most complex system --- life.
He himself provides one of the best examples of such a contribution.
In his essay ``What is life?'' Schr\"odinger  used his concepts from statistical mechanics in order to discuss the structure of the hereditary substance~\cite{Schroedinger1992}.
His conclusion was that the latter must have the form of an ``aperiodic crystal'' --- ten years before Watson and Crick unveiled the structure of DNA.

Another contribution from physics to biology comes in the form of simplified models for complex systems.
The models for small systems formulated within the framework of ST is just one example.
In his pioneering contributions, Hill already focused on models of the molecular machinery of life~\cite{Hill1977}.
At that time, however, fluctuations in those systems where not yet accessible by measurements.

More recent experiments provide ways to compare model predictions with actual data.
Hence, a good model must yield more than only a correct account for the average behaviour of a system.
Additionally, it must capture the behaviour of the random fluctuating trajectories observed in single realizations of an experiment.

For biological systems, which have evolved for millions of years, this is particularly important.
In the light of evolution \cite{Dobzhansky1973}, it is fair to assume that fluctuation may play a crucial role for the \emph{function} of living systems.
Consequently, we start the present chapter with a discussion of functional fluctuations.

After that, we focus on the fluctuations arising in the model of one of the most well-studied small biological systems, namely the molecular motor \emph{kinesin}.
Using the tools we have established throughout the previous Chapter~\ref{chap:cycles}, we perform a quantitative study of the fluctuating properties in various models for kinesin.
In particular we give a detailed account on kinesin's phase diagram, which is spanned by the external chemical and mechanical forces driving the motor.

We are further interested in the comparison of more complicated models with both simpler \emph{reduced} models and \emph{minimal} models obtained from first principles.
Simplified models can be obtained via a coarse-graining procedure.
Minimal models are more fundamental and in a sense maximally non-committal with respect to the physical information we have about a system:
The topological properties of a physically motivated network of states, the thermodynamic forces exerted by the medium and observable physical currents.

Using a classical model for kinesin \cite{Liepelt+Lipowsky2007} as an example, we find that the simpler models are extremely good in capturing the fluctuating behaviour of the original model --- for values ranging over more than twenty logarithmic decades.
We end with a discussion of the visible and hidden structure in phase-diagrams.
In particular, we discuss the benefits of considering a certain signal-to-noise ratio, which has a direct interpretation in the theory of non-linear response.

Some of the content of the present chapter is discussed in the pre-print of a manuscript prepared by A.~Wachtel, J.~Vollmer and the present author \cite{Wachtel_etal2014}.
The non-linear response theory of ST is the topic of another publication in preparation by the same authors \cite{Wachtel_etal2014-2}.
The plots of kinesin's phase diagram where compiled by A.~Wachtel using \emph{Mathematica}\texttrademark.
As such, they are obtained from analytical expressions.

%
%

\section{Fluctuations in small systems}
\label{sec:fluctuations-intro}
Systems in thermodynamic environments are always subject to noise.
As we have discussed in detail in Chapter~\ref{chap:entropy}, noise is the result of the interaction of the system's degrees of freedom with its environment.
There we also introduced the notion of the medium which comprises any unobserved degrees of freedom.
Generally speaking, fluctuations originate from the unpredictable behaviour of these degrees of freedom.

On macroscopic scales, however, we do not resolve this microscopic behaviour and fluctuations are not visible.
The smaller the systems we observe become, the larger the effect of fluctuations.
Within the last twenty years or so, novel microscopy techniques allowed probing of systems on ever smaller length and time scales.
Though these systems may be \emph{small}, most of them are by no means \emph{simple}.

\subsection{Thermodynamic aspects of molecular motors}

Throughout this chapter, we consider the thermodynamic properties of molecular motors.
Generally speaking, a molecular motor is a macromolecular machine involved in some intra-cellular activity.
Such activities can vary widely from intracellular transport to the locomotion of the whole cell.

To perform these tasks a molecular motor needs an energy source.
The most common fuel used by the machinery of life is \emph{adenosine triphosphate} (ATP).
It is produced by another molecular motor called ATP synthase which itself is powered by a gradient in proton concentration through a membrane.
In green plants, this gradient is provided by \emph{photosynthesis} in the chloroplasts.
For eukaryotic organisms, ATP is produced in the mitochondria, as a part of the cellular respiration, \ie the conversion of energy provided by nutrients like sugars or fat.
For a detailed account on the molecular biology of the cell \cf Ref.~\cite{Alberts1994}.

Many molecular motors have catalytic sites where the exothermic hydrolysis of ATP into \emph{adenosine diphosphate} (ADP) and inorganic phosphate (P) provides the motor's energy.
Given their fundamental role for life, thermodynamic aspects of such small engines like their \emph{efficiency} or \emph{power} have recently attracted much attention of both the physics and biology community \cite{Howard1997,Qian2005,Seifert2005,Liepelt+Lipowsky2007,Lipowsky_etal2009,Seifert2011,Schmiedl+Seifert2007,Esposito_etal2009}.

The notions of the efficiency and power of an engine have their origins in the thermodynamics of macroscopic engines.
Both small and macroscopic engines operate in a cyclic fashion.
Upon the completion of a cycle the system returns to its initial state, while in the mean time some form of energy is converted into another.

However, the difference in size also yields important differences in their functioning.
For a macroscopic motor the environment is can usually be divided into several heat reservoirs.
The role of the fuel is simply to provide a temperature gradient in which the motor operates.
In contrast, molecular motors usually work in an isothermal environment which acts both as a heat bath and as a source of (chemical) energy.
Moreover, the mechanical motion of a molecular machine is strongly influenced by its surroundings.

In addition, inertia plays a major role in sustaining the mechanical motion of a macroscopic motor.
In contrast, for small micro-organisms and even smaller molecular motors inertial effects are damped out by a viscous environment.
A comprehensive account of the fundamental concepts of locomotion in overdamped situations can be found in Purcell's work on ``Life at low Reynolds number'' \cite{Purcell1977}.

Another difference of molecular motors and macroscopic engines is the importance of thermal fluctuations.
While these fluctuations are negligible for macroscopic engines, they lead to unexpected events in the case of small systems.
For instance, rather than hydrolysing an ATP molecule and performing work against some force, a molecular motor may perform this task while \emph{synthesizing} ATP.
Surely such an event can never break the first law of thermodynamics, which states that energy is conserved.
Consequently, the energy needed to synthesize ATP, which is an exothermic process, must have been taken \emph{from the environment}.
Such \emph{anti-dissipative} events are very uncommon on the macroscopic scale.
However, rather than being impossible, they are just extremely improbable.
The fluctuation relations we have briefly discussed in Chapter~\ref{chap:entropy} express a symmetry between dissipative ($Q:=\kb T \Delta \totentrv >0$) and anti-dissipative ($Q:=\kb T \Delta \totentrv <0$) events:
The ratio of observing the latter over the former becomes exponentially small upon increasing the magnitude $\abs{Q}$ of an (anti-)dissipative event.
The typical scale of $Q$ for a molecular motor is of the order of the thermal energy $\kb T$, which is much lower than typical macroscopic energy scales.
Hence, anti-dissipative events are much more probable for small systems than for macroscopic engines.

\subsection{Functional fluctuations}
\label{sec:functional}
In the light of evolution, it is not surprising that fluctuations in biological systems may be \emph{functional}.
Evolution itself provides us with a first example of such a functionality:
Random beneficial mutations (fluctuations) are selected and fixated by external factors, thus increasing the so-called \emph{fitness} (reproduction rate).
Theoretical evolution itself has become a major field of research and is built on the tools of statistical physics \cite{Hallatschek2010,Geyrhofer2014,Mustonen+Laessig2010}.
However, in the rest of this section we discuss another example of the functionality of fluctuations occurring in a special kind molecular motor.

\begin{figure}[t]
  \centering
  \includegraphics{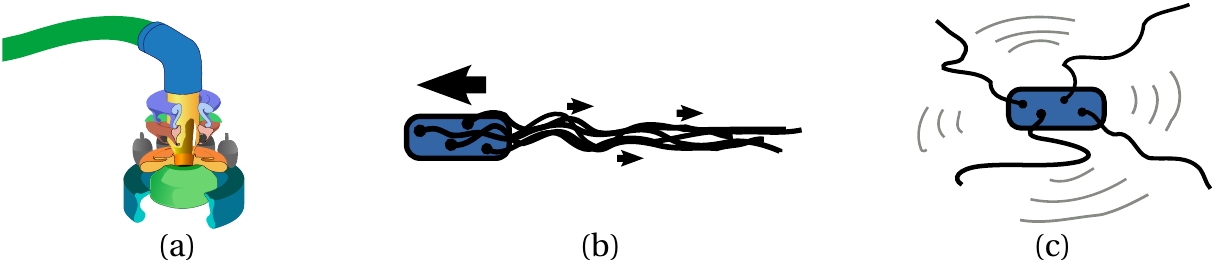}
  \caption{a) Schematic representation of the molecular motor that drives the flagellum of the bacterium E.~Coli.
  Its base is immersed in the cell membrane.
  Flagella are homogeneously distributed over the body of the bacterium.
b) If all flagella rotate in the same direction, they entangle to form a large propeller.
This enables the bacterium to ``run'' in a straight line.
c) Fluctuations in the rotation direction of some of the motors lead to a disentangling of the propeller.
The bacterium tumbles and reorients its direction of motion.}
  \label{fig:run-tumble}
\end{figure}
Bacteria and other micro-organisms that live in a liquid medium, propel themselves by the motion of extracellular filaments called flagella.
In many cases the appendages are themselves passive and their motion is the result of the motion of a molecular motor at the base of the flagellum.
For the case of the bacterium \emph{E. Coli}, flagella are attached all over the cell's body, as depicted schematically in Figure~\ref{fig:run-tumble}.
On average, the motors are synchronized, \ie all the motors and thus the flagella move in the same counter-clockwise direction.
In that case, the flagella entangle and form a screw-like propeller and the bacterium performs a movement in a straight line.
We say it performs the ``run'' motion depicted in Figure~\ref{fig:run-tumble}b.
In order to change its direction of motion, the bacterium goes into a ``tumbling'' movement, \cf \ref{fig:run-tumble}c.

After the tumbling motion, the bacterium performs another run in a new direction.
Alternating run and tumble motions result in a random walk of the organism in its spatial environment.
The origin of this ``run and tumble'' movement has been found to be a random fluctuation of one or few of the molecular motors driving the flagella~\cite{Larsen_etal1974}:
Every now and then one motor may perform a counter-clockwise motion which leads to a disentanglement of the screw propeller.
During the process of disentanglement and reformation of the propeller the bacterium tumbles.
Moreover, the probability of tumbling to occur is dependent on the chemical environment of the motors.
On average, this may lead to a bias in the direction of motion called \emph{chemotaxis}~\cite{Macnab+Koshland1972}.
Chemotaxis enables the bacterium to actively move towards sources of food and thus is an essential \emph{function} of the cell.
Hence we see that fluctuations may play a pivotal role in the functionality of living matter.

In Section~\ref{sec:large-deviations} of the previous chapter we have introduced large-deviations theory as way to quantify fluctuations.
Moreover, we have developed an analytical tool-kit to calculate the fluctuation spectrum.
In the following, we use our formalism in order to characterize the fluctuating statistics of biologically relevant physical observables.
However, before treating multiple models of the molecular motor kinesin in detail, we introduce a general method of model reduction which is sensitive to fluctuations.

\section{Fluctuation-sensitive model reduction}
\label{sec:coarse-graining}
We have discussed the general idea of model reduction by \emph{coarse graining} already in Chapter~\ref{chap:entropy}.
There, we were mostly concerned in the coarse graining of continuous degrees of freedom $x$ into a finite number of effective variables $\omega$.
The present section is concerned with the coarse graining of Markov chains, \ie we already start with a finite state space $\ospace$.
After the coarse-graining, we obtain a new state space $\tilde \ospace$ with $\abs{\tilde \ospace} < \abs{\ospace}$.

As we have seen, coarse graining is often based on the separation of time scales between the original and the coarse-grained level of description.
Several coarse-graining procedures for Markov jump processes which are based on time-scale separation have been presented in the recent literature \cite{Pigolotti+Vulpiani2008,Puglisi_etal2010,Esposito2012}.
In general, a coarse-grained view of a Markovian model is not Markovian any more, due to some memory contained in the probabilities of the fundamental states that make up a coarse-grained state.
Only in the limit of an infinite separation of time-scales, Markovianity is rigorously recovered on the coarse-grained level.

Esposito's approach \cite{Esposito2012} explicitly discusses this limit. 
In contrast, in the work of Pigolotti and co-workers \cite{Pigolotti+Vulpiani2008,Puglisi_etal2010} Markovianity is already built into the coarse-grained model.
In order to compare the original with the coarse-grained model, they performed stochastic simulations of both models.
From these simulations they obtained finite-time approximations to the large deviation function for the motance, \ie the observable identified with the \emph{entropy variation in the medium}.
They found that while certain reduction steps hardly change the fluctuations, others do so in a rather drastic way.

In Ref.~\cite{Altaner+Vollmer2012}, we presented an alternative coarse-graining procedure.
Our approach is in the spirit of Pigolotti's work in the sense that reduced description is formulated as a Markovian model.
However, rather than being based on the separation of time scales, it is based on topological considerations and Schnakenberg's decomposition for the entropy.
Further thermodynamic consistency arguments where used to close the equations.

In contrast to Pigolotti's work \cite{Pigolotti+Vulpiani2008}, we found that our approach yields a fluctuation-sensitive coarse-graining procedure \cite{Altaner+Vollmer2012}.
Moreover, this result seemed to be independent of the concrete model or observable under consideration.
In the previous Chapter~\ref{chap:cycles} we presented an algebraic-topological approach to the large deviation theory of Markov jump processes.
In fact, these methods were developed in order to quantify the small differences between the original and reduced models.
In this section, we briefly review the ideas of Ref.~\cite{Altaner+Vollmer2012} and discuss them in the light of the new results obtained in Section~\ref{sec:cycle-ldt}.

\subsection{Heuristic motivation}
In this subsection we present the heuristic argument for a coarse-graining.
We have seen before (\cf Section~\ref{sec:discrete-st}) how entropies are identified in stochastic models of physical systems in thermodynamic environments.
Moreover, we have seen in Chapter~\ref{chap:information-st} how entropies and the abstract definitions of motance/affinity also arise without the need for a thermodynamic model.
Hence, we consider the anti-symmetric motance and (steady state) affinity matrix $\mot,\aff \in\currents$ with elements $\mota{\omega}{\omega'} = \log \frac{\tprob{\omega}{\omega'}}{\tprob{\omega'}{\omega}}$ and $\affa{\omega}{\omega'} = \log \frac{\fluxa{\omega}{\omega'}}{\fluxa{\omega'}{\omega}}$ as the relevant physical observables.
Our method is based on the following (heuristic) requirements:
\begin{enumerate}
  \item[(i)a] {\it Cycle topology:} The number and mutual connections of cycles are preserved.
  \item[(i)b] {\it Cycle affinities:} The algebraic values of the affinity of any cycle is preserved.
  \item[(ii)] {\it Locality}: Fluxes and probabilities may only change locally
  \item[(iii)] {\it Trajectories:} The system's entropy variation along single trajectories is the same between the models.
\end{enumerate}
Requirement (i) describes consistency within the stochastic models and with macroscopic thermodynamics.
It is motivated by the role of cyclic currents as the fundamental building blocks of Markov jump processes, \cf also the discussion in Sec.~\ref{sec:relevance}.
The Schnakenberg decomposition of the entropy production in the steady state is based on the cycle affinities, are directly related to the thermodynamic forces driving the system.

Requirement (ii) is motivated in the light of the connection to an underlying microscopic phase space:
Imagine that the coarse-graining acts only on a part of the (observable, mesoscopic) states and leaves others untouched.
Then, the dynamics on the elements of the phase-space partition that correspond to untouched states should not change.
This is the notion of \emph{locality} expressed in requirement (ii).

The last requirement (iii) is motivated in the light of stochastic thermodynamics.
The system's entropy variation along trajectories is preserved.
It constitutes the final relation needed to close our equations.

\subsection{Target topologies and coarse-grained observables}
\begin{figure}[thb]
  \centering
  \includegraphics{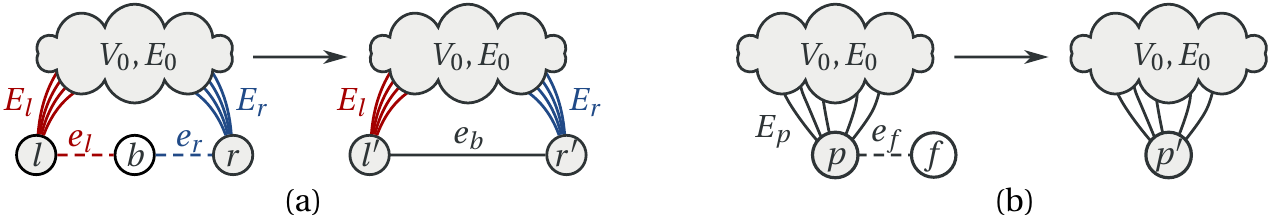}
  \caption{
    The reduction of vertices (white) appearing in certain subgraphs does not change the cycle structure of the graph:
    a) Bridges $(b)$ are vertices which have exactly two neighbours $(l,r)$, which are themselves not connected.
    b) Leaves $(v)$ have only one neighbour $(p)$.
    Note that trees are composed out of bridges and leaves only.
  }
  \label{fig:target-topologies}
\end{figure}
The requirement (i) restricts our coarse-graining procedure to target only vertices with a certain topology of its neighbours.
More precisely, we apply it only to states appearing in certain subgraphs.
Our target topologies are called \emph{bridges} and \emph{leaves}, which we illustrate in Figure~\ref{fig:target-topologies}.
A bridge is defined as a vertex that is only connected to two neighbouring vertices, which are not themselves adjacent.
The latter requirement is important, because the smallest non-trivial cycle consists of three states.
A \emph{leaf} is a vertex that has only a single neighbour.

Note that a tree is a subgraph that can be reduced to a single vertex by subsequently removing leaves.
Further, for any connected graph one can find a spanning tree that includes any bridge and leaf states together with their adjacent edges.
In that regard, dropping leafs and edges corresponds to dropping edges of the spanning tree, while preserving the chords of a graph.
In the light of the discussion in the Section~\ref{sec:cycle-ldt}, this appears to be a good choice in order to preserve the (asymptotic) fluctuations.

\subsubsection{Observables and bridges}
By changing the mesoscopic state space in our model reduction, we also need to define the new coarse-grained  observables $\tilde \obs \in \tilde \currents$ for the observables $\obs \in \currents$ of the original model.
In the following, we denote the ``left'' and ``right'' neighbours of a bridge state $b$ by $l$ and $r$, respectively.
The remaining vertices are summarized in the vertex subset $V_0$, \cf Fig.~\ref{fig:target-topologies}.
\begin{subequations}%
  \begin{align}%
    \widetilde \obs^{l'}_{r'} &= \obs^l_r + \obs^r_l + (d_l - d_r),\\ 
    \widetilde \obs^{n'}_i &=\obs^n_i + d_n \quad \mathrm{for}~i \in V_0, n \in \left\{ l,r \right\},\\
    \widetilde \obs^i_j &= \obs^i_j\quad\mathrm{for}~i,j\in V_0.
  \end{align}%
  \label{eq:newCurrentObservable}%
\end{subequations}%
The constants $d_l$ and $d_r$ act as gauges that do not change the macroscopic observations.
They only change the part of $O$ that lies in the co-cycle space.
In the phase-space picture, they depend on the microscopic dynamics and the chosen re-partitioning of phase space.
As we do not know these details, we choose $d_l \equiv d_r \equiv 0$ for simplicity.

\subsubsection{Observables and leaves}
We also need a rule for coarse-graining the observables if we drop leaves.
As on average the system jumps into the leaf state as often as it leaves it again.
Hence, we set:
\begin{subequations}
  \begin{align}%
    \widetilde \obs^{p'}_i &=\obs^p_i + d_p \quad \mathrm{for}~i \in V_0,\\
    \widetilde \obs^i_j &= \obs^i_j\quad\mathrm{for}~i,j\in V_0.
  \end{align}%
  \label{eq:newCurrentObservable-leaf}%
\end{subequations}
Again, $d_p$ is a co-cyclic gauge and for simplicity we choose $d_p\equiv0$.

\subsection{The coarse-graining algorithm}
In Ref.~\cite{Altaner+Vollmer2012} we presented the coarse-graining prescription for bridge states.
In the CG procedure we absorb the bridge into its neighbours leading to new states $l'$ and $r'$.
Thus, the transition rates for the sets of edges $\edges_l$ and $\edges_r$ connecting $l$ and $r$ to the rest of the network have also to be adjusted.

This has to be done in accordance with requirement (i)b, \ie $\aff_\alpha = \aff'_\alpha$ where $\aff_\alpha$ is the cyclic affinity for any cycle $\alpha$.
Demanding the conservation of fluxes along any edge $e \in \edges \setminus \left\{ e_l,e_r \right\}$ that is not adjacent to the bridge state $b$, Eq.~(\ref{eq:cycle-affinity}) yields
\begin{subequations}%
\begin{equation}%
  \frac{\fluxa{l'}{r'}}{\fluxa{r'}{l'}}\stackrel!= \frac{\fluxa{l}{b} \fluxa{b}{r}}{\fluxa{r}{b} \fluxa{b}{l}}. 
  \label{eq:CGflux}
\end{equation}%
Any trajectory passing through the two edges $(l,b)$ and $(b,r)$ in the original model will be a trajectory through $(l',r')$ in the coarse-grained model.

In ST, the change of entropy associated to the system's state is the difference of the logarithms, \cf~Eq.~\eqref{eq:defn-sysentrv}.
In the following, $\bvec p = \bvec p\tind{\infty}$ always denotes the stationary distribution.
Hence, for brevity of notation we write $p_\omega$ instead of $p\tind{\infty}(\omega)$, as we have done in previous chapters.
Condition (iii) demands that the change in system's entropy along a trajectory entering the bridge on one side and re-emerging on the other side is the same between the models.
In terms of the invariant probabilities, this amounts to
\begin{equation}%
  \frac{p_l}{p_r}\stackrel{!}{=}\frac{p_{l'}}{p_{r'}}.
  \label{eq:CGprob}
\end{equation}%
\label{eq:CG}%
\end{subequations}%
A priori, other closures of the form $p_{l'}/p_{r'} = c$ are also possible but lack the advantage of an interpretation in ST.

Without loss of generality we assume that there is a positive net current $\curr\tsub{bridge} = \curr^l_b = \curr^b_r>0$ flowing through the bridge from the left to the right neighbour state.
Together with the steady-state balance condition (\ref{eq:master-equation}) and the locality assumption, the Eqs.~(\ref{eq:CG}) uniquely determine all rate constants of the coarse-grained model~\cite{Altaner+Vollmer2012}:
\begin{subequations}%
  \begin{align}%
    w^i_{n'} &= w^i_n\quad \mathrm{for}~i \in V_0, n \in \left\{ l,r \right\},\\
    w^{n'}_i &= w^n_i/g\quad \mathrm{for}~i \in V_0, n \in \left\{ l,r \right\},\\
    w^{l'}_{r'} &= (\curr\tsub{bridge}+m)/(g p_l),\\ 
    w^{r'}_{l'} &= m/(g p_r),
  \end{align}%
\label{eq:newrates}%
\end{subequations}%
where
\begin{subequations}%
  \begin{align}%
    g &= (p_l + p_r + p_b)/(p_l + p_r),\\
    m &= \phi^r_b \phi^b_l/(\curr\tsub{bridge} + \phi^r_b + \phi^b_l).
  \end{align}%
  \label{eq:mf_defn}%
\end{subequations}%

The determination of the new rates can also be expressed as a reduction procedure for the motance:
\begin{subequations}%
  \begin{eqnarray}%
    \widetilde \mot^{l'}_{r'} &=& \mot^l_r + \mot^r_l, \label{eq:Bbridge}\\ 
    \widetilde \mot^{n'}_i &=& \mot^n_i - \log g \quad \mathrm{for}~i \in V_0, n \in \left\{ l,r \right\}, \label{eq:Bneighbour}\\
    \widetilde \mot^i_j &=& \mot^i_j\quad\mathrm{for}~i,j\in V_0. \label{eq:Brest}
  \end{eqnarray}%
  \label{eq:newElectromotance}%
\end{subequations}%
Eq.~(\ref{eq:Bbridge}) expresses that the dissipation along a trajectory which passes through the bridge is conserved.
Eq.~(\ref{eq:Bneighbour}) states that along the edges $\mathcal E_n,~n \in \left\{ l,r \right\}$, there is an additional contribution $- \log(g)$ to $\widetilde \mot^n_i$, which is the same for both neighbours due to the closure (\ref{eq:CGprob}).
Eq.~(\ref{eq:Brest}) expresses locality and is independent of the closure.

For the case of leaves, locality of probability demands that $p_{p'} \shouldbe p_p + p_f$.
In order to preserve the fluxes, the new rates need to obey
\begin{subequations}
  \begin{align}
    w^i_{p'} &= w^i_p\quad \mathrm{and}\\
    w^{p'}_i &= \frac{p_p w^p_i}{p_p+p_f}\quad \mathrm{for}~i \in V_0.
  \end{align} 
  \label{eq:newrates-leaf}
\end{subequations}

\section{Applications to kinesin}
\label{sec:motors}

In order to be concrete, we illustrate our methods using the well-established model of the molecular motor kinesin \cite{Liepelt+Lipowsky2007,Liepelt+Lipowsky2009}.
Kinesin is one of the most well-studied molecular motors.
It is responsible for the transport of cargo in eukaryotic cells.
As such, it plays a major role also in the reproductive cycle of cells, \cf Refs.~\cite{Alberts1994,Yildiz_etal2004,Carter+Cross2005} and the references therein for further reading.

Figure~\ref{fig:kinesin-walking}a) shows a schematic representation of kinesin, which is a macromolecule composed of several subunits.
The ``feet'' or \emph{motor domains} are its active sites constituting the \emph{head end} of the molecule.
The ``body'' of the molecule is a stalk connecting the head end to the tail end.
At the latter one finds binding site for kinesin's payload, which is a vesicle containing other chemical compounds.

The active sites bind to intracellular filaments called \emph{microtubules}, which are one on the central structural elements of the cytoskeleton.
Microtubules are themselves hollow cylinders which consist of polymerized dimers called \emph{tubulin}.
They further show \emph{polarity} due to the helical arrangement of the tubulin components.
Hence, the two ends of a microtubule are distinguishable.

The active sites act as a catalyst for the hydrolysis of adenosine triphosphate (ATP) into adenosine diphosphate (ADP) and inorganic phosphate (P).
This exothermic reaction provides the energy for a mechanical transition along the microtubule.
The polarity of the microtubule thereby ensures a preferred direction of motion.
The mechanics of kinesin's mechanical step have only recently been understood \cite{Yildiz_etal2004}.
They arise from a change of the active sites' binding affinity to the microtubule that goes along with the hydrolysis reaction.
Additionally, during this reaction the molecule undergoes a conformational change.
The complex interplay of this mechanisms leads to the motion depicted in Figure~\ref{fig:kinesin-walking}b).

\begin{figure}[ht]
  \centering
    \includegraphics{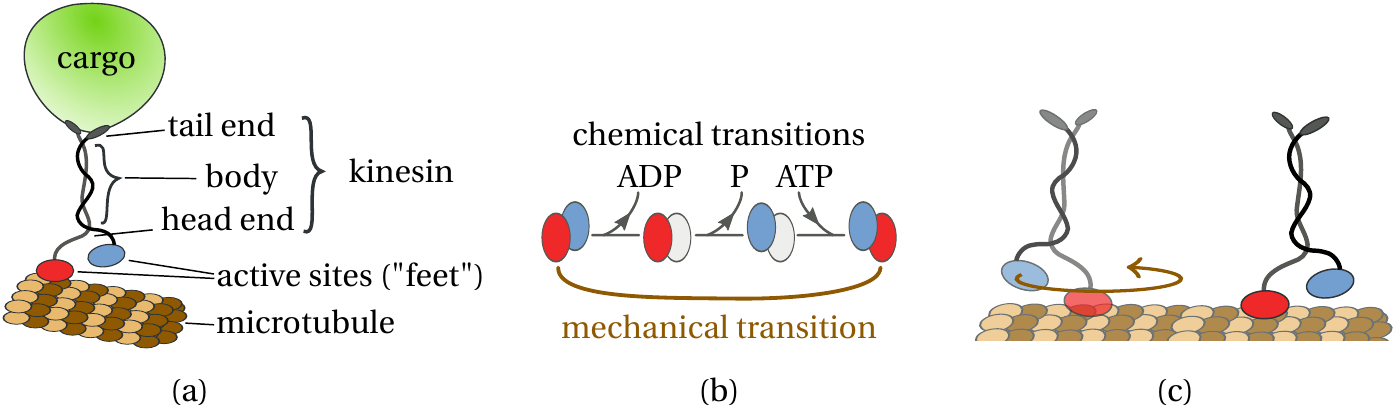}
  \caption{
    Kinesin is a molecular motor featuring two chemically active sites (``feet'').
    Different colours at kinesin's active sites represent the different chemical compositions.
    Here, we display the typical behaviour of the motor where forward movement is linked to the hydrolysis of ATP (red) into ADP (blue).
    (a) Schematic representation of kinesin attached to a microtubule.
    (b) Dominant mechano-chemical pathway of kinesin under physiological conditions:
    The net attachment, hydrolysis and release of one ATP molecule drives a conformational change, leading to a mechanical step.
    (c) Kinesin walks in a ``hand-over-hand'' motion, where the mechanical transition exchanges the leading with the trailing active site.
  }
  \label{fig:kinesin-walking}
\end{figure}

\subsection{A model for kinesin}
\label{sec:kinesin}

Liepelt and Lipowsky~\citep{Liepelt+Lipowsky2007} have constructed a chemo-mechanical Markov model for kinesin, which is shown in Figure~\ref{fig:kinesinreduction}a).
Their minimal model included six states, each of which corresponding to a different chemical composition of the active sites.
Upon a mechanical step of distance $l\approx \SI{8}{\nano\meter}$, the position of the leading and trailing ``foot'' is interchanged.

In order to explore kinesin's response to different non-equilibrium conditions, Carter and Cross designed an experiment where a mechanical force $F$ acts on the motor \cite{Carter+Cross2005}.
A positive force $F>0$ thereby indicate that it acts \emph{against} kinesin's natural direction of motion.

Under physiological chemical conditions and in the absence of an external force, on average one molecule of ATP is hydrolysed per mechanic step.
However, the chemical and the mechanical cycle are not \emph{tightly coupled} \cite{Liepelt+Lipowsky2009}.
It is possible to have a futile hydrolysis of ATP, with no mechanical step taking place.
In the model, this is reflected by the presence of three cycles, two of which are independent, \cf Fig.~\ref{fig:kinesinreduction}.
The \emph{forward cycle} $\mathcal F$ corresponds to the standard motion under physiological conditions.
The \emph{backward cycle} $\mathcal B$ describes the situation where fuel consumption leads to a backward rather than a forward step.
Upon completion of the \emph{dissipative slip cycle} $\mathcal D$, two molecules of ATP are hydrolysed while no mechanical transition takes place.

The construction of the six-state model shown in Figure~\ref{fig:kinesinreduction}a) is found in Ref.~\cite{Liepelt+Lipowsky2007}.
For the determination of the transition rates and their dependence on chemical concentration and mechanical load, experimental data was used.
Appendix~\ref{app:construction} reviews the arguments brought forward in Ref.~\cite{Liepelt+Lipowsky2007} to construct a simpler four-state model.

In the following Sections~\ref{sec:kinesin-fpcg} and~\ref{sec:analytical-phase-space}, we illustrate our methods using different models for kinesin.
At first, we demonstrate the power of the coarse-graining procedure introduced in the previous Section~\ref{sec:coarse-graining} as published in Ref.~\cite{Altaner+Vollmer2012}.
For both the original and several reduced models we plot the rate function $I_{\obs}(x)$ obtained for the time-averages of several observables $\obs$.
In addition, we demonstrate the convergence of the scaling form of sampled probability distributions to $I_\obs$.
Remarkably, not only the asymptotic statistics (characterized by the rate function) but also the sampled distributions for finite times agree to a very high degree between the models.

Secondly, we investigate the full \emph{phase diagram} of kinesin, which is spanned by the values of the driving parameters $\mu$ and $F$.
Motivated by the success of our coarse-graining procedure, we conclude that a minimal model with four-states is enough.
In order to get rid of the ambiguity when it comes to choosing \emph{which} bridge states are removed, we introduce a simple four-state model, \cf Appendix~\ref{app:construction}.
\begin{figure}[t]
  \centering
  \includegraphics{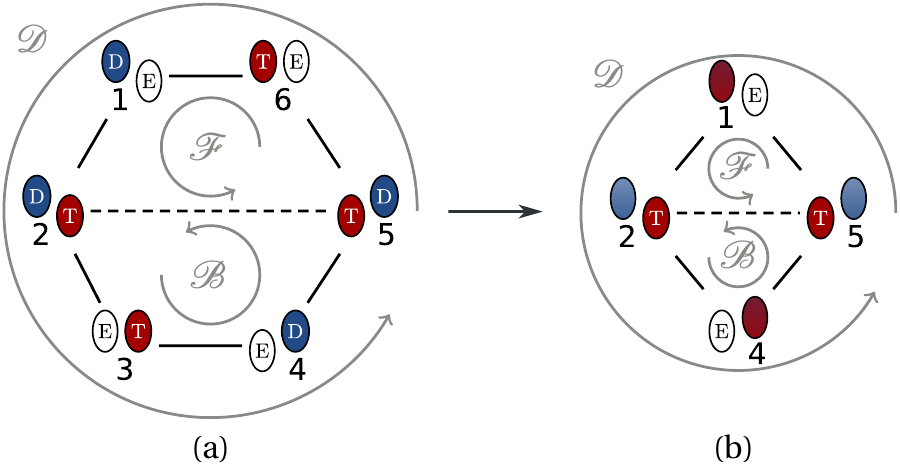}
  \caption{(a) 6-state model of kinesin \cite{Liepelt+Lipowsky2007} with its cycles.
  The dashed line represents the mechanical transition which allows the motor to move along the microtubule.
  All other transition are chemical.
  The cycle $\mathcal F$ is called the \emph{forward cycle}.
  It corresponds to the process that is usually found under physiological conditions shown in Fig.~\ref{fig:kinesin-walking}b:
  The hydrolysis of one ATP molecule leads to a forward step.
  In the lower cycle $\mathcal B$, an ATP molecule is hydrolysed and the motor performs a backward step.
  The outer cycle $\mathcal D$ describes the futile hydrolysis of two ATP molecules.
  (b) Coarse-grained description with bridges $3$ and $6$ reduced.
  Note that all cycles are preserved.
}
  \label{fig:kinesinreduction}
\end{figure}

\subsection{Fluctuations in the coarse-grained model}
\label{sec:kinesin-fpcg}
In Ref.~\cite{Altaner+Vollmer2012} we applied the coarse-graining procedure outlined in Section~\ref{sec:coarse-graining} to the bridge states of the six-state model.
Note that both the forward and the backward cycle contain two bridges that lie next to each other.
Upon reduction of one of the neighbouring bridge states, the resulting state is not a bridge any more, as its neighbours are mutually connected.
Hence, there is an ambiguity when it comes to the choice, which states to remove.
As an example, in Fig.~\ref{fig:kinesinreduction}b we show the removal of the bridge states $6$ and $3$, yielding a coarse-grained model on four states.

Figure \ref{fig:kinesin-fluctuations} shows the effect of the coarse-graining procedure for a set of parameters that correspond to physiological conditions.
More precisely, we considered the time averages of three observables: 
The hydrolysis rate, the velocity and the dissipation rate.
The former two were obtained by counting the transitions along the respective edges in the graph, whereas the latter is the time-average \eqref{eq:time-average-ldt} associated with the motance matrix.
We will discuss the \emph{chord representations} of these observables in detail below.

\begin{figure}[tb]
  \includegraphics{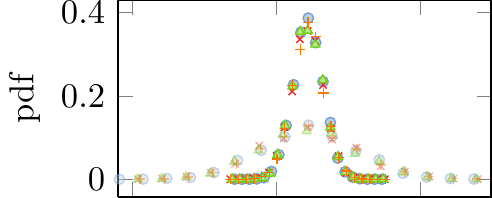}%
  \includegraphics{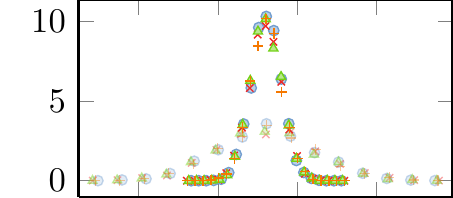}%
  \includegraphics{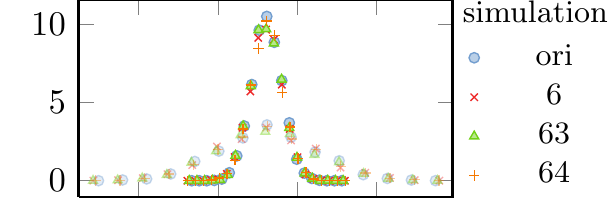}\\%
  \includegraphics{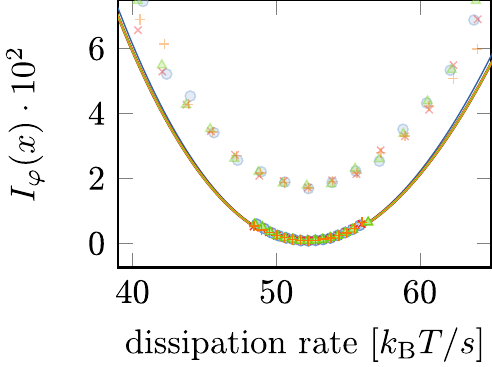}%
  \includegraphics{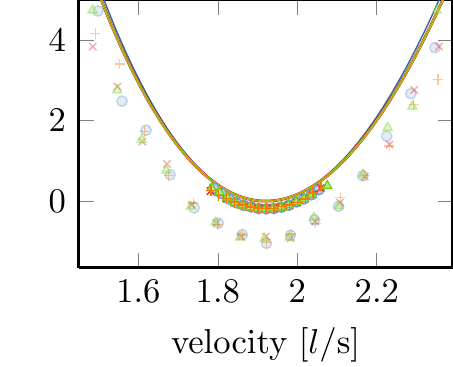}%
  \includegraphics{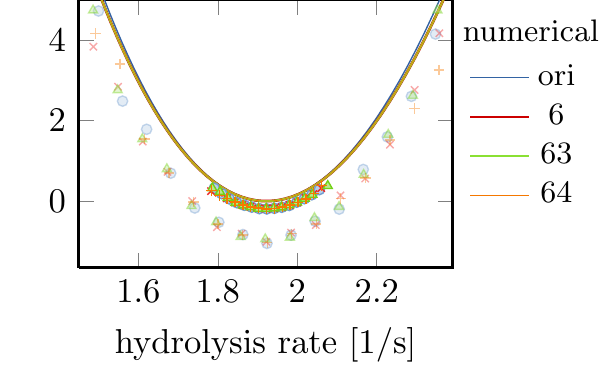}%
  \caption{Simulation and numerical results for dissipation rate, moving velocity and hydrolysis rate of the kinesin model taken from Ref.~\cite{Altaner+Vollmer2012}.
  Data is shown for the original 6-state model (``ori''), a 5-state model with state $6$ reduced (``6'') and two 4-state models with state $6,3$ or $6,4$ reduced (``63'' and ``64''). 
  The rate constants for the original model are taken from Ref.~\cite{Liepelt+Lipowsky2007} describing the data in Ref.~\cite{Carter+Cross2005} for chemical concentrations $c\tsub{ADP}=c\tsub{P}=c\tsub{ATP}=$\SI{1}{\micro\Molar} and stepping size $l\approx$\SI{8}{\nano\meter}.  
  The top row shows the sampled pdf for $\tau \approx $\SI{1200}{\second} (opaque symbols) and $\tau \approx$ \SI{120}{\second} (transparent symbols).
  The bins with the width of half an empirical standard deviations are centred around the empirical mean.
  For the simulation we sampled $N = 5000$ trajectories. 
  The bottom row shows convergence of rescaled data (\cf Eq.~(\ref{eq:ldt-pden-approx})) to  the rate function $I_\obs(x)$ (solid lines) obtained via Legendre transform (implemented using Ref.~\cite{Lucet1997}) of the SCGF $\lambda_\obs(q)$.
  }
  \label{fig:kinesin-fluctuations}
\end{figure}

The data presented in Figure~\ref{fig:kinesin-fluctuations} illustrates several results.
First of all, the data obtained for different implementations of the coarse-graining lie on top of the data obtained in the original model.
This holds true for both the asymptotic statistics characterized by a rate function $I_\obs(x)$ as well as for the finite-time probability distributions $\pden_\obs\rlind{\tau}(x)$.
While the former were calculated using Equation~\eqref{eq:gaertner-ellis}, the latter are sampled from the simulation of a large number of stochastic trajectories.

The results show the efficiency of our fluctuation-preserving coarse-graining:
Firstly, it preserves the whole fluctuation spectrum (encoded in the rate function) to a very good degree.
This is true not only for the dissipation, but also for other observables.
The good agreement is a consequence of the generalized Schnakenberg decomposition~\eqref{eq:gen-schnakenberg}.
Secondly, the results do not change between the original model on six states, and the different five- and four-state models obtained from the removal of bridge states.
This underlines the importance of the cycle topology, which we emphasized in our heuristic motivation of the coarse-graining procedure.

In conclusion, a minimal model on four states that respects the cycle topology and the affinities of the six-state model is sufficient to capture the fluctuations of a more complicated model.
Because we do not allow multiple transitions between states, a non-tightly coupled model featuring multiple cycles cannot be formulated on only three states.
This motivates the construction of minimal four-state model directly from experimental data which is shown in Appendix~\ref{app:construction}.

This minimal model has several advantages over a four-state model obtained from the coarse-graining procedure.
Firstly, there is no ambiguity with respect to the choice of the removed bridge states.
Secondly, when removing a bridge state, one distributes it among its two neighbours.
In that process, the neighbours lose their former interpretation as being the equivalence classes belonging to a certain observable result.
This is also reflected by the new transition rates:
They lose their direct interpretation as the kinetic rates of a specific chemical reaction.

\subsection{Analytical treatment of kinesin's phase diagram}
\label{sec:analytical-phase-space}
In the previous section, we discussed the fluctuation spectrum of the 6-state kinesin model and its reduction to 5-state and 4-state models for a special set of (physiological) parameters, \cf Figure~\ref{fig:kinesin-fluctuations}.
In Ref.~\cite{Liepelt+Lipowsky2009}, Liepelt and Lipowsky analysed the ``phase diagram'' of their model obtained by varying the forces that drive the system.
The effective parameters of the model are the thermodynamic affinities characterized by the (non-dimensionalized) chemical potential difference \(\mu\) and a mechanical load  \(f\) \cite{Liepelt+Lipowsky2007}.
With the equilibrium constant $K\tsub{eq}$ for the hydrolysis of ATP, one defines
\begin{align}
  \mu := \log\frac{K\tsup{eq} [\mathrm{ATP}]}{[\mathrm{ADP}][\mathrm{P}]},
  \label{eq:defn-mu}
\end{align}
where $[\mathrm{X}]$ stands for the chemical concentration of compound X in the solution.
The non-dimensionalized force is defined as
\begin{align}
  f:= \frac{l F}{\kb T}
  \label{eq:defn-force}
\end{align}
where $l$ is the distance of kinesin's mechanical step and $F$ is the applied external force.
The non-dimensionalized driving parameters constitute the affinities of the cycles of the model.
One also refers to this thermodynamic consistency requirement as the ``Schnakenberg conditions'' \cite{Andrieux+Gaspard2007,Faggionato+dPietro2011}.

The Schnakenberg decomposition~\eqref{eq:schnakenberg} ensures that the affinities of all other cycles are obtained as linear combinations of the affinities of the fundamental cycles \cite{Schnakenberg1976}.
Note that these affinities in turn are linear combinations of the (dimensionless) chemical and mechanical drivings acting on the system.
They correspond to net (free) energy difference in the environment (divided by $\kb T$) upon the completion of a thermodynamic cycle, \cf also Ref.~\cite{Seifert2011}.

After a mechanical step, the work performed by the external force on the system amounts to $f\kb T = F\cdot l$.
Similarly, after a hydrolysis reaction the chemical potential in the environment has changed by a factor $\kb T \mu = \kb T \log\frac{K\tsub{eq}[\text{ATP}]}{[\text{ADP}][\text{P}]}$.
To the present author's knowledge, Hill first realized and appreciated this fact.
He calls this phenomenon the ``transduction of (free) energy'' \cite{Hill1977}.
Consequently, we prefer the term Hill--Schnakenberg conditions or simply \emph{thermodynamic balance conditions}.
Formally, one may express this fact as
\begin{align}
  \mot_\alpha := \sum_{\edge \in \alpha} \mot_\edge \shouldbe \sum_{\nu}n_\alpha^\nu (\Delta A)^\nu,
  \label{eq:thermodynamic-balance}
\end{align}
where $\mot_\alpha$ is the motance of a cycle, $\Delta A^\nu$ refers to a difference in a (thermodynamic) potential and $n^\alpha_\nu \in \integers$ is an integer.

\subsubsection{Chord representation of the observables}
Throughout the present Section~\ref{sec:analytical-phase-space} we plot results obtained using the six-state model.
In the next Section~\ref{sec:minkinesin} we compare these results to minimal and coarse-grained models formulated on only four states.

For the sake of simplicity, however, already in the present section we exemplify the application of our method using a four-state model.
A detailed treatment can be found in Ref.~\cite{Wachtel_etal2014}.
Here, we focus on the chord representation~\eqref{eq:chord-representation} of biologically relevant physical observables.
Moreover, using the four-state model we ensure continuity with respect to the previous chapter, where we exemplified cycles and co-cycles using a graph on four vertices.

Figure~\ref{fig:minkinesin} shows an abstract representation of a minimal four-state model alongside its abstract representation as a directed graph.
We chose the enumeration of the vertices in Fig.~\ref{fig:minkinesin}a) to be consistent with Fig.~\ref{fig:minkinesin}b and hence with all the examples used in Chapter~\ref{chap:cycles}.
This allows us to identify a basis of fundamental cycles basis of fundamental cycles $(\zeta_2,\zeta_5)$ which corresponds to the dissipative slip and forward cycle, respectively.
\begin{figure}[htb]
  \centering
  \includegraphics[scale=1]{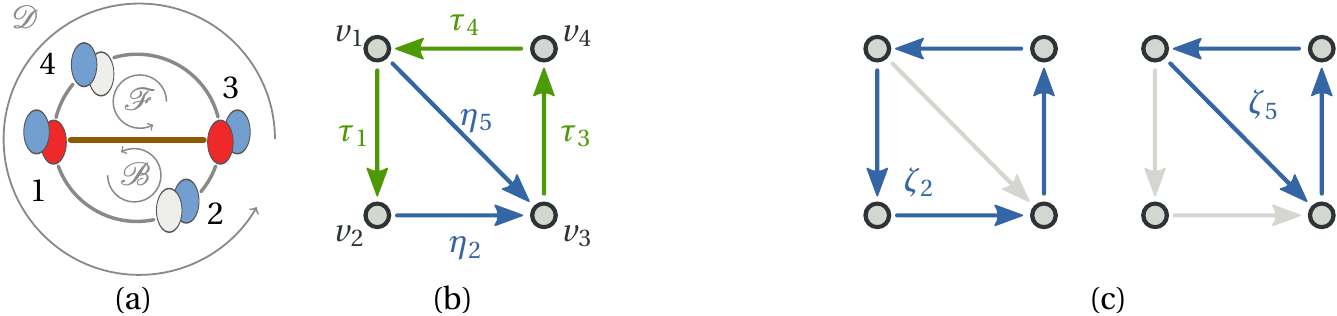}
  \caption{
  (a) Our minimal 4-state model for kinesin.
  Details of the parametrization can be found in Appendix~\ref{app:construction}.
  Note that the enumeration is different than in Figure~\ref{fig:kinesinreduction}.
  However, it is consistent with the graph shown in (b).
  Note that this graph was used as an example throughout Chapter~\ref{chap:cycles}.
  The chosen spanning tree \( \left( \verts, \left\{ \tau_1,\tau_3,\tau_4 \right\} \right)\) is marked in green.
  The fundamental chords \(\eta_2,\eta_5\) are shown in blue.
  (c) The fundamental cycles \(\zeta_2\) and \(\zeta_5\) correspond to the dissipative slip cycle $\mathcal D$ and the forward cycle $\mathcal F$, respectively.
  }
  \label{fig:minkinesin}
\end{figure}

The chord representation~\eqref{eq:chord-representation} $\obs_\chords$ of a physical observable $\phi$ simplifies the calculation the fluctuations.
We demonstrate our method using the same observables considered already in Figure~\ref{fig:kinesin-fluctuations}:
The displacement rate $d:=l\,\eta_5$ and the hydrolysis rate $h:=\tau_1 + \tau_3$ count the centre of mass movement ($l=\SI{8}{\nano\meter}$) and the numbers of ATP molecules hydrolysed, respectively.
The dissipation rate is calculated as the time-average of the motance $\mota{\omega}{\omega'} \mapsto \log\frac{\tprob{\omega}{\omega'}}{\tprob{\omega'}{\omega}}$.

The chord representations $\obs_\chords = \sprod{\zeta_2}{\obs}\eta_2 + \sprod{\zeta_5}{\obs}\eta_5$ for those observables read
\begin{subequations}
\begin{align}
  d_\chords &= l\sprod{\zeta_2}{\eta_5} + l\sprod{\zeta_5}{\eta_5} \eta_5 = l\,\eta_5\,,\\
  h_\chords &= \sprod{\zeta_2}{\tau_1+\tau_3}\eta_2 + \sprod{\zeta_5}{\tau_1+\tau_3} \eta_5 = 2\eta_2 + \eta_5\,,\\
  \mot_\chords &= \sprod{\zeta_2}{\mot}\eta_2 + \sprod{\zeta_5}{\mot} \eta_5 = 2\mu\eta_2 +(\mu-f) \eta_5\,.
\end{align}  
  \label{eq:chord-reps-obs}
\end{subequations}
The first two lines are directly evident from Figure~\ref{fig:minkinesin}.
The third line reflects the thermodynamic balance conditions \eqref{eq:thermodynamic-balance}.
These conditions enter explicitly in the construction of the minimal four-state model as well as the construction of the original six-state model, \cf Appendix~\ref{app:construction} and Ref.~\cite{Liepelt+Lipowsky2007}, respectively.

\subsubsection{A phase diagram and beyond}
\label{sec:phase-diagram}
Based on Equations~\ref{eq:gen-schnakenberg} derived in Section~\ref{sec:cycle-ldt}, we can calculate the asymptotic fluctuation spectrum of the (time-averages) of these currents analytically.
All that we need are the chord representations~\eqref{eq:chord-reps-obs} of the observables we are interested in and the scaled cumulants for the currents on the fundamental chords.
For brevity, we will only deal with the dominant part of the spectrum, \ie with the first and second scaled cumulants.

\begin{figure}[htb]
  \centering
  \begin{tabular}{cc}
    \includegraphics[width=0.35\columnwidth]{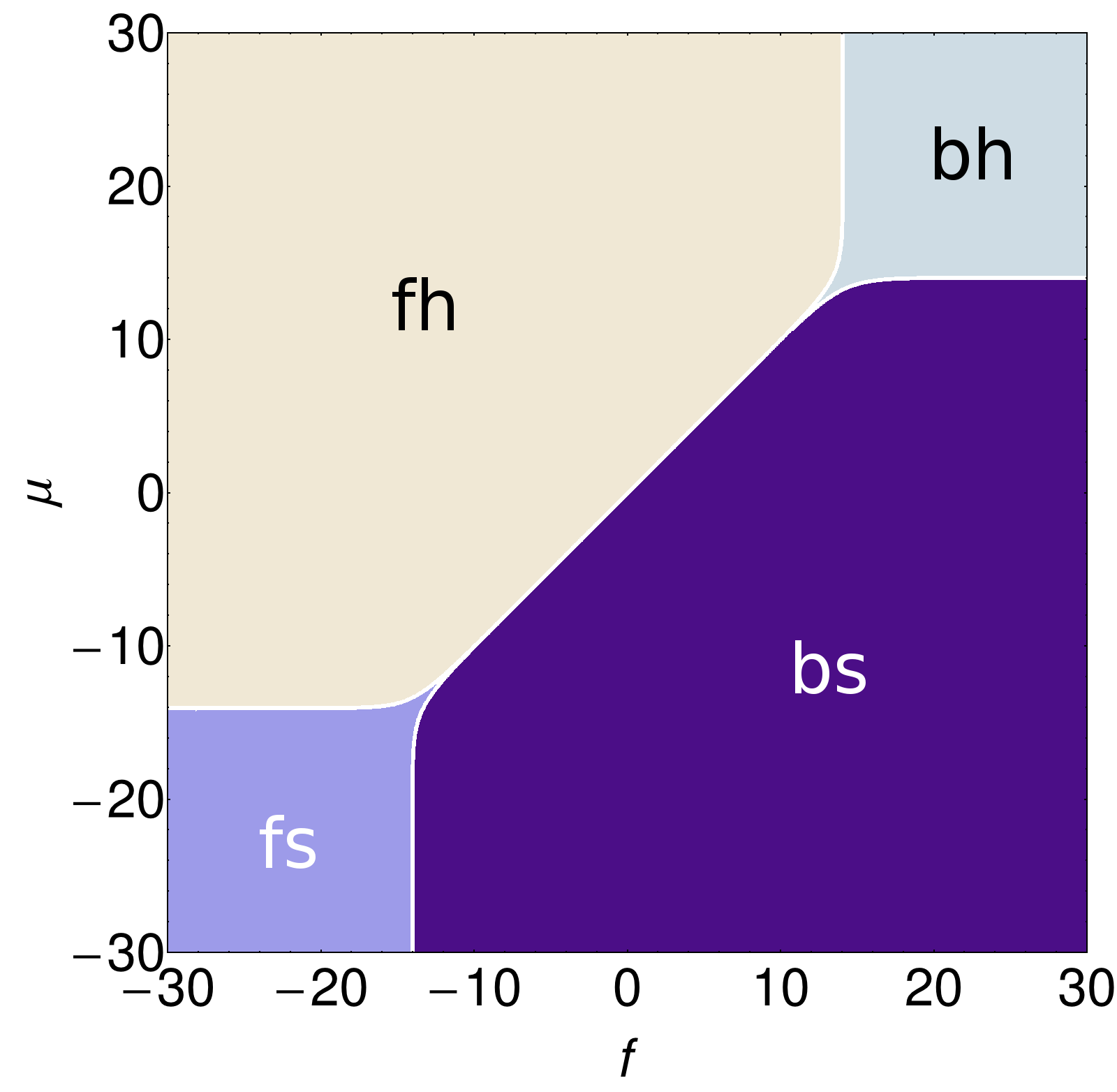}\qquad\qquad & \includegraphics[width=0.35\columnwidth]{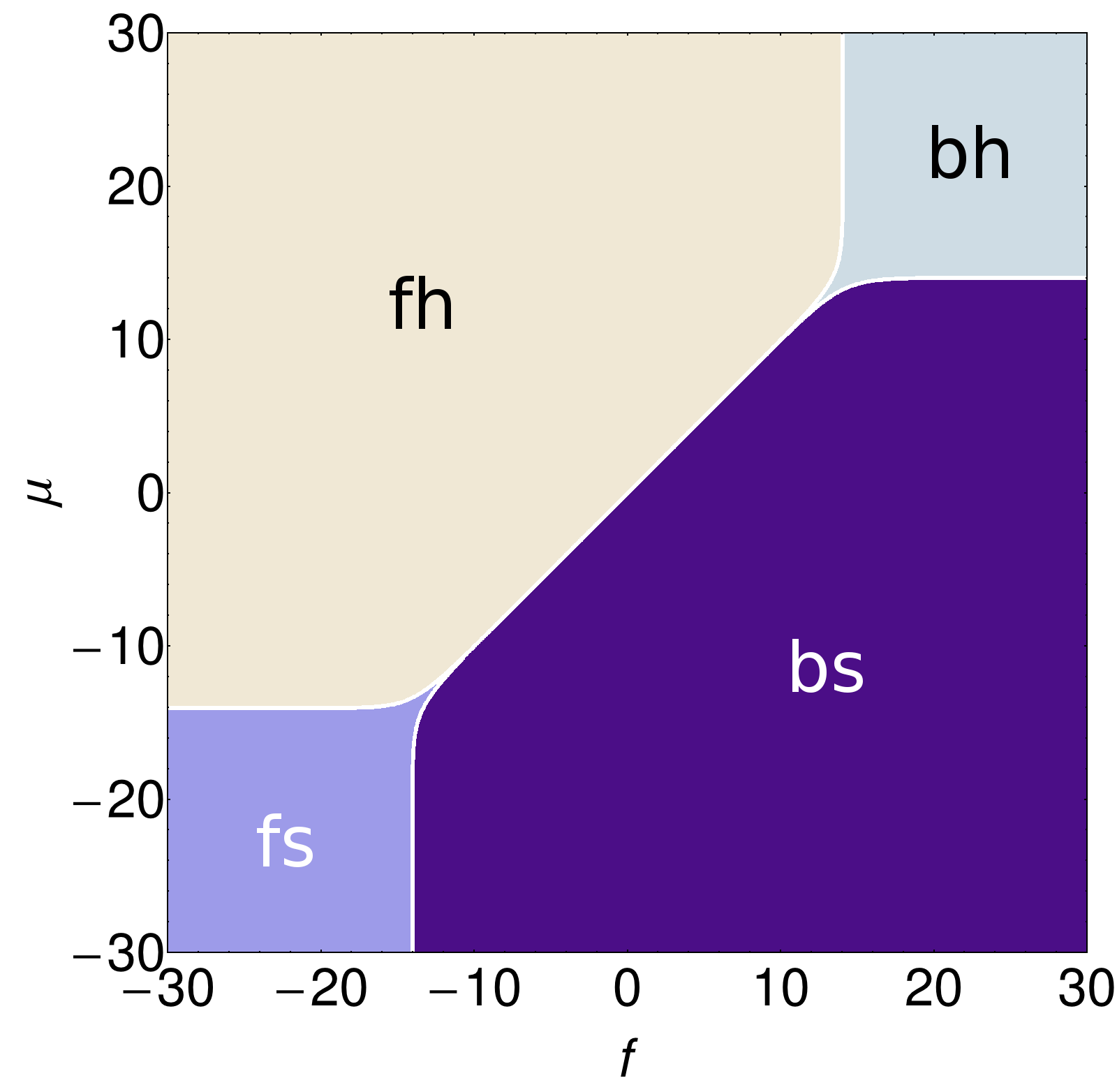}\\
    (a) &\qquad (b) 
  \end{tabular}
  \caption{
    Phase diagram of the molecular motor kinesin in the \( (f,\mu) \)-plane.
    The left an right plots are analytical results obtained for the well-established six-state model \cite{Liepelt+Lipowsky2007} and a minimal four-state model, \cf Appendix~\ref{app:construction}, respectively.
    The different regions in the diagram reflect the different tuples $\left(\sgn\left( c_1(d) \right),\sgn\left(c_1(h)\right)\right)$, \ie the signs of the average motor velocity and the average hydrolysis rate. 
    They correspond to the four operational modes of kinesin: forward stepping and hydrolysis (\,$\mathrm{fh} \equiv (+,+)$), backward stepping and hydrolysis (\,$\mathrm{bh} \equiv (-,+)$), forward stepping and synthesis (\,$\mathrm{fs} \equiv (+,-)$), and backward stepping and synthesis (\,$\mathrm{bs} \equiv (-,-)$), \cf also Fig.~\ref{fig:velo-hydro-6-state}.
   In this phase diagram, no difference is visible between the two models.
 }
  \label{fig:modes}
\end{figure}

Using the first scaled cumulants, \ie the steady-state averages, in Figure~\ref{fig:modes}a) we reproduce a result by Liepelt and Lipowsky, regarding the operational modes of kinesin~\citep{Liepelt+Lipowsky2009}.
The modes are defined by the signs of the first cumulants \(c_1(d)\) and \(c_1(h)\) of the displacement and hydrolysis rate.
They distinguish the direction of the motor's velocity along the microtubule and whether ATP is produced or consumed, respectively.
The top left part of the diagram corresponds to the usual operation mode of kinesin under physiological conditions:
The motor moves forward on the microtubule while hydrolysing ATP.
Even if we apply a force ($f>0$) pulling the motor back this motion is sustained.
However, for sufficiently large forces the motor runs backward while still hydrolysing ATP (top right).
In the other large region on the lower right, kinesin synthesizes ATP from ADP and P.
In the part of this region where the force is positive, we can say that kinesin acts as a ``chemical factory'':
Mechanical work is converted into chemical energy stored in ATP.
Finally, the lower left region corresponds to ATP synthesis during forward motion.
We see that for this behaviour there must be both a force that pushes the motor while also the chemical concentrations in the environment highly favour ATP synthesis.

As a comparison, Figure~\ref{fig:modes}b shows the phase diagram obtained with our minimal four-state model. 
To the eye, it is indistinguishable from Figure~\ref{fig:modes}a) obtained for the six-state model. 
Thus in order to compare the models in more detail, we need additional information than just the sign of the first cumulants.

Our method provides this additional information.
Before we discuss the actual \emph{values} of the first and second cumulants for the displacement and the hydrolysis, we look at the (non-dimensionalized) dissipation provided by the motance matrix $\mot$.
The dissipation is in a sense the most natural scalar observable to look at.
It is completely determined by the logarithmic ratio of the rates alone.
Hence, it is well-defined for \emph{any} dynamically reversible Markovian model and does neither require a physical interpretation of the transition rates nor of the system as such. 

\begin{figure}[ht]
  \centering
  \includegraphics[width=.35\textwidth]{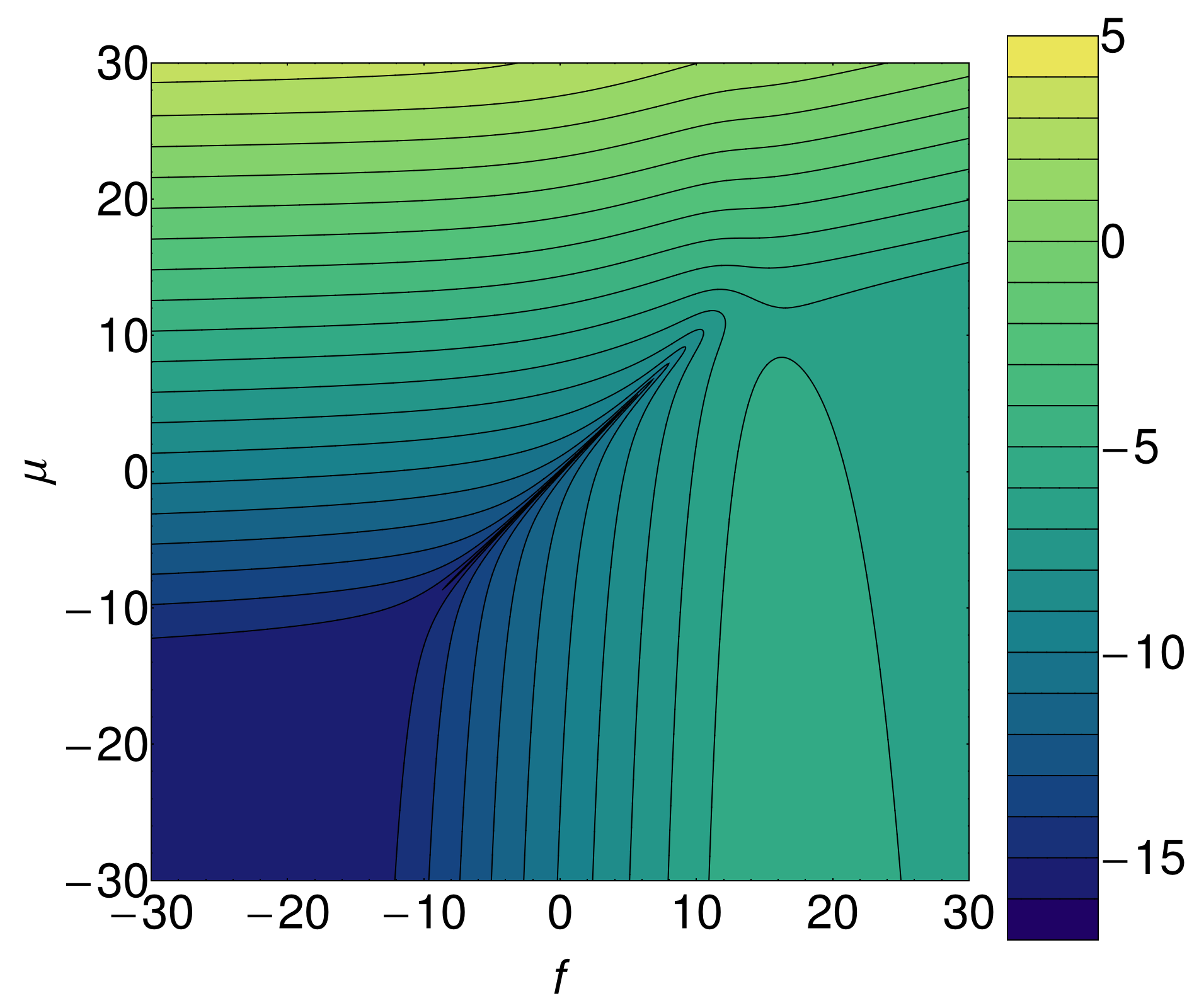}%
  \includegraphics[width=.35\textwidth]{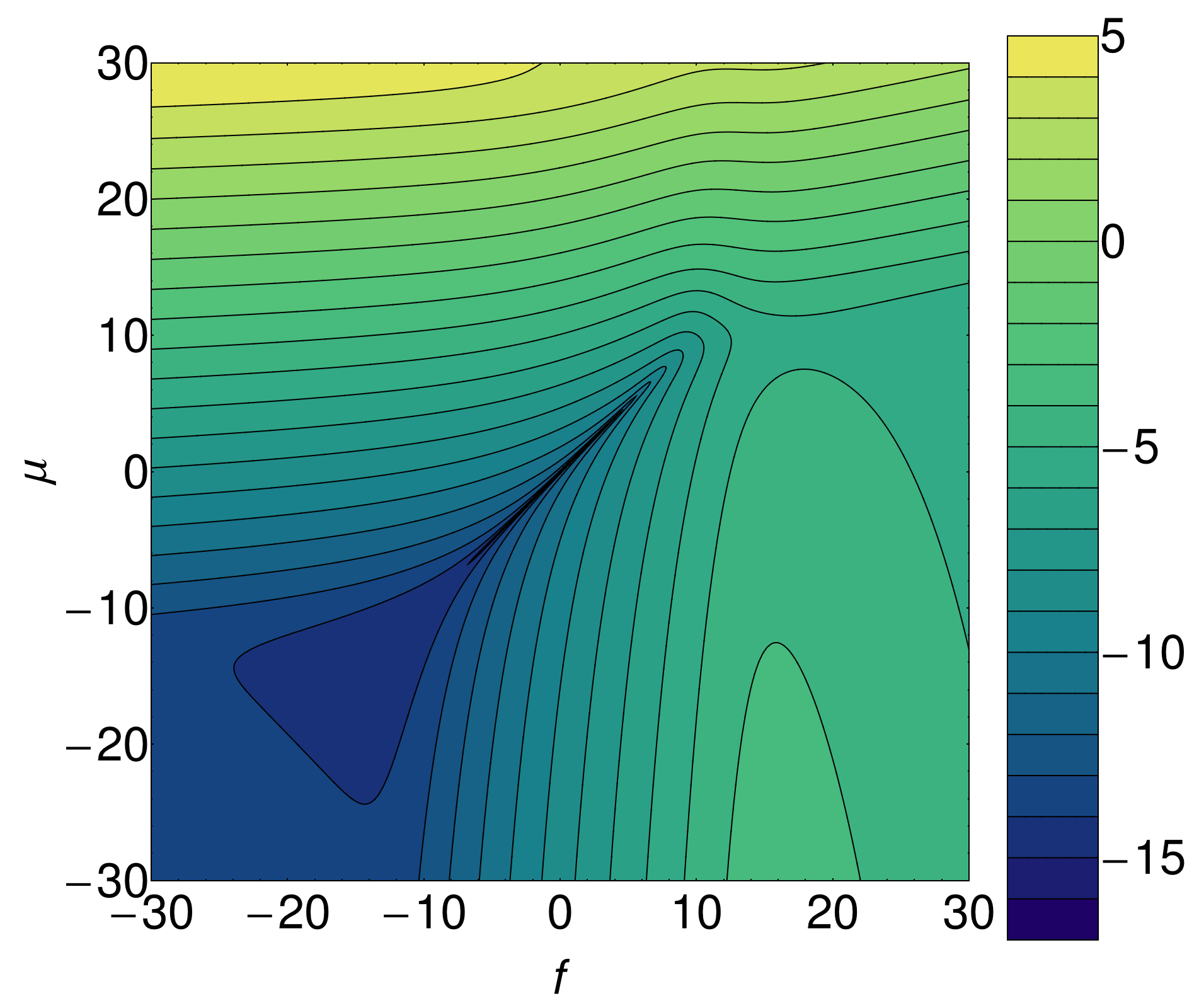}%
  \includegraphics[width=.35\textwidth]{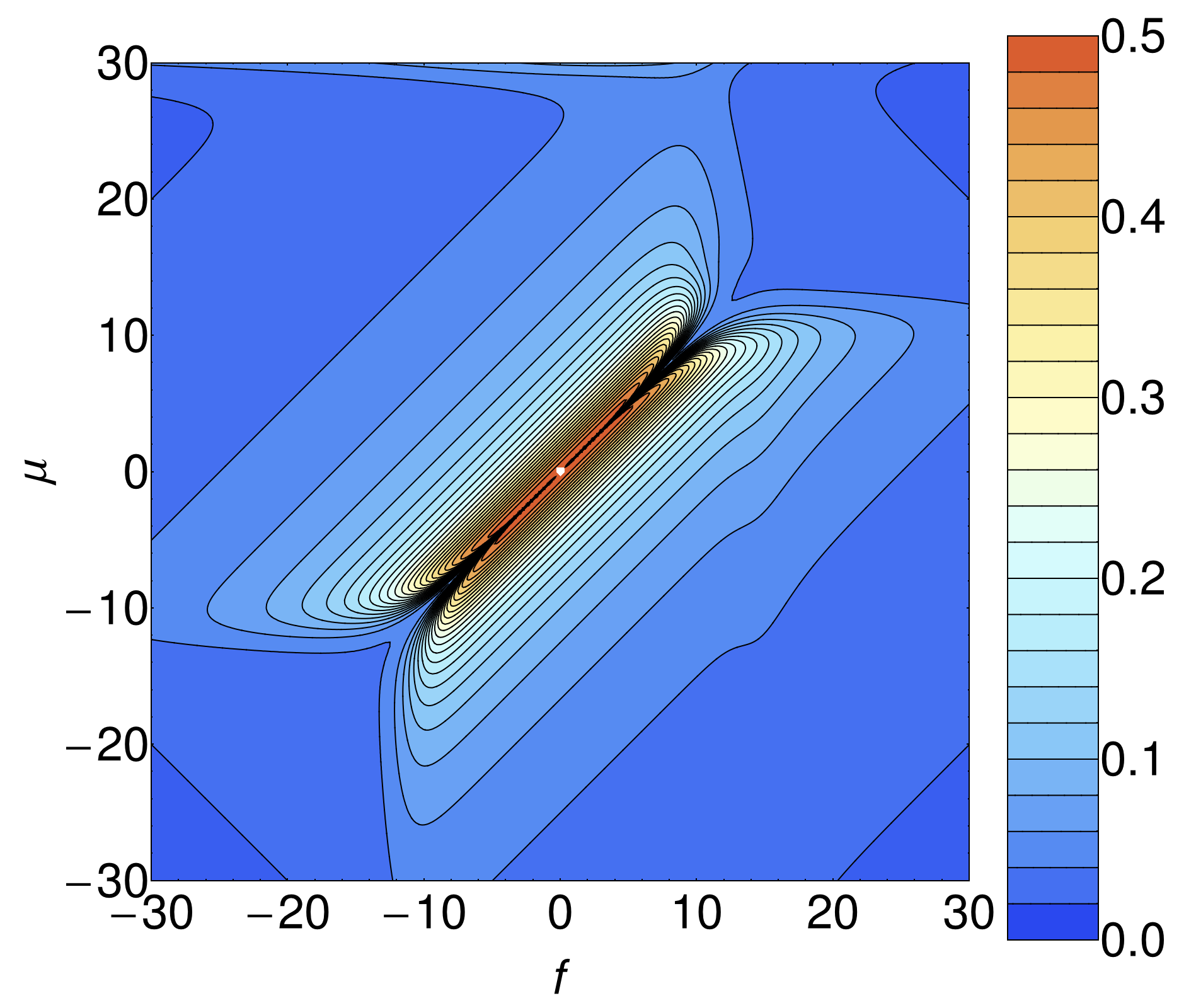}
  \caption{Dissipation in the six-state model corresponding to the physical observable $\mot$.
    The axis the plots correspond to the chemical and mechanical driving parameters $\mu$ and $f$, respectively.
    Left: The decadic logarithm $\log_{10}c_1(\mot) = \log_{10}\eav{\totentrv}\tind{\infty} = \log_{10}\eav{\medentrv}\tind{\infty}$.
    Centre: The decadic logarithm $\log_{10}c_2(\mot)$ characterizing the strength of fluctuations in the dissipation.
    Right: The signal-to-noise ratio $c_1(\mot)/c_2(\mot)$.
  } 
  \label{fig:dissipation-6-state}
\end{figure}

Figure~\ref{fig:dissipation-6-state} shows our results.
The plot on the left shows the decadic \emph{logarithm} of the first scaled cumulant $\log_{10} c_1(\mot)$, \ie the expectation value of the total entropy production in the steady state.
The black contour lines thus correspond to values separated by one order of magnitude.
Hence, the values we plot range over more than twenty decades.
The centre plot for the logarithm of the second scaled cumulant $\log_{10} c_2(\mot)$, which characterizes the (scaled) variance in the dissipation rate, ranges over similar values.
Qualitatively, it looks rather similar to the fist cumulant.
In order to find more structure,  the rightmost plot shows a signal-to-noise ratio (SNR) calculated as the ratio $c_1(d)/c_2(d)$.
Note that we use a \emph{linear} scale here.

\begin{figure}[t]
  \centering
  \includegraphics[width=.35\textwidth]{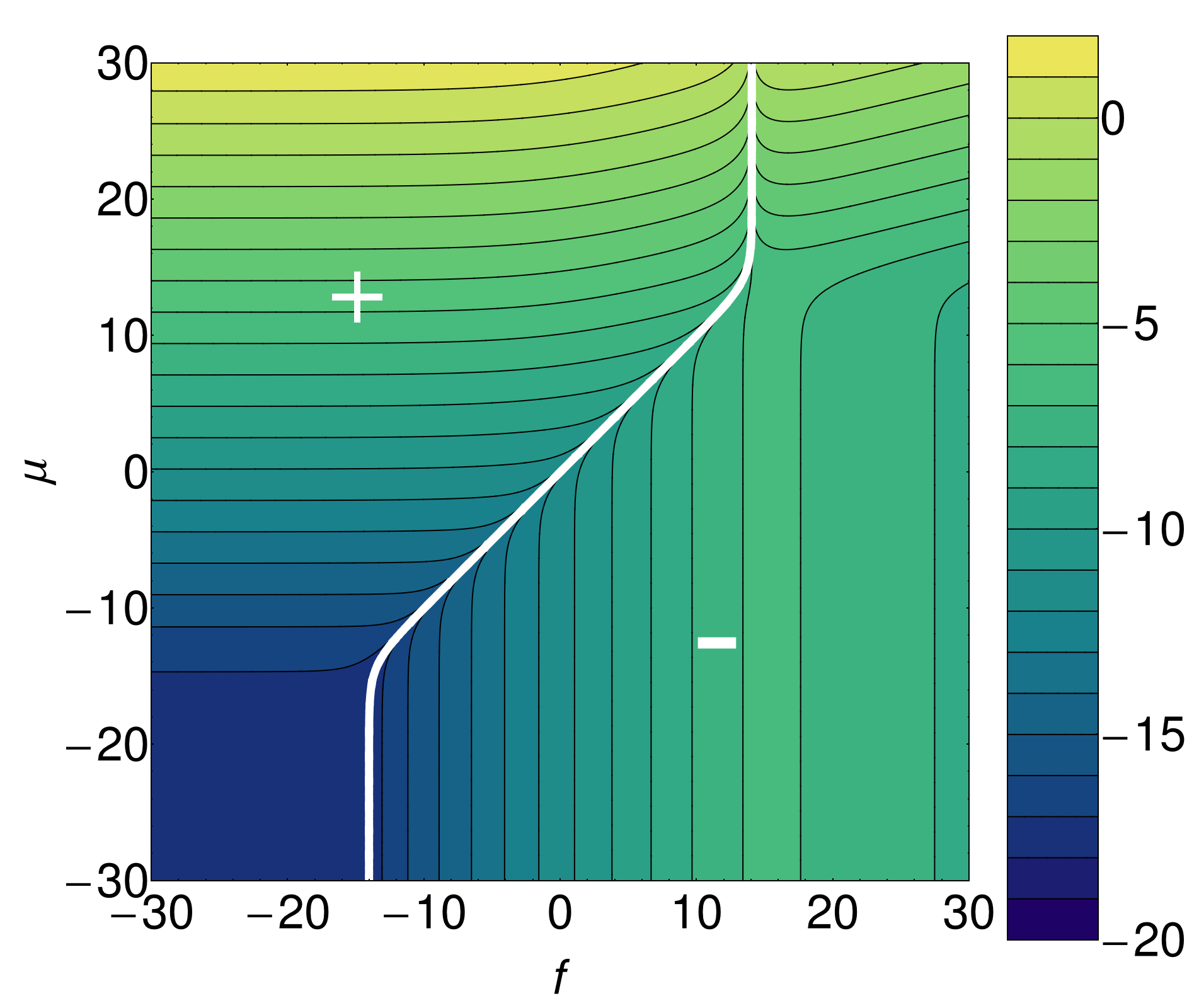}%
  \includegraphics[width=.35\textwidth]{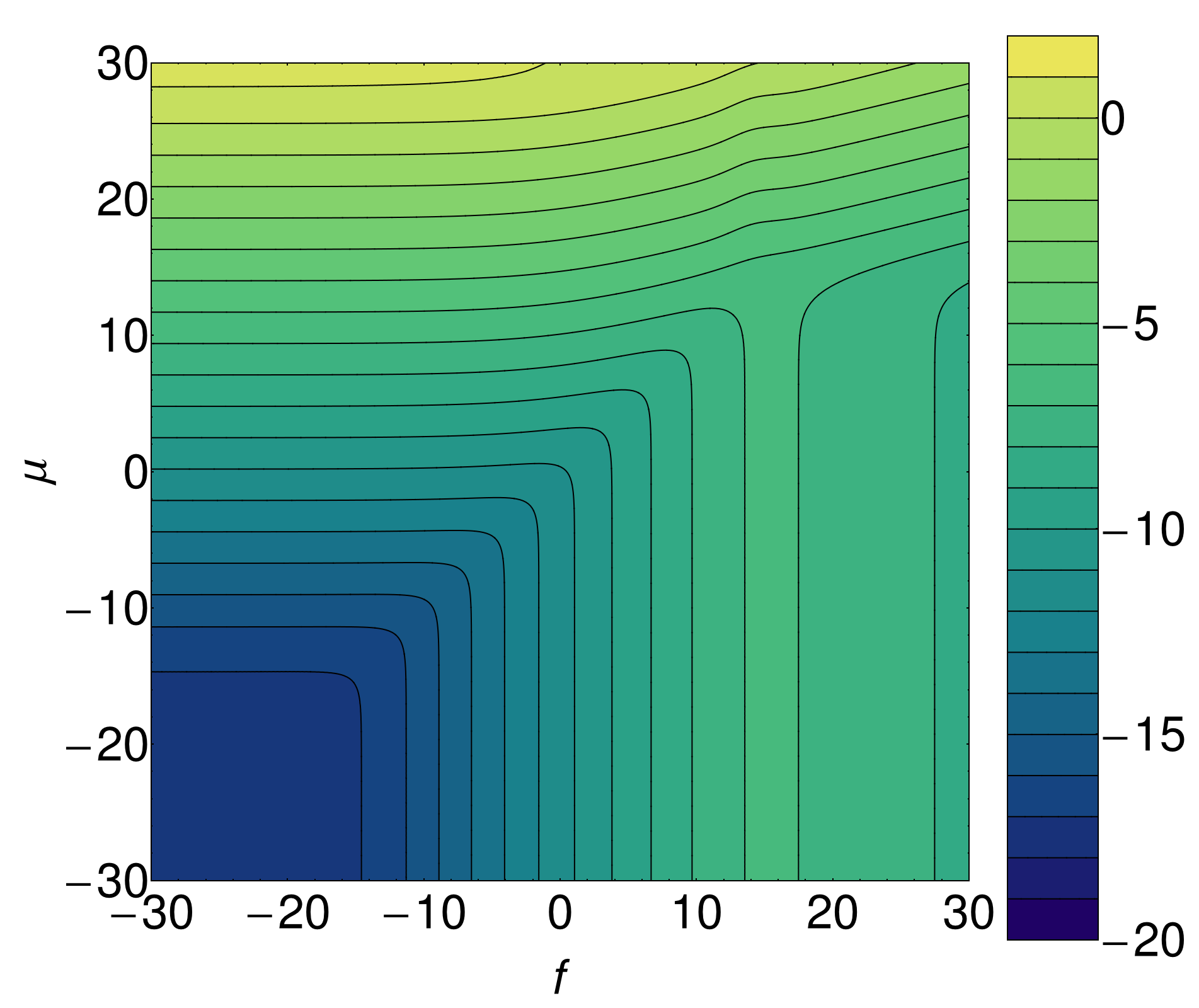}%
  \includegraphics[width=.35\textwidth]{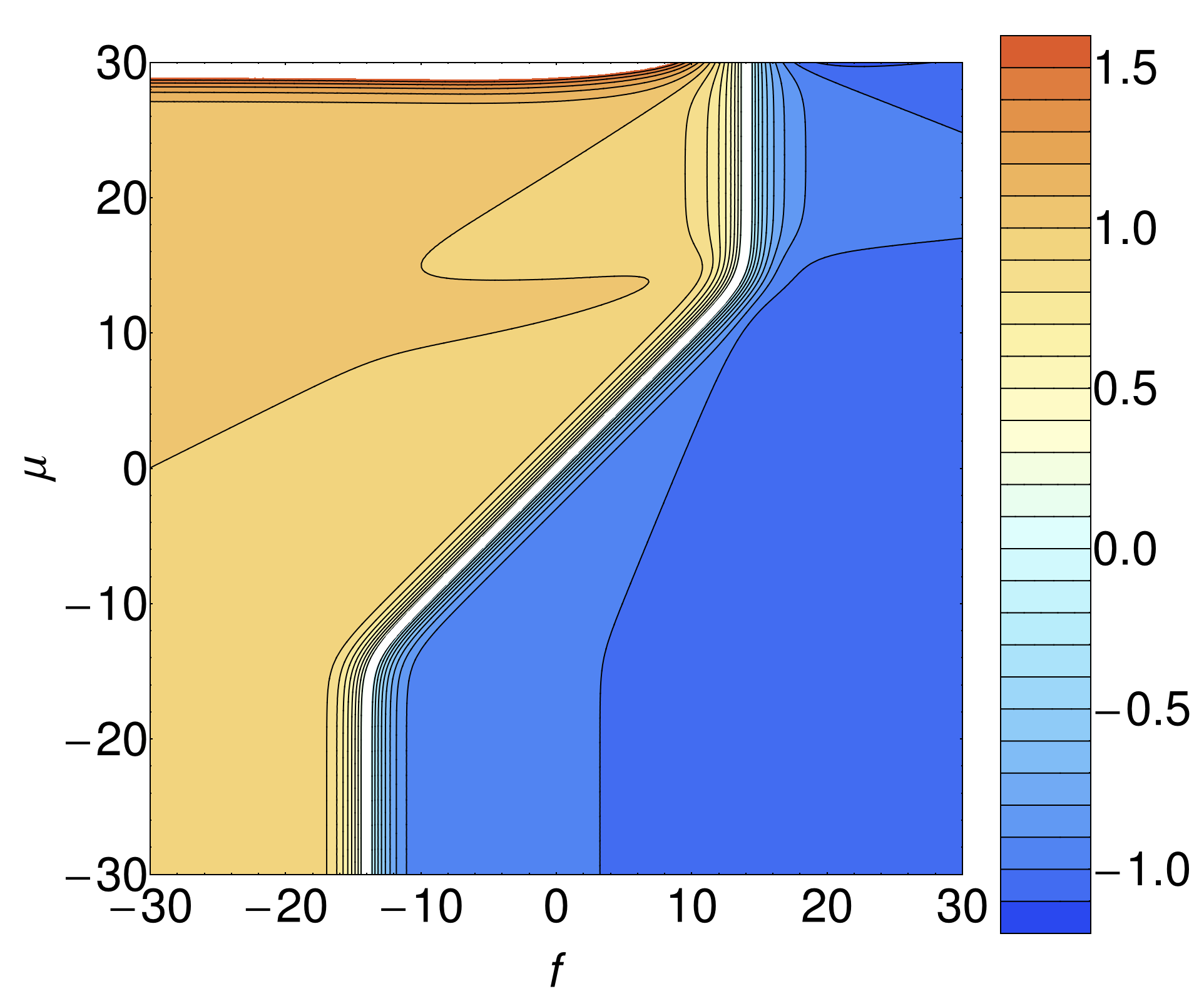}\\
  \includegraphics[width=.35\textwidth]{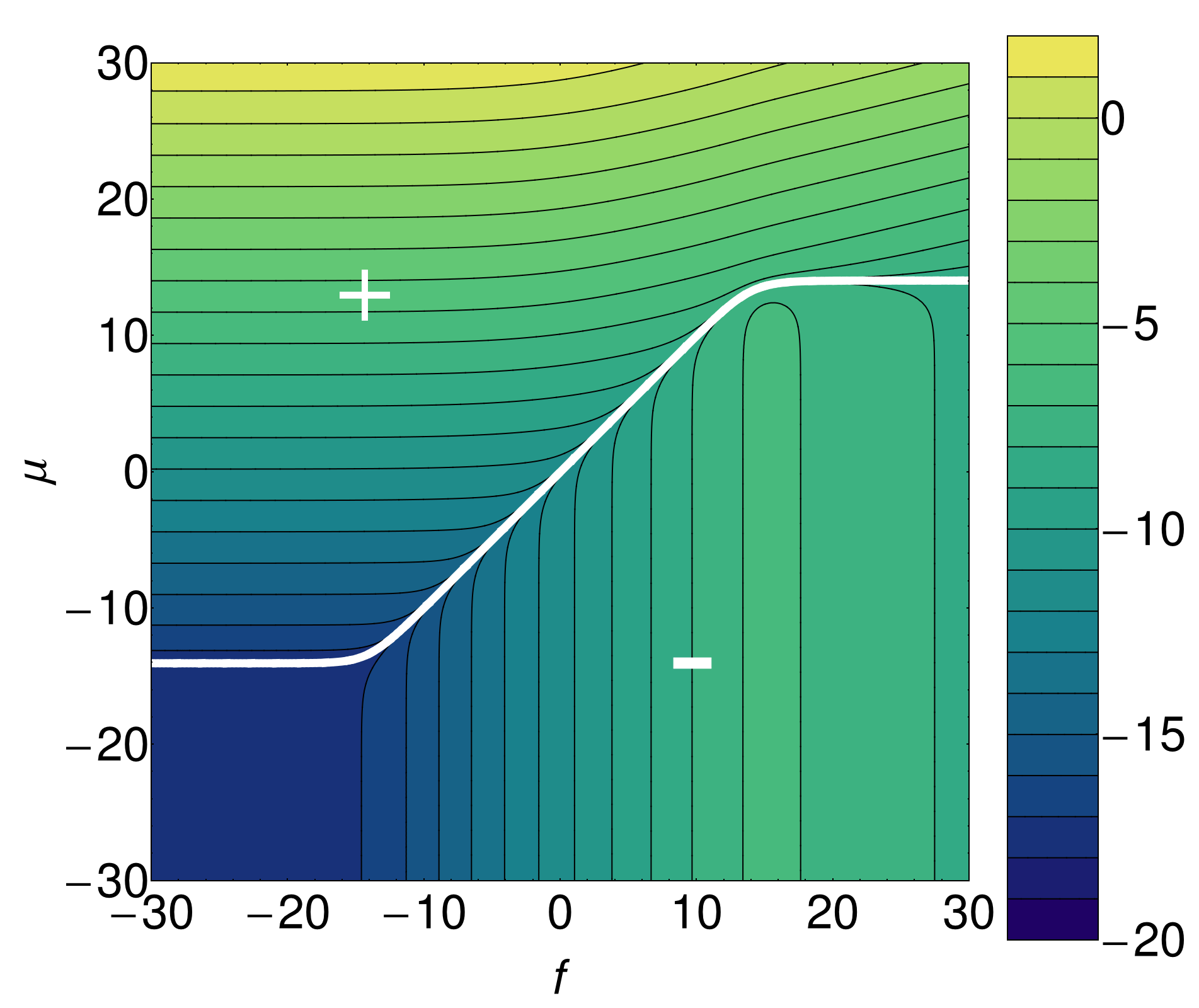}%
  \includegraphics[width=.35\textwidth]{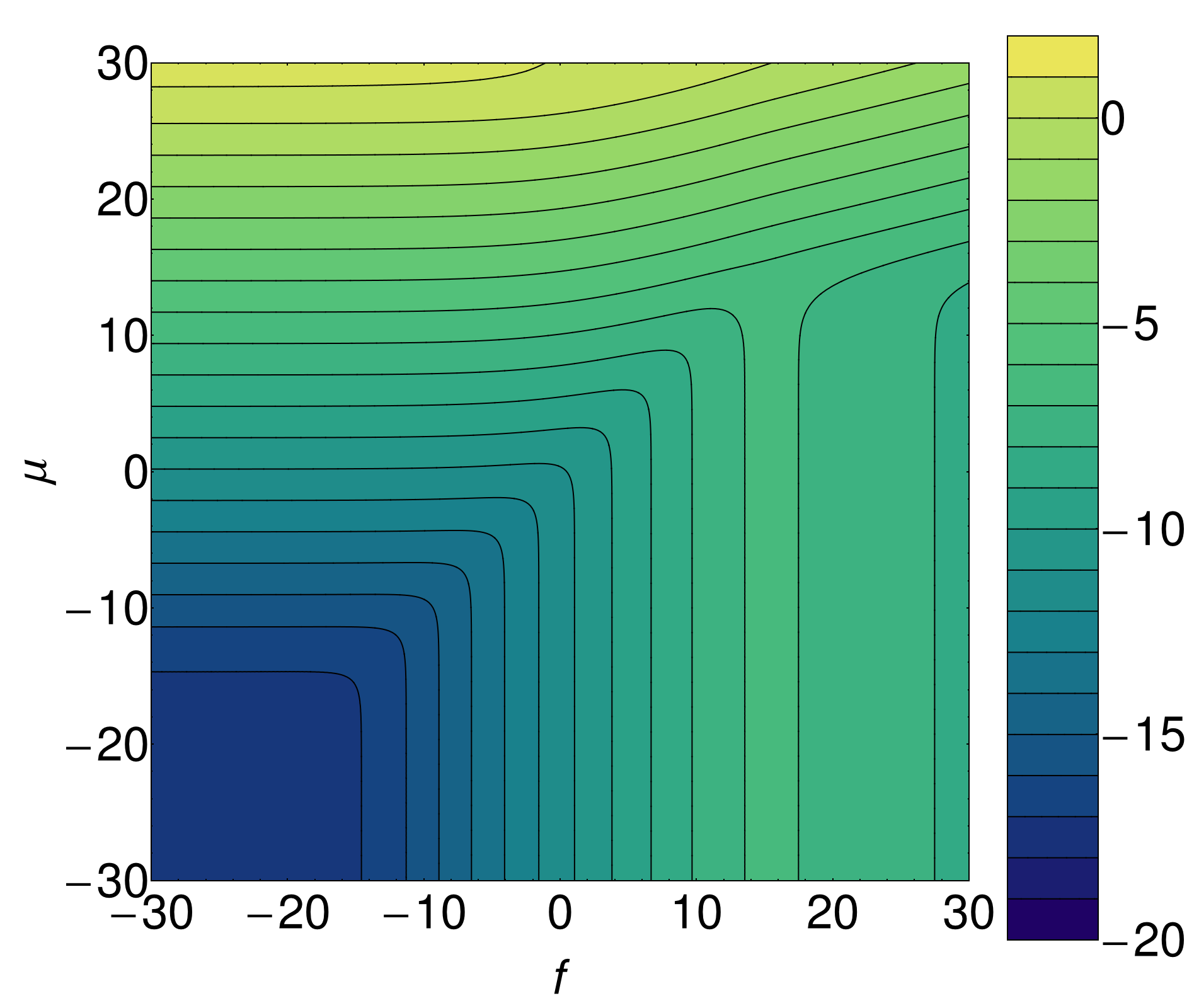}%
  \includegraphics[width=.35\textwidth]{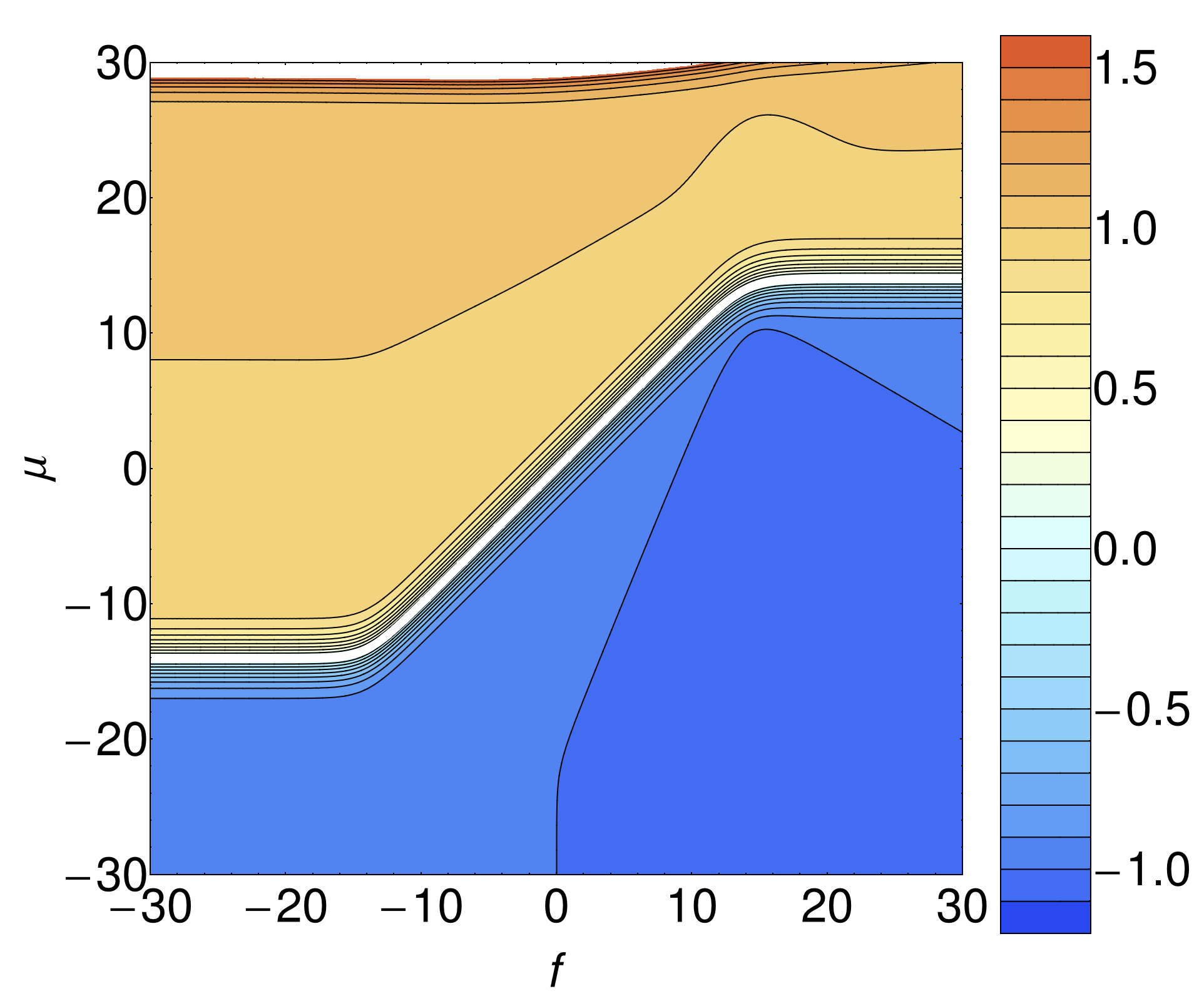}
  \caption{
    First two cumulant of the mechanical displacement $d$ (top row) and hydrolysis $h$ (bottom row) in the six-state model.
    The first column shows the absolute value of the first scaled cumulants $c_1(d)$ and $c_1(h)$, corresponding to the average motor velocity and hydrolysis rate, respectively.
    The sign is indicated as an inset, with the two regions separated by the think solid line corresponding to parameters such that $c_1 =0$, \cf also the phase diagram Fig.~\ref{fig:modes}a).
    The presentation of the data is similar to the presentation in Fig.~\ref{fig:dissipation-6-state}, \ie the middle and right columns show the second cumulant $c_2$ and the SNR $c_1/c_2$, respectively.
  }
  \label{fig:velo-hydro-6-state}
\end{figure}

Before we interpret these results, let us first take a look at the corresponding plots for the observables that define the operation modes of the motor. 
Figure~\ref{fig:velo-hydro-6-state} shows similar plots for the logarithms of first and second cumulants $\abs{c_1},c_2$ as well as the SNR $c_1/c_2$ for the displacement $d$ (top row) and hydrolysis rate $h$ (bottom row), respectively.

On first sight, the plots for the first and second cumulants look qualitatively similar to the ones shown in Fig.~\ref{fig:dissipation-6-state}:
The values range over the large range spanning more than twenty logarithmic decades.
Moreover, the overall features illustrated by the contour lines (\eg the existence of a local  maximum in the absolute value of the cumulants around $f\approx 15$) are common to all plots.

However, there is a caveat to this statement.
Unlike the average entropy production, the average displacement rate (\ie the velocity) and the average hydrolysis rate change their signs.
Consequently, the logarithm of the absolute value of the first cumulant $\abs{c_1}$ shows a continuous line of singularities.
The position of these singularities is marked by a white line in the first column of Fig.~\ref{fig:velo-hydro-6-state}.
Note that the white lines are by definition the lines separating the different modes of kinesin, shown in the phase-diagram in Fig.~\ref{fig:modes}a).

Let us discuss the structure found in the cumulants in more detail.
The non-negative dissipation rate vanishes at the centre of the diagram, where we have thermodynamic equilibrium $f = \mu = 0$ and the motor does neither move nor catalyse any chemical reactions on average.
Further, the dissipation is very small along the first diagonal $f=\mu$ in a neighbourhood around the origin.

It is a consequence of the quasi-tight coupling  mechanical and chemical transitions for kinesin:
It is very natural for the motor to behave in the way shown in Fig.~\ref{fig:kinesin-walking}b, as expressed by the succession of states along the forward cycle.
To appreciate this fact, note that the affinity of the forward cycle $\zeta_5$ is given by $\mot(\zeta_5) = \mu-f$.
Hence, the diagonal in the phase diagram corresponds to a vanishing affinity along that cycle.
Similarly, in the centre region the values of the velocity and hydrolysis rate vanish somewhere very close to this diagonal.

Moreover, for not too high driving forces we have an (approximate) reflection symmetry along the diagonal, which holds for the first and second cumulant of all the observables we considered here.
In conclusion, the (affinity of the) forward cycle dominates the dynamics of kinesin --- at least in regions where the absolute values of the non-dimensionalized drivings are smaller than about ten.

Another prominent feature is the region of low dissipation in the lower left corner of the phase diagram.
It is the region where the system on average runs along the reversed backwards cycle, \ie on average it moves forwards while synthesizing ATP.
It does so, however in an extremely slow fashion, as indicated by the small values of the velocity and hydrolysis rate (\cf the left column of Fig.~\ref{fig:velo-hydro-6-state}).
Hence, the reason for the small dissipation rate in that part of the phase diagram is the slowness of the kinetics.
Note that this feature also shows in the plots for the second cumulants:
In the kinetically hindered region in the lower left, the system is essentially ``frozen''.
Upon increasing the chemical driving, the system starts moving again. 

More structure is visible in the third column where we plot the signal-to-noise-ratio (SNR) $c_1/c_2$.
Nevertheless, we stop the discussion of the structure of kinesin's  phase-diagram at this point.
Additional remarks regarding the significance of the SNR are made in the discussion in Section~\ref{sec:discussion-kinesin}.
Before we come to that point, however, let us see whether these features are preserved in simplified models.

\subsection{Simplified models}
In the present section, we discuss the structure of kinesin's phase diagram in simplified models.
The main focus thereby lies on the minimal model constructed in Appendix \ref{app:construction}, where we carefully made sure to follow the physical arguments for the construction of the six-state model in Ref.~\cite{Liepelt+Lipowsky2007}.
After that, we quickly discuss simplified models obtained by the fluctuation-sensitive coarse-graining approach presented in Section~\ref{sec:coarse-graining}.

\subsubsection{Comparison to the minimal model}
\label{sec:minkinesin}
We have already seen in Figure~\ref{fig:modes} that the phase diagram of a minimal four-state model agrees very well with the phase diagram of the more involved six-state model.
In order to see whether we also have an agreement in the more detailed structure, we consider relative errors between the models.
For any quantity $X$ that takes the values $X_6$ and $X_4$ in the six and four-state model, respectively, we define the relative error $\delta{X}:= \frac{X_4-X_6}{X_6}=\frac{X_4}{X_6}-1$.
For our case, $X$ represents the (decadic logarithms of the) first and second cumulants as well as the SNR.
For these observables, we have discussed $X_6$ already in Figures~\ref{fig:dissipation-6-state} and~\ref{fig:velo-hydro-6-state}.

\begin{figure}[tp]
  \centering
  \includegraphics[width=.35\textwidth]{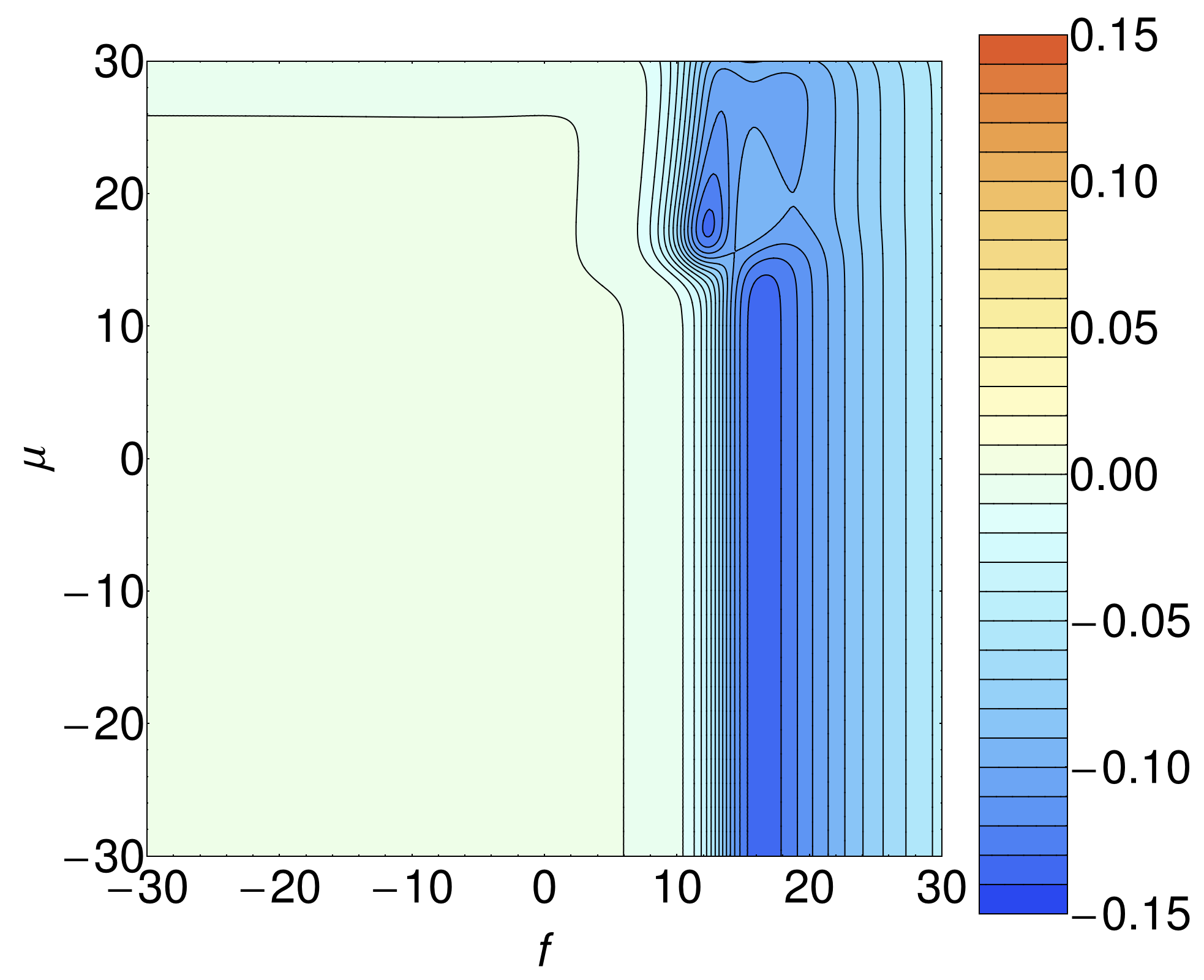}%
  \includegraphics[width=.35\textwidth]{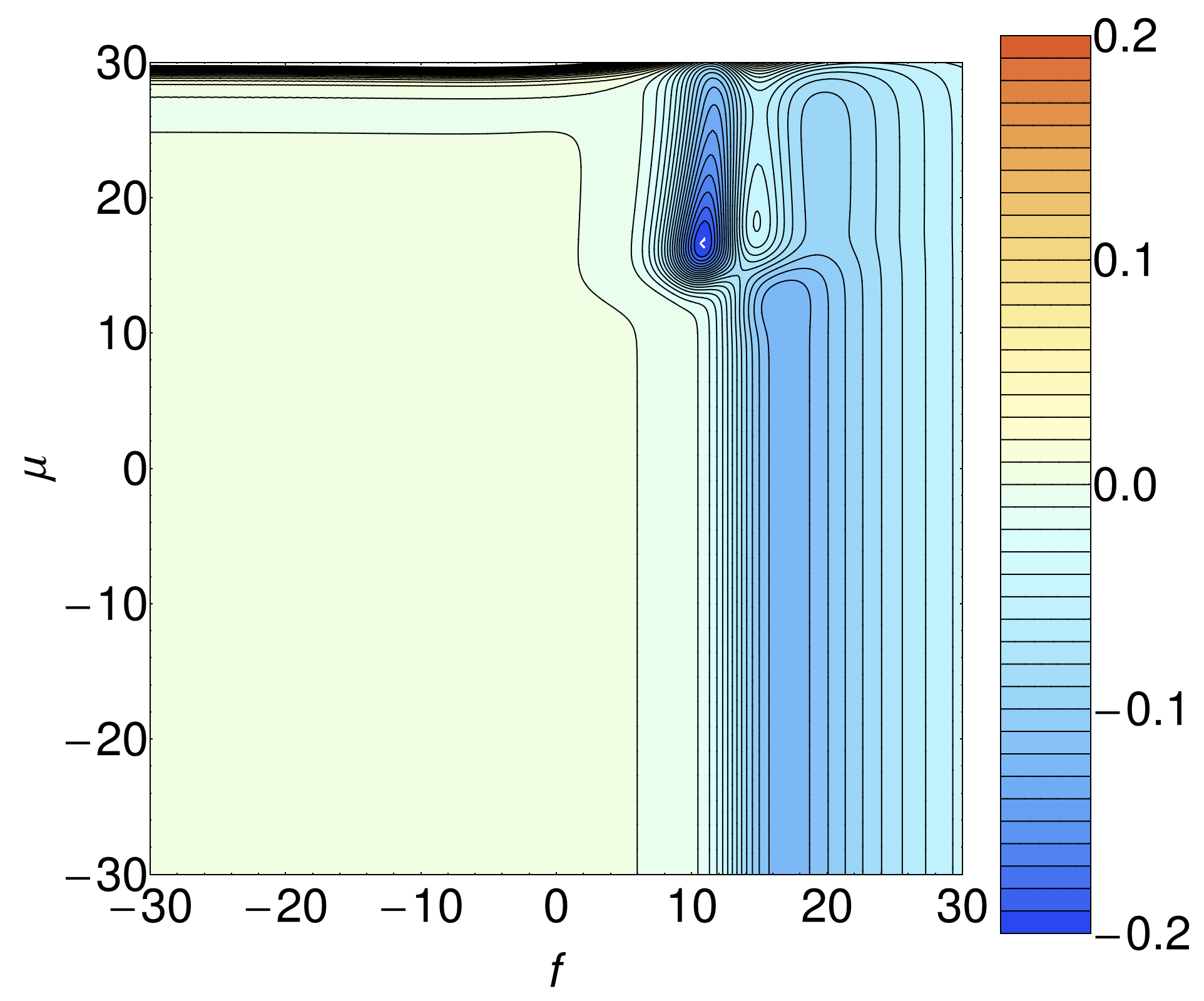}%
  \includegraphics[width=.35\textwidth]{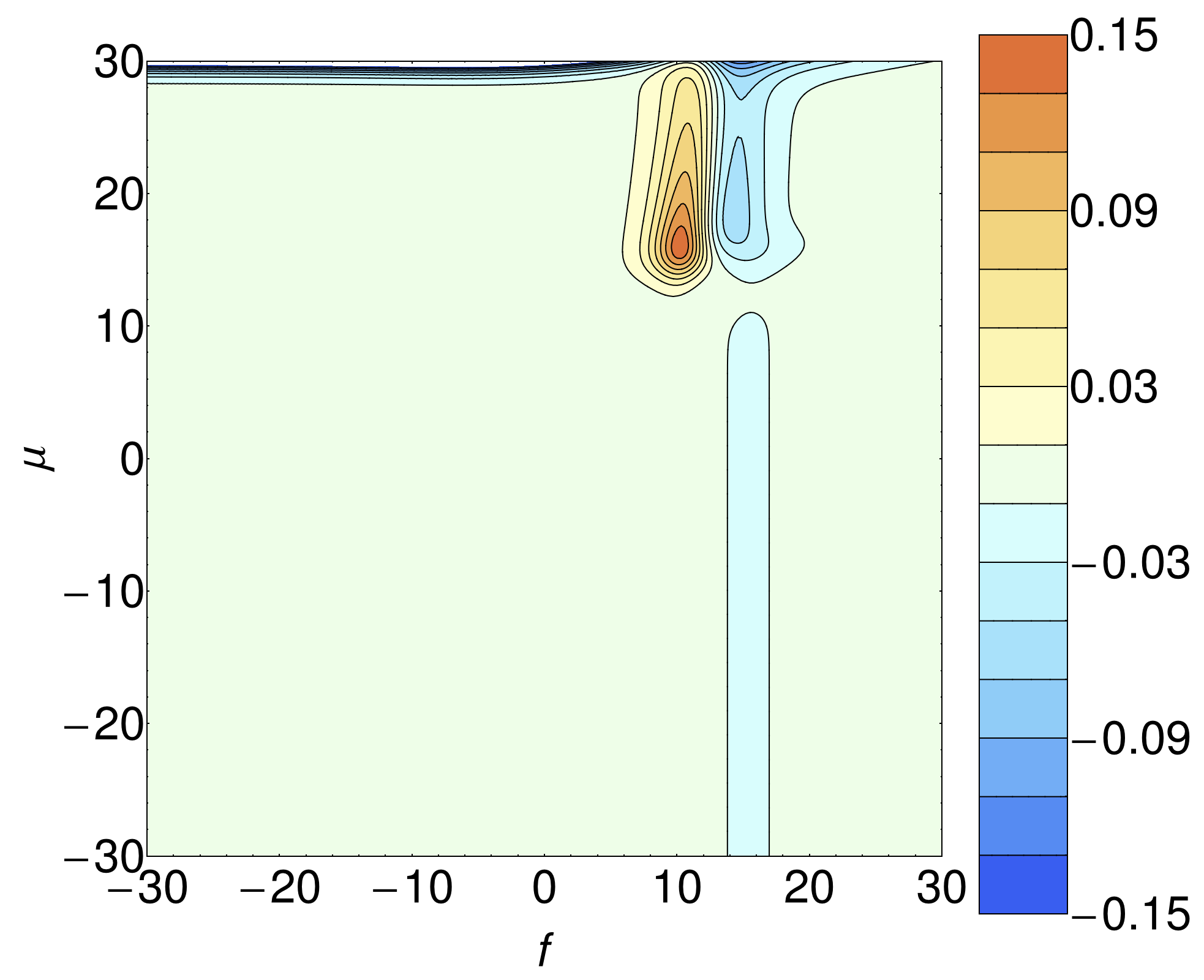}\\
  \includegraphics[width=.35\textwidth]{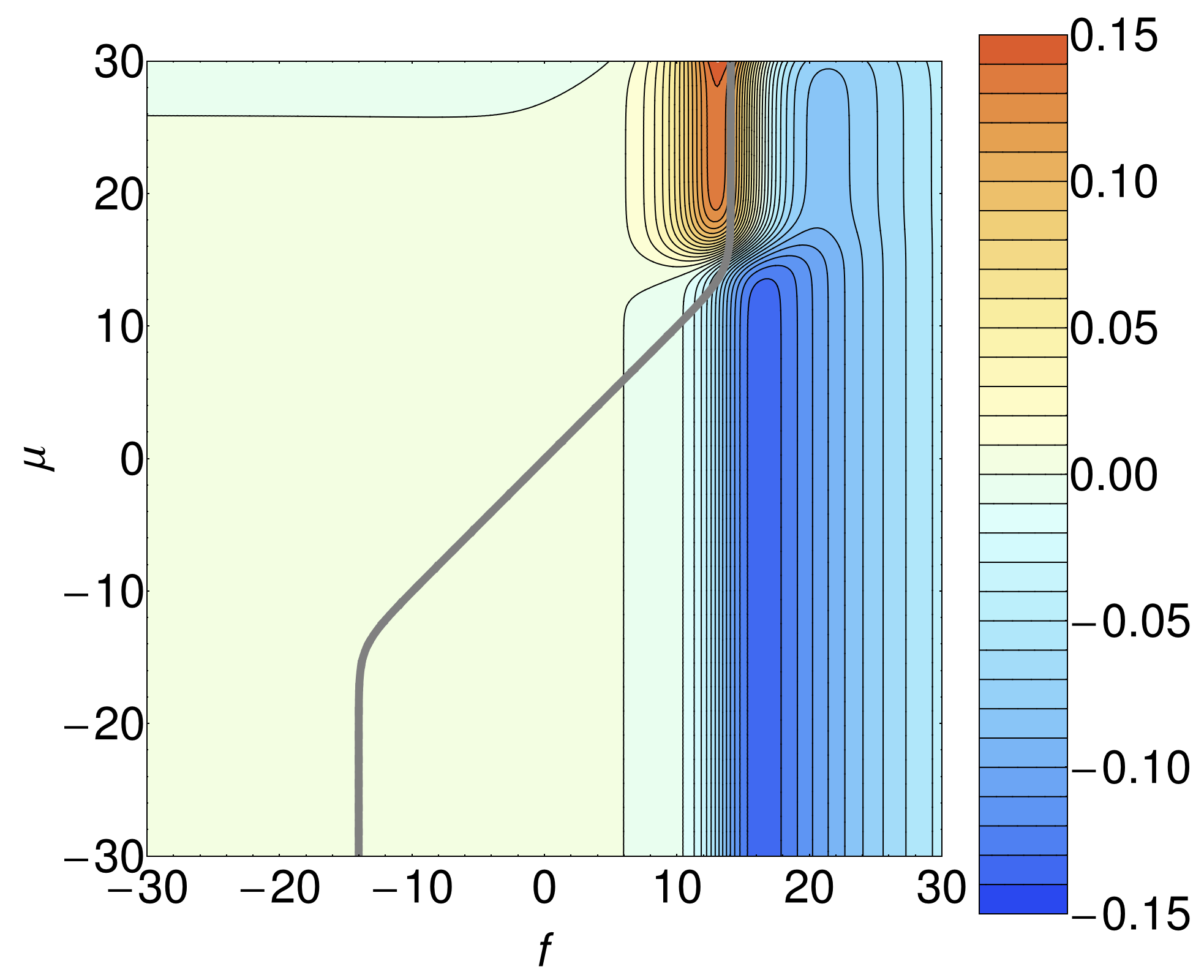}%
  \includegraphics[width=.35\textwidth]{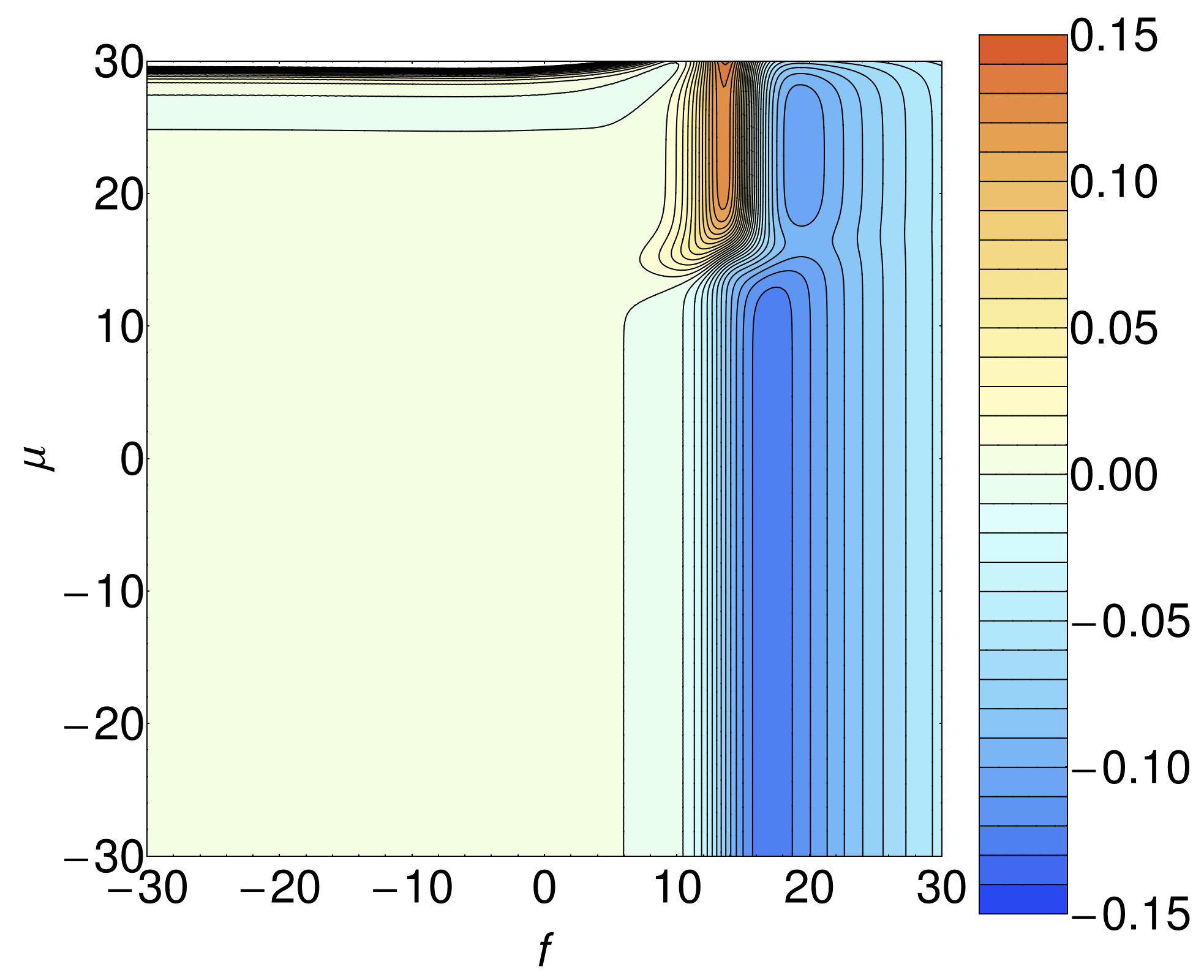}%
  \includegraphics[width=.35\textwidth]{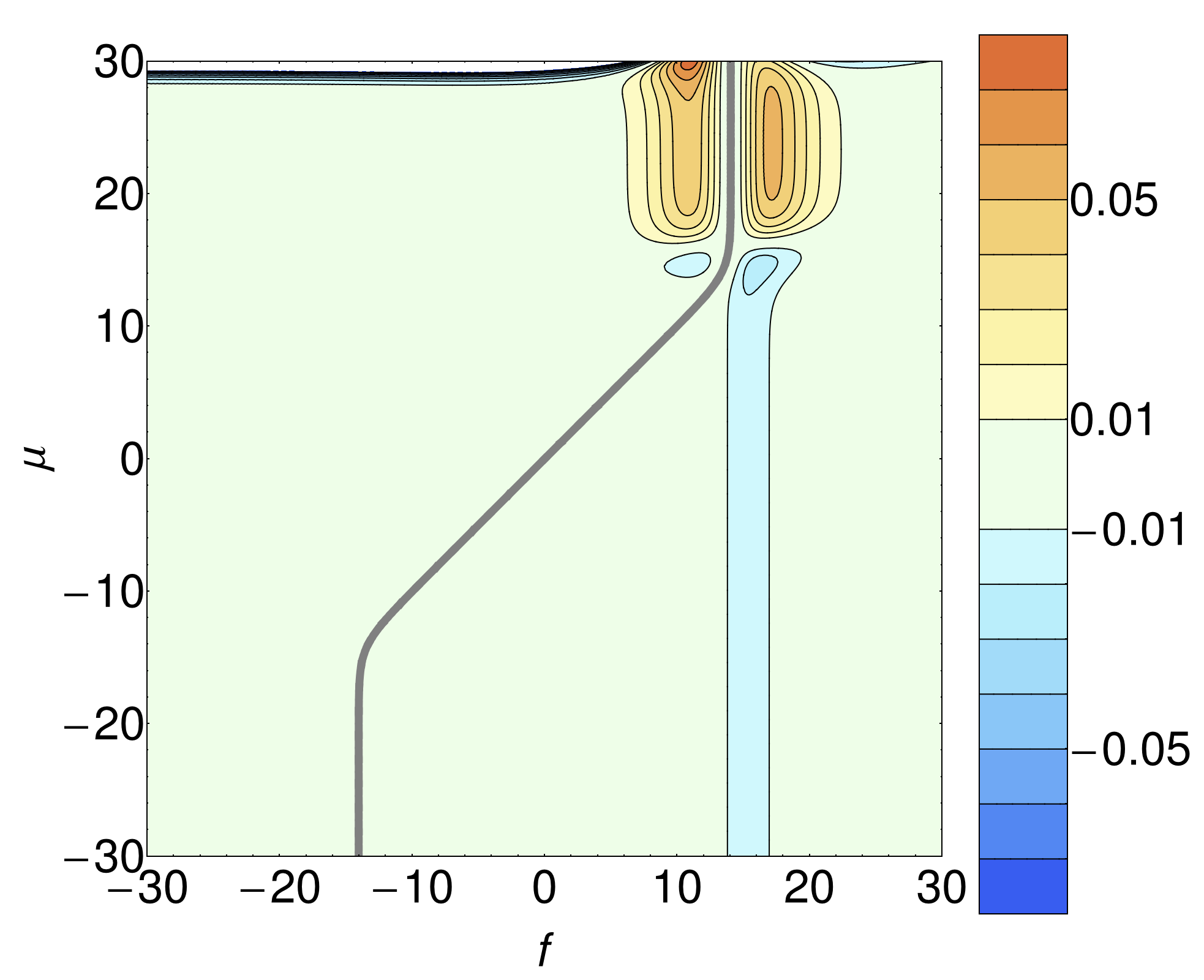}\\
  \includegraphics[width=.35\textwidth]{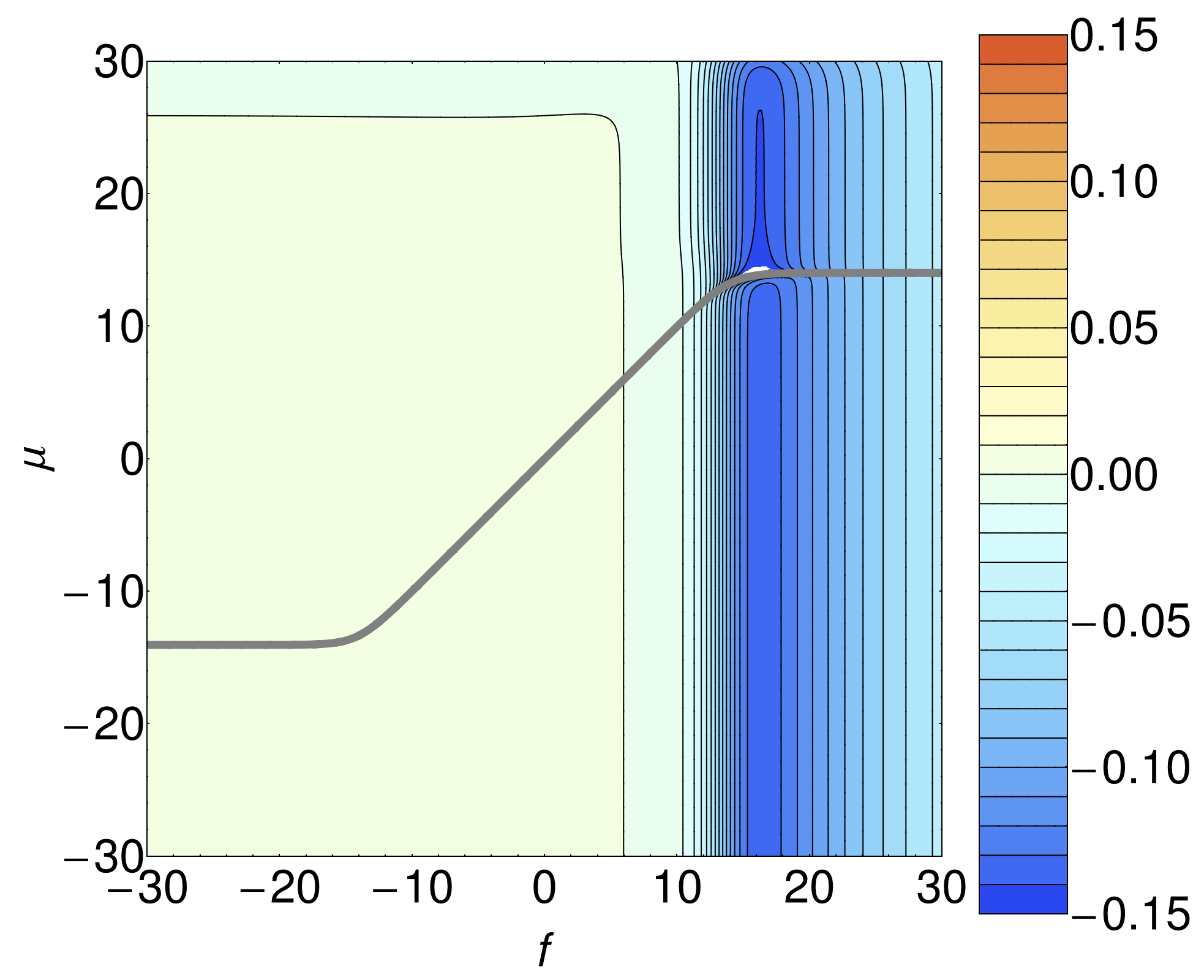}%
  \includegraphics[width=.35\textwidth]{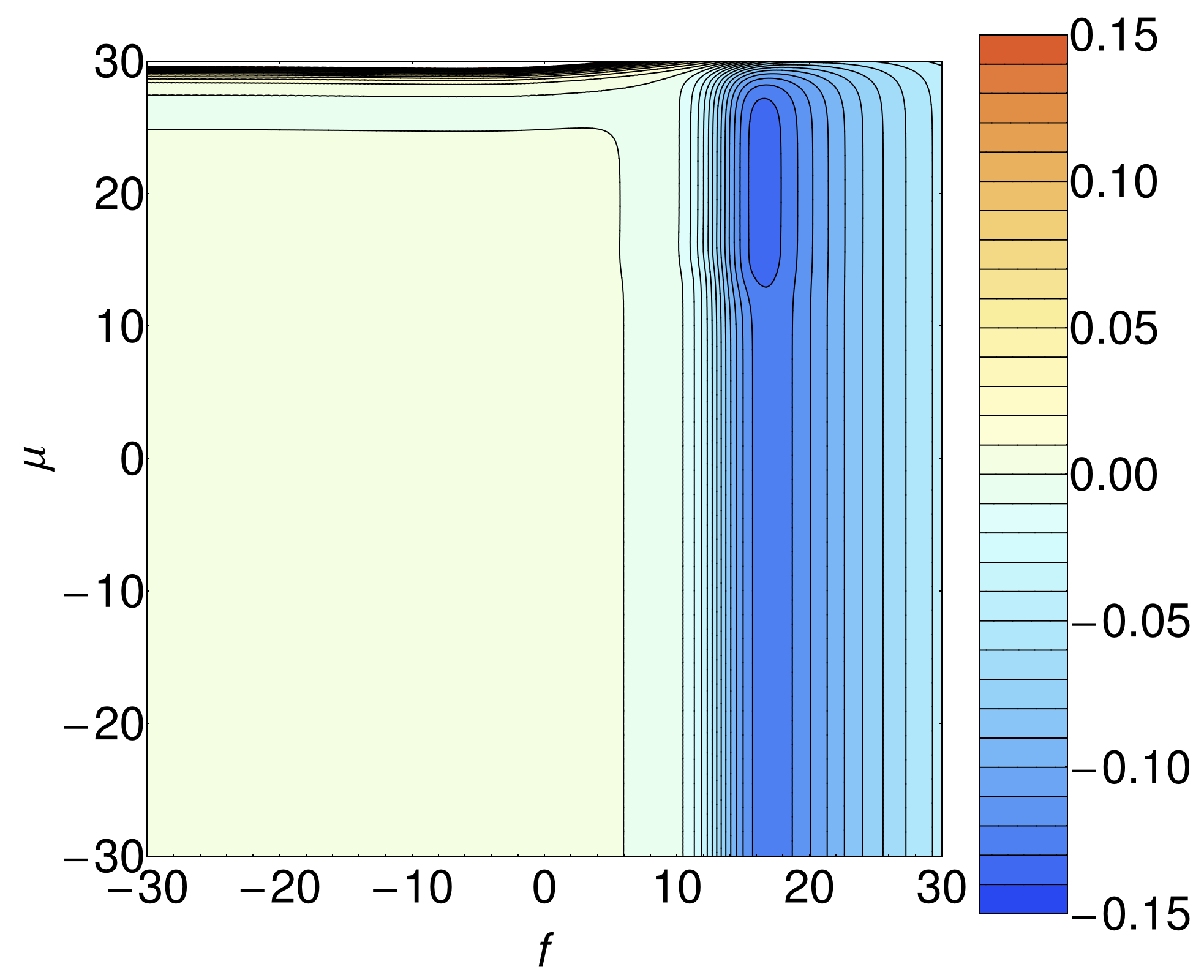}%
  \includegraphics[width=.35\textwidth]{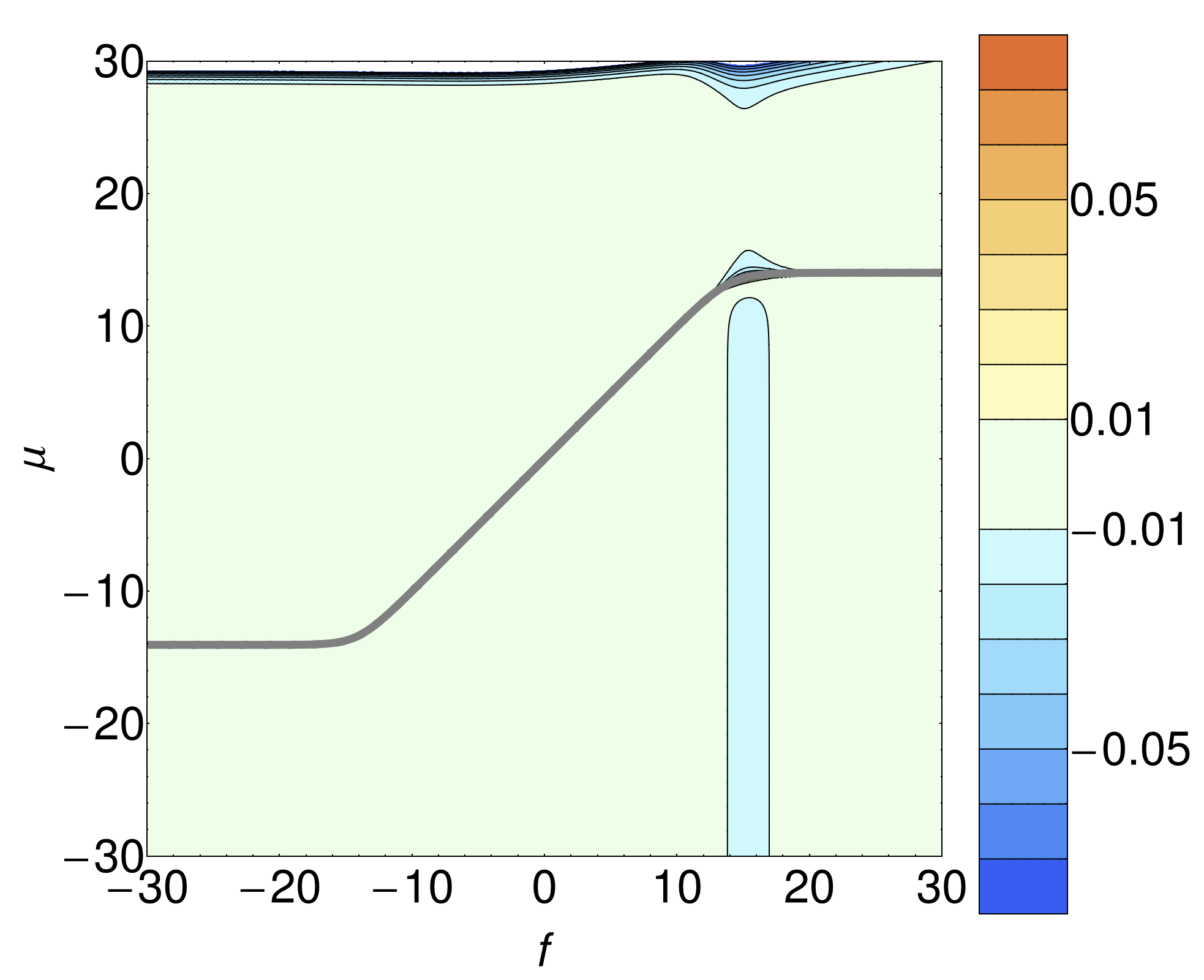}
  \caption{Relative errors $\delta X:= \frac{X_4-X_6}{X_6}$ for the quantities $X_4$ calculated in the minimal four-state model with respect to their values $X_6$ in the six-state model.
  From top to bottom we show the errors for the dissipation $\mot$, the displacement $d$ and the hydrolysis $h$, respectively.
  The columns (left to right) correspond to the first cumulant $c_1$, the second cumulant $c_2$ and the SNR $c_1/c_2$.
  The relative error for all quantities is bounded below $20\%$ everywhere.
  In large parts of the phase diagram, it is less or of the order of a couple of percent.
}
  \label{fig:min4state-relerr}
\end{figure}

Figure \ref{fig:min4state-relerr} shows the relative errors $\delta X$ for those quantities.
In the entire region depicted in the phase diagram, the relative error is bounded below approximately 15\%.
This alone is a remarkable result, given the fact that the values range over more than twenty logarithmic decades.

Even more spectacular is the fact that in a large part of the phase-diagram the mismatch is even less than one percent.
Especially for the SNRs this low error is achieved almost everywhere.
An exception to this very low error can be observed in the vicinity of the \emph{stalling force}, which is defined by the force necessary to stop the motor from moving.
For physiological chemical conditions, \ie a (non-dimensionalized) chemical potential difference $\mu$ between 20 and 30, we observe mismatches of around 10 to 20\%.
This is well within the expected uncertainty in the experimental data both models are built on, \cf Ref.~\cite{Carter+Cross2005}.
In conclusion, we see that a four-state model is as effective as a more involved six-state model --- if the physics are the same.

\subsubsection{Comparison to the coarse-grained model}
We omit showing similar plots for the five- and four-state models obtained using our approach to model reduction presented in Section~\ref{sec:coarse-graining}.
Qualitatively the results are the same, as we would expect from the good agreement for a typical set of physiological parameters shown in Fig.~\ref{fig:kinesin-fluctuations}.
Further, we have no physical arguments that would prefer the reduction of one particular bridge state over any other.

Moreover, note that \emph{by construction} our coarse-graining procedure preserves the currents on all chords and hence in the entire network.
Then, the Schnakenberg decomposition \eqref{eq:schnakenberg} ensures that the first cumulant of \emph{any} physical observables agrees between the models.
Or differently put:
The relative error $\delta{c_1}$ identically vanishes for any physical observable.

\section{Discussion}
We close this chapter with a discussion of our results.
At first, we give some additional remarks on the role of the SNR obtained as the ratio $c_1/c_2$ of the first two (scaled) cumulants of a physical observable.
In particular, we show how it reveals additional structure hidden in the phase diagram of non-equilibrium systems modelled by the means of ST.
After that,  we formulate ``take-home messages'' regarding the formulation of Markovian models for small (biological) systems.

\subsection{The significance of the SNR}
\label{sec:discussion-kinesin}

In the following we consider the SNR, \ie the ratio $\frac{c_1(g)}{c_2(g)}$ of the first and second cumulant of a physical observable $g\in\currents$.
First, we motivate why the SNR reveals more of the hidden structure contained in phase diagrams of non-equilibrium situations.
After that, we briefly discuss its role in the theory of non-linear response.
Finally, we discuss the special role of the motance $\mot$ for arbitrary dynamically reversible Markov processes and that of the displacement for molecular motors performing a linear motion.

\subsubsection{Revealing hidden structure}

We have seen that plotting the SNR shows features which are not directly visible in the plots of the cumulants.
We observed that the order of magnitude of the first and second cumulants are roughly the same throughout the phase diagram.
This fact can be easily understood from the scaling of the SCGF with respect to the transition matrix.

To that end, note that the thermodynamic balance conditions~\eqref{eq:thermodynamic-balance} ensure that  $\exp(\mota{\omega}{\omega'})$, \ie the ratio of forward and backward transition rates scales with the driving forces.
Disregarding correct mathematical notation, one can say $\tmat \propto \exp{\abs{\mot}}$ where $\abs{\mot}$ reflects (the magnitude) of external driving.
It is an easy mathematical exercise to show that the SCGF and hence the entire fluctuation spectrum is homogeneous of first order in $\tmat$.
Hence, a rescaling of $\tmat$ by a common factor $\exp{\abs{\mot}}$ amounts to a rescaling of all the cumulants by the same factor.
Plotting the SNR (and thus hiding the general exponential trend), we are more sensitive to the detailed structure of the phase diagram (like hidden symmetries, \cf the next paragraph).

%
%
%
%
%

\subsubsection{The role of the motance}
As an example for the additional structure found in the SNR, consider Figure~\ref{fig:dissipation-6-state}, where we show our results for the dissipation.
The observable corresponding to the dissipation is the motance $\mot$.
The latter plays a special role: 
It solely depends on the transition rates of the stochastic process and thus is a \emph{neutral} observable:
One only requires that the Markov process under consideration is dynamically reversible, but not necessary that it constitutes a \emph{model} for any physical system.
It is well-defined by the transition rates for any dynamically reversible Markov process used in ST, \cf also Ref.~\cite{Polettini2012}.

In Section~\ref{sec:phase-diagram} we have seen that for moderate values of the driving there is an (approximate) reflection symmetry at the first diagonal of the phase diagram.
We argued that this symmetry originates from the quasi-tight coupling of kinesin's chemical and mechanical transitions.
However, this symmetry is broken for sufficiently high driving parameters.
We explained this fact with a change of kinesin's dominant cycles, \cf also Refs.~\cite{Altaner_etal2012} and~\cite{Hill1979}.

The SNR for the dissipation exhibits this symmetry also for much higher values of the driving.
Moreover, we find an additional second mirror symmetry in the phase diagram.
We have the following explanation for this fact:
The symmetry along the first diagonal still expresses the tight-coupling of ATP hydrolysis with forward stepping.
The existence of the additional symmetry arises from the existence of a second fundamental cycle.
More precisely, it indicates the existence of two loosely coupled fundamental cycles. 
In the case of completely uncoupled cycles, the symmetry $\mot_\alpha \mapsto - \mot_\alpha$, $\curr_\alpha \mapsto -\curr_\alpha$ is obeyed for each individual cycle $\alpha$.
As kinesin's cycles \emph{are} coupled, the symmetry only holds approximately.

\subsubsection{Displacement, drift and diffusion}

Many molecular motors perform a one-dimensional motion along a linear track.
Kinesin is just one example, but there are many others like dynein or RNA-polymerase \cite{Chemla_etal2008}.
In Markovian models, one can always consistently identify the step length (which may be zero) associated to any transition.

As for kinesin, one can then write the displacement $d\in\currents$ as an anti-symmetric observable.
Usually one is interested and the drift velocity $V$ and the diffusion constant $D$ of the motor.
The drift velocity $V=c_1{d}$ is just the first cumulant of the displacement.
The diffusion constant is, up to a factor of two, the second cumulant, \cf also Ref.~\cite{Derrida1983}.
This fact is directly visible from the scaling of the \emph{mean square displacement} $\kappa_2$ along a trajectory $\traj \omega\rlind{\tau}$ of run-length $\tau$, which amounts to 
\[l^2 c_2(\eta_5)= c_2(d) = \lim_{\tau\to\infty} \frac{1}{\tau} \kappa_2\left( \tau\, \overline{d}\rlind{\tau} \right) = \lim_{\tau\to\infty} \frac{l^2}{\tau} \kappa_2\left( \sum_{e\in\traj\omega\rlind{\tau}} \eta_5(e) \right)= 2D,\]
where $e \in \traj\omega\rlind{\tau}$ sums over the directed edges passed by $\traj \omega\rlind{\tau}$.
There is plenty of literature on the calculation of velocity $V$ and diffusion constant $D$ in Markovian models~\cite{Chemla_etal2008,Boon+Hoyle2012}.
Our approach generalizes the treatments presented in these references.
In our method, one does not need to bother with the combinatorial complexity due to the topology of the graph:
It is hidden in the coefficients $a_n$ of the characteristic polynomial of the tilted matrix~$\tmat_d(q)$, \cf Equation~\eqref{eq:skewed-generator}.

Moreover, for the kinesin model, the displacement $d$ constitutes an example where the choice of an appropriate spanning tree can simplify calculations, \cf the discussion in Sec.~\ref{sec:relevance}.
Recall that the choice of spanning tree defines the fundamental cycles.
In the present case, the fundamental cycles $\zeta_2$ and $\zeta_5$ correspond to the dissipative slip and the forward cycle, respectively.
The former involves only chemical transitions and hence $d$ vanishes on $\zeta_2$.
Consequently, $\eta_2$ does not appear in its chord representation $d_\chords= d \equiv l \eta_5$.
Note that this is not the case for the spanning trees depicted in Figure~\ref{fig:trees}b/c.
In that case, the fundamental cycles are the forward and backward cycle, which both involve a mechanical transition.

Another interesting quantity is given by the inverse of the SNR for the displacement~$d$:
It yields the typical length scale above which drift dominates diffusion.
In the more general context, the inverse SNR is also known in the literature as the \emph{Fano factor}.
It was introduced originally for particle detection in high-energy physics \cite{Fano1947}.

Recently, the notion of a Fano factor has also been used in the context of stochastic transport and chemical systems \cite{Roche_etal2005,Fange_etal2010,Qian+Kou2014}.
As the inverse of the SNR, it diverges when the signal (expressed by the first cumulant) in the denominator passes through zero.
In contrast, the SNR seems to show no singularities indicating that the denominator, \ie the second cumulant is always positive.

Interpreted as a transport coefficient like the diffusion constant, the positivity of the second (self-)cumulant $c_2(\obs)$ of a physical observable $\obs$ is clear:
A negative diffusion constant (indicating a negative mean \emph{square} displacement) is not possible.
However, there is a caveat to this statement regarding the origin of the phase diagram, where all driving forces vanish:
At equilibrium, the symmetry imposed by detailed balance ensures that \emph{all} scaled cumulants vanish identically, due to the fact that the SCGF $\lambda_\phi(q)\equiv 0$ is constantly zero.
However, this non-generic feature is a mathematical peculiarity that has its origin in the formulated of the theory.
Outside of (and arbitrarily close to) equilibrium, transition-rate independent physical observables $\phi$ have a positive second moment and the SNR is well-defined, \cf also Figures~\ref{fig:dissipation-6-state} and~\ref{fig:velo-hydro-6-state}.

\subsubsection{Linear and non-linear response in ST}
The lack of divergence of the SNR even at thermodynamic equilibrium (where both cumulants vanish) is evident from the interpretation of the second cumulant as a \emph{transport coefficient} or generalized susceptibility.
Let us briefly comment on this interpretation.
A more detailed discussion can be found in the (yet unpublished) manuscript~\cite{Wachtel_etal2014-2}.

For any observable, the second cumulant has a direct interpretation in terms of response theory.
In particular, the SNR allows us to characterize fluctuation-dissipation relations in situations far from equilibrium.
In Section~\ref{sec:langevin} we have discussed the Einstein relation, which connects the strength of a transport coefficient (in that case, the mobility or inverse drag $\zeta$) to the strength of fluctuations (in that case given by the diffusion constant).
The general relation is formulated in Equation~\ref{eq:white-noise-general}, which states that the noise correlations amount to the elements of the mobility matrix multiplied by temperature and a numerical factor of 2.
It is a consequence of the theory of linear response.

For abstract ST, the mobility is given as the derivative of the first cumulant $c_1$ with respect to a driving force.
Recall that all physical currents can be obtained as linear combinations of the currents associated to the family $\ifam{\eta_\alpha}_{\alpha \in (1,2,\ldots \abs{\chords})}$ of fundamental chords.
The associated abstract driving force is the motance $\mot_\alpha$ of the fundamental cycle $\zeta_\alpha$.
Consequently, we define the mobility matrix:
\begin{align}
  \mobmat_{\alpha,\beta} := \frac{\partial c_1(\eta_\alpha)}{\partial \mot_\beta}
  \label{eq:mobility-general}
\end{align}
Note, that $\mu_{\alpha,\beta}$ amounts to an entry of a generalized Onsager matrix $\onsmat$, \cf Eq.~\eqref{eq:onsager}.

The fluctuation-dissipation theorem \cite{Callen+Scott1998} ensures that close to equilibrium we have linear response, \ie
\begin{align}
  \mobmat_{\alpha,\beta} = 2 c_2(\eta_\alpha,\eta_\beta).
  \label{eq:linear-response}
\end{align}

Moreover, we can apply the fluctuation relations discussed in Sec.~\ref{sec:stochastic-fr} to determine the region in phase space where we expect this relation to hold:
In the limit of vanishing driving, the distribution of the dissipation is a centred Gaussian one.
The fluctuation relations \cite{Lebowitz+Spohn1999,Seifert2005} for the steady-state dissipation then ascertains that the scaled variance of the limiting distribution must approach two times the mean.
Hence, a value close to $\frac{1}{2}$ in the SNR of the dissipation $\mot$ amounts to the region of linear response.
Results in the same spirit have been obtained for deterministic dynamics \cite{Gallavotti1998,Ruelle1999}.
For the current set-up, linear response has been discussed in a paper by Lebowitz and Spohn on the fluctuation relation for stochastic dynamics \cite{Lebowitz+Spohn1999}.

\subsection{Consequences for models}
\label{sec:modeling}
Equations~\eqref{eq:thermodynamic-balance} relate the affinities of cycles to (linear combinations) of the thermodynamic forces.
The relate the microscopic rates to the macroscopic conditions found in the medium.
Hence, they must hold in any model describing the same physical situation, independent of the mesoscopic resolution.

However, thermodynamic balance is actually a thermo\emph{static} statement about (local) equilibrium distributions.
As such, they are insufficient to specify any kinetic, \ie dynamic properties of the model.
The observable dynamic quantities are the average currents associated with physical observables.
The Schnakenberg decomposition~\eqref{eq:schnakenberg} ensures that these currents are completely determined by the average probability currents flowing on the fundamental chords.

Hence, for thermodynamic consistency on the level of the (average) thermodynamic currents one must always take the probability currents into account.
If one fails to do so while inferring a simpler from a more complex model, one gets inconsistent results.
The next paragraph discusses an example.

\subsubsection{Na\"ive coarse-graining of bridge states}
\begin{figure}[ht]
  \begin{center}
    \includegraphics[width=\textwidth]{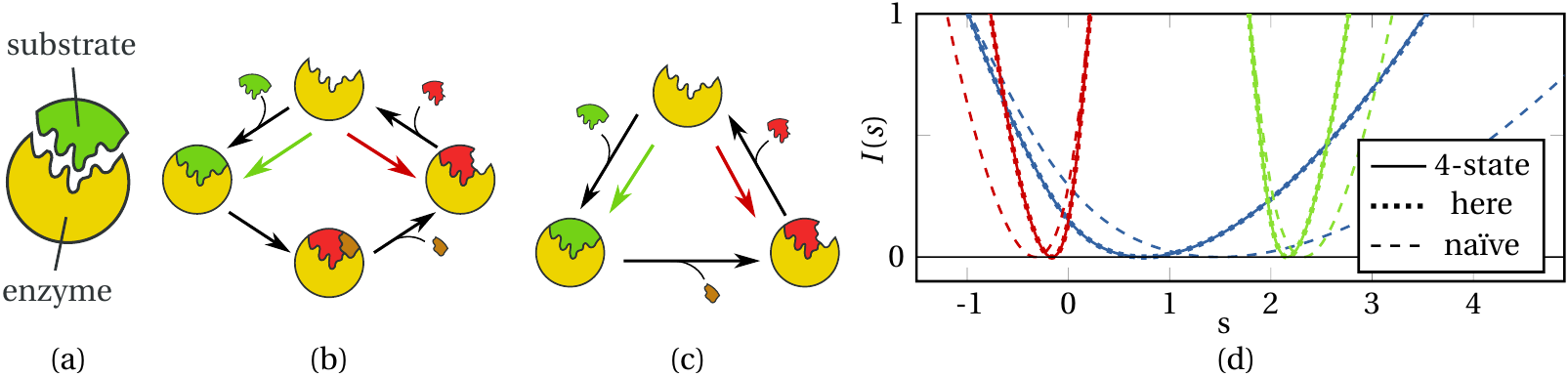}
  \end{center}
\caption{a) Schematic representation of an enzyme that binds a substrate.
  b) A simple one-cycle catalysis model. Upon binding the substrate, it is catalytically split into two subunits.
  The release of the small subunit happens almost instantly, whereas the large subunit stays attached for a longer time.
  c) A reduced three-state model. 
  Catalytic splitting and release of the small subunit yield a combined transition.
  d) Rate functions for the dissipation (blue, center), the association of the large subunit (red, left) and the association of the substrate rate (green, shifted to the right by $s_{0}=2$). 
  As a parametrization we choose all transition rates as unity, with exception of the fast transition $\tprob{\Theta}{D}=100$ for the release of the small subunit.
  The method presented in Sec.~\ref{sec:coarse-graining} preserves fluctuation to a high degree.
  The na\"ive choice \eq{naive-cg}, however, results in severe changes even in the first cumulant.
}
  \label{fig:enzyme}
\end{figure}

The catalytic cycle of ATP hydrolysis on kinesin's active sites is an example of a general enzymatic activity.
Figure~\ref{fig:enzyme} shows the catalytic splicing of a \emph{substrate} molecule into two subunits at the active site of an enzyme.
We assume that one subunits are always released in the same order, for instance due to steric effects.
Often the substrate splits into a small and a large subunit, similar to the case for the hydrolysis of ATP in P and ADP.
The small subunit is released immediately after splicing, whereas the larger part stays bound to the enzyme for a longer time.

In Ref.~\cite{Lipowsky+Liepelt2008}, the authors consider the catalytic cycle at kinesin's active sites.
Denote by $D$, $T$ and $E$ the states where ADP, ATP and nothing, respectively are bound to the active site.
In addition, we use the symbol $\Theta$ to denote the state where ATP is already split but both ADP and P are still attached to kinesin's active site.

Kinesin provides an example for a very fast of the small subunit P.
Consequently, it is natural to remove the (bridge) state $\Theta$ in a coarse-graining procedure.
For the rates of the new transition connecting $T$ and $D$ the choice 
\begin{align}
  \frac{\tprob{T'}{D'}}{\tprob{D'}{T'}} = \frac{\tprob{T}{\Theta}\tprob{\Theta}{D}}{\tprob{\Theta}{T}\tprob{D}{\Theta}}
  \label{eq:wrong-choice}
\end{align}
preserves the affinity of the single cycle, if all rates are left constant.

For three linearly connected states Hill proposed to choose 
\begin{subequations}
  \begin{align}
    \tprob{T'}{D'} &= \frac{\tprob{T}{\Theta}\tprob{\Theta}{D}}{\eav{\tau_\Theta}},\\
    \tprob{D'}{T'} &= \frac{\tprob{\Theta}{T}\tprob{D}{\Theta}}{\eav{\tau_\Theta}},
  \end{align}
  \label{eq:naive-cg}
\end{subequations}
if the staying time $\eav{\tau_\Theta}$ is small \cite{Hill1977}.
However, a linear chain always fulfils detailed balance, and hence does not carry any currents.

Out of equilibrium one has to be careful with using Eq.~\eqref{eq:naive-cg} as a prescription for the new rates.
Generically, adapting these rates for the new transition while leaving all others constant massively changes the currents running through the network.
Consequently, we expect the entire fluctuation spectrum to change rather drastically, which is in fact the case.

In Figure~\ref{fig:enzyme}d the solid lines show the rate function obtained for different physical observables in a simple parametrization of the enzymatic reaction network depicted in Figure~\ref{fig:enzyme}b.
Dashed and dotted lines correspond to reduced three-state models, \cf Figure~\ref{fig:enzyme}c.
More precisely, the dashed lines show the na\"ive choice prescribed by \eq{naive-cg}.
Similarly, the dotted lines amount to a reduced model obtained with our fluctuation-sensitive coarse-graining procedure.

The change in the currents is generic for the choice given by \eqref{eq:wrong-choice}, if the other transition rates are \emph{unmodified}.
Recall that in the heuristic derivation in of our coarse-graining algorithm Sec.~\ref{sec:coarse-graining} we demanded locality.
Locality amounts to the requirement that the steady-state probabilities are not changed in the part of the network which is unaffected by the coarse-graining.
However, locality cannot be achieved, unless some other transition rates are changed.

Though our fluctuation-sensitive uses further constraints than only locality, the moral from the data presented Fig.~\ref{fig:enzyme} is the following:
The thermodynamic balance conditions \ref{eq:thermodynamic-balance} for the (fundamental) cycles of the network ensure thermodynamic consistency, which is a \emph{static} requirement. 
A model's dynamic properties are characterized by the steady-state currents of physical (and hence in principle measurable) currents.
The generalized Schnakenberg decomposition~\eqref{eq:schnakenberg} for their expectation values relates them to the steady-state probability currents $\curr$.
The preservation of the currents of \emph{all} physical observables is given if probability current $\curr$ agrees on a fundamental set of chords.

Consequently, the static and dynamic requirements specify $2\abs{\chords}$ independent constraints, where $\abs{\chords}$ is the cyclomatic number.
Any two models for the same physical situations, should (at least approximately) obey these constraints.
A direct consequence of these constraints is the preservation of the dissipation in the steady state.
We will discuss the significance of that statement in more detail in the final Section~\ref{sec:personal}.

\subsubsection{Some modest advice for model construction}
A main result of the present chapter is that a four-state model is sufficient for the modelling of kinesin.
More precisely, we need a four-state model that shares the same cycle topology than the original model.
We have also seen that the cycle topology determines how different currents are coupled to each other.
A non-tightly coupled model thus always needs at least two (independent) cycles.

The minimal four-state model constructed in Appendix~\ref{app:construction} qualitatively captures all the features of the more complicated model.
In the parametrization of the model we followed the same physical arguments than Liepelt and Lipowsky adapted for constructing their model in Ref.~\cite{Liepelt+Lipowsky2007}.
The quantitative agreement between the models is remarkable:
Relative errors of a few percent are typically much smaller than the experimental uncertainty in the measurement of the (chemical) transition rates.

Hence, we propose the following paradigm for constructing good models:
At first, determine how experimentally accessible currents couple to the external driving forces.
To that end, try to measure or estimate the coupling matrix $L_{\alpha\beta}$ between different measurable currents and external forces.

After that, design a network of (observable) states where changes between states can be attributed to observable currents.
This step is the hardest: it requires a certain degree of physical, chemical or biological knowledge or intuition.
The coupling matrix $L_{\alpha\beta}$ will be useful in that regard.

After the construction, reduce the network to the minimal topology be removing all bridges and leaves.
The topology of the network then tells you about the structure in the \emph{abstract} matrix $\mu_{\alpha,\beta}$ introduced in Equation~\eqref{eq:mobility-general}.
The physical forces and observable currents which are assigned to transitions between states connect the measured matrix $L_{\alpha}{\beta}$ to $\mu_{\alpha,\beta}$.
This can be used as another consistency check.

Finally, try to measure, predict or by means of thermodynamic balance \eqref{eq:thermodynamic-balance} infer the transition rates between states.
Symmetry considerations may help in that approach.
When applying the thermodynamic balance, make sure that you have a \emph{physical reason} why to use the constraint to infer a particular transition and not another.\footnote{
We stress this here because of the following reason:
In both Ref.~\cite{Liepelt+Lipowsky2007} and in App.~\ref{app:construction} one \emph{chooses to infer} the rate for the transition modelling ATP release from the trailing active site by thermodynamic balance.
If any other rate on the backward cycle is inferred in that way rather than determined by symmetry, the physics described by the model are different.
The application of our method than shows that in that case other cycles become dominant for high values of the driving, and the phase diagram changes drastically.
}

A (Markovian) model constructed in this way minimizes the amount of additional assumptions.
It is thus maximally non-committal with respect to our missing knowledge.
Moreover, one can use this paradigm in order to systematically \emph{test} assumptions about small (biological) systems in fluctuating environments.
Though not more than a sketch, we hope that the ideas outlined here may guide scientists to design more and more accurate models.
In a sense we are back with Schr\"odinger's quote:
We hope that our abstract physical-mathematical considerations might contribute to improving our understanding of life --- if even just a tiny bit.

\section{Summary}

Fluctuations, \ie a behaviour away from the expected average, are non-negligible in small systems in thermodynamic environments due to the influence of thermal noise.
For biological systems, which have evolved over millions of years, such fluctuations are often functional.
The molecular machinery of life, which includes so-called molecular motors, ensures the function of living cells.
ST can be understood as a paradigm for designing models of such systems as well as for their thermodynamic interpretation.

Static properties of a model are prescribed by the thermodynamic properties in the medium.
The requirement of thermodynamic balance manifests in the Hill--Schnakenberg conditions, \cf Eq.~\eqref{eq:thermodynamic-balance}.
The dynamic properties of physical currents on an ensemble level are completely determined by the probability currents running in the network.
The latter in turn are determined by the currents on a set of fundamental chords, reflected in the Schnakenberg decomposition~\eqref{eq:schnakenberg}.
Together these requirements yield a set of consistency criteria regarding the asymptotic ensemble properties of different models.

The (approximate) preservation of fluctuations between two models requires additional conditions.
A heuristic motivation of the latter yields the fluctuation-sensitive coarse-graining procedure presented in Section~\ref{sec:coarse-graining}.
Additional constraints can also be obtained as the result of experimental data.
Minimal models are constructed using thermodynamic consistency and all other available information, while ensuring the simplest required cycle topology of a network.

A main emphasis of this chapter was on the application of our ideas to molecular motors.
Using the tools from the previous Chapter~\ref{chap:cycles}, we compared an established model of the motor protein kinesin with simplified models.
Our results are remarkable:
Our reduced models captured the fluctuations of the original model to a very high degree.
Mismatches in relevant quantities are bounded below 15\% for the all the values of the external driving we considered.
This is even more spectacular given the fact that the \emph{values} of observable currents change over about twenty logarithmic decades.

In the discussion, we were concerned with the physical relevance of the SNR of an observable, \ie ratio of its first and second cumulant.
We mentioned its role in thermodynamic response theory and showed how the SNR of the motance can be used to infer the linear response regime.
We further discussed how our results simplified previous approaches to the calculation of drift and diffusion in models of molecular motors which perform a one-dimensional motion.
Finally, we formulated a set of ideas to be applied in the future design of models.

At this point, we conclude the main part of the present work.
We give a summary and a final discussion in the following final Chapter~\ref{chap:discussion}.
In addition, we also give an outlook on further research and take a final bird's eye perspective on the topics discussed in the present thesis.

  \chapter{Conclusion and outlook}
  \label{chap:discussion}
  \begin{fquote}[C.~H.~Bennett][The thermodynamics of computation][1981]
Computers may be thought of as engines for transforming free energy into waste heat and mathematical work.  
\end{fquote}

The present chapter intends to bridge the gap to our introductory remarks in Chapter~\ref{chap:intro}.
In Section~\ref{sec:summary}  we start by telling the reader what we have told him in the previous Chapters~\ref{chap:entropy}--\ref{chap:fluctuations}.
This summary can be understood as a brief synopsis of the discussions and summaries that were provided at the of each of these chapters.

After that, Section~\ref{sec:outlook} provides a review as to how our results are placed amongst the current literature --- at least from the author's knowledge of the latter.
As an outlook, we discuss promising directions for follow-up work.

Finally, we conclude this thesis with an author's perspective in Section~\ref{sec:personal}.
Instead of focussing on the rigorous results obtained so far, this personal perspective provides the ``bigger picture'' as perceived by the author, who views complex systems as information processing devices --- as already hinted at in the epigraph of the present chapter.

\section{Summary}
\label{sec:summary}
Chapters~\ref{chap:intro} and~\ref{chap:entropy} provided the necessary background for this thesis.
They reviewed the state of the art of modern statistical physics in general and ST in particular.
Even though many fascinating aspects of this active field of research could not be covered, the cited references should provide a good starting point for further reading.

The main part of the thesis split into two major parts.
Chapters~\ref{chap:marksymdyn} and~\ref{chap:information-st} discussed deterministic microscopic foundations of ST.
In Chapters~\ref{chap:cycles} and~\ref{chap:fluctuations} we focused on the structure of Markovian ST on finite state spaces.

Before discussing anything new, let us summarize what we have discussed so far.

\subsection{Microscopic foundations of stochastic thermodynamics}
\subsubsection{The foundations of Markovian stochastic thermodynamics}
A crucial assumption for statistical physics in general and stochastic thermodynamics in particular is the \emph{Markovian postulate} \cite{Penrose1970}.
It states that observable time-series corresponding to measurements on thermodynamic systems obey Markovian statistics.
It is equivalent to the claim that observable states do not contain any memory of their past.

From the point of a deterministic evolution for observationally inaccessible microscopic states, this is a highly non-trivial statement.
In thermodynamics, the Markovian postulate is closely linked to the hypothesis of local equilibrium (LE).
The latter assumes that on the time-scale of observations, the distribution of the microscopic degrees of freedom has relaxed to a constrained equilibrium ensemble --- at least for all practical purposes.

In Chapter~\ref{chap:marksymdyn} we investigated the microscopic implications of the  Markovian postulate.
We considered deterministic dynamics $\Phi$ in discrete time which map phase space $\pspace$ bijectively onto itself.
An observable $\partmap\colon\pspace \to \ospace$ maps microscopic states $x\in\pspace$ to a finite number of observable states $\omega \in \ospace$.
The observable thereby partitions phase space into equivalence classes $\cell_\omega$ which are indexed by the values of the measurement result.

An observable time-series $\traj \omega\rlind{\tau}=\ifam{\omega_0,\omega_1,\cdots,\omega_\tau}$ of length $\tau$ summarizes subsequent observations.
Different microscopic initial conditions generically yield different time-series.
The main result of the chapter states the requirements on the observable $\partmap$ and a probability distribution $\pden$ for the microscopic states, such that the time-series $\traj \omega\rlind{\tau}$ obey Markovian statistics.

We found that for a given dynamics $\Phi$ and an appropriate observable $\partmap$, there are many distributions $\pden$ that fulfil this requirement.
Generically, these distributions are non-stationary and yield Markovian statistics only after some distinct point in time $t_0$, which corresponds to the time of the preparation of the system.
This statement provides an example for an operational interpretation of our abstract results.
In addition, we discussed our results from the perspective of modern ergodic theory.

\subsubsection{Stochastic thermodynamics as a theory of statistical inference}
In 1957, Jaynes proposed the view of statistical mechanics as a theory of statistical inference, formulated in the language of information theory \cite{Jaynes1957}.
Indeed, his ideas provide a self-consistent foundational framework, which formalizes Gibbs' approach to statistical ensembles \cite{Gibbs1948}.
However, he stresses that the probability densities $\pden$ which characterize phase space ensembles must not be interpreted in a frequentist way.
Rather, they should be understood as  maximally non-committal statements about our expectations of the probability of microscopic states --- given our previous knowledge about the microscopic physics that govern their dynamics.

In Chapter~\ref{chap:information-st} we propose a tentative information-theoretic framework for ST.
The formal background is provided by the mathematical treatment of observed time-series in Chapter~\ref{chap:marksymdyn}.
We use Jaynes' notion of maximum-entropy priors to infer the distribution of microstates $x\in\cell_\omega$.

From these priors we construct the coarse-grained ensemble $\cgden\tind{t}$ to expresses our expectation about the distributions of microstates $x\in\pspace$.
Subsequent measurements provide new information about the state of the system.
Consequently, we obtain an update rule that specifies the temporal evolution of $\cgden\tind{t}$.
The coarse-grained ensemble is inferred without a detailed knowledge of the microscopic dynamics~$\Phi$.
However, a given dynamics~$\Phi$ specifies a deterministic evolution rule for the initial ensemble, which is expressed by the fine-grained ensemble $\fgden\tind{t}$.

In our information-theoretic perspective, the uncertainty expressed in the coarse- and fine-grained ensemble is quantified by their differential entropy.
Their relative entropy, \ie the Kullback--Leibler divergence of $\fgden$ from $\cgden$ quantifies the mismatch between the microscopic and an inferred description based on coarser models or measurements.
We found that these entropies can be obtained as time-series averages of entropic $\tau$-chains, \ie random variables that depend on finite time-series $\traj\omega\rlind{\tau}$.
Motivated by (non-equilibrium) molecular dynamics simulation, we related phase-space contraction and dissipation with the entropy change in the hidden, unobservable degrees of freedom.
A corollary is the identification of the relative entropy with the total entropy.

As an example for the application of the mathematical framework we introduced network multibaker maps (NMBM) as a versatile model dynamics.
Reversible NMBM share the mathematical properties of the equations of motion used in molecular dynamics.
We explicitly showed how ST emerges in the context of NMBM.
After a discussion of the results of Chapter~\ref{chap:marksymdyn} in the light of NMBM, we postulate that our results apply more generally.
In particular, we conjecture that under the right assumptions, ST naturally emerges from physical microscopic dynamics, \ie dynamics that are time-reversible with a measure-preserving time-reversal involution.

\subsection{Structure and models of stochastic thermodynamics}

\subsubsection{Kirchhoff's laws, cycles and fluctuations}
In Chapter~\ref{chap:cycles} we dealt with the structure of Markovian jump processes on finite state spaces.
The topology of the network of states is determined by the transition probabilities and can be visualized as a graph.
In the steady state, probability currents on that graph resemble electrical currents in an electrical circuit.
For the latter, Kirchhoff's first and second law state that (i) the current balances at each vertex and (ii) that the integrated difference of a reference voltage vanishes along a loop.
We find that Kirchhoff's laws equally apply for Markov processes and provide a complete electrical analogy.

The reason for the applicability of Kirchhoff's results is the algebraic structure of the network.
In an abstract sense, electrical currents and voltage drops are anti-symmetric observables defined on the edges of a graph.
In the context of ST, ensemble averages of anti-symmetric observables correspond to physical currents.

For such physical observables algebraic graph theory provides powerful tools for abstract structural investigations.
For instance, the abstract space of all physical observables $\currents$ decomposes into two orthogonal components $\cycles$ and $\cocycles$.
The former contains so-called \emph{cycles}, whereas the latter contains the \emph{co-cycles}.
Cycles are the analogue of divergence-free currents in field theory, hence they correspond to the currents of conserved quantities.
Co-cycles are obtained as the discrete gradients of potentials $\potentials$ defined on the vertices of the graph.
An abstract formulation of Kirchhoff's first law holds for all abstract cycles $z \in \cycles$, whereas Kirchhoff's second law generalizes to observables $y \in \cocycles$.

We are particularly interested in the fluctuations of physical currents, \ie the stochastic deviations from their averages.
In mathematical terms, we consider the probability distribution associated to the finite-time averages $\bar{\obs}\rlind{\tau}$ of physical observables $\obs\in\currents$.
A time-average is obtained from integrating $\obs$ along the transitions defined by a stochastic time-series $\traj\omega\rlind{\tau}$.
For ergodic Markov processes, time-averages converge to a distinct value $\eav{\obs}\tind{\infty}$ in the asymptotic limit $\tau \to \infty$.
This value agrees with the steady-state ensemble average of the corresponding physical current.

Consequently, the asymptotic distribution is a $\delta$-peak.
For any finite time $\tau$, the probability distribution of $\bar{\obs}\rlind{\tau}$ has a finite width.
For Markov processes, its convergence to the $\delta$-distribution is governed by a large deviation principle, which amounts to a scaling form characterized by a \emph{rate function} $I_\obs(x)$.

The rate function is completely determined by a set of \emph{scaled cumulants} $c_n(\obs)$, which we call the \emph{fluctuation spectrum} of $\obs$.
Two of the main results of Chapter~\ref{chap:cycles} regard the latter:
Firstly, we showed that the fluctuation spectrum of $\obs=z_\obs + y_\obs$ only depends on its component $z_\obs \in \cycles$ in cycle space.
Secondly, we demonstrated that the spectrum of any physical observable $\obs$ is completely determined by the cumulants of the probability currents.

The importance of cycles in the network of states of Markovian processes yields several cycle decompositions.
Probably the most well-known is the Schnakenberg decomposition for the average steady state probability currents.
We generalized this decomposition to the entire fluctuation spectrum of arbitrary physical observables.
Moreover, we introduced the chord representation $\obs_\chords$ of $\obs\in\currents$ and explain its benefits for the purpose of calculations.

\subsubsection{Models of molecular motors}
In Chapter~\ref{chap:cycles} we developed the necessary tools for the quantification of \emph{fluctuations} in models of small systems under non-equilibrium conditions.
Chapter~\ref{chap:fluctuations} applied these methods to models of actual (bio-)physical systems.
Currently, the study of molecular motors is attracting a lot of interdisciplinary attention.
In biology, fluctuations are not only relevant but often even important for the \emph{function} of living systems.
In quantifying fluctuations, statistical physics provides biologists and modellers with new tools for the study of life.

More precisely, our methods help us to separate good from less good models.
The minimal requirement on any model used in ST is that the calculated average currents agree with ensemble measurements.
Consequently, two different models of the same physical system must in this respect also agree \emph{with each other}.
%
The results of the previous chapter emphasized the importance of the topology of cycles in the network of states.
The latter is characterized by the connections of a set of (fundamental) cycles.
The motance or affinity of a cycle amounts to the logarithm of forward to backward transition rates integrated along its edges.
Physically, the motance of any cycle is a linear combination of the \emph{thermodynamic forces} that drive the system out of equilibrium.
This thermodynamic balance requirement is also known as the Hill--Schnakenberg conditions.

The results of Chapter~\ref{chap:fluctuations} can be summarized as follows:
The ensemble behaviour of a model is fully characterized by the topology of its fundamental cycles together with their affinities and associated steady-state currents.
We say that two models are equivalent on the ensemble level, if they share these properties.

Based on this \emph{necessary} requirement, we presented a coarse-graining algorithm for Markov processes.
In addition to the currents $\curr$ and affinities, this approach additionally preserves the fluxes $\flux$.
With these additional constraints, we find that in addition to the \emph{exact} preservation of the averages of any physical observable, also its fluctuations are preserved to a very good degree.

In order to be concrete, we exemplified our abstract results using a well-known model for the molecular motor \emph{kinesin}.
We explore its phase-diagram, \ie its dependence on chemical and mechanical forces that drive the system out of equilibrium.
In that context we illustrate the ability of our coarse-graining procedure to preserve fluctuations.
We further construct a minimal model for kinesin based on the same data and the same physical assumptions used in the construction of the original model \cite{Liepelt+Lipowsky2007}.
Remarkably, the relative error of the simplified versions with respect to the original one is only a few percent throughout the phase diagram --- in most parts it is even much lower.

In comparison with the typical uncertainties in an experimental measurements of kinetic rates, this mismatch is negligible.
In conclusion, we propose a guiding principle for the construction of physical models from available data.

\section{Outlook}
\label{sec:outlook}

\subsection{Possible generalizations}
In the first part of the present thesis we discussed the microscopic deterministic foundations of ST.
For simplicity, we assumed a stroboscopic picture, \ie a dynamics evolving in discrete time.
Much of the ergodic theory for measurable dynamical systems was first formulated using discrete-time maps and later generalized to flows.
For instance, the SRB measure was first formulated for Anosov maps \cite{Sinai1968} and then extended for continuous Axiom-A flows \cite{Sinai1972,Bowen+Ruelle1975}.
We expect that this is also the case for our set-up.
However, given the limited temporal resolution of any real experiment, the discrete-time case might be more natural.

We further demanded discreteness (in fact finiteness) of the space of observations, \ie the space of possible measurement outcomes.
In that case, an observable time-series is generated by a jump process. 
In Section~\ref{sec:stochastic-models} we have briefly considered ST for continuous state spaces using Langevin and Fokker--Planck equations.
Consequently, an extension of the algebraic framework introduced in Chapters~\ref{chap:cycles} and~\ref{chap:fluctuations} to continuous state spaces is desirable, but seems rather involved.

In contrast, we expect the formulation of the results in Chapters~\ref{chap:cycles} and~\ref{chap:fluctuations} for the case of discrete \emph{time} to be straightforward.
The algebraic-topological treatment in Chapter~\ref{chap:cycles} is independent of the temporal evolution.
Also the time-discrete formulation of the large deviation principle resembles the continuous case.
The main difference is that the SCGF is obtained as the largest eigenvalue of the tilted stochastic matrix, rather than as the logarithm of the corresponding tilted transition matrix \cite{Touchette2011}.

\subsection{Network multibaker maps as a versatile tool for ergodic theory}
Let us briefly comment on the role of network multibaker maps (NMBM) as a model dynamics.
In our opinion, NMBM constitute a sufficient representation of the dynamics between elements of Markov partitions for generic uniformly hyperbolic maps.
Consequently, they can provide a good pictorial representation of any system that satisfies the ``Chaotic Hypothesis'' of Gallavotti--Cohen \cite{Gallavotti+Cohen1995}.

This is interesting for several reasons.
Network multibaker maps are both analytically tractable and formulated on a two-dimensional phase space, which can easily be sketched.
Moreover, a two-dimensional phase space is sufficient to exhibit transversal stable and unstable manifolds.
Transversality is an important aspect of the topology of a Markov partition~\cite{Adler1998}.

Further, NMBM are extremely versatile:
Not only can we design them to be (uniformly) conservative or dissipative, but we can also make them time-reversible.
Throughout this thesis we have argued that time-reversal is a hallmark of a \emph{physical} microscopic dynamics.
In fact, the authors of Ref.~\cite{Maes+Verbitskiy2003} formulated the need for a uniformly hyperbolic model dynamics with time-reversal.
NMBM provide this example.

Finally, for any NMBM which is based on a \emph{simple} graph, the cells representing the vertices form a generating Markov partition.
That means that every symbol sequences generated by a Markov jump process has a corresponding phase-space points.
After choosing the parameters $s^i_j$ which define a NMBM, one has full information about its symbolic (equivalent) dynamics.
Hence, one immediately knows whether or not to expect certain features of chaotic dynamics (like homo- and heteroclinic orbits) and where to find them in phase space.
Further, NMBM constitute a constructive example for Theorem~3.2.~of Ref.~\cite{Blank+Bunimovich2003} about the existence of a deterministic representation of stochastic cellular automata.

\subsection{Towards a dynamical picture of local equilibrium}
We argued how NMBM serve as a representation for systems that fall under the chaotic hypothesis.
The chaotic hypothesis implies the existence of a Markov partition $\parti =\ifam{\cell_\omega}_{\omega \in \ospace}$ for physical systems.
However, the partition $\mathcal V = \ifam{\mathcal V_k}$ which is induced by a real physical observable is usually much coarser.
Let us assume that each $\mathcal V_k = \bigcup_{\omega \in \ospace_k} \cell_\omega$ is comprised of a subset of ``microscopic''%
\footnote{Note that in this context ``microscopic'' does \emph{not} refer to a point in the phase space of a deterministic dynamical system, but to an element $\cell_\omega$ of a Markov partition.}
cells $\ifam{\cell_{\omega}}_{\omega \in \ospace_{\omega}}$ indexed by $\omega \in \ospace_k \subset\ospace$.
In order to distinguish them from the microscopic cells $\cell_\omega$, we call $\mathcal V_k$ an observable cell.
Further, we consider an observable time-scale $\tau\tsub{obs} \gg \tau\tsub{mic}$, where we use the stroboscopic time interval $\Delta\tau \equiv \tau\tsub{mic}$ as the microscopic time-scale.

Transitions $k \to k'$ between observable cells $\mathcal V_k$ on observable time-scales are generally non-Markovian.
However, there might be situations where ``internal'' transitions $\omega_k \to \omega'_{k}$ with $\omega_k,\omega'_k \in \ospace_k$ happen much faster than transitions $\omega_k \to \omega'_{k'}$, $k\neq k'$ between different cells. 
If these time-scales are properly separated, one can approximate observable transitions $k \to k'$ on a coarse-grained time scale $\tau\tsub{obs}\gg\tau\tsub{mic}$ as a Markov process.

In Ref.~\cite{Esposito2012}, Esposito investigates this situation for Markov jump process in continuous time.
He derives a renormalized form for the total entropy production $\delta \totent=\delta\totent\tsub{hom}+\delta\totent\tsub{inhom}$, which consists of homogeneous and an inhomogeneous term.
The homogeneous term $\delta\totent\tsub{hom}$ is formally identical with the usual expression \eqref{eq:transient-totent} used in ST.
The probabilities $p_k$ appearing in $\delta\totent\tsub{hom}$ are the probabilities of observable states $k$ rather than the finer microscopic states $\omega$.
Similarly, the microscopic transition rates $\tprob{\omega}{\omega'}$ are replaced by the observable transition rates $V^k_{k'}$.
In the limit of infinite time-scale separation these probabilities become time-independent and homogeneous and thus define the observable Markov process.

However, the inhomogeneous term $\delta\totent\tsub{inhom}$ does not vanish in this limit.
The following additional requirements are needed:
\begin{enumerate}
  \item Internal transitions $\omega_k \to \omega'_{k}$ obey detailed balance.
  \item Transitions between different cells $\omega_k \to \omega'_{k'}$ happen in a certain regular way.
\end{enumerate}
The ``regular way'' can be formulated as a time-reversal symmetry for the transitions between microscopic states conditioned on an observable transition.

Network multibaker maps allow us to translate these notions into the terms of an underlying phase space dynamics.
A sufficient separation of time-scales allows us to formulate an approximate autonomous fast dynamics $\Phi_k$ for the microstates $x \in \mathcal V_k$.
Requirement 1) then amounts to $\Phi_k$ being a conservative map, \ie a map which features an equilibrium distribution as its steady state.

The phase space formulation of requirement 2) is not that obvious.
However, the existence of such an additional requirement is already interesting on its own.
It emphasizes the requirement of an additional symmetry regarding the microscopic realization of transitions between observable states:
To the author's knowledge, this has not been mentioned anywhere in the literature yet.

In conclusion, we sketched how Esposito's coarse-graining procedure together with the NMBM perspective provides a possible dynamical picture of LE.
Moreover, the inhomogeneous term allows \emph{quantitative} statements about how well the LE hypothesis is satisfied in a given physical system.
Working out the details of this dynamical perspective on LE will be an objective for future work by the present author.

\subsection{Deterministic foundations of stochastic fluctuations relations}
\label{sec:master-fr}
In addition to a more detailed picture of LE, our work provides a new perspective on the fluctuation relations, both in the deterministic as well as in the stochastic setting.
In order to appreciate this fact, note that dynamically reversible Markov processes and time-reversal symmetric NMBM are equivalent.
For the sake of brevity of the following argument, consider a NMBM with cells $\cell_i$ of equal size $\vcell_i =1$.
Then, a dynamically reversible rate matrix $\tmat$ containing transition probabilities $\tprob{\omega}{\omega'}$ equivalently defines a reversible network multibaker with relative strip widths $\tprob{i}{j} = s^i_j = \hat s^j_i$.

For dynamically reversible Markov processes, a number of fluctuation relations are known.
The most general one holds for the total entropy.
In the NMBM setting, it translates into a fluctuation theorem for the relative entropy.
The change in relative entropy, \eq{tentative-diss-fun}, can be written as an integral over two contributions:
One that accounts for the negative of the phase space expansion rate and one that describes the ratio of two inferred densities.
Hence, it has the form of a phase space average of the generalized dissipation Evans--Searles dissipation function \cite{Evans+Searles2002}.
Thus, we have established a tentative connection between the latter and the general transient fluctuation relation for stochastic dynamics.

In addition, reversible NMBM fulfil the chaotic hypothesis.
Moreover, they can be made dissipative and as such have a natural (SRB) measure as a \emph{microscopic} steady state.
Trivially, the stochastic processes is also stationary and obeys the fluctuation relation for the steady state where $\Delta\totent = \Delta\medent$.
In that setting, it is a constructive example for a system where the Gallavotti--Cohen fluctuation relation holds \cite{Gallavotti+Cohen1995}.

In general, the existence of a time-reversal involution which factors on the elements of an absolutely $\invo$-invariant partition implies an abstract fluctuation theorem \cite{Wojtkowski2009}.
In fact, this abstract fluctuation theorem holds in the more general case where we have two bijective phase space dynamics $\Psi$ and $\Phi$, which are conjugate to each other.
Conjugacy in this general setting means that there is a time-reversal involution $\invo$ such that $\Psi = \invo \circ \Phi \circ \invo$.
Here, we have always demanded that the conjugate dynamics $\Psi = \Phi^{-1}$ is also the inverse one.

The more general setting opens up a way to study the microscopic foundations of the ``Master fluctuation relation'' proposed by Seifert~\cite{Seifert2012}.
In the Master fluctuation relation, one is not required to consider reversed trajectories generated by the \emph{same} stochastic process.
Rather, the reversed process is any process obtained from the original one by applying an involution-symmetry to the model.
Typically, the action of the time-reversal symmetry on control parameters includes the inversion of electric or magnetic fields.
Further, if $\lambda(t)$ specifies a protocol (\ie a deterministic change of the systems parameters in a non-autonomous model), $\lambda(t)$ may be reversed in a generalized time-reversed, \emph{conjugate stochastic process}.
Consider now any observable $\obs$ with a defined parity, \ie where the action of the time-reversal $\invo(\obs) = \pm \obs$ is an involution.
The master fluctuation relation then holds for all such observables $\obs$ such that $\invo(\obs)$ has the same physical interpretation in the conjugate process.
Examples of such quantities are (besides the total entropy production) are the work or the heat \cite{Seifert2012}.

Wojtkowski's conjugate map $\Psi$ then acts as the deterministic dynamics that gives rise to the conjugate stochastic process.
Hence, we conjecture that the abstract fluctuation theorem provides a microscopic basis for the Master fluctuation relation.
The appropriate two-sided Markov measure in that case needs to be formulated with respect to the natural measure $\mu_{\Psi}=\mu_{\Phi}\circ\invo$ of $\Psi$ rather than with respect to $\mu_{\Phi^{-1}}$.

\subsection{Information, complexity and neuroscience}
Finally, let us draw a connection to neuroscience.
The Lyapunov spectrum quantifies the average phase space expansion in ergodic systems.
In Section~\ref{sec:topent-ksent} we mentioned how Pesin's formula connects Kolmogorov--Sinai entropy to the Lyapunov spectrum and thus to phase space contraction.
We finally showed how this connects to a stochastic description on the level of a Markov partition, \cf also Ref.~\cite{Gaspard+Wang1993}.

These notions have just recently been investigated from the perspective of theoretical neuroscience \cite{Monteforte+Wolf2010,Lajoie_etal2013}.
Mathematically, the cortex can be thought of as a high-dimensional systems capable of performing complex computations \cite{vVreeswijk+Sompolinsky1996}.
A statistical mechanics perspective in the spirit of Jaynes' ideas can be found in Ref.~\cite{Tkacik_etal2013}.

We expect huge developments in this very modern field in the years to follow.
The dynamical systems framework (\eg the usage of SRB measures and KS entropy) applied to neuronal computation promises novel qualitative and quantitative results.
An overview about the models used by theoretical neuroscientists can be found in a recent review \cite{Wolf_etal2014}.
Network multibaker maps may also provide a novel perspective on these information-processing non-equilibrium systems.

\section{A summarizing (personal) perspective}
\label{sec:personal}
Finally, let us come back to Bennett's initial quote on computers as machines that turn available free energy into waste heat and mathematical work.
We slightly generalize its statement in the following way:
In our opinion, an interesting thermodynamic system (like a computer or a molecular motor) turns free energy into some useful work and some waste heat.
Surely, this opinion is as subjective as the meaning of the notion of \emph{useful} work is ambiguous. 
We just vaguely understand it as the work necessary to generate some form of \emph{pattern}.
A pattern is something that is created (and may be sustained) in spite of entropic decay of structure dictated by the second law, \cf also Ref.~\cite{Nicolis+Prigogine1977} by Nicolis and Prigogine.
It can be as complex as a living being or as mundane as the output of a trivial mathematical calculation.

A reverse statement of the initial quote then reads:
A thermodynamic system under sustained non-equilibrium conditions is an engine that converts (free) energy in waste heat and the generation of patterns.
If we understand a pattern as encoding some form of information, we might even say:
An (interesting) non-equilibrium thermodynamic system is an information processing device, which computes patterns and produces waste heat \textit{via} the consumption of (free) energy.


\subsection{On the ubiquity of Markovian statistics}
\label{sec:ubiquity}
In Section~\ref{sec:operational} we have already commented on the ubiquity of Markovian statistics for measured time-series.
We hinted at a certain anthropic principle expressing that scientist observe Markovianity because they look for it.
After all, Markovian stochastic processes are the most well-understood ones and are readily used for the purpose of modelling --- as we have seen throughout this work.

In the present section, we pick up on that discussion.
In his book on the foundations of (classical) statistical mechanics \cite{Penrose1970}, Penrose stresses that the Markovian postulate ensures reproducibility of experimental results.
In order to appreciate this fact, suppose that an initial preparation procedure always leaves the system in some observational state $\omega \in \ospace$.
Note that these states are defined with respect to the measurement apparatus that a scientist A uses to record time series $\traj\omega\rlind{\tau}$.

Now imagine another scientist B, who works on the same system with the same measurement apparatus.
Scientist B wants to reproduce some results reported by scientist A.
Hence, she needs to be able to prepare initial conditions that yield experiments showing the same statistical properties.
However, this is only possible if the initial condition $\omega$ reported by A does not carry knowledge of their preparation protocol.

This is the main argument behind the Markovian postulate:
If the dynamics on the level of the observable states shows Markovian behaviour, the system admits statistically regular and hence reproducible experimental trials.
In a sense, Markovian observable states are maximally non-committal with respect to the preparation procedure.

From that perspective the ``Markovian anthropic principle'' is nothing else than the Scientific Method:
Scientific statements need to be reproducible and thus experimentally falsifiable \cite{Popper2002}.
Consequently, they must be statistically regular, in the sense that the prepared initial state fully specifies the probabilities of future observed time-series.

Local equilibrium, which comes with the time-scale separation between microscopic and observable dynamics, is one way to ensure this statistical regularity.
However, in a real experiment the temporal resolution might just be ``too good'' for the system to relax between two subsequent observations:
Observable time series with entries $\omega \in \ospace$ are not (1-step) Markovian.
In that case, one commonly summarizes $k$ subsequent measurement results $(\omega_t,\omega_+1,\cdots,\omega_{t+k-1})$ into a new observable state $\traj\omega\rlind{k-1} \in \ospace^k$.

Often, time-series recorded in terms of these new observables are (at least to a very good approximation) described by Markovian statistics.
Mathematically, this is equivalent to a refinement $\bigvee_{t=0}^{k-1}\left[\Phi^{-t}(\parti)\right]$ of the partition $\parti$ induced by the measurement observable~$\partmap$.

A discussion of observable states which points in the same direction was attempted by Shalizi and Moore in the framework of ``Computational Mechanics'' \cite{Shalizi+Moore2003}.
More precisely, the concept of refining partitions by summarizing subsequent observations into new observable states is the idea behind the construction of \emph{causal states} \cite{Shalizi+Crutchfield2001,Shalizi+Moore2003}.
These states allow for an \emph{optimal} predictions of future events, \cf~Ref.~\cite{Shalizi+Crutchfield2001}.
Hence, the idea of a causal state is somewhat analogue to our information-theoretical discussion of ST in Chapter~\ref{chap:information-st}.
Interestingly, causal states have been argued to be the observable states of minimal ``thermodynamic depth'', \ie the process of their preparation is the least complex~\cite{Crutchfield+Shalizi1999}.

In analogy to deterministic Turing machines, the network of causal states is called an $\epsilon$-machine.
These machines can be constructed from data or a probabilistic specification of a dynamics.
The stochastic transitions between causal states are Markovian.
Some argue that an ``$\epsilon$-machine'' with a finite number of causal states is simple another name for a Markov chain.

In our opinion, $\epsilon$-machines are a way to interpret Markov chains as \emph{computing devices}.
This interpretation then allows for a connection to measures of (computational) complexity, like the Chaitin--Kolmogorov complexity or the notion of Kolmogorov--Sinai entropy \cite{Shalizi+Crutchfield2001}.
These measures of complexity in turn can be used for the characterization of patterns generated by stochastic or deterministic algorithm as outlined in a recent review article by Crutchfield \cite{Crutchfield2012}.

\subsection{On information processing systems}
Motivated by the perspective of Markovian ST as an $\epsilon$-machine, we discuss thermodynamic systems as information processing devices.
In particular, we focus on information processing in living systems: 
Molecular motors and other biological devices that are involved in the cell's regulatory feedback processes.

In the introductory Chapter~\ref{chap:intro} we have mentioned Landauer's principle \cite{Landauer1961}.
It states that any irreversible (\ie non-invertible) logical operation dissipates energy.
Bennett summarizes the argument as follows \cite{Bennett2003}: 
Reversible operations on the memory of a computer yield a decrease of entropy in its information-bearing degrees of freedom (IBDF).
Consider for instance the erasure of a piece of memory, facilitated by resetting every bit to a neutral binary state, say ``$1$''.
If we assume the binary data to be essentially random with ``$0$'' and ``$1$'' appearing at the same frequencies, the average information per bit is $\log 2=1\text{bit}$.
After erasure,  we know that any bit is in state ``$1$'' and thus the uncertainty (or information) per bit is exactly zero.
Consequently, the entropy of the IBDF has decreased by an amount of $\log 2$ per bit.
The second law of thermodynamics then ensures that the entropy of the medium (\ie the non-information bearing degrees of freedom, NIBDF) has to be increased by at least the same amount.
In essence, this is Landauer's principle for the thermodynamic cost of irreversible computation.
In an isothermal environment, Landauer's limit on the entropy $S\tsub{L}=\log 2$ yields a lower bound $Q\tsup{L} := \kb T \log 2$ for the dissipated heat per bit.

Although it is possible to design \emph{reversible} computers in a gedankenexperiment \cite{Fredkin+Toffoli1982}, this is neither practically possible nor useful:
A reversible computer cannot delete anything stored in its its memory.
As such, it will be useless as soon as the latter has reached its storage capacity --- unless this memory is erased and Landauer's principle holds.

For real (irreversible) computers engineered by humans, Landauer's limit for the minimum heat is not of technical relevance:
It is completely negligible against the dissipated power due to resistivities in the electrical circuits and the energy lost as heat in other conversion processes.
However, a recent work inspired by Landauer's original setting demonstrated that his limit can be reached experimentally \cite{Berut_etal2012} --- if the experimental set-up is prepared carefully enough.

Even before that experiment, scientists have achieved an experimental realization of Maxwell's demon, \ie a device that turns information about a system into useful work \cite{Toyabe_etal2010}.
Abstractly, a Maxwell demon is a kind of feedback control.
Similarly, Landauer's principle can be understood as a special case of a generalized second law for systems with feedback \cite{Sagawa+Ueda2010,Horowitz+Parrondo2011,Sagawa+Ueda2012,Sagawa2012}.

In the experiments mentioned above, the set-ups of the measurement device and its feedback control are very elaborate.
Yet, there are machines that work very closely to Landauer's limit.
In contrast to digital computers and experimental set-ups, they have not been designed by humans, but by evolution:
In every living cell, \emph{polymerases} are enzymes that copy, transcribe and replicate genetic information.
They can be understood as molecular motors that run along single-stranded pieces of RNA or DNA, while copying information from the template strand onto the new strand.

A letter in the genetic alphabet corresponds to one of four nucleic acids.
Hence, the information is $\log{4} =2\log{2} =2\text{bit}$ per copying event.
In Ref.~\cite{Bennett1982}, Bennett calculated that the thermodynamic cost of genetic copying maybe as small as $20\kb T$ per nucleotide, \ie $10\kb T$ per bit.
This is not so far away from Landauer's limit which amounts to $1\kb T$ per bit.

In biological processes, not only thermodynamic efficiency but also the error rate is important \cite{Bennett1982}.
Relatively high error rates are admissible when genetic information is used as the template for protein synthesis.
However, for DNA replication much lower error rates are needed in order to prevent too many unwanted mutations.
In real cells, the error rate is reduced by additional molecular motors which run along the freshly synthesized strand and check for errors.
This proof-reading scheme has recently been treated in a model of stochastic thermodynamics \cite{Sartori+Pigolotti2013}.

The copying of genetic information is just one example where biological systems perform computations.
Generally, any regulatory mechanism can be understood as information processing through some feedback loop.
One example which has been recently studied along these lines is sensory adaptation in the chemotactic response of E.~Coli~\cite{Lan_etal2012,Sartori_etal2014,Barato_etal2014}, \cf also Section~\ref{sec:functional} of the present thesis.
From an abstract perspective, an adapting system learns how to predict and react accordingly to changes of its environment.
Prediction and learning have also been been discussed in the framework of stochastic thermodynamics \cite{Still_etal2012,Hartich_etal2014}.

\subsection{On dissipation}
\begin{figure}[t]
  \centering
  \includegraphics{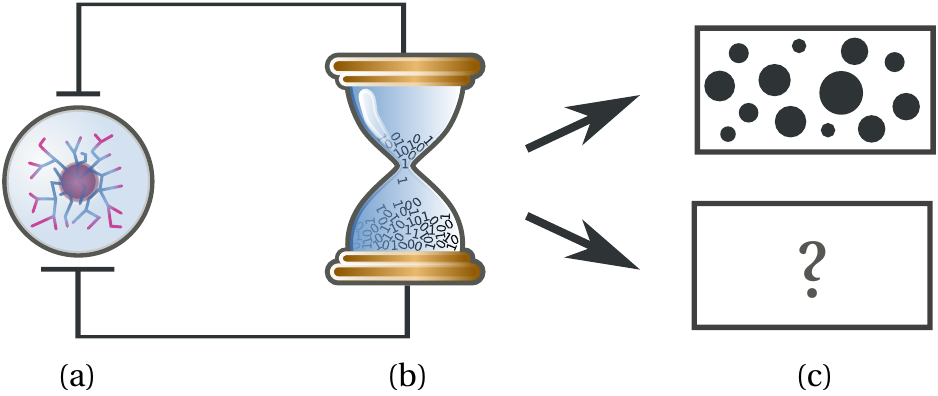}
  \caption{A (non-equilibrium) thermodynamic system as an information processing device. A source of (free) energy (a) enables a thermodynamic system (b) to produce patterns in information-bearing, accessible degrees of freedom (c, top) at the cost of information dumped to the unobservable medium (c, bottom).
    The latter process is called dissipation. 
    Here, we interpret it as information written to non-information bearing degrees of freedom.
  }
  \label{fig:ips}
\end{figure}

In conclusion, recent work emphasizes the intimate connection between entropy and information --- and thus between entropy production and information processing.
In thermodynamics, entropy is connected to heat and entropy production to dissipation.
After all, the intuitive notion of heat is information-theoretical in the first place:
Heat is energy stored in inaccessible and thus non-information bearing degrees of freedom.
Dissipation is information lost to the latter.

Useful computations need to be irreversible, because information has to be deleted at some point.
Landauer's principle and its extensions state that this comes at the cost of dissipation, which is the transfer of information into the realm of inaccessibility.
Maxwell's demon is a hypothetical apparatus which by some means has access to that realm.
Experimental realizations of Maxwell's demon stress that the definition of ``inaccessibility'' is operational rather than objective --- as is the distinction between system and medium, which we have adapted throughout this thesis.

We argue that this apparent subjectivity is not a problem for physics or science in general.
In contrast, it reflects the role of \emph{observations} and \emph{models} in the method of scientific discovery.
A model is a mathematical formalization of empiric rules about the evolution of the world around us.
Hence, it enables us to make predictions about the future state of systems, given a certain knowledge about the system's present state.

The knowledge about this state is described by an ensemble.
Its information content is described by the corresponding differential entropy.
Thereby one gets different numerical values for the entropy, depending on the ensemble and for instance, the dimensionality of the state space of the model.
However, the exact value of the entropy is not important.
What counts, is how this information is changing in the course of time --- either due to insights from measurements or due to the evolution of the system.
ST is a framework to process this information.

Dissipation is a key quantity in this information processing.
It formalizes the information written to unobservable degrees of freedom --- in contrast to the information contained in the pattern formed by the statistics of the information-bearing observable degrees of freedom. 
An (irreversible) computation performed on the latter requires some information being dumped to the former, \cf Figure~\ref{fig:ips}.

In order to compare different \emph{models} of thermodynamic systems, one usually refers to its observable \emph{steady state}.
Then, the (subjective) choice of an initial condition does not play any role.
However, in this situation we do not observe any \emph{changes} in the statistics of observable patterns described by these models.
Consequently, from looking at the observable steady state it is impossible to formulate \emph{dynamical} statements about the information \emph{processed} by the system. 
We can, however, make dynamical statements regarding the change of entropy in the \emph{unobservable} degrees of freedom, formalized by the dissipation rate.
%

Ultimately, this is the argument that leads us to say that different models of the same system are consistent if they share the value of the dissipation rate.
In the first part of the present thesis we have witnessed how the correct identification of dissipation between to levels of description leads to the emergence of ST. 
The second part emphasized how dissipation, which is obtained as a bi-linear form involving thermodynamic currents and forces, acts as a consistency criterion between stochastic models on different resolutions.

This is the final conclusion for this work:
Viewing dissipation in a complex physical system as information processing is more than an analogy.
It formalizes the predictive capabilities of the mathematical models we use to describe nature. 
Ultimately, we see it as a consequence of the mathematical logic of scientific discovery.

  \clearpage
  \chapter*{Acknowledgements}
  \addcontentsline{toc}{chapter}{Acknowledgements}
  Many people were involved in the perception of this thesis.
This is the place to thank them.

First and foremost, I want to thank J\"urgen Vollmer for his supervision during the last years.
His physical intuition and perfect amount of mathematical rigorosity have substantially influenced both the style of this thesis and my way of my physical thinking.
By sharing his connections to scientists working in various fields, he facilitated my entry in the world of scientific research.
I am also grateful for the amount of freedom and responsibility I have enjoyed working in his group.

Many thanks go to the members of this group and generally to all the people on the Yellow Floor.
I found my scientific home in the Department of Complex Fluids at the Max Planck Institute for Dynamics and Self-Organization.
As the head of the department, Stephan Herminghaus always supported my theoretical work and enabled me to share it with people all over the world.
In that context let me also thank the other funding agencies, namely the German Academic Exchange Service (DAAD), the G\"ottingen Graduate School for Neurosciences, Biophysics, and Molecular Biosciences (GGNB) and the Studienstiftung des Deuschen Volkes.

Throughout my time as a Ph.D.~candidate I could rely on my thesis committee and especially my second supervisor Marc Timme for both good collaborations and career advices.
Further thanks go to all the members of my thesis examination committee for listening to my story.
Let me explicitly thank Stefan Kehrein for agreeing to act as a second referee for this thesis.

I am further indebted to the poeple who helped me to organize myself.
Without Barbara Kutz, Monika Teuteberg, Antje Erdmann and Frauke Bergmann I would have dissipated much more energy in the buerocratic medium surrounding institutional research in G\"ottingen.

Throughout the last years I was lucky to enjoy discussions with many people in and outside science.
Lukas Geyrhofer and David Hofmann where often the first people to hear about my most recent thoughts over one of many dinners we have prepared together.
Talking to fellow scientists has constantly changed and challenged my perspective.
I particularly acknowledge discussions with Christian Bick, Guillaume Lajoie, Hugo Touchette, Jonathan Dawes, Matteo Polettini, Nigel Goldenfeld, Pablo Sartori, Susanne Still and Udo Seifert.

Two people are still missing in this admittedly incomplete list.
The first one is Lamberto Rondoni, who I especially want to thank for many vivid discussions and his warm hospitality in Torino.
Ultimatively, he has challenged me to pursue the programme that has eventually become the first part of the present thesis.

Secondly, I am indebted to Artur Wachtel in so many ways.
His knowledge of physics and mathematics as well as his incredible motivation made him one of my most valued collaborators.
Moreover, his technical skills with \textit{Mathematica\texttrademark}, \LaTeX~ and computers in general contributed majorly to the final shape of the present work.

The latter was also influenced by the many people that provided me with feedback on the manuscript, namely Artur, David, Guillaume, Jakob, J\"urgen, Laura and Lukas.
Thank you all very much for your constructive criticism.

I would also like to express my gratitude towards my many friends in G\"ottingen and elsewhere.
Thank you so much for making these last years a wonderful time, which I will never forget.
Thank you also for your support on every front; you were always there for me when I needed you.

Finally, thanks go to my family, who has always supported and encouraged me on my way in and outside of science --- even if that meant that I would spent much less time with them than they deserved.
I love you and I will always be grateful for everything you have done for me.
\vspace{2cm}

\par{\hfill Bernhard Altaner}
\par{\hfill G\"ottingen, June 2014}

  \appendix
  \label{chap:appendix}
  \chapter{Mathematical appendix}

\section{Construction of the Markov measure}
\label{app:construction-mms}
The author of this thesis is not aware of an explicit construction of non-stationary Markov measure in the literature.
Usually, the one-sided or stationary Markov measures are introduced like we did in the main text, \cf Defs.~\ref{def:one-sided-mms} and \ref{def:stationary-mms}.
However, the proof of existence and uniqueness is omitted with a reference to the ``usual extension theorems''.
In this appendix we explicitly prove existence and uniqueness of the measures defined in definitions \ref{def:one-sided-mms}, \ref{def:stationary-mms} and \ref{def:two-sided-mms}.

We first show the existence and uniqueness of certain measures (whose definition seems rather complicated) using the Daniell--Kolmogorov extension theorem.
Then, we prove that these measures are the same as the ones defined in the main text, by using the following lemma:
\begin{lemma}
  Let $\fsets \subset \powset(X)$ be a $\pi$-system, \ie a non-empty family of subsets of $X$ which is closed under finite intersections.
  Then, a measure on $\sigma(\fsets)$ is uniquely defined by its values on $\fsets$.
  \label{theo:pi-system}
\end{lemma}

\subsection{The basic theorems}

We start with some preliminary definitions:
A \emph{semi-algebra} on a space $\spc$ is a family of sets $\alg \subset \powset(\spc)$ that contains the empty set $\emptyset$ and is closed under intersections.
Further it is \emph{semi-closed} under the formation of complements, \ie $\forall A,B \in \alg\colon\, A \setminus B = \bigcup_{k=1}^K C_k$, where
$\ifam{C_k}_{k \in \{1,2,\cdots,K\}}\subset\alg$ is a finite family of mutually disjoint subsets.
A probability \emph{pre-measure} on a semi-algebra $\alg$, is a $\sigma$-additive map $\widehat \pms\colon\alg\to[0,1]$ that obeys $\widehat \pms(\emptyset) = 0$.

The first theorem we need is a generalized version of Carath\'eodory's extension theorem~(\cf Ref.~\cite{Capasso+Bakstein2012}, Proposition A 25).
It states that a pre-measure $\widehat \pms$ on a semi-algebra can be uniquely extended to a measure on the generated $\sigma$-algebra:

\begin{theorem}[Carath\'eodory]
  Let $\alg \subset \powset(\spc)$ be a semi-algebra and $\widehat \pms$ a probability pre-measure on $\alg$.
  Then, there exists a unique probability measure $\pms\colon\alg\to[0,1]$ on $\sigma(\alg)$, obeying $\pms\vert_\alg = \widehat \pms$, \ie $\pms$ restricted to $\alg$ agrees with $\widehat \pms$.
  \label{theo:caratheodory}
 \end{theorem}

The other tool we need is the Daniell--Kolmogorov extension theorem for inner regular measures.
Inner regularity is not a very restrictive property and states that measurable sets can be approximated from within by compact sets:

\begin{definition}[Inner regular measure]
  Let $\left( \left( \spc,\alg\right),\topo \right)$ be a measurable space $(\spc,\alg)$ with topology $\topo$, such that any open set is measurable, \ie $\topo \subset \alg$.
  A measure $\ms$ on $\left( \spc,\alg\right)$ is called \emph{inner regular}, if for any set $A \in \alg$
  \begin{align*}
    \ms(A) = \sup \set{\ms(K) \,\middle\vert\, K \subset A\text{ is compact}}.
  \end{align*}
\end{definition}

Now we can formulate the version of the Daniell-Kolmogorov theorem, which can be found in T.\,Taos introductory textbook \cite{Tao2011}:
\begin{theorem}[Daniell--Kolmogorov]
  Let $F$ be an arbitrary set and $\left( \left( X_t,\alg_t\right),\topo_t   \right)_{t \in F}$ be an indexed family of measurable spaces  $\left( X_t,\alg_t\right)$ with respective topologies $\topo_t$.
  For all nested finite subsets $T'\subset T\subset F$, let $\pms_T$ be an inner regular probability measure on the product $\sigma$-algebra $\alg_T := \bigotimes_{t\in T}\alg_t$ which obeys
  \begin{align*}
    \left( \proj_{T'\leftarrow T} \right)_*\pms_T = \pms_{T'}, 
  \end{align*}
  where $ \proj_{T'\leftarrow T}$ is the projection map. 
  Then, there exists a unique probability measure $\pms_F$ on $\alg_F = \bigotimes_{t\in F}\alg_t$ such that $\left( \proj_{T\leftarrow F} \right)_*\pms_F = \pms_T$ for all $T\subset F$.
  \label{theo:daniell-kolmogorov-extension}
\end{theorem}

\subsection{Equivalent definitions of the Markov measures}

We start with a minimal structure on finite products of $\alphabet$:
\begin{lemma} 
  \label{theo:markov-pre-measure-setup}
  Let $\alphabet$ and $T \subset \timdom$ be finite sets.
  Let $\alphabet^T = \prod_{t\in T}\alphabet$ be the product space consisting of generalized tuples $\traj \omega_T$ with components $\omega_t = \proj_t\traj\omega_T \in \alphabet$.
  Let $\powset_T = \powset(\alphabet^T)$ and  $\mathcal R_T \subset \powset_T$ be the set containing the empty set $\emptyset$, the entire set $\alphabet^T$ and all singleton subsets $\set{\traj\omega_T}\subset\alphabet^T$.
  Then, $\mathcal R_T$ is a semi-algebra and $\sigma(\mathcal R_T) = \powset_T$.
\end{lemma}
\begin{proof}
  Because $\alphabet$ and $T$ are finite, so are $\alphabet^T$ and $\powset_T$.
  Hence, each element in $\powset_T$ can be constructed as the union of a finite number of singleton sets $\set{\traj\omega_T}$.
  In consequence, $\mathcal R_T$ is semi-closed under the formation of complements, closed under intersections and contains the empty set and the entire set by definition.
  Therefore $\mathcal R_T$ is a semi-algebra which generates the whole power set as its $\sigma$-algebra.
\end{proof}

On these (finite) sets we define a families of pre-measures:

\begin{definition}[Markov pre-measure]
  Let $T\subset \naturals$ be a finite set and $\mathcal R_T$ be as in lemma \ref{theo:markov-pre-measure-setup}.
  Let $\tmx$ be the largest integer in $T$ and $T_{\text{max}} := \set{0,1,\cdots,t_{\text{max}}}$.
  Let $\sset \in \mathcal R_T$.
  Further, let $\tmat$ be a $N \times N$ stochastic matrix with entries $\tprob{\omega}{\omega'}$ and $\bvec p \equiv \bvec p\tind{0}$ a stochastic vector. 
  We define functions $\fmms_T\colon\mathcal R_T \to [0,1]$ as follows:
  \begin{align*}
    \fmms_T\left(\sset\right)
    := \begin{cases}
      0, & \sset=\emptyset,\\
      1, & \sset=\alphabet^T,\\
      \displaystyle \tsum{\omega}{T_{\mathrm{max}}\setminus T} \left[ \p{0}{\omega_0}\prod_{t=1}^{\tmx} \tprob{\omega_{t-1}}{\omega_t}\right], &\sset = \set{\traj \omega_T}.
    \end{cases}
  \end{align*}
  If $T = T_{\text{max}}$, the set of summation variables is empty and in the sum there appears only one term, which has all $\omega_t$ specified by $\sset = \set{\traj\omega_T}$.
  \label{def:markov-pre-measure}
\end{definition}


We need to show that these set functions are indeed (probability) pre-measures.
Then, theorem \ref{theo:caratheodory} ensures that they can be uniquely extended to measures on $\salg_T:=\powset(\alphabet^T)$.

\begin{lemma}
  The set function $\mms_T=\mms_T(\tmat, \bvec p)$ introduced in definition \ref{def:markov-pre-measure} is a pre-measure on $\mathcal R_T$.
  \label{theo:markov-pre-measure}
\end{lemma}
\begin{proof}
  By definition $\mms_T(\emptyset) =0$.
  Hence, we only need to show $\sigma$-additivity, which is the same as additivity because $\mathcal R_T$ is finite.
  Then, the Carath\'eodory extension theorem \ref{theo:caratheodory} yields the required result.

  Additivity is also easy to see:
  Let $\fsets = \ifam{\sset^i}_{i\in \indset}$ be a finite, disjoint family of sets $\sset^i \in \mathcal R_T$ whose union, $\sset=\bigcup_i A^i$, is an element of $\mathcal R_T$.
  If $\sset = \emptyset$, the family must consist only of the empty set and additivity is trivial.
  If $\sset = \set{\traj\omega_T}$ is a singleton, $\fsets$ contains exactly one element which is $\set{\traj\omega_T}$ and again, additivity is trivial.
  The only remaining case is that $\sset = \alphabet^T$ and hence $\fmms_T(\sset) = 1$.
  In that case, $\fsets$ consists of \emph{all} singleton sets and hence $\fsets$ can be indexed by $\traj \omega_T \in \alphabet^T$.
  Therefore,
  \begin{align*}
    \sum_{\sset^i \in \fsets} \mms_T(\sset^i) &= \tsum{\omega}{T}\mms_T\left( \set{\traj \omega_T} \right)\\
    &= \tsum{\omega}{T}\quad \tsum{\omega}{T_{\text{max}}\setminus T} \left[ \p{}{\omega_0}\prod_{t=1}^{t_{\text{max}}} \tprob{\omega_{t-1}}{\omega_t} \right]\\
    &= \sum_{\omega_0 \in \alphabet}\sum_{\omega_1 \in \alphabet}\dots\sum_{\omega_{t_{\text{max}}} \in \alphabet} \left[ \p{}{\omega_0}\prod_{t=1}^{t_{\text{max}}} \tprob{\omega_{t-1}}{\omega_t} \right]
    = 1.
  \end{align*}
  In the last line, we just used the fact that $\bvec p$ is a stochastic vector and $\tmat$ is a stochastic matrix.
  Thus, we have shown additivity which ensures that $\mms_T$ is a pre-measure on $\mathcal R_T$.
\end{proof}

At this stage, we constructed a measure space $( \alphabet^T,\powset_T,\fmms_T )$ for any finite $T \subset \naturals$.
To apply the Daniell--Kolmogorov theorem we have to show inner regularity of the measures and the compatibility criterion:

\begin{lemma}
  Let $T \subset \integers$ be a finite subset.
  Then, the measure space $(\alphabet^T,\powset_T,\fmms_T)$ is inner regular with respect to the discrete topology $\topo_T = \powset_T$.
  \label{theo:markov-measure-inner-regular}
\end{lemma}
\begin{proof}
  Any measure defined on the power set of any finite set is inner regular, because all sets are compact in the discrete topology on a finite set.
\end{proof}

\begin{corollary}
  \label{theo:markov-measure-exists}
  There is a unique measure $\fmms := \fmms(\tmat, \bvec p)$ on $(\alphabet^{\naturals}, \salg^\naturals)$, such that its restriction to $\powset_T$ is $\fmms_T(\tmat, \bvec p)$ for all finite $T \subset \naturals$.
\end{corollary}
\begin{proof}
  Lemmata \ref{theo:markov-pre-measure-setup}--\ref{theo:markov-measure-inner-regular} ensure that all of the assumptions for the Daniell--Kolmogorov theorem \ref{theo:daniell-kolmogorov-extension} with the exception of the compatibility criterion are already satisfied.
  The only thing left to show is that 
  \begin{align*}
    \left( \proj_{T'\leftarrow T} \right)_*\fmms_T = \fmms_{T'}, \quad \forall T'\subset T
  \end{align*}
  for any finite $T \subset \naturals$.
  In our consideration we can restrict ourselves to $\sset \in \mathcal R_{T'}$.
  The Carath\'eodory extension ensures that this agreement on the semi-algebras $\mathcal R_{T'}$ carries on to the family of measures defined on the generated $\sigma$-algebras.
  First, let $\sset = \emptyset$ and observe that $\left( \proj_{T'\leftarrow T} \right)_*\fmms_T(\emptyset) =0 = \fmms_{T'}(\emptyset)$.
  Now let $\sset = \alphabet^{T'}$ and hence $\proj_{T'\leftarrow T}^{-1} \sset = \alphabet^{T}$ and therefore $\left( \proj_{T'\leftarrow T} \right)_*\fmms_T(\alphabet^{T'}) =1 = \fmms_{T'}(\alphabet^{T'})$.
  
  The interesting case is $\sset = \set{\traj\omega_{T'}}$ where $\traj\omega_{T'} \in \alphabet^{T'}$.
  For convenience, we introduce the following notation:
  Let $T \subset \naturals$ be finite and $T_1,T_2 \subset T$ be disjoint subsets.
  Let $\traj \omega_{T_1} \in \alphabet^{T_1}$ and $\traj \omega_{T_2} \in \alphabet^{T_2}$.
  The expression $\traj \omega_{T_1}\oplus\traj \omega_{T_2} \in \alphabet^{T}$ denotes the unique element satisfying $ \proj_{T_1\leftarrow T}\left( \traj \omega_{T_1}\oplus\traj \omega_{T_2}\right)=\traj \omega_{T_1} $ and $ \proj_{T_2\leftarrow T} \left(\traj \omega_{T_1}\oplus\traj \omega_{T_2}\right)=\traj \omega_{T_2} $.
  Let $T'' = T\setminus T'$ and find that
  \begin{align*}
    \left( \proj_{T'\leftarrow T} \right)_*\fmms_T(\set{\traj\omega_{T'}}) &\equiv \fmms_T\left(\proj_{T'\leftarrow T}^{-1}(\set{\traj\omega_{T'}})\right) \\
    & = \fmms_T\left( \bigcup_{\traj \omega_{T''} \in \alphabet^{T''}}\set{\traj\omega_{T''}\oplus\traj\omega_{T'}} \right)\\
    & {=}  \tsum{\omega}{T''} \left[\fmms_T\left(\set{\traj\omega_{T''}\oplus\traj\omega_{T'}} \right)\right]\\
    & =  \tsum{\omega}{T''}\quad \tsum{\omega}{T_{\text{max}}\setminus T}\left[ \p{}{\omega_{0}}\prod_{t=1}^{t_{\text{max}}} \tprob{\omega_{t-1}}{\omega_t} \right]\\
    &= \tsum{\omega}{T'_{\text{max}}\setminus T'}\left[ \p{}{\omega_0}\prod_{t=1}^{t'_{\text{max}}} \tprob{\omega_{t-1}}{\omega_t} \right]
    \underbrace{\tsum{\omega}{T_{\mathrm{max}}\setminus T'_{\text{max}} } \left[\prod_{t=\tmx'+1}^{t_{\text{max}}} \tprob{\omega_{t-1}}{\omega_t} \right]}_{=1}\\
    &= \fmms_{T'}(\set{\traj\omega_{T'}}).
  \end{align*}
  Since we have shown compatibility, the Daniell--Kolmogorov extension theorem finally yields the desired result.
  The rearrangement of the sums over index sets can be seen from
  \begin{align*}
    T''\cup( T_{\mathrm{max}}\setminus T) &= (T\setminus T') \cup (T_{\mathrm{max}} \setminus T)\\ &= T_{\mathrm{max}} \setminus T'\\ &=(T_{\mathrm{max}}\setminus T_{\mathrm{max}}'\cup T_{\mathrm{max}}')\setminus T' = (T_{\mathrm{max}}'\setminus T')\cup((T_{\mathrm{max}}\setminus T_{\mathrm{max}}')\setminus T')\\&=(T_{\mathrm{max}}'\setminus T') \cup (T_{\mathrm{max}}\setminus T_{\mathrm{max}}')
  \end{align*}
\end{proof}

By definition, this measure agrees with the set function defined in Def.~\ref{def:one-sided-mms} on all $0$-shifted $\tau$-cylinders.
It is easy to see that these sets together with the empty set are closed under intersection:
Just note that for any given $\tau$, the family $\cyls\tind{0}\rlind{\tau}$ is a partition of $\alphabet^\naturals$.
Further, the set $\cyls\tind{0}\rlind{\tau'}$ is a refinement of $\cyls\tind{0}\rlind{\tau}$ whenever $\tau'>\tau$.
Consequently, if one takes two arbitrary elements of $\set{\cyls\tind{0}\rlind{\tau}}_{\tau \in \naturals}$, they are either disjoint or one is contained within the other.
Hence, $\set{\cyls\tind{0}\rlind{\tau}}_{\tau \in \naturals}$ is a $\pi$-system and uniqueness is guaranteed by Lemma~\ref{theo:pi-system}.

\subsection{Measures on the bi-infinite sequences}

So far we have only shown the argument for the one-sided case of Def.~\ref{def:markov-pre-measure}.
Due to the fact that the construction always proceeds using finite $T \subset \timdom$, the proofs for $\timdom = \integers$ proceed along the same lines.
Throughout this subsection let $T \subset \integers$ be finite, $\mathcal R_T$ be the semi-algebra on $\alphabet^T$ that contains the singletons, the full set and the empty set.
Further, let $t_{\text{max}}= \max(\set{\max(T), 0})$ and $t_{\text{min}} = \min(\set{\min(T), 0})$ and $T_{\text{def}} := \set{t_{\text{min}},t_{\text{min}}+1,\cdots,t_{\text{max}}}$.
We define the set functions corresponding to Definition~\ref{def:stationary-mms} in the main text:
\begin{definition}[Stationary Markov pre-measure for the bi-infinite sequences]
  Let $\bvec p\tind{\infty}$ be a stochastic left eigenvector for the unity eigenvalue of a stochastic matrix $\tmat$.

  For $\sset \in \mathcal R_T$ we define functions $\fmms_T\colon\mathcal R_T \to [0,1]$ as follows:
  \begin{align*}
    \fmms_T\left(\sset\right)
    := \begin{cases}
      0, & \sset=\emptyset,\\
      1, & \sset=\alphabet^T,\\
      \sum_{\traj \omega_{T_{\text{def}}\setminus T}} \left[ \p{\infty}{\omega_0}\left(\prod_{t=\tmn+1}^{t_{\mathrm{max}}} \tprob{\omega_{t-1}}{\omega_t}\right)\right], &\sset = \set{\traj \omega_T}.
    \end{cases}
  \end{align*}
  \label{def:stationary-markov-pre-measure}
\end{definition}

Similarly, the set functions corresponding to definition \ref{def:two-sided-mms} read:
\begin{definition}[Two-sided Markov pre-measure]
  Let $\bvec p$ be a stochastic vector and let $\tmat$ and $\btmat$ be $N \times N$ stochastic matrices compatible with some adjacency matrix $\adjm$ respectively its transpose $\adjm^{\mathrm T}$.

  For $\sset \in \mathcal R_T$ we define functions $\fbmms_T\colon\mathcal R_T \to [0,1]$ as follows:
  \begin{align*}
    \fbmms_T\left(\sset\right)
    := \begin{cases}
      0, & \sset=\emptyset,\\
      1, & \sset=\alphabet^T,\\
      \sum_{\traj \omega_{T_{\text{def}}\setminus T}} \left[ \p{}{\omega_0}\left(\prod_{t=1}^{t_{\mathrm{max}}} \tprob{\omega_{t-1}}{\omega_t}\right)  \left(\prod_{t=1}^{-{t_{\mathrm{min}}}} \btprob{\omega_{-t+1}}{\omega_{-t}}\right)\right], &\sset = \set{\traj \omega_T}.
    \end{cases}
  \end{align*}
  \label{def:two-sided-markov-pre-measure}
\end{definition}

As said above, the proof of existence is directly analogous to the one-sided Markov measure.
It remains to be shown that the measure defined here is the equivalent to the one in the main text.
We start with the two-sided measure.
It is defined on the set of $0$-shifted forward and backward cylinders, as well as on intersections of elements of the two.
Thus, it is defined on a $\pi$-system and yields a unique measure.
Further, it is easy to check that Def.~\ref{def:two-sided-markov-pre-measure} is consistent with Def.~\ref{def:two-sided-mms} and thus the measures are equivalent.
Finally, the same argument together with the fact that backward cylinders can be rewritten as shifted forward cylinders for the reversed sequences ensures that Def.~\ref{def:stationary-markov-pre-measure} is consistent with Def.~\ref{def:stationary-mms}.
The same argument is used to prove Proposition~\ref{theo:fbmms-alternative}.

It remains to be shown that $s^t_*\fmms(\tmat,\bvec p\tind{0}) = \fmms(\tmat,\bvec p\tind{t})$ for $t \in \integers$, which was the statement of Proposition~\ref{theo:pushforward}.
But this is clear from definition \ref{def:markov-pre-measure} and the fact that $s^{-t}\cyl\tind{0}[\traj \omega\rlind{\tau}] = \cyl\tind{t}[\traj \omega\rlind{\tau}]$.

\section{Sequences of $\tau$-chains and their variations}
\label{app:tau-chains}
Stochastic thermodynamics uses the notion of time-series dependent observables for the identification of fluctuating entropies, \cf Sec.~\ref{sec:discrete-st}.
For the case of of stochastic jump processes in discrete time, these observables are examples of \emph{$\tau$-chains}.
Formally, a $\tau$-chain is a measurable function
\begin{align*}
  \obs\rlind{\tau}\colon\alphabet^{\tau+1}\times \integers &\to\reals,\\
  (\traj \omega\rlind{\tau},t) &\mapsto \obs\tind{t}[\traj\omega\rlind{\tau}].
\end{align*}
The $(\tau+1)$\emph{-point average} or \emph{time-series average} is defined as
\begin{align}
  \trava{\obs\rlind{\tau}}{t_0}{\tau} := \sum_{\traj \omega\rlind{\tau} \in \alphabet^{\tau+1}} \prob\tind{t_0}\left[\traj \omega\rlind{\tau}\right] \obs\tind{t_0}\left[\traj \omega\rlind{\tau}\right].
    \label{eq:ts-average}
\end{align}
where $\prob\tind{t_0}\left[\traj \omega\rlind{\tau}\right] $ is the probability for a time-series $\traj \omega\rlind{\tau}$ to occur at time $t_0$.
For brevity of notation we usually write $\trava{\obs}{t_0}{\tau} $ instead of $\trava{\obs\rlind{\tau}}{t_0}{\tau}$, when no ambiguity can arise.
In Section \ref{sec:discrete-st} we saw that jump averages reduce to state averages, if the observable is a one-chain, \ie if it depends only on the initial and final state of a jump $\omega\to\omega'$.
The general result is the following:
\begin{lemma}
  Let  $\obs\rlind{\tau}_t = \obs_{t+k} \left(\omega_k\right) $ be a $\tau$-chain, which depends only on the $k$th component $\omega_k$ of a block $\traj \omega\rlind{\tau}$.
  Then, the time-series average reduces to the state average 
  \begin{align*}
    \trava{\obs}{t}{\tau} = \eava{\obs}{t+k}.
  \end{align*}
  \label{theo:trav-to-eav}
\end{lemma}
\begin{proof}
  Let $T = \set{0,1,\cdots,\tau} \setminus \set{k}$.
  \begin{align*}
    \trava{\obs\rlind{\tau}}{t}{\tau}
    &= \sum_{\traj \omega\rlind{\tau}} \prob\tind{t}[\traj \omega^{\left( \tau \right)}]\obs\tind{t}[\traj \omega\rlind{\tau}]\\
    &= \sum_{\omega_k}\obs\tind{t+k}(\omega_k)  \tsum{\omega}{T} \prob\tind{t}[\traj \omega^{\left( \tau \right)}]\\
    &= \sum_{\omega_k} \obs\tind{t+k}(\omega_k)  \p{t+k}{\omega_k}= \eava{\obs}{t + k}
  \end{align*}
  To get to the last line we used the rule of marginal probability, \ie the fact that $\pms$ is a measure and that $\p{t}{\omega_k} = \pcyl_t[(\omega_k)] =   \tcup{\omega}{T}  \cell_t[\left(\traj \omega^{\left( \tau \right)}\right)]$.
\end{proof}

In the following we consider the relation between state averages $\eav{\obs\tind{t}}$ and time-series averages $\trav{\obs\tind{t}}$.
More precisely, we are interested in averages of certain \emph{sequences} of $\tau$-chains.

Note that any state observable (\ie any $0$-chain) $\obs\tind{t}(\omega)$ induces such a \emph{canonical sequence}, \cf Sec.~\ref{sec:entropic-tau-chains}:
\begin{definition}[Canonical sequence]
  \label{def:induced-sequence}
  Let $\obs\tind{t} \colon\ospace \to \reals$ be a $0$-chain parametrized by a time index $t$.
  Let $t_0,\tau \in \naturals$ and denote by $\omega_\tau$ the final state of $\traj \omega\rlind{\tau}$.
  The sequence $\ifam{\obs\tind{t_0}\rlind{\tau}}_{\tau \in \naturals}$ with elements
  \begin{align*}
    \obs\tind{t_0}\rlind{\tau}[ \traj \omega\rlind{\tau}] := \obs\tind{t_0+\tau}(\omega_\tau),
  \end{align*}
  is called the \emph{canonical sequence of time-series observables} for $\obs\tind{t}$.
\end{definition}
For the canonical sequence, Lemma~\ref{theo:trav-to-eav} ensures that 
\begin{align}
  \trava{\obs\rlind{\tau}}{t_0}{\tau} = \eava{\obs}{t_0+\tau}.
  \label{eq:trav-to-eav-induced}
\end{align}
The notion of a sequence of $\tau$-chains is needed for the consistent definition of the \emph{temporal variation} associated with a $\tau$-chain.
Before we discuss the \emph{sequence of variations} associated to a sequence of $\tau$-chains, we define the \emph{temporal variation} $\Delta F$ of a function $F(\tau)\colon\naturals \to \reals\,$ as
\begin{align}
   \Delta F(\tau) := F(\tau) - F(\tau-1).
  \label{eq:defn-temporal-variation-function}
\end{align}
We define a similar notion for sequences of $\tau$-chains:
\begin{definition}
  Let $\ifam{\obs\rlind{\tau}}_ {\tau \in \naturals} $ be a sequence of $\tau$-chains.
  The sequence of time-series observables $(\delta\tind{t}\rlind{\tau}\obs)_{\tau \in \naturals^+}$ with elements
  \begin{align*}
    \left(\delta\rlind{\tau}\obs\tind{t}\right) \left[(\traj\omega\rlind{\tau-1},\omega_\tau)\right] := \obs\rlind{\tau}\tind{t}\left[(\traj\omega\rlind{\tau-1},\omega_\tau)\right] - \obs\rlind{\tau-1}\tind{t}\left[\traj\omega\rlind{\tau-1}\right]
  \end{align*}
  is called the \emph{temporal variation} of the sequence $\ifam{\obs\rlind{t}}_ {t \in \naturals} $.
\end{definition}
Again, for a more compact notation we may drop the superscript $(\tau)$ on $\delta\rlind{\tau}\obs\tind{t}$ if the meaning is clear from the context.

With these definition we are able to state and proof the following result.
It relates the variations of (explicitly time-dependent) averages for sequences of $\tau$-chains with the average of the variation of the $\tau$-chain:
\begin{lemma}[Variations and averages]
  Let $\ifam{\obs\tind{t}\rlind{\tau}}_{\tau \in \naturals}$ be a sequence of $\tau$-chains.
  Let $\ifam{\delta\rlind{\tau}\obs\tind{t}}_{\tau \in \naturals^+}$ be its temporal variation.
  Define $F\tind{t_0}(\tau) := \trav{\obs\tind{t}\rlind{\tau}}\rlind{\tau}\tind{t_0}$.
  Then, we have
  \begin{align*}
    \Delta\tind{t_0+\tau-1}\rlind{1}F := \left(\Delta F\tind{t_0}\right)(\tau)=\trav{\delta\rlind{\tau} \obs\tind{t}}\rlind{\tau}\tind{t_0}.
  \end{align*}
  \label{theo:variations}
\end{lemma}

\begin{proof}
  Let $\traj\omega\rlind{\tau-1} := \proj_{ \set{0,1,\cdots,\tau-1} \leftarrow \set{0,1,\cdots,\tau} }(\traj\omega\rlind{\tau})$ denote the first $\tau$ components of $\traj\omega\rlind{\tau}$.
  First observe that
  \begin{align*}
    \trava{\obs\rlind{\tau-1}}{t_0}{\tau}& =\sum_{\traj \omega\rlind{\tau}\in\ospace^{\tau+1}}\left[ \proba{t_0}{\omega}{\tau} \obs\rlind{\tau-1}\tind{t_0}\left[\traj\omega^{\left( \tau-1 \right)}\right] \right]\\
    &=\sum_{\traj \omega\rlind{\tau-1}\in\ospace^{\tau}} \sum_{\omega_{\tau}} \left[ \proba{t_0}{\omega}{\tau-1}\cdot \prob\tind{t_0+\tau-1}\left[\omega_{\tau-1}\to\omega_\tau  \right]\obs\rlind{\tau-1}\tind{t_0}\left[\traj\omega^{\left( \tau-1 \right)}\right] \right]\\
    &=\sum_{\traj \omega\rlind{\tau-1}\in\ospace^{\tau}} \left[ \proba{t_0}{\omega}{\tau-1}\obs\rlind{\tau-1}\tind{t_0}\left[\traj\omega^{\left( \tau-1 \right)}\right]\right] \cdot  \sum_{\omega_{\tau}}\left[ \prob\tind{t_0+\tau-1}\left[\omega_{\tau-1}\to\omega_\tau  \right]\right]\\
    &=\sum_{\traj \omega\rlind{\tau-1}\in\ospace^{\tau}} \left[ \proba{t_0}{\omega}{\tau-1}\obs\rlind{\tau-1}\tind{t_0}\left[\traj\omega^{\left( \tau-1 \right)}\right]\right] = \trava{\obs\rlind{\tau-1}}{t_0}{\tau-1},
  \end{align*}
  where $ \prob\tind{t_0+\tau-1}\left[\omega\to\omega'  \right] := \frac{\prob\tind{t_0+\tau-1}\left[ (\omega,\omega')\right]}{\p{t_0+\tau-1}{\omega}}$ is the conditional probability for symbol $\omega'$ to occur at time $t_0+\tau$ given  $\omega$ has occurred at time $t_0+\tau-1$.
  With that we obtain the desired result:
  \begin{align*}
    (\Delta F\tind{t_0})(\tau) &= F\tind{t_0}(\tau) - F\tind{t_0}(\tau-1)   = \trava{\obs\tind{t}\rlind{\tau}}{t_0}{\tau} -  \trava{\obs\tind{t}\rlind{\tau-1}}{t_0}{\tau-1}&&\\
    & = \trava{\obs\tind{t}\rlind{\tau}}{t_0}{\tau} -  \trava{\obs\tind{t}\rlind{\tau-1}}{t_0}{\tau}  = \trava{\obs\tind{t}\rlind{\tau}- \obs\tind{t}\rlind{\tau-1}    }{t_0}{\tau}\\&=\trava{\delta\rlind{\tau}\obs\tind{t}}{t_0}{\tau} %
  \end{align*}
\end{proof}
Note that if $\ifam{\obs\tind{t}\rlind{\tau}}_{\tau \in \naturals}$ is a canonical sequence for a $0$-chain $\obs\tind{t}(\omega)$, the average depends only on the sum of $t_0$ and $\tau$, \cf Eq.~(\ref{eq:trav-to-eav-induced}).

\chapter{Construction of a minimal model for kinesin}
\label{app:construction}

In this section we review the construction and parametrization of the four-state model for the molecular motor kinesin.
We follow the arguments of Liepelt and Lipowsky~\cite{Liepelt+Lipowsky2007} to construct the model, and use the experimental data from Ref.~\cite{Carter+Cross2005} to obtain numerical values for the parameters.
Further we rely on Hill--Schnakenberg balance conditions~\eqref{eq:thermodynamic-balance} to infer some rates from the others.
Even with this consistency requirements, we still lack one first order rate constant in order to specify all numerical rates of the model.
This parameter is obtained by a comparison to the six-state model for physiological values of the chemical concentrations at vanishing mechanical driving.
Comparing the results obtained with that choice to previous results \cite{Liepelt+Lipowsky2009}, in Section~\ref{sec:phase-diagram} we find that this choice --- made for physiological conditions --- is good \emph{globally}, that is everywhere in the phase diagram.

\section{State space of the four-state model}
Kinesin is a molecular motor ``walking'' along a one-dimensional ``track'' by hydrolysing adenosine triphosphate (ATP) into adenosine diphosphate (ADP) and inorganic phosphate (P).
For a more detailed exposition of the mechanics of the kinesin step, see Ref.~\cite{Yildiz_etal2004,Carter+Cross2005}.
It is important that mechanical stepping and chemical hydrolysis are \emph{not} tightly coupled.
Physically, this means that it is possible to have ``futile'' hydrolysis events in which the motor does not take any step.
Mathematically, this means that there must be at least two (fundamental) cycles in any discrete stochastic model.
Disregarding multiple transitions between states, the simplest compatible topology, is given by the four-state network used as an example throughout Chapter~\ref{chap:cycles}.

\begin{figure}[t]
  \centering
  \includegraphics{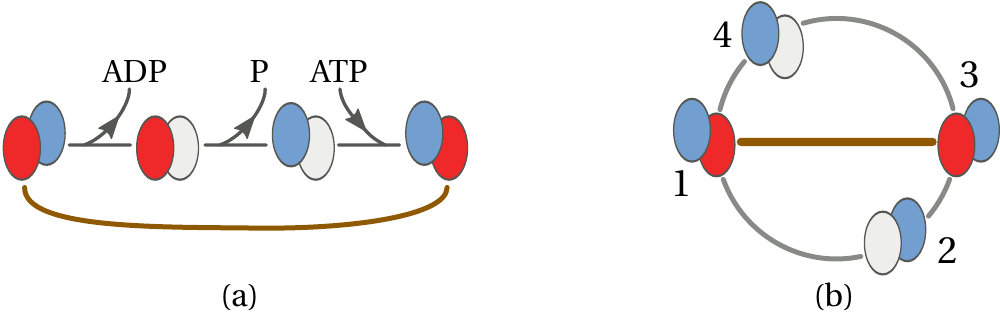}
  \caption{a) Chemical pathway involving (from left to right) detachment of ADP (blue) from the leading foot, hydrolysis of ATP (red) at the trailing foot with release of P, and finally attachment of ATP to the leading foot.
  From the final configuration the system is likely to undergo a mechanical transition (brown) which reverts the order of the feet yielding kinesin's mechanical step.
  b) In our model, we summarize the first two steps (attachment of ADP, hydrolysis of ATP and release of P) into one transition.
  Symmetry and the demand for a non-tightly coupled model give rise to the four-state model (right).
}
  \label{fig:kinesin-model}
\end{figure}

In that model, the states are labelled by the chemical composition of kinesin's active groups:
They can either be empty (E) or have ATP or ADP attached, see Figure~\ref{fig:kinesin-model}.
In the following, we repeat the arguments given by Liepelt and Lipowsky \cite{Liepelt+Lipowsky2007,Liepelt+Lipowsky2009} in order to arrive at a parametrization of our four-state model.
In general, transition rates \(w^i_j\) (unit: \(\frac{1}{s}\)) are parametrized by first-order rate constants \(\kappa^i_j\) which are multiplied concentration and/or force-dependent factors \(C^i_j\) and \(\Phi^i_j\), respectively.

\subsubsection*{Chemical transitions}
For a chemical transition, we have
\begin{align}
  w^i_j := \kappa^i_j\cdot C^i_j \cdot \Phi^i_j(f).  
\end{align}
Here, \(\kappa^i_j\) is the first order rate constant (see table \ref{tab:parameters}) for the transition \((i\to j)\).
Further,
\begin{align}
  C^i_j := 
  \begin{cases}
    \prod_X[X] & \text{ if compound \(X\) is attached during the transition, }\\
    1 & \text{ else,}
  \end{cases}
\end{align}
where \([X]\) denotes the (fixed) concentration of compound \(X\) in the surrounding medium.
The factor
\begin{align}
  \Phi^i_j(f) = \frac{2}{1+\exp{[\chi^i_j f]}}
  \label{}
\end{align}
depends on the (non-dimensional) applied force \(f = l F /(k_\mathrm{B} T)\) and symmetric mechanical parameters \(\chi^i_j = \chi^j_i\).
Values for the mechanical parameters are the ones used in Ref.~\cite{Liepelt+Lipowsky2007} to account for the data of Ref.~\cite{Carter+Cross2005}, see table \ref{tab:parameters}.

\subsubsection*{Mechanical transition}
The mechanical transition lacks the chemical attachment factor \(C^i_j\) and has a slightly different force-dependent factor \(\tilde\Phi(f)\).
The combined rates are thus:
\begin{align}
  w^i_j := \kappa^i_j \cdot \tilde\Phi^i_j(f), 
\end{align}
with the mechanical (load distribution) factor
\begin{align}
  \tilde\Phi^i_j(f) :=
  \begin{cases}
	  \exp{(-\theta f)}\,, & \text{if }(i\to j) = (1\to 3)\\
	  \exp{( (1- \theta) f)}\,, & \text{if }(i\to j) = (3 \to 1).
  \end{cases}
  \label{}
\end{align}
Again, the choice for the parameter \(\theta=0.65\) corresponds to the experimental data in Ref.~\cite{Carter+Cross2005}.

\begin{table}[t]
  \centering
  \begin{tabular}{lr@{\,=\,}lr@{\,=\,}l}
    \toprule
    Mechanical transition& \(\kappa^1_3\)& \num{3e5}& \(\kappa^3_1\)& \num{0.24}\\
    \midrule
    Chemical transitions& \(\kappa^1_4\)& \num{100}& \(\kappa^4_1\)& \num{2.0}\\
    (forward cycle) &\(\kappa^4_3=c\)& \num{2.52e6}& \(\kappa^3_4=\frac{K_{\mathrm{eq}}\kappa^4_3\kappa^1_4\kappa^3_1}{\kappa^4_1 \kappa^1_3}\)& \num{49.3}\\
    \midrule
    Chemical transitions&\(\kappa^3_2 =
\left(\frac{\kappa^3_1}{\kappa^1_3}\right)^2\,\kappa^1_4\)& \num{6.4e-11}& \(\kappa^2_3=\kappa^4_1\)& \num{2.0}\\
    (backward cycle)&\(\kappa^2_1=c\)& \num{2.52e6}& \(\kappa^1_2=\kappa^3_4\)& \num{49.3}\\
    \midrule
    Mechanical load &\(\chi^3_4=\chi^4_3=\chi^1_2=\chi^2_1\)& \num{0.15}& \(\chi^4_1=\chi^1_4=\chi^2_3=\chi^3_2\)& \num{0.25}\\
    \bottomrule
  \end{tabular}
  \caption{Parameters of the 
    four-state model for kinesin.
    All first-order reaction rates \(\kappa\) are given in units of \si{\per\second} or, if attachment of chemicals is involved, \si{\per\second}\(\si{\micro}\si{\Molar}^{-1}\).
    The equilibrium constant of the ATP hydrolysis reaction is \protect{\(K_{\tsub{eq}} = \SI{4.9e11}{\micro\Molar}\)}.
    The parameter \(\theta=\num{0.65}\) enters the mechanical factor of the transition rates.
  }
  \label{tab:parameters}
\end{table}

\section{Choice of parameters}
The parametrization of the mechanical transition is taken from the work of Liepelt and Lipowsky \cite{Liepelt+Lipowsky2007} to reflect experiments \cite{Carter+Cross2005}.
However, the choice of first-order rate constants for the chemical transitions requires adaption for our simpler model.
The transition \((\text{ADP},\,\text{ATP}) \to (\text{ADP},\,\text{E}) \) is present in both our and the original six-state model.
For this transition we use the same values as in Ref.~\cite{Liepelt+Lipowsky2007}.
Thus, we only need to find a good parametrization of the rate \(w^4_3\) for the transition \((\text{ADP},\,\text{E}) \to (\text{ATP},\,\text{ADP})\) and its reverse.
In the force-free case the mechanical parameters drop out.
Hence, the first-order constant for one of the transitions \(3\leftrightarrows 4\) can be obtained from the Hill--Schnakenberg conditions~\eqref{eq:thermodynamic-balance} \cite{Schnakenberg1976}.
In equilibrium, \ie when there are no driving forces acting on the system, these conditions ensure that the (chemical) affinity of every cycle vanishes.
For the upper (forward) cycle \((1\to3\to4)\) this statement can be cast into the expression
\begin{align}
  \frac{\kappa^3_4 \kappa^4_1\kappa^1_3}{\kappa^4_3\kappa^1_4\kappa^3_1} \stackrel{!}{=} K_{\mathrm{eq}},
  \label{eq:balance}
\end{align}
where \(K_{\mathrm{eq}} = \SI{4.9e11}{\micro\Molar}\) is the chemical equilibrium constant for the ATP hydrolysis reaction.
Thus, we still have the freedom to choose one of the rate constants of the forward cycle.
Hence, \(c\equiv\kappa^3_4\) is the above mentioned fit parameter.
It is fixed by demanding the motor's velocity \(V\) in the four-state model to be identical to that of the six-state model in the force-free case (\(f=0\)) and at the physiological concentrations \mbox{\([\text{ATP}]=[\text{ADP}]=[\text{P}]=\SI{1}{\micro\Molar}\).}

%
By symmetry, the chemical first-order rates of the lower cycle \( \left( 3\to1\to2 \right)\) are chosen to be the same as that of the upper cycle, with the exception of \(\kappa^3_2\).
The latter determines the likelihood for the system to let go of the ATP molecule rather than (after releasing ADP from the other head) hydrolysing it.
Because the balance condition (\ref{eq:balance}) is required to hold also for the lower cycle, we use it to determine the missing  rate \(\kappa^3_2\).
This is the same reasoning as in Ref.~\cite{Liepelt+Lipowsky2007}.

Finally, we have to estimate the missing parameter \(\chi^3_4 = \chi^4_3\) for the combined transition \( (\text{ADP},\,\text{ATP}) \to (\text{ADP},\,\text{E})\).
Liepelt and Lipowsky used the same parameter (\(\chi = 0.15\)) for both of its chemical substeps.
We take this as an argument to use \(\chi^3_4=\chi^1_2 = 0.15\) for the combined rate and its reversed counter-part in the lower cycle.
All model parameters are summarized in table \ref{tab:parameters}.

%

  \newpage 
  \printbibliography[heading=bibintoc]%

  \newpage
\thispagestyle{empty}
\null
\vfill
This work is licensed under the Creative Commons Attribution-ShareAlike 4.0 International License. To view a copy of this license, visit~ \url{http://creativecommons.org/licenses/by-sa/4.0/}.

\vfill
\par{\hfill\includegraphics[scale=.7]{./figures/by-sa}}


\end{document}